\documentclass[11pt,a4paper]{article}
\pdfoutput=1

\usepackage{jcappub}
\usepackage{bm}
\usepackage{lmodern}
\usepackage[T1]{fontenc}
\usepackage{relsize}
\usepackage{booktabs}
\usepackage{array}
\usepackage[table]{xcolor}
\usepackage{tcolorbox}
\usepackage[frozencache]{minted}
\usepackage{etoolbox}
\usepackage{upquote}
\usepackage{abraces}
\usepackage{refcount}
\makeatletter
\let\rc@refused\refused
\makeatother

\tcbuselibrary{skins}
\tcbuselibrary{breakable}

\renewcommand{\texttt}[1]{{\ttfamily\fontseries{l}\selectfont{#1}}}

\newcounter{advancedbox}[section]

\BeforeBeginEnvironment{minted}{\begin{tcolorbox}[enhanced,breakable,colback=white,boxrule=0pt,frame hidden,top=-0.75mm,bottom=-0.75mm,left=-0.75mm,right=-0.75mm]}
\AfterEndEnvironment{minted}{\end{tcolorbox}\noindent}

\newenvironment{example}{\begin{tcolorbox}[enhanced,colback=white,colbacktitle=black!20,colframe=black!40,coltitle=black,title=Example,fonttitle=\sffamily\fontseries{b}\selectfont]}{\end{tcolorbox}}

\newenvironment{advanced}[1]{\stepcounter{advancedbox}\begin{tcolorbox}[enhanced,breakable,colback=red!10,colbacktitle=red!20,colframe=red!40,coltitle=black,title={Advanced usage: {#1}},fonttitle=\sffamily\fontseries{b}\selectfont]}{\end{tcolorbox}}

\newenvironment{warning}{\begin{tcolorbox}[enhanced,breakable,colback=red!10,colbacktitle=red!20,colframe=red!40,coltitle=black,title={Warning},fonttitle=\sffamily\fontseries{b}\selectfont]}{\end{tcolorbox}}

\renewcommand{\d}{\mathrm{d}}

\newcommand{\e}[1]{\mathrm{e}^{{#1}}}

\newcommand{\Mp}{M_{\mathrm{P}}}

\newcommand{\kmax}{k_\text{max}}
\newcommand{\ellmax}{\ell_\text{max}}

\newcommand{\Pzeta}{P_\zeta}
\newcommand{\Bzeta}{B_\zeta}

\newcommand{\DimlessSigma}{\mathcal{P}}
\newcommand{\DimlessB}{S}
\newcommand{\fNL}{f_{\mathrm{NL}}}
\newcommand{\fNLlocal}{\fNL^{\text{local}}}
\newcommand{\fNLortho}{\fNL^{\text{ortho}}}
\newcommand{\fNLequi}{\fNL^{\text{equi}}}

\newcommand{\vect}[1]{\bm{\mathrm{{#1}}}}

\newcommand{\Ninit}{N_{\text{init}}}
\newcommand{\Nexit}{N_{\text{exit}}}
\newcommand{\Nstar}{N_{\ast}}
\newcommand{\Npre}{N_{\text{pre}}}
\newcommand{\Nzero}{N_{0}}
\newcommand{\kstar}{k_{\ast}}
\newcommand{\texit}{t_{\text{exit}}}
\newcommand{\tmassless}{t_{\text{massless}}}

\newcommand{\astar}{a_{\ast}}
\newcommand{\Hstar}{H_{\ast}}

\newcommand{\kcom}{k_{\text{com}}}

\newcommand{\repoobject}[1]{{\ttfamily\bfseries\small #1}}

\newcommand{\packagefont}{\sffamily}

\newcommand{\MadGraph}{{\packagefont MadGraph}}

\newcommand{\FormCalc}{{\packagefont FormCalc}}
\newcommand{\GiNaC}{{\packagefont GiNaC}}
\newcommand{\xloopsginac}{{\packagefont XLOOPS-GiNaC}}
\newcommand{\CLN}{{\packagefont CLN}}

\newcommand{\CompHEP}{{\packagefont CompHEP}}
\newcommand{\CalcHEP}{{\packagefont CalcHEP}}

\newcommand{\HELAC}{{\packagefont HELAC}}
\newcommand{\SHERPA}{{\packagefont SHERPA}}
\newcommand{\Whizard}{{\packagefont Whizard}}

\newcommand{\ModeCode}{{\packagefont ModeCode}}
\newcommand{\MultiModeCode}{{\packagefont MultiModeCode}}
\newcommand{\FieldInf}{{\packagefont FieldInf}}
\newcommand{\PyFlation}{{\packagefont PyFlation}}

\newcommand{\BINGO}{{\packagefont BINGO}}

\newcommand{\CppTransport}{{\packagefont CppTransport}}

\newcommand{\Xcode}{{\packagefont Xcode}}
\newcommand{\Python}{{\packagefont Python}}

\newcommand{\Matplotlib}{{\packagefont Matplotlib}}
\newcommand{\seaborn}{{\packagefont seaborn}}

\newcommand{\MPI}{{\packagefont MPI}}
\newcommand{\OpenMPI}{{\packagefont OpenMPI}}
\newcommand{\MPICH}{{\packagefont MPICH}}
\newcommand{\IntelMPI}{{\packagefont Intel MPI}}
\newcommand{\SQLite}{{\packagefont SQLite}}
\newcommand{\MacPorts}{{\packagefont MacPorts}}
\newcommand{\Homebrew}{{\packagefont Homebrew}}
\newcommand{\Graphviz}{{\packagefont Graphviz}}
\newcommand{\Mathematica}{{\packagefont Mathematica}}
\newcommand{\Maple}{{\packagefont Maple}}
\newcommand{\CLion}{{\packagefont CLion}}
\newcommand{\DataGrip}{{\packagefont DataGrip}}
\newcommand{\Sqliteman}{{\packagefont Sqliteman}}
\newcommand{\DBeaver}{{\packagefont DBeaver}}
\newcommand{\OpenSSL}{{\packagefont OpenSSL}}
\newcommand{\SPLINTER}{{\packagefont SPLINTER}}
\newcommand{\JsonCpp}{{\packagefont JsonCPP}}
\newcommand{\Eigen}{{\packagefont Eigen}}

\newcommand{\Boost}{{\packagefont Boost}}
\newcommand{\odeint}{{\packagefont odeint}}

\newcommand{\jQuery}{{\packagefont jQuery}}
\newcommand{\DataTables}{{\packagefont DataTables}}

\newcommand{\CMake}{{\packagefont CMake}}

\newcommand{\file}[1]{\texttt{{#1}}}
\newcommand{\envvar}[1]{\mintinline{bash}{#1}}
\newcommand{\cmakevar}[1]{\texttt{\textbf{\footnotesize #1}}}
\newcommand{\block}[1]{\mintinline{text}{#1}}
\newcommand{\attribute}[1]{\mintinline{text}{#1}}
\newcommand{\descfile}[1]{\mintinline{text}{#1}}
\newcommand{\option}[1]{{\ttfamily\bfseries\small #1}}
\newcommand{\token}[1]{{\footnotesize\color{orange}\texttt{\textbf{{\$}#1}}}}
\newcommand{\indexset}[1]{{\footnotesize\color{orange}\texttt{\textbf{[#1]}}}}

\newcommand{\Gb}{\,\mathrm{Gb}}

\newcommand{\second}{\,\mathrm{s}}

\DeclareMathOperator{\Or}{O}
\newcommand{\iprod}[2]{\langle\!\langle {#1}, {#2} \rangle\!\rangle}

\newcommand{\semibold}[1]{{\fontseries{b}\selectfont{#1}}}
\newcommand{\para}[1]{\par\vspace{2mm}\noindent\semibold{{#1.}---}\ignorespaces}
\newcommand{\grouptag}[1]{\textsf{{#1}}}

\newcommand\CC{C\nolinebreak\hspace{-.05em}\raisebox{.4ex}{\relsize{-3}{\textbf{+}}}\nolinebreak\hspace{-.10em}\raisebox{.4ex}{\relsize{-3}{\textbf{+}}}}

\newcolumntype{s}{>{$\displaystyle}l<{$}}
\newcolumntype{t}{>{$\displaystyle}c<{$}}
\newcolumntype{u}{>{$\displaystyle}r<{$}}
\newcolumntype{v}{>{$\displaystyle}m{4cm}<{$}}

\newcolumntype{d}{D{!}{\;\pm\;}{-1}}

\newenvironment{sqltablelist}{\renewcommand{\arraystretch}{1.3}\small}{}

\renewcommand{\geq}{\geqslant}

\DeclareMathOperator{\realpart}{Re}

\renewcommand{\Re}{\realpart}

\setminted[bash]{
    autogobble=true,
    bgcolor=blue!10,
    fontfamily=tt,
    fontseries=l,
    fontsize=\scriptsize,
    frame=none,
    linenos=true,
    xleftmargin=0.05\linewidth,
    style=tango,
    breaklines=true,
    breakbytoken=true
}

\setminted[text]{
    autogobble=true,
    bgcolor=green!10,
    fontfamily=tt,
    fontseries=l,
    fontsize=\scriptsize,
    frame=none,
    linenos=true,
    xleftmargin=0.05\linewidth,
    style=tango,
    breaklines=true,
    breakbytoken=true
}

\setminted[python]{
    autogobble=true,
    bgcolor=red!10,
    fontfamily=tt,
    fontseries=l,
    fontsize=\scriptsize,
    frame=none,
    linenos=true,
    xleftmargin=0.05\linewidth,
    style=tango,
    breaklines=true,
    breakbytoken=true
}

\setminted[c++]{
    autogobble=true,
    bgcolor=yellow!10,
    fontfamily=tt,
    fontseries=l,
    fontsize=\scriptsize,
    frame=none,
    linenos=true,
    xleftmargin=0.05\linewidth,
    style=tango,
    breaklines=true,
    breakbytoken=true
}

\setminted[cmake]{
    autogobble=true,
    bgcolor=olive!10,
    fontfamily=tt,
    fontseries=l,
    fontsize=\scriptsize,
    frame=none,
    linenos=true,
    xleftmargin=0.05\linewidth,
    style=tango,
    breaklines=true,
    breakbytoken=true
}

\setminted[sql]{
    autogobble=true,
    fontfamily=tt,
    bgcolor=magenta!5,
    fontseries=l,
    fontsize=\scriptsize,
    frame=none,
    linenos=true,
    xleftmargin=0.05\linewidth,
    style=tango,
    breaklines=true,
    breakbytoken=true
}
\setmintedinline{fontsize=\footnotesize}

\begin{document}
\title{\fontseries{s}\selectfont {\CppTransport}: a platform to
automate calculation of inflationary correlation functions
\vspace{5mm}\hrule}
\author{\fontseries{s}\selectfont\large David Seery}

\affiliation{\vspace{2mm}\small
Astronomy Centre, University of Sussex, Falmer, Brighton BN1 9QH, UK}

\emailAdd{D.Seery@sussex.ac.uk}

\abstract{{\CppTransport} is a numerical platform that can automatically
generate and solve the evolution equations for the 2- and 3-point correlation
functions
(in field space and for the curvature perturbation $\zeta$)
for any inflationary model with canonical kinetic terms.
It makes no approximations beyond the applicability of tree-level
perturbation theory.
Given an input Lagrangian,
{\CppTransport} performs symbolic calculations to determine the
`Feynman rules' of the model
and generates efficient {\CC} to integrate the correlation functions of interest.
It includes a visualization suite
that automates extraction of observable quantities from the raw
$n$-point functions and generates high quality plots with minimal
manual intervention.
It is intended to be used as a collaborative platform,
promoting the rapid investigation of models and
systematizing their comparison with observation.
This guide describes how to install and use the system,
and illustrates its use through some simple examples.
\par\vspace{2cm}
\mbox{}\hfill
\includegraphics[scale=0.1]{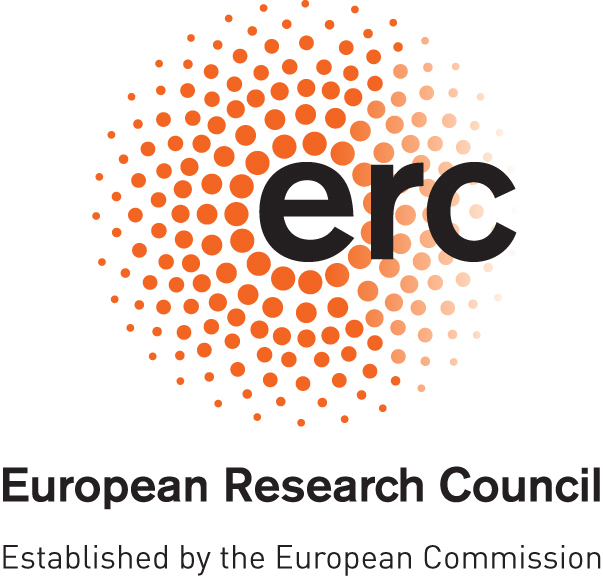}
\quad
\includegraphics[scale=0.5]{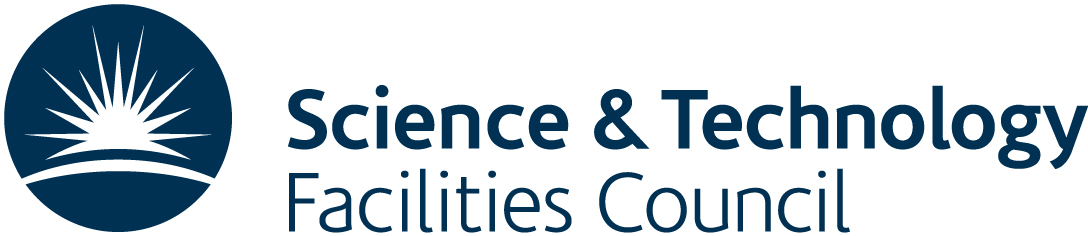}}

\maketitle
\newpage

\section{Introduction}
There is now broad agreement that the inflationary
scenario~\cite{Guth:1980zm,Starobinsky:1980te,Albrecht:1982wi,
Hawking:1981fz,Linde:1981mu,Linde:1983gd}
provides an acceptable framework within which to interpret
observations of the very early universe.
In this scenario, all structure arises from a primordial distribution
of gravitational potential wells laid down by quantum
fluctuations during an early phase of accelerated expansion.
After inflation the universe is refilled with a sea of cooling radiation,
and matter begins to condense within these potential wells.
This generates a network of structure inheriting its statistical
properties from those of the seed quantum fluctuations.
By measuring these properties we can hope
to infer something about the microphysical modes whose fluctuations
were responsible.

Information about the pattern of correlations visible within our Hubble patch
can be extracted from any observable that traces the condensed matter
distribution.
To determine the viability of some particular inflationary model
we must compare these observations with predictions
that carefully account for the precise character of quantum
fluctuations given the field content, mass scales and coupling constants of the model.

Unfortunately these calculations are challenging.
Various
approximate schemes to compute a general $n$-point function are known,
but even where these are available they require numerical methods
except in special cases.
Such schemes typically break the calculation into two pieces:
a `hard' contribution---that is, involving comparatively
large momenta---characterized by wavenumbers near the scale
of the sound horizon $k/a \sim H / c_s$,
and a `soft' contribution involving comparatively low wavenumbers
$k/a \ll H / c_s$~\cite{Dias:2012qy}.
The soft contribution can be computed using the classical
equations of motion and is normally the only one to be handled exactly.
The hard contribution is estimated by assuming
all relevant degrees of freedom are massless and non-interacting.
A typical example is the $\delta N$ formula for the
equal-time two-point function
of the curvature perturbation $\zeta$,
\begin{equation}
    \langle \zeta(\vect{k}_1) \zeta(\vect{k}_2) \rangle_t
    = \aunderbrace[l1r]{N_\alpha(t, t_\ast) N_\beta(t, t_\ast)}_{\text{soft part}}
    \aoverbrace[L1R]{\langle \delta\phi^\alpha(\vect{k}_1) \delta\phi^\beta(\vect{k}_2) \rangle_{t_\ast}}^{\text{hard part}} .
    \label{eq:deltaN-approx}
\end{equation}
The indices $\alpha$, $\beta$, \ldots, label species of scalar
field (with summation implied over repeated indices)
and the subscript attached to each correlation function
denotes its time of evaluation.
The times are ordered so that $t \geq t_\ast$.
Taking $t_\ast$ to label an epoch when
$k_1 / a = k_2 / a \sim H / c_s$
makes $N_\alpha N_\beta$
correspond to the soft piece
and the $t_\ast$ correlation function
correspond to the hard piece.
This division is entirely analogous to the factorization
of hadronic scattering amplitudes into a hard subprocess
followed by soft hadronization.

Any scheme of this type will break down if the
hard initial condition
can not be approximated by the `universal'
massless, non-interacting case.
In recent years it has been understood that there is
a rich phenomenology associated with this possibility,
including `gelaton-like'~\cite{Tolley:2009fg}
or
`QSFI-like'~\cite{Chen:2009we,Chen:2009zp,Chen:2012ge}
effects.
With sufficient care these effects can be captured
in an approximation scheme such as~\eqref{eq:deltaN-approx},
but the approach becomes more complex---and
even if this is possible we have only exchanged the problem
for computation of the hard
component
$\langle \cdots \rangle_{t_\ast}$.
If this hard component is not universal,
then the problem
is no easier than
calculation with which we started.

A different way in which~\eqref{eq:deltaN-approx}
loses its simplicity
occurs when there is not a single hard scale,
but a number of widely separated scales.
For example, this can occur in an $n$-point function
with $n \geq 3$
where the external wavenumbers $\vect{k}_i$
divide into groups characterized by typical
magnitudes $\mu_1$, $\mu_2$, \ldots, $\mu_N$
and $\mu_1 \ll \mu_2 \ll \cdots \ll \mu_N$.
In this case the factorization in~\eqref{eq:deltaN-approx}
becomes more involved~\cite{Kenton:2015lxa},
and must be modified in a way depending on the precise
hierarchy of groups.

Taken together, these difficulties generate a significant
overhead
for any analysis where accurate predictions of
$n$-point functions are important.
The form of this overhead varies from model to model,
and even on the range of wavenumbers under consideration.
If we choose to pay the overhead
and pursue this approach, we encounter three
significant obstacles.
First, a sizeable investment may be required---due to field-theory
calculations
of the hard component---%
before analysis can commence for each new model.
Second, because each hard component is model-specific, there
may be limited opportunities for economy by re-use.
Third, if we implement the hard component of each model
individually (perhaps using a range of different
analytic or numerical methods), we must painfully test and validate
the calculation in each case.

\subsection{Automated calculation of inflationary correlation functions}
To do better we would prefer a completely general method
that could be used to obtain accurate predictions for each $n$-point
function, no matter what mass spectrum
is involved or what physical
processes contribute to the hard component.
Such a method could be used to compute each correlation function directly,
without imposing any form of approximation.

The same problem
is encountered in any area of physics for which
observable predictions depend on the methods of quantum field theory.
The paradigmatic example is collider phenomenology,
where the goal is
to compare theories of beyond-the-Standard-Model physics
to collision events recorded at the Large Hadron Collider.
In both early universe cosmology and collider phenomenology the challenge is to
obtain sufficiently accurate predictions from a diverse and growing range
of physical models---and,
in principle, the obstacles listed above apply
equally in each case.
But, in collider phenomenology, the availability of sophisticated tools
to \emph{automate}
the prediction process has allowed models to be developed and investigated
at a remarkable rate.
Examples of such tools include
\href{http://theory.sinp.msu.ru/~pukhov/calchep.html}{{\CompHEP}/{\CalcHEP}}~\cite{Pukhov:1999gg,Boos:2004kh,Pukhov:2004ca,Belyaev:2012qa},
\href{http://www.feynarts.de/formcalc/}{\FormCalc}~\cite{Hahn:1998yk,Hahn:2000kx,Hahn:2006qw,Agrawal:2011tm},
\href{http://helac-phegas.web.cern.ch/helac-phegas/helac-phegas.html}{\HELAC}~\cite{Kanaki:2000ey,Cafarella:2007pc},
\href{http://madgraph.hep.uiuc.edu}{\MadGraph}~\cite{Maltoni:2002qb,Alwall:2007st,Alwall:2011uj,Alwall:2014hca},
\href{https://sherpa.hepforge.org/trac/wiki}{\SHERPA}~\cite{Gleisberg:2003xi,Gleisberg:2008ta} and
\href{https://whizard.hepforge.org}{\Whizard}~\cite{Moretti:2001zz,Kilian:2007gr}.
(For an early review of the field, see the
\emph{Les Houches Guide to MC Generators}~\cite{Dobbs:2004qw}.)
Their
common feature is support for automatic
generation of LHC event rates
directly from a Lagrangian
by mixing three components:
(1) symbolic calculations
to construct suitable Feynman rules,
(2) automatically-generated compiled code
to compute individual matrix elements,
and (3) Monte Carlo event generators to convert these matrix
elements into
measurable event rates.
This strategy of automation has successfully
overcome the difficulties encountered in
developing cheap, reliable,
model-dependent predictions.
In addition,
reusable tools
bring
obvious advantages of simplicity
and reproducibility.
There have also been indirect benefits.
For example, the existence of widely-deployed
tools has provided a common language
in which to express not only the models but also the
elements of their analysis.

In early universe cosmology the available toolbox is substantially less developed.
A number of public codes are available to assist computation of the
two-point function,
including
\href{http://theory.physics.unige.ch/~ringeval/fieldinf.html}{\FieldInf}
\cite{Ringeval:2005yn,Martin:2006rs,Ringeval:2007am},
\href{http://modecode.org}{\ModeCode} and
\href{http://modecode.org}{\MultiModeCode}
\cite{Mortonson:2010er,Easther:2011yq,Norena:2012rs,Price:2014xpa},
and
\href{http://pyflation.ianhuston.net}{\PyFlation}~\cite{Huston:2009ac,
Huston:2011vt,Huston:2011fr}.
But although these codes are \emph{generic}---they can handle any model
within a suitable class---%
they are not \emph{automated} in the sense described above,
because expressions for the potential and its derivatives must be obtained
by hand and supplied as subroutines.
For the three-point function the situation is more restrictive.
Currently the only public code is
\href{https://sites.google.com/site/codecosmo/bingo}{\BINGO}~\cite{Hazra:2012yn,
Sreenath:2014nca}
which is limited to single-field canonical models.

Partially, this difference
in availability of solvers for the two- and three-point functions
has arisen because
a direct implementation of the Feynman calculus is not straightforward
for $n$-point functions with $n \geq 3$.
In these cases, conversion of formal Feynman integrals into concrete
numerical results
usually
depends on techniques such as Wick rotation
that are difficult to implement without an analytic
expression for the integrand.
Such expressions are seldom available
for the time-dependent backgrounds required by cosmology,
making
integration over the time variable more
demanding than for Minkowski-space scattering amplitudes.
For this reason it would be considerably more convenient to work with an
explicitly real-time formulation.

Recently, Dias et al. described a
formulation of field theory with these properties.
It can be applied to time-dependent backgrounds more straightforwardly
than the traditional machinery of Feynman diagrams~\cite{DiasFrazerMulryneSeery}.
This formulation is based on direct computation of the $n$-point functions
by an evolution or `transport' equation,
allowing most of the complexities of field theory to be absorbed into
calculation of suitable initial conditions.
In inflation these initial conditions \emph{are} universal,
\emph{provided} the calculation is started at sufficiently early times where
all scalar fields can be approximated as massless.
Therefore, obtaining suitable initial conditions becomes a one-time
cost, the results of which are easy to compute numerically.
The remaining challenge is to implement the evolution equations that
bring these initial conditions
to the final time of interest.
These also have a universal form,
parametrized by coefficient matrices
that depend only algebraically on the model at hand.

Using this scheme it becomes possible to implement automated
calculation of inflationary correlation functions in the same sense
as the tools used in collider phenomenology.
By performing suitable symbolic calculations we can
compute the necessary coefficient matrices for any model,
and given knowledge of these matrices it is straightforward to generate
specialized code that implements the necessary evolution equations.
When compiled this code will take advantage of any opportunities for
optimization detected by the compiler, making evaluation of
each correlation function as rapid as possible.
Finally, by
mapping each correlation function over a range of wavenumbers we
place ourselves in a position to determine any
late-universe observable of interest.

\subsection{The {\CppTransport} platform}

{\CppTransport} is a platform that implements
this prescription.
It is the result of three years of development, amounting to roughly 60,000
lines of {\CC},
and consists of three major components:
\begin{enumerate}
    \item
    A \emph{translator} (17,000 lines)
    converts `model description files' into
    custom {\CC} implementing suitable evolution equations and
    initial conditions.
    The model description file enumerates the field content of the model,
    lists any parameters required by the Lagrangian, and
    specifies the inflationary potential.
    It is also possible to document the model by providing
    a rich range of metadata.

    \item
    Once this specialized {\CC} code is available it must be compiled
    together with a runtime support system to produce a finished product capable
    of integrating the transport equations and producing correlation
    functions.
    The \emph{management system} (29,000 lines) is the largest component
    in the runtime support
    and
    has responsibility for coordinating integration jobs
    and
    handling parallelization.
    It also provides database services.

    \item The remaining component is a \emph{visualization and reporting suite}
    (15,000 lines)
    that can process the raw integration data to produce observable
    quantities
    and present the results as plots or tables.
    The reporting component generates interactive HTML
    documents
    containing these outputs
    for easy reading or sharing with collaborators.
\end{enumerate}
\begin{figure}
    \begin{center}
        \includegraphics[scale=0.75]{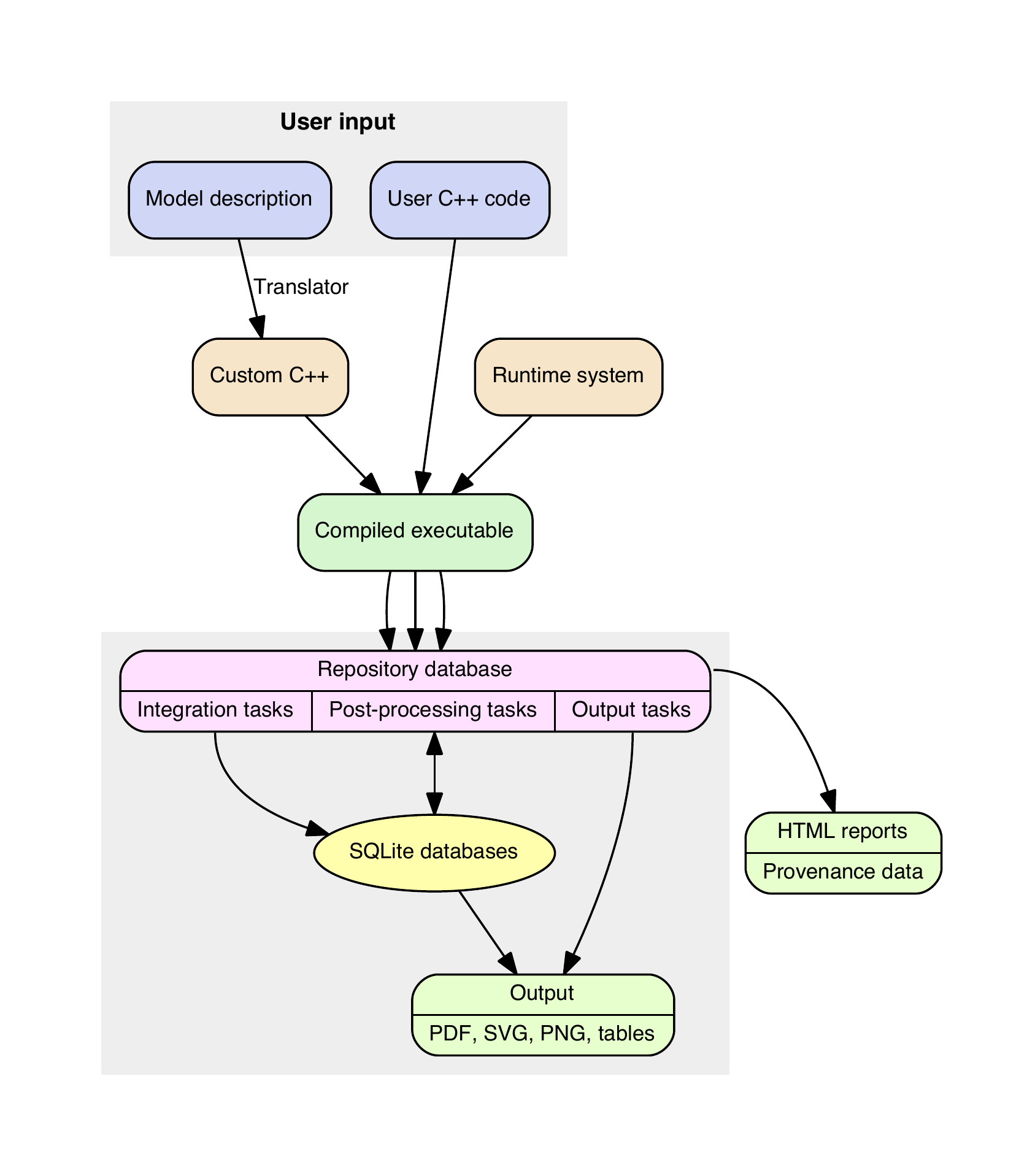}
    \end{center}
    \caption{\label{fig:organization}Block diagram showing relation
    of {\CppTransport} components.}
\end{figure}
A block diagram showing the interaction among {\CppTransport} components is given in
Fig.~\ref{fig:organization}.
To investigate some particular model normally requires the following steps:
\begin{enumerate}
    \item Produce a suitable model-description file and process it
    using the {\CppTransport} translator.

    \item Produce a short {\CC} code
    that couples the runtime system to some number of model implementations produced by
    the translator---at least one, but up to as many as required.
    Each implementation is pulled in as a header file via the \mintinline{c++}{#include}
    directive.
    The code can define
    any number of \emph{integration tasks},
    \emph{post-processing tasks}
    and \emph{output tasks}
    that describe the work to be done:
    \begin{itemize}
        \item \emph{Integration tasks} associate a single model with a fixed choice
        for the parameters required by its Lagrangian, and initial conditions for
        the fields and their derivatives.
        The task specifies a set of times and
        configurations (assignments for the wavenumbers
        $\vect{k}_i$ characterizing each correlation function)
        at which samples should be stored.

        \item \emph{Post-processing tasks} act on the output from an integration
        task or other post-processing task. They are typically used to convert
        the field-space correlation functions generated by integration tasks into
        observable quantities, such as the correlation functions of the
        curvature perturbation $\zeta$.
        Further post-processing tasks can compute inner products of
        the $\zeta$ three-point function with commonly-defined
        templates.

        \item \emph{Output tasks} draw on the data produced by integration
        and post-processing tasks to produce summary plots and tables.
    \end{itemize}

    \item When compiled and executed, the code writes all details of its tasks into
    a \emph{repository}---an on-disk database that is used to aggregate
    information about the tasks and the numerical results they produce.

    \item The runtime system uses the information stored in a repository to
    produce output for each task on demand.
    The results are stored in the repository and information about them
    is collected in its database.

    \item Once predictions for the required correlation functions have
    been obtained, they can be converted into science outputs:
    \begin{itemize}
        \item If relatively simple observables are required, such as
        a prediction of the amplitude or spectral indices for
        the $\zeta$ spectrum or bispectrum,
        then it may be sufficient to set up an output task
        to compute these observables directly.
        The result can be written as a set of
        publication-ready plots in
        some suitable format such as PDF, SVG or PNG,
        or as ASCII-format tables listing numerical values.

        \item Output tasks support a limited range of observables.
        For more complex cases, or to produce plots by hand,
        the required data can be exported from the databases stored inside
        the repository.

        {\CppTransport} does not itself provide this functionality,
        but because its databases are of the industry-standard SQL type
        there is a wide selection of powerful tools to choose from. Many
        of these are freely available.

        \item To share information about the results that have been generated,
        {\CppTransport} can produce a report in HTML format
        suitable for exchanging with collaborators.
        These reports include a summary analysis of content
        generated by integration tasks.
        They also
        embed the plots and tables produced by output tasks.

    \end{itemize}

\end{enumerate}

\subsection{Summary of features}
The remainder of this paper will describe these steps
in more detail
and illustrate how the numerical results they generate can be
used to study inflationary models.
Acting together, the components
of {\CppTransport} provide much more than a bare implementation
of the evolution equations for each $n$-point function.
The main features of the platform are:
\begin{itemize}
    \item Numerical results including \semibold{all relevant
    field-theory effects at tree-level}.
    The method correctly accounts for
    a hierarchy of mass scales,
    interactions among different field species,
    and correlation or interference effects
    around the time of horizon exit.
    It makes no use of approximations
    such as the separate universe method or the slow-roll
    expansion.%
        \footnote{In order to obtain accurate estimates of
        the initial conditions,
        the slow-roll approximation should be
        approximately satisfied at the initial time.
        For more details, see the accompanying technical
        paper~\cite{DiasFrazerMulryneSeery}.
        This also contains a discussion of the validity
        of the tree-level approximation.}

    \item An \semibold{SQL-based workflow} based on
    the \href{http://sqlite.org}{\SQLite} database
    management system,%
        \footnote{`SQL' is the \emph{Structured Query Language},
        a set-based language used nearly universally
        to express queries acting on the most common `relational' type
        of database. It is useful because, to extract
        some subset from a large database, one need only \emph{describe}
        the subset rather than give an explicit algorithm to search for it;
        it is the responsibility of the database management system
        to devise a strategy to read and collate the required records.
        This is very convenient for
        scientific purposes
        because it allows a dataset to be analysed in
        many different ways, by many different tools,
        with only modest effort.}
    which
    {\CppTransport} uses for its data storage.
    Because SQL is an industry-standard technology
    there is a rich ecosystem of existing
    tools that can be used to
    read SQLite databases and perform real-time SQL queries.
    This enables powerful GUI-based workflows that allow
    \semibold{scientific exploitation and analysis without
    extensive programming}.

    \item A \semibold{fully parallelized} {\MPI}-based implementation that
    scales from laptop-class hardware up to many cores, using
    \semibold{adaptive load-balancing} to keep all cores fed with work.
    A \semibold{transactional design} means it is safe to run multiple
    jobs simultaneously,
    and automated \semibold{checkpointing and recovery} prevent
    work being lost in the event of a crash.
    If modifications are required then the messaging implementation is
    \semibold{automatically instrumented} to assist with debugging and
    performance optimization.

    \item Manages the \semibold{data lifecycle} by
    linking each dataset
    to a repository
    storing all information
    about the parameters, initial conditions and sampling points
    used for the calculation.
    The repository also collects \semibold{metadata about the integration},
    such as the type of stepper used and the tolerances applied.
    Together, this information
    ensures that each dataset is properly documented and
    has long-term archival value.
    (All repository data is stored in human-readable
    JSON documents in order that this information is accessible, if necessary,
    without requiring the {\CppTransport} platform.)

    \item The repository system
    supports \semibold{reproducible research} by providing
    an unambiguous means to regenerate
    each dataset, including any products derived
    from it.
    This already provides clear benefits at the analysis stage, because it
    is not possible to confuse when or how each
    output was generated.
    But if shared with the community, the information stored
    in a repository
    enables every step of an analysis to
    be audited.

    \item When derived products such as plots or tables are produced,
    their dependence on existing datasets is recorded.
    This means that the platform can be provide \semibold{a detailed
    provenance for any data product tracked by the repository}.
    The reporting suite generates
    HTML documents containing a hyperlinked
    audit trail summarizing this provenance.
    Notes can be attached to each repository record, meaning that
    the report functions as a type of \semibold{electronic
    laboratory notebook}.

    \item \semibold{Leverages standard libraries},
    including the \href{http://www.boost.org}{\Boost} {\CC} library.
    Integrations are performed using high-quality
    steppers taken from
    \href{http://www.boost.org/doc/libs/release/libs/numeric/odeint}{{\Boost}.{\odeint}}~\cite{2011AIPC.1389.1586A}.
    These steppers are interchangeable, meaning that they can be customized
    to suit the model in question.
    For difficult integrations, very high-order adaptive steppers
    are available.

    \item The translator is a full-featured tool in its own right,
    capable of customizing arbitrary template code
    for each model using sophisticated replacement rules.
    It understands a form of Einstein summation convention,
    making generation of \semibold{specialized template code}
    rapid and convenient.
\end{itemize}

\subsection{Notation and conventions}
This document includes examples of computer code written in a variety
of languages.
To assist in understanding the context of each code block, its background
is colour-coded according to the language:
\begin{itemize}
	\item Shell input or output, blue background: \mintinline{bash}{export PATH=/usr/local/bin:$PATH}
	\item Configuration files, green background: \mintinline{text}{input = /usr/local/share/cpptransport}
	\item {\CC} source code, yellow background: \mintinline{c++}{class dquad_mpi;}
	\item Python source code, red background: \mintinline{python}{def plot:}
	\item {\CMake} scripts, olive background: \mintinline{cmake}{TARGET_LINK_LIBRARIES()}
	\item SQL code, magenta background: \mintinline{sql}{SELECT * FROM}
\end{itemize}
{\CppTransport} uses units where $c=\hbar=1$ but the
reduced Planck mass $\Mp = (8\pi G)^{-1/2}$
can be set to an arbitrary value.

Each inflationary model can have an arbitrary number of scalar fields. These are
all taken to be singlets
labelled by indices $\alpha$, $\beta$, \dots,
and are written $\phi^\alpha$; their perturbations are $\delta\phi^\alpha$.

{\CppTransport} does not use the slow-roll approximation, and therefore it is necessary
to deal separately with the scalar field derivatives
$\dot{\phi}^\alpha$ and $\delta\dot{\phi}^\alpha$.
We often write these generically as $\pi^\alpha$ and collect them
into a larger
set of fields indexed by labels $a$, $b$, \ldots:
\begin{equation}
    X^a = (\phi^\alpha, \pi^\alpha)
    \qquad
    \text{or}
    \qquad
    \delta X^a = (\delta\phi^\alpha, \delta\pi^\alpha) .
\end{equation}

\section{Installation}

\subsection{Minimum requirements}

\para{Compiler}
{\CppTransport} is written in modern {\CC} and requires a relatively recent compiler
with support for {\CC}14.
It has been confirmed to build
correctly with the three major {\CC} toolchains---%
\href{http://clang.llvm.org}{Clang}
(including Apple Clang),
\href{https://gcc.gnu.org}{gcc} and the
\href{https://software.intel.com/en-us/c-compilers}{Intel compiler}.
The minimum recommended versions
are $\geq$ gcc 5.0
and $\geq$ Intel 16.0.
Versions of gcc prior to 5.0 have insufficient standard library support,
and versions of the Intel compiler prior to 16.0
contain bugs that prevent a successful build.%
    \footnote{On Linux, the Intel compiler normally depends on the
    standard library supplied with gcc.
    This means that gcc $\geq$ 5.0 should also be available.}
Any moderately recent version of Clang should work correctly.

In the absence of specialized requirements
it is usually simplest to build with the default
toolchain on your platform.
On Linux the default compiler
will normally be gcc, and for versions of OS X later than 10.7
it will be Clang.
Testing has shown that there is little to be gained by switching between
different compilers, although the Intel compiler can give better performance
under certain circumstances.
Where this occurs
it is sometimes possible to obtain the performance improvement
by building executables for individual models using the Intel compiler,
even if the base {\CppTransport} system is built with the default system
toolchain.%
    \footnote{This can go wrong if there are binary incompatibilities
    between compiled code generated by different compilers.}

\para{Dependencies}
{\CppTransport} is packaged to minimize its pre-requisites and
dependencies.
Nevertheless, there are inevitably some libraries and tools that must be present
before installation can be attempted.
These dependencies have been organized by splitting them into two groups.
The
first
group contains those that must be installed system-wide
(and therefore may be not be installable by individual users in a cluster environment),
or which are very commonly available from package management systems.
The second contains more specialized libraries.
{\CppTransport}
expects dependencies in the first group to be pre-installed by a system
administrator,
or via a package-management system on a personal computer. (Some examples are discussed
below.)
This approach uses system resources economically by
promoting use of shared libraries.
Dependencies in the second group are managed internally
and
do not require user intervention.

\para{Pre-requisite dependencies}
The dependencies that must be installed prior to
building {\CppTransport} are:

\begin{itemize}

\item \semibold{{\CMake} build system.}
The build process for {\CppTransport} is managed by the
\href{https://cmake.org}{\CMake} tool,
which
is responsible for finding the various libraries and system files
needed by {\CppTransport}.
It is also responsible for downloading and installing those dependencies
that {\CppTransport} manages internally.
Once all resources are available, CMake automatically builds and installs
the {\CppTransport} platform.

{\CppTransport} requires {\CMake} version 3.0 or later.

\item \semibold{A working {\MPI} installation.}
{\CppTransport} uses the standard
{\MPI} message-passing system
to coordinate parallel calculations.
A suitable implementation must therefore be installed.
Any standards-compliant choice should work,
including
\href{https://www.open-mpi.org}{\OpenMPI},
\href{https://www.mpich.org}{\MPICH}
(or its derivatives)
or the
\href{https://software.intel.com/en-us/intel-mpi-library}{\IntelMPI} libraries.

\item \semibold{The Boost {\CC} libraries.}
{\CppTransport} uses
a suite of {\CC} libraries called
\href{http://www.boost.org}{Boost}.
Most Boost libraries are header-only and do not require
shared libraries to be pre-built.
However, some do require a build step.
Those required by {\CppTransport}
are
\href{http://www.boost.org/doc/libs/1_60_0/doc/html/date_time.html}{\packagefont Date{\_}Time},
\href{www.boost.org/doc/libs/1_60_0/libs/filesyste…}{\packagefont Filesystem},
\href{http://www.boost.org/doc/libs/1_60_0/libs/log/doc/html/index.html}{\packagefont Log},
\href{http://www.boost.org/doc/libs/1_60_0/doc/html/mpi.html}{\packagefont MPI},
\href{http://www.boost.org/doc/libs/1_60_0/doc/html/program_options.html}{\packagefont ProgramOptions},
\href{http://www.boost.org/doc/libs/1_60_0/doc/html/boost_random.html}{\packagefont Random},
\href{www.boost.org/doc/libs/1_60_0/libs/regex/do…}{\packagefont RegEx},
\href{http://www.boost.org/doc/libs/1_60_0/libs/system/doc/index.html}{\packagefont System},
\href{http://www.boost.org/doc/libs/1_60_0/libs/serialization/doc/}{\packagefont Serialization},
\href{www.boost.org/doc/libs/1_60_0/doc/html/thre…}{\packagefont Thread}
and
\href{http://www.boost.org/doc/libs/1_60_0/libs/timer/doc/index.html}{\packagefont Timer}.

{\CppTransport} will function with any version of Boost later than 1.56,
but is more efficient with version 1.58 or later.

\item \semibold{The {\GiNaC} computer algebra library.}
\href{http://www.ginac.de}{\GiNaC} is a library for performing symbolic computations
in {\CC}. It was originally developed as part of the
\href{http://wwwthep.physik.uni-mainz.de/~xloops/}{\xloopsginac} project
to develop an automated 1-loop particle physics code.
However, the library itself is independent of any particular application.
{\GiNaC} has a further dependence on the
\href{http://www.ginac.de/CLN/}{\CLN} project, but this will
be handled automatically if installation is managed by
a packaging system.

\item \semibold{The {\SQLite} database library.}
This is almost certain to be installed on every Linux or OS X machine,
but some extra developer files may be required.

\item \semibold{The {\OpenSSL} library.}
This is needed to calculate MD5 hashes, which {\CppTransport} uses
to uniquely identify models. The MD5 algorithm is used to give consistent results
on all platforms.

\end{itemize}
Typically, some of
these dependencies
(such as {\CMake}, an {\MPI} implementation, Boost, {\SQLite}, {\OpenSSL})
will already
be available in a managed HPC environment
such as a compute cluster.
All of them
are widely packaged for convenient installation on personal
machines:
they are included in the most common Linux distributions,
and are available using the
\href{https://www.macports.org}{\MacPorts}
or
\href{http://brew.sh}{\Homebrew}
package-management systems for OS X.
{\CppTransport} is agnostic about
how these libraries are installed,
but
unless there are compelling reasons to install in some other way
the
packaged versions normally represent
the most convenient approach.

In addition {\CppTransport} can use certain external programs
if they are available, but does not depend on them for its
core functionality:
\begin{itemize}
    \item Using output tasks to generate plots
    depends on \href{https://www.python.org}{\Python} and the
    \href{http://matplotlib.org}{\Matplotlib} library.
    If one or both is unavailable then it is instead possible
    to generate Python scripts which can be processed to
    product plots at a later date.
    (This provides a means to customize the plot format, if desired.)

    If the
    \href{https://stanford.edu/~mwaskom/software/seaborn}{\seaborn}
    statistical library is available, {\CppTransport} can use
    it to style its plots.

    \item If the \href{http://www.graphviz.org}{\Graphviz} tools are
    available, the HTML report generator will produce a dependency diagram
    showing how each product generated by an output task
    depends on content produced by earlier integration and
    post-processing tasks.
\end{itemize}

\subsection{Downloading the {\CppTransport} platform}
\label{sec:downloading}

The {\CppTransport} sources can be downloaded from various locations:
\begin{itemize}
    \item Citeable archives tagged with a unique
    DOI and containing the source code are deposited
    at the CERN/OpenAIRE Zenodo repository
    \begin{center}
        \url{https://zenodo.org/record/61237}
    \end{center}

    \item The same tar archives can be downloaded from
    \url{http://transportmethod.com}, or
    from {\CppTransport}'s GitHub homepage:
    \begin{center}
        \url{https://github.com/ds283/CppTransport/releases}
    \end{center}

    \item If you wish to install a pre-release version of {\CppTransport},
    or contribute to its development, you can fork or clone the
    git repository from GitHub.
\end{itemize}

\para{Reporting issues}
Bug reports, feature requests or other issues are best reported
using the issue tracker on the GitHub page:
\begin{center}
    \url{https://github.com/ds283/CppTransport/issues}
\end{center}

\subsection{Building the translator and installing the runtime system}
\label{sec:build-install}

This section describes how to install {\CppTransport},
with explicit summaries for handling dependencies in a number of
common cases---OS X with {\MacPorts} or {\Homebrew}, and Ubuntu.
Users with experience building and installing software may wish
to skip directly to~\S\ref{sec:build-translator}
which gives instructions for building the {\CppTransport}
once all dependencies have been installed.

\subsubsection{Installing dependencies on OS X}

As explained above, to build on OS X it is usually convenient to use
the {\MacPorts} or {\Homebrew} packaging systems to simplify installation
of its dependencies.
Whichever package manager is chosen, the first step is to install
{\Xcode} and its associated command-line tools.
\begin{enumerate}
    \item Download {\Xcode} from the App Store. It is a large download
    (roughly $\sim$ 6 Gb) so this may take a while.

    \item Install the command-line tools associated with {\Xcode}
    by opening the Terminal application and typing:
    \begin{minted}{bash}
        xcode-select --install
    \end{minted}

    \item Agree to the {\Xcode} license by typing:
    \begin{minted}{bash}
        sudo xcodebuild -license
    \end{minted}
    You will need to page to the end of the license or chose
    \mintinline{bash}{q} to quit, followed by typing
    \mintinline{bash}{agree} to confirm that you accept the license.
\end{enumerate}

\para{Using {\MacPorts}}
To install {\CppTransport}'s dependencies using {\MacPorts}:
\begin{enumerate}
    \item Install the {\MacPorts} system from
    \url{http://www.macports.org}. Installers are available
    for each recent release of OS X.

    \item Once {\MacPorts} is installed, open a new Terminal.
    ({\MacPorts} makes some changes to your configuration files
    in order to make its packages available.
    These changes are only picked up when you open a new Terminal.)

    The dependencies for {\CppTransport}
    can be installed simultaneously by typing
    \begin{minted}{bash}
        sudo port install cmake openmpi boost +openmpi ginac openssl
    \end{minted}
    (Note that the combination
    \mintinline{bash}{boost +openmpi} is a single item and instructs
    {\MacPorts} to install the Boost libraries using {\OpenMPI}
    as the {\MPI} implementation.)
    Each of these packages has further dependencies which {\MacPorts}
    will download and install automatically.
    This process can take some time.

    If you want to use {\Python} to produce plots and
    {\Graphviz} for dependency diagrams then this can be followed with
    \begin{minted}{bash}
        sudo port install py-matplotlib py-seaborn graphviz
    \end{minted}
    Alternatively you can combine all these packages together
    in a single \mintinline{bash}{sudo port install} instruction.

    \item When all packages have installed, issue the command
    \begin{minted}{bash}
        sudo port select --set mpi openmpi-mp-fortran
    \end{minted}
    This selects {\OpenMPI} as the default {\MPI}
    implementation,
    which will enable {\CppTransport} to find its libraries
    while it is being built.
\end{enumerate}

\para{Using {\Homebrew}}
The procedure is similar for Homebrew.

\begin{itemize}
    \item Install {\Homebrew} by following the instructions at
    \url{http://brew.sh}.

    \item To install the major {\CppTransport} dependencies, execute
    \begin{minted}{bash}
        brew install cmake openmpi ginac openssl
        brew install boost --c++11 --with-mpi --without-single
    \end{minted}

    \item Although {\Homebrew} includes {\Python} and {\Graphviz} it does not
    include {\Matplotlib}, which must be installed separately.
    First install {\Python} and {\Graphviz}:
    \begin{minted}{bash}
        brew install python graphviz
    \end{minted}
    We will also want two further dependencies:
    \begin{minted}{bash}
        brew install pkg-config pip
    \end{minted}
    It is now possible to install {\Matplotlib} and {\seaborn}:
    \begin{minted}{bash}
        pip2 install matplotlib seaborn
    \end{minted}

\end{itemize}

\subsubsection{Installing dependencies on Ubuntu 16.04}
Most Linux distributions will include packages for all
{\CppTransport} dependencies. For illustration we describe the process for
Ubuntu 16.04, but the process will be nearly unchanged for any
Debian-based distribution.

From a terminal, issue the command:
\begin{minted}{bash}
    sudo apt-get install libsqlite3-dev libboost-all-dev libginac-dev libopenmpi-dev libssl-dev cmake python-matplotlib python-seaborn graphviz git texlive texlive-latex-extra texlive-fonts-recommended
\end{minted}
This will download and install all required packages and their dependencies.
Depending what is already available on your machine, this may be a sizeable
download and could take some time.
The large \mintinline{bash}{texlive} dependencies are needed only if you plan to
use {\LaTeX} typesetting with {\Matplotlib}.

Ubuntu provides a tool called ubuntu-make which can conveniently
\href{https://wiki.ubuntu.com/ubuntu-make}{install
development platforms} and their dependencies.
Although alternatives exist, you may wish to investigate the
\href{https://www.jetbrains.com/clion/}{\CLion}
and
\href{https://www.jetbrains.com/datagrip/}{\DataGrip}
platforms which are available through ubuntu-make.
These are commercial products, but free licenses are available
to researchers with an academic email address.
In particular, {\DataGrip} is a good candidate
for a tool to manage or interrogate the SQL databases that
{\CppTransport} produces (see~\S\ref{sec:examine-k-database}).

\subsection{Building the translator}
\label{sec:build-translator}

Once all dependencies are installed it is possible to build {\CppTransport}.
Assuming you have downloaded the source code
from \href{http://zenodo.org}{zenodo.org},
\href{http://transportmethod.com/cpptransport}{transportmethod.com}
or as a specific release from
\href{https://github.com/ds283/CppTransport/releases}{GitHub},
it will be packaged as a \file{.tar.gz} archive containing the source tree.
Place this archive in a suitable directory, then unpack the
archive by typing
\begin{minted}{bash}
	tar xvf CppTransport_2016_03.tar.gz
\end{minted}
The name of the archive may be different if you are using a more recent
version.
The {\CppTransport} source code will be unpacked into a directory
with the name \file{CppTransport}.
The build process proceeds by entering this directory,
creating a \emph{new} directory called
\file{build} that will hold temporary files,
and then configuring {\CMake} to use your preferred compiler
and install to your preferred location.

{\CppTransport} can be installed \emph{system-wide}, making it available
to all users on a machine. Alternatively it can be installed locally,
just for a single user. For example, if installing
system wide we might choose to locate it in \file{/usr/local}.
This usually requires administrator privileges.
Single-user installation would usually locate {\CppTransport} within the
user's home directory
and does not require administrator privileges. This may be the only option
if you are building on a managed system such as a cluster.

In what follows we shall assume that installation is happening locally, but the
changes required for system-wide installation are minimal.
First, enter the {\CppTransport} directory and create a new directory
for temporary files:
\begin{minted}{bash}
	cd CppTransport
	mkdir build
	cd build
\end{minted}
The next step is to configure {\CMake}. If you are building with the default
compiler you can enter
\begin{minted}{bash}
	cmake .. -DCMAKE_BUILD_TYPE=Release -DCMAKE_INSTALL_PREFIX=~/.cpptransport-packages
\end{minted}
This instructs {\CMake}
to build using a release configuration
(some debugging code is suppressed)
and install to the directory \file{\textasciitilde{}/.cpptransport-packages}.
The precise name of this directory is arbitrary and can be freely changed,
although it is wise to avoid the names
\file{\textasciitilde{}/.cpptransport}
and
\file{\textasciitilde{}/.cpptransport\_runtime}
which {\CppTransport} expects to be associated with configuration files.
(See the discussion on p.\pageref{page:config-files} below.)
If you are installing system-wide the install prefix should
be set using
\mintinline{bash}{-DCMAKE_INSTALL_PREFIX=/usr/local}
or similar.

If you wish to build with a different compiler then {\CMake} will require
further information.
For example, if the Intel compiler is available on your \envvar{PATH}
and you wish to build with it, you should use
\begin{minted}{bash}
	cmake .. -DCMAKE_BUILD_TYPE=Release -DCMAKE_INSTALL_PREFIX=~/.cpptransport-packages -DCMAKE_C_COMPILER=icc -DCMAKE_CXX_COMPILER=icpc
\end{minted}
More generally, you should pass the location of the C compiler
as the value of
\cmakevar{CMAKE\_C\_COMPILER}
and the location of the {\CC}
compiler as the value
of \cmakevar{CMAKE\_CXX\_COMPILER}.

If configuration is successful, build the translator
and then install:
\begin{minted}{bash}
	make CppTransport -j4
	make install
\end{minted}
Adjust the argument \mintinline{bash}{-j4}
to correspond to the number of cores available on your machine;
for example, on a dual-core machine you should
use \mintinline{bash}{-j2}
and on a quad-core machine with hyperthreading
you could use \mintinline{bash}{-j8}.
If you don't wish to use parallelized builds then it is possible to
omit the \mintinline{bash}{-j} argument altogether,
although the process may take substantially longer.

\subsection{Configuring your environment}
\label{sec:environment}
\begin{figure}
	\begin{center}
		\includegraphics[scale=0.65]{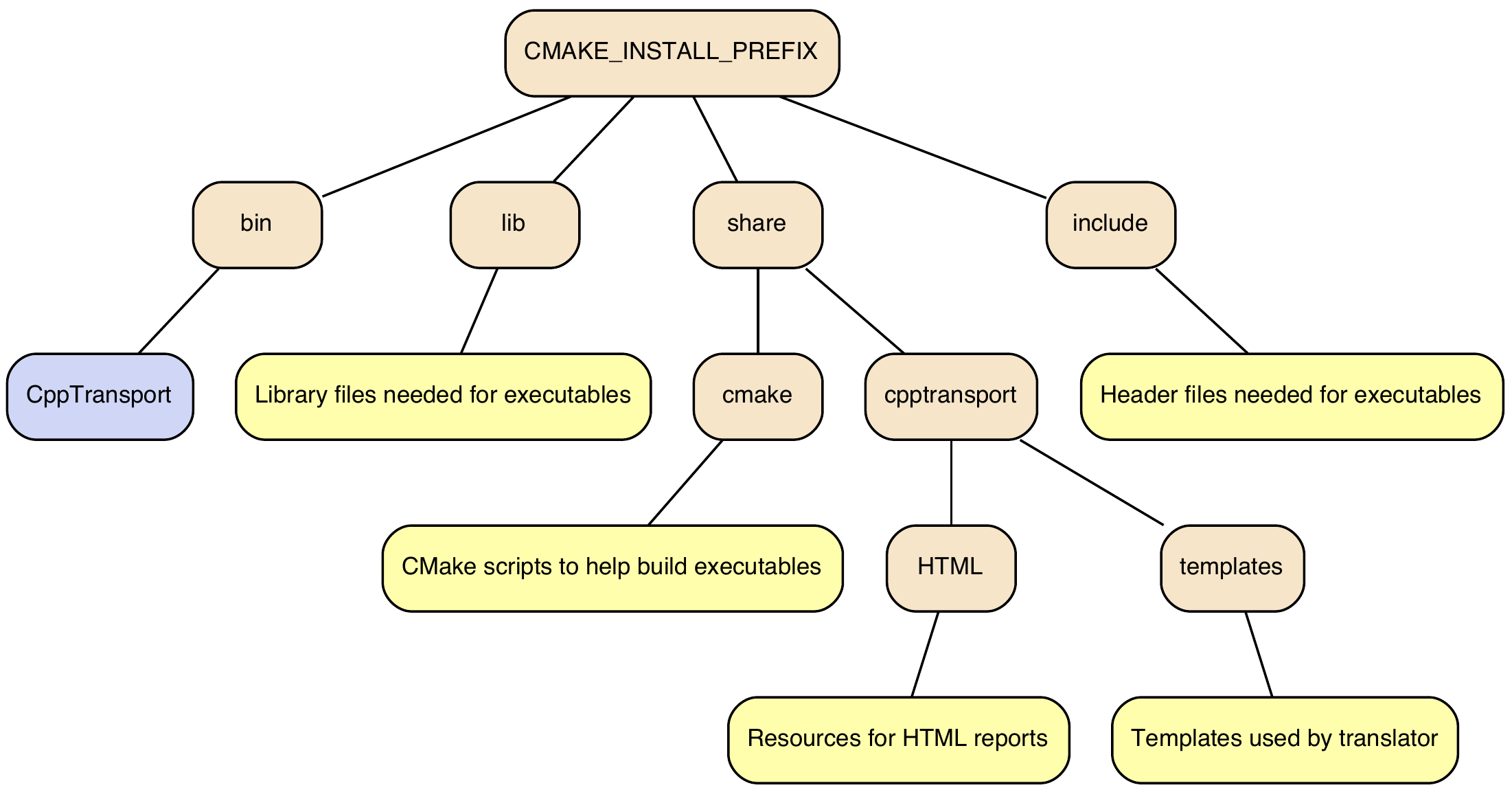}
	\end{center}
	\caption{\label{fig:directory-structure}Directory structure created by
	{\CppTransport} installation process.}
\end{figure}

\para{\envvar{PATH} variable}
{\CppTransport} is now installed, but is not yet usable.
The install procedure writes a large number of files and
resources into directories
under the installation prefix
specified in
\cmakevar{CMAKE\_INSTALL\_PREFIX};
see Fig.~\ref{fig:directory-structure}.
One of these files is the translator,
called \file{CppTransport},
which is installed under \file{bin}.
The operating system needs to know where to find this
when we ask it to process a model file,
and this means adding its parent directory to the
\envvar{PATH} variable.
We also have to inform {\CppTransport}
where its supporting files have been installed;
for example,
the translator requires access to its templates,
and the runtime environment requires access to various assets
that are used when writing HTML reports.

The first step is to add the \file{bin} directory to your path.
Typically this would be set in a configuration script such as
\file{.profile}.%
	\footnote{There are several possible locations where
	\envvar{PATH} can be set, but
	\file{.profile} is a good choice because it will typically be read
	for non-interactive shells. This can be important if you will be running
	{\CppTransport} via {\MPI} in a cluster environment.}
You may find that this file already contains a line of the form
\begin{minted}{bash}
	export PATH=/opt/local/bin:/opt/local/sbin:$PATH
\end{minted}
although the precise list of colon-separated paths may be different.
If not, or there was no existing
\file{.profile} script,
add a new entry that points to the \file{bin}
directory under your installation prefix.
For example, for a user named \texttt{ds283}
the resulting line might be
\begin{minted}{bash}
	export PATH=/Users/ds283/.cpptransport-packages/bin:$PATH
\end{minted}
Ensure that you \emph{add} to the list of
colon-separated paths rather than replacing any existing ones,
or you may find that you lose access to some of your installed software.

\para{{\CppTransport} resources}
At this stage it should be possible to invoke
the {\CppTransport} translator simply by typing
\file{CppTransport} at the command line.
and it is worth opening a new terminal
(causing your \file{\textasciitilde{}/.profile} script to be read)
to check that this happens.
\begin{minted}{bash}
	CppTransport --version
\end{minted}
The translator should respond by printing information about
the installed version, such as
\begin{center}
	\ttfamily\fontseries{l}\selectfont
	CppTransport 2016.3 (c) University of Sussex 2016
\end{center}

If this has worked successfully
then nothing else need to be done to translate model files
or build them into executables.
The only step that is still required
is to inform the runtime system
where it can find the files installed under \file{share}.
There are two ways to do this:
\begin{itemize}
    \item \semibold{Use the \envvar{CPPTRANSPORT_PATH} environment variable.}
    {\CppTransport} will search a list of filesystem locations when looking
    for files. One option is to supply this information as a colon-separated list
    in the environment variable \envvar{CPPTRANSPORT_PATH},
    which functions very like the variable \mintinline{bash}{PATH}
    used above to inform the shell where it should search for executable files.

    If set, \envvar{CPPTRANSPORT_PATH}
    should point to the subdirectory \file{share/cpptransport}
    of the installation prefix.
    For example, your \file{\textasciitilde{}/.profile} could include
    a line such as
    \begin{minted}{bash}
        export CPPTRANSPORT_PATH=~/.cpptransport-packages/share/cpptransport
    \end{minted}
    The \envvar{CPPTRANSPORT_PATH} variable is used by both the
    translator and the runtime library.

    \item \label{page:config-files}\semibold{Use configuration files.}
    Alternatively, options may be supplied to {\CppTransport}
    using configuration files in the top level of your home directory.
    The translator will look for a configuration file named
    \file{\textasciitilde{}/.cpptransport},
    and the runtime environment will look for a file named
    \file{\textasciitilde{}/.cpptransport\_runtime}.
    The use of separate files
    enables different options to be passed
    to each component.

    This method allows you to avoid adding extra material to
    your \file{\textasciitilde{}/.profile} script
    (or related files), if that is desirable.
    To use configuration files for this purpose,
    create a \file{\textasciitilde{}/.cpptransport} file
    containing a the line such as
    \begin{minted}{text}
        include = /Users/ds283/.cpptransport-packages/share/cpptransport
    \end{minted}
    where the path on the right-hand side should be adjusted to have
    the correct prefix and home directory.
    Notice that
    although is the same path that would appear in \envvar{CPPTRANSPORT_PATH},
    the symbol \file{\textasciitilde{}} cannot be used to represent
    the path to the home directory.

    The runtime system requires a separate configuration
    file called \file{.cpptransport\_runtime}
    which should include the same line:
    \begin{minted}{text}
        include = /Users/ds283/.cpptransport-packages/share/cpptransport
    \end{minted}

\end{itemize}

\para{Using {\Python} to produce plots}
Provided the {\Python} interpreter is available on your
\envvar{PATH} it will be automatically detected.
{\CppTransport} will also detect whether {\Matplotlib} is available.
Therefore it is not necessary to adjust any settings in order to use these tools.

By default {\CppTransport} will produce plots in {\Matplotlib}'s own default style.
This was designed to mimic the appearance of MatLab and is not ideal for publication-quality
results.
If the installed version of {\Matplotlib} is sufficiently recent to support
\href{http://matplotlib.org/users/whats_new.html#style-package-added}{style sheets},
or if the {\seaborn} package is available,
then {\CppTransport} can use these features to produce more attractive output.
These features are enabled using the
\option{plot-style} option.
This can be provided on the command line
(see {\S}\ref{sec:option-summary}), but it is usually more convenient to include it in
the \file{\textasciitilde{}/.cpptransport\_runtime} configuration file.
This file should include a line such as
\begin{minted}{text}
    plot-style = seaborn
\end{minted}
Currently, the available styles are
\mintinline{text}{ggplot},
\mintinline{text}{ticks}
(corresponding to the {\Matplotlib} style sheets with the same name; for example, see
\href{https://tonysyu.github.io/raw_content/matplotlib-style-gallery/gallery.html}{here})
and
\mintinline{text}{seaborn}.
If you wish to use a different style it is possible to generate {\Python} scripts
from an output task and insert any required customization by hand.

If you are running {\CppTransport} via SSH or a similar remote login, you may need
to force {\Matplotlib} to use a noninteractive backend.
By default {\CppTransport} will use whatever backend has already been configured;
for more details, \href{http://matplotlib.org/faq/usage_faq.html#what-is-a-backend}{see
the {\Matplotlib} documentation}.
If {\Matplotlib} has been configured to use an interactive backend this will usually
fail for jobs started via remote login.
To fix this you should force {\CppTransport} to use a noninteractive backend
using the \option{{-}{-}mpl-backend} option.
The allowed backends are
\mintinline{text}{Agg},
\mintinline{text}{Cairo},
\mintinline{text}{PDF}
and
\mintinline{text}{MacOSX}.
The Cairo renderer sometimes has problems with {\LaTeX}-formatted text. The
\mintinline{text}{Agg} backend is a good choice unless you know you need something
different.
To set this permanently, add line line
\begin{minted}{text}
    mpl-backend = Agg
\end{minted}
to your
\file{\textasciitilde{}/.cpptransport\_runtime} file.

\para{Using {\Graphviz}}
As for {\Python},
{\CppTransport} will automatically detect the {\Graphviz} tools, provided
they are available on your \envvar{PATH}.

\section{The translator: generating custom code for a specific model}
The first step in using {\CppTransport} to perform practical calculations
is to generate a \emph{model description file}.
As explained above, this describes details of the inflationary model
such as its field content and Lagrangian.
It is used by the translator to generate specialized code
capable of computing the required initial conditions and transport equations.
This code is constructed from a supplied template by applying
well-defined replacement rules.
The template can be modified if required, but in practice this is
not normally necessary.

\begin{warning}
    \noindent
    \semibold{Will {\CppTransport} work for my model?}---%
    Before using {\CppTransport} to study a model, you should
    carefully evaluate whether it satisfies the criteria for
    applicability of the underlying numerical scheme.
    For more details see the accompanying technical paper~\cite{DiasFrazerMulryneSeery}.
    The key considerations are:
    \begin{itemize}
        \item \semibold{Is it possible to find an initial time
        at which the slow-roll conditions approximately apply?}
        {\CppTransport} needs initial estimates of the two- and three-point
        correlation functions, which are obtained from analytic methods
        that use the slow-roll approximation.

        Although the slow-roll conditions normally do not have to be strongly satisfied
        at the initial time, they should be \emph{approximately} satisfied in order
        that the analytic estimates fall within the basin of attraction of the true
        numerical solution. Usually the initial conditions will safely relax
        to this true value, although there is no guarantee that this will always happen.

        \item \semibold{Does the tree approximation apply?}
        {\CppTransport} implements a numerical scheme that computes tree-level
        estimates of each correlation function. For many models this is safe,
        but you should exercise caution if:
        \begin{itemize}
            \item loop corrections are already important for S-matrix processes such
            scattering or decays
            \item copious particle production could allow significant contributions to
            the curvature perturbation $\zeta$ from multiparticle production channels.
            Multiparticle channels such as $n \rightarrow 1$ decay for $n \geq 2$
            make loop-level contributions to expectation values, even if the
            $n \rightarrow 1$ process itself is tree-level when considered as
            an S-matrix element.
            For more details, see~{\S}3.2 of Ref.~\cite{DiasFrazerMulryneSeery}.
            \item production of finite-wavenumber modes can significantly drain energy
            from the zero-mode, as in warm inflation or trapped inflation.
        \end{itemize}
    \end{itemize}
\end{warning}

The model description file consists of a number of \emph{blocks}
that declare properties and attributes for the model.
They generally take the
form
\begin{center}
    \slshape\ttfamily\fontseries{l}\selectfont
    block-name tag {\upshape \{} attribute-list {\upshape \}}
\end{center}
Here, \texttt{\textsl{block-name}} is a keyword indicating
what kind of attributes are being declared;
\texttt{\textsl{tag}} is a label used to identify the block;
and
\texttt{\textsl{attribute-list}} is a list of
assignments in the form
\begin{center}
    \slshape\ttfamily\fontseries{l}\selectfont
    property = value;
\end{center}

\para{Breaking the model description into files}
If desired, it is possible to spread the model description
over several files
by using the directive
\mintinline{text}{#include "string"}.
The effect is as if the contents of the file
whose name matchs
\mintinline{text}{string} had been included at the same point.
The included file can itself contain
further \mintinline{text}{#include} directives.

\subsection{Adding model metadata}
\label{sec:model-block}
Blocks can come in any order, but
it is generally preferable to place
the\block{model} block at or near the top of
the file because it
contains a range of useful summary information.
Most of the fields are optional,
but if provided
they are embedded within the custom {\CC}
output and subsequently attached to any data products
that it is used to generate.
By specifying this data we reduce the risk of
`orphaned' code or data that cannot be traced back
to a specific combination of model,
parameters and initial conditions.

The attributes available within the \block{model} block are:
\begin{itemize}
	\item The tag: this is used to construct the names of the
	{\CC} classes build by the translator,
	and also the names of the output files it generates.
	For this reason it should be fairly short and obey the
	rules for constructing valid {\CC} identifiers and
	filenames.,

    \item \attribute{name = "string";} \\
    Sets the model's textual name to \attribute{string}. The textual name
    is generally used only when producing reports;
    automatically-generated code normally refers to the model
    using the tag associated with the block.

    \item \attribute{description = "string";} \\
    Adds a short description of the model. This should briefly identify
    its origin and major features.

    \item \attribute{citeguide = "string";} \\
    Give short guidance about how to cite this model and its description file.

    \item \attribute{license = "string";} \\
    If you intend to make your description file publicly available (eg. on the
    arXiv or via a data repository service such as
    \href{http://www.zenodo.org}{zenodo.org}),
    you may wish to explicitly set a license that allows re-use
    such as the
    \href{https://creativecommons.org/licenses/}{Creative Commons Attribution license}.
    Some funding agencies may express a preference
    (or even specific requirements) for the licensing of research outputs.

    \item \attribute{revision = integer;} \\
    A model description may evolve through multiple iterations during its
    lifetime.
    Where significant changes occur it can be helpful to indicate this
    unambiguously by changing the model's textual name and the tag
    used to identify it in generated {\CC}.
    However, for minor changes it may be less confusing to retain
    the same identifiers. In these circumstances the
    \attribute{revision} field can be used to
    distinguish between different versions
    of the model file.

    The runtime system will not allow code generated using
    an earlier revision of a model description file to
    handle tasks prepared using a later revision.

    \item \attribute{references = [ string, string, ... ];} \\
    Attach a comma-separated list of strings
    that reference publications associated with this model.
    The string are free-format and can be used for any suitable
    purpose. For example, you may wish to identify papers by their
    arXiv number or by DOI.

    \item \attribute{urls = [ string, string, ... ];} \\
    Attach a comma-separated list of strings
    corresponding to URLs associated with this model.
    These should be internet locations to which an end-user
    can refer to obtain more details about the model or
    its implementation.
\end{itemize}

\begin{example}
    To illustrate the process of preparing a model file and
    constructing tasks,
    in these panels we will work through the steps needed
    for the model of double quadratic inflation---%
    eventually building up to an analysis of its
    bispectrum.

    This model was introduced by Rigopoulos, Shellard \& van
    Tent~\cite{Rigopoulos:2005xx,Rigopoulos:2005us}
    and later studied by
    Vernizzi \& Wands~\cite{Vernizzi:2006ve}.
    It has been widely used as a test case for numerical methods;
    see eg. Refs.~\cite{Mulryne:2009kh,Mulryne:2010rp}.

    {\CppTransport} does not expect any particular naming convention
    for model description files, and they do not need to have a fixed
    extension.
    However, to keep them readily recognizable it may help to
    apply a uniform extension such as \file{.model} or
    \file{.mdl}.
    In this case we will write the model description into a file
    named \file{dquad.model}.
    The first step is to construct a suitable \block{model} block.
	We use the tag \attribute{"dquad"},
	and provide links to the original literature.
	We also assign the model file
	a specific license by tagging it with the abbreviation
	``CC BY'', which indicates the Creative Commons
	Attribution License.

    \begin{minted}{text}
    model "dquad"
      {
        name        = "Double quadratic inflation";
        description = "A two-field model with quadratic potentials";
        citeguide   = "Example from the CppTransport user guide";
        license     = "CC BY";
        revision    = 1;

        references  = [ "astro-ph/0504508",
                        "astro-ph/0511041",
                        "astro-ph/0603799",
                        "arXiv:160x.yyyy" ];
        urls        = [ "http://transportmethod.com" ];
     };
    \end{minted}
\end{example}

\subsection{Specifying a template}
\label{sec:template-block}
The translator produces customized {\CC}
output by reading the model description file,
using it to construct all the information needed
for concrete calculations, and then writing this information
into a \emph{template}. We will examine this
process in more detail in \S\ref{sec:run-translator}.
{\CppTransport} allows arbitrary templates to be used,
although there will not normally be any need to
modify the supplied examples.
The purpose of the \block{model} block
is to tell {\CppTransport} which templates are intended
for use. It does not have a tag.

To use the standard templates, the
\block{templates} block should read:
\begin{minted}{text}
	templates
	  {
	    core           = "canonical_core";
	    implementation = "canonical_mpi";
	  };
\end{minted}
Notice that two templates are required.
The \emph{core} template
writes an output file called
\file{tag\_core.h}
and
defines a {\CC} class
called \mintinline{c++}{tag_core}, where
\mintinline{c++}{tag} is the tag used to declare the
\block{model} block.
This class provides common services such as computation
of initial conditions and mass matrices, which are
the same no matter how we choose to solve the
equations of motion for each correlation function.
The \emph{implementation} class defines a {\CC}
class that integrates these equations.

In principle {\CppTransport} can support many different
implementations. For example, these could use different resources
to carry out the calculation, perhaps by
splitting the work across a range of CPUs and
offload processors such as GPUs or Xeon Phis.
Currently only an {\MPI}-based CPU integrator is supplied
because
testing has shown that---for the specific
system of differential equations
that {\CppTransport} needs to solve---
it is not straightforward to extract
good performance from GPUs.
This difficulty is partially driven by memory requirements,
and may change in future.
The CPU integrator is supplied as a template called
\file{canonical\_mpi.h}
and
writes an output file called
\file{tag\_mpi.h}.
It
defines a {\CC} class
called \mintinline{c++}{tag_mpi}.

\subsection{Choosing a stepper}
\label{sec:stepper-block}
The translator customizes the integration
template to use
a stepper drawn from
the {\Boost}.{\odeint}
\href{http://www.boost.org/doc/libs/1_60_0/libs/numeric/odeint/doc/html/boost_numeric_odeint/getting_started/overview.html}{collection}.
Not all the steppers provided by {\odeint} are available,
and the selection may expand in future.
Currently, the supported steppers are:
\begin{itemize}
	\item \option{runge\_kutta\_dopri5}.
	This is a $4^{\mathrm{th}}$/$5^{\mathrm{th}}$-order
	Dormand--Prince solver, and a good general purpose stepper.
	It is capable of efficiently interpolating the solution between
	sample points, meaning that the step-size can often be kept large
	even when high accuracy is required.
	This gives the method good overall performance.
	It should be regarded as the default unless the model requires
	special treatment.

	\item \option{runge\_kutta\_fehl78}.
	This is a $7^{\mathrm{th}}$/$8^{\mathrm{th}}$ order
	Fehlberg solver.
	It is higher-order than the Dormand--Prince algorithm,
	but cannot interpolate the solution between sample points
	and
	therefore
	sometimes struggles to control its step-size.
	Nevertheless,
	it remains a useful alternative.

	\item \option{bulirsch\_stoer\_dense\_out}.
	This is a Bulirsch--Stoer algorithm that adapts both its
	step-size and the order of the method, currently up to 8th order.
	It can interpolate the solution, enabling the same good control
	of step-size exhibited by the Dormand--Prince algorithm.
	It is typically slower than the other algorithms
	but
	is a good choice where high precision is required.
	It may be the only practical choice if the solution exhibits
	sharp features, which the adaptive order control can handle
	quite effectively.
\end{itemize}
No matter which stepper is selected, it is a good idea to check
that features in a solution
are stable to changes in the stepper and sample mesh.

Because the background equations seldom require a stepper with advanced
capabilities it is possible to specify separate steppers
for the background and perturbations.
At present this has limited utility because background integrations
usually constitute a negligible proportion of the runtime,
but it may have more impact in future.

The background stepper is specified with a \block{background}
block, and the stepper for perturbations is specified with
a \block{perturbations} block.
Each block accepts the same attributes, and neither has a tag.
\begin{itemize}
    \item \attribute{stepper = "string";} \\
    Sets the stepper for this block to be one of the supported
    steppers listed above.

    \item \attribute{stepsize = number;} \\
    Sets the initial step-size.
    All steppers supported by {\CppTransport} are adaptive
    and will adjust their step-size depending on the structure
    of the solution, but an initial estimate is needed.
    The step-size is measured in e-folds.
    Typically values in the range $10^{-12}$ to $10^{-15}$
    are reasonable.
    The stepsize will rapidly be adjusted upwards if the solution
    makes this practicable.

    \item \attribute{abserr = number;} \\
    Sets the absolute tolerance for the stepper.

    \item \attribute{relerr = number;} \\
    Sets the relative tolerance for the stepper.
\end{itemize}
Suitable values for the tolerances are typically in the
range
$10^{-8}$
to
$10^{-12}$,
although in some cases they need to be smaller.
A reasonable default choice is $10^{-8}$, followed by reduction
to $10^{-10}$ or $10^{-12}$ if the stepper fails to keep the solution
under control.

\begin{example}
    Nothing special is needed for the double-quadratic model,
    so we can use the default
    \option{runge\_kutta\_dopri5}
    solver and conventional values for the step-size and tolerances.
    In addition we are using the standard templates,
    so we should add the following lines to the description:
    \begin{minted}{text}
        templates
          {
            core           = "canonical_core";
            implementation = "canonical_mpi";
          };

        background
          {
            stepper  = "runge_kutta_dopri5";
            stepsize = 1E-12;
            abserr   = 1E-12;
            relerr   = 1E-12;
          };

        perturbations
          {
            stepper  = "runge_kutta_dopri5";
            stepsize = 1E-12;
            abserr   = 1E-12;
            relerr   = 1E-12;
          };
    \end{minted}

\end{example}

\subsection{Adding author metadata}
\label{sec:author-block}
The authors of the model description file can be
identified by including one or more \block{author}
blocks.
In principle these are intended to identify the authors
of the \emph{model description} rather than to assign
credit for the original model,
which can be done via the
\attribute{references} attribute
of the \block{model} block.

The tag for each block should be a string giving the
author's textual name.
The available attributes are:
\begin{itemize}
    \item \attribute{email = "string";} \\
    Attaches an email address for this author. Only one
    address is allowed per author.
    If multiple email attributes are given then the
    translator will issue a warning.

    \item \attribute{institute = "string";} \\
    Attach an institutional affiliation.
    As with email addresses, only one affiliation
    is allowed per author.
\end{itemize}

\begin{example}
    A suitable author block for our example file might be:
    \begin{minted}{text}
        author "David Seery"
          {
            institute = "Astronomy Centre, University of Sussex";
            email     = "D.Seery@sussex.ac.uk";
          };
    \end{minted}
\end{example}

\subsection{Specifying field content and Lagrangian parameters}
\label{sec:field-param-block}
The final step is to specify the Lagrangian of the model.
Because {\CppTransport} is currently restricted to models with
canonical kinetic terms it is only necessary to specify the potential.
Before doing so, we must enumerate the fields used by the model
and any parameters appearing in the Lagrangian.
This is done by giving
a \block{field} block for each field, and a \block{parameter} block
for each parameter.
The tag for each block is a symbolic name that can be used to refer
to the corresponding quantity in the potential.
Currently, only one attribute is available which is used to give
a {\LaTeX} name for the quantity:
\begin{itemize}
    \item \attribute{latex = "string";} \\
    Set the {\LaTeX} name of the quantity to be
    \attribute{string}.
    The {\LaTeX} name is available for use when generating derived
    products such as plots.
\end{itemize}

\begin{example}
    In double-quadratic inflation there are two
    fields, conventionally $\phi$ and $\chi$,
    and the potential is
    \begin{equation}
        V(\phi, \chi) = \frac{1}{2} M_\phi^2 \phi^2
            + \frac{1}{2} M_\chi^2 \chi^2 .
        \label{eq:double-quadratic-V}
    \end{equation}
    This means there are two parameters, $M_\phi$
    and $M_\chi$.
    To declare all of these objects we would write
    \begin{minted}{text}
        field phi
          {
            latex = "\phi";
          };

        field chi
          {
            latex = "\chi";
          };

        parameter Mphi
          {
            latex = "M_\phi";
          };

        parameter Mchi
          {
            latex = "M_\chi";
          };
    \end{minted}
\end{example}

\subsection{Specifying the Lagrangian}
\label{sec:lagrangian-block}
Once all fields and parameters have been declared
we can use them to give an expression for the potential.
The syntax for this is
\attribute{potential = expression;}
where \attribute{expression} is a mathematical
expression written using the same kind of syntax
one would employ in
{\Mathematica} or {\Maple}.
{\CppTransport} understands the standard mathematical
operators, including
\descfile{+} for addition,
\descfile{-} for subtraction,
\descfile{*} for multiplication,
\descfile{/} for division
and
\descfile{^} for exponentiation.
Nested brackets
\descfile{(} $\cdots$ \descfile{)}
can be used to indicate precedence.
It also understands the mathematical functions listed in
Table~\ref{table:funcs}.
\begin{table}

    \begin{center}

        \small
    	\heavyrulewidth=.08em
    	\lightrulewidth=.05em
    	\cmidrulewidth=.03em
    	\belowrulesep=.65ex
    	\belowbottomsep=0pt
    	\aboverulesep=.4ex
    	\abovetopsep=0pt
    	\cmidrulesep=\doublerulesep
    	\cmidrulekern=.5em
    	\defaultaddspace=.5em
    	\renewcommand{\arraystretch}{1.5}

        \rowcolors{2}{gray!25}{white}

        \begin{tabular}{ll}

            \toprule
            \semibold{function} & \semibold{meaning} \\
            \texttt{abs(x)} & absolute value $|x|$ \\
            \texttt{sqrt(x)} & square root $\sqrt{x}$ \\
            \texttt{sin(x)} & sine $\sin x$ \\
            \texttt{cos(x)} & cosine $\cos x$ \\
            \texttt{tan(x)} & tangent $\tan x$ \\
            \texttt{asin(x)} & inverse sine $\sin^{-1} x$ \\
            \texttt{acos(x)} & inverse cosine $\cos^{-1} x$ \\
            \texttt{atan(x)} & inverse tangent $\tan^{-1} x$ \\
            \texttt{atan2(y,x)} & inverse tangent $\tan^{-1} y/x$ using signs of $x$, $y$ to determine quadrant \\
            \texttt{sinh(x)} & hyperbolic sine $\sinh x$ \\
            \texttt{cosh(x)} & hyperbolic cosine $\cosh x$ \\
            \texttt{tanh(x)} & hyperbolic tangent $\tanh x$ \\
            \texttt{asinh(x)} & inverse hyperbolic sine $\sinh^{-1} x$ \\
            \texttt{acosh(x)} & inverse hyperbolic cosine $\cosh^{-1} x$ \\
            \texttt{atanh(x)} & inverse hyperbolic tangent $\tanh^{-1} x$ \\
            \texttt{log(x)} & natural logarithm $\ln x$ \\
            \texttt{pow(x, y)} & exponentiation $x^y$ \\
            \bottomrule

        \end{tabular}

    \end{center}

    \caption{\label{table:funcs}Mathematical functions understood by {\CppTransport}}

\end{table}

In simple cases it is easy to specify the potential in just one line.
For example, in single-field $\phi^2$ inflation we could write
\begin{minted}{text}
    potential = m^2 * phi^2 / 2;
\end{minted}
Fields are assumed to have dimension $[\mathrm{M}]$,
and the pre-defined symbol \descfile{M_P} is available to represent
the Planck mass.

In more complex cases, single-line expressions become difficult to read
or debug and it is preferable to break the potential down into subexpressions.
{\CppTransport} provides a \block{subexpr} block for this purpose.
Its tag is the symbol that will be used to refer to the subexpression,
and the block accepts two attributes:
\begin{itemize}
    \item \attribute{latex = "string";} \\
    Specifies a {\LaTeX} symbol associated with this subexpression.

    \item \attribute{value = expression;} \\
    Defines the symbolic expression
    associated with this quantity.
\end{itemize}
For example, consider the potential studied
by Gao, Langlois \& Mizuno~\cite{Gao:2012uq},
\begin{equation}
    V(\phi, \chi)
    =
        \frac{1}{2} M^2
        \big[
            \chi - (\phi - \phi_0) \tan \Xi
        \big]^2
        \cos^2 \frac{\Delta\theta}{2}
        +
        \frac{1}{2} m_\phi^2 \phi^2 ,
\end{equation}
where $\Xi$ is defined by
\begin{equation}
    \Xi \equiv \frac{\Delta\theta}{\pi}
    \tan^{-1}
    \frac{s(\phi - \phi_0)}{\Mp^2} .
\end{equation}
This potential is designed to contain an inflationary
valley with a turn.
The quantities $M$ and $m_\phi$ are mass scales, and
$\phi_0$, $\Delta\theta$ and $s$ are
constants that parametrize the turn.
Assuming we have defined
suitable fields \descfile{phi}, \descfile{chi}
and parameters \descfile{M}, \descfile{mphi}, \descfile{phi0}
\descfile{Delta} and \descfile{s},
the potential can be broken down into subexpressions:
\begin{minted}{text}
    subexpr Xi
      {
        latex = "\Xi";
        value = (Delta/pi) * atan(s*(phi-phi0) / M_P^2);
      };

    subexpr V1
      {
        latex = "V_1";
        value = (1/2) * M^2 * (chi - (phi-phi0)*tan(Xi))^2 * cos(Delta/2)^2;
      };

    subexpr V2
      {
        latex = "V_2";
        value = (1/2) * mphi^2 * phi^2;
      };

    potential = V1 + V2;
\end{minted}

\begin{example}
    The potential for double-quadratic inflation is simple.
    Given the fields and parameter definitions
    described above it can be written in one line:
    \begin{minted}{text}
        potential = Mphi^2 * phi^2 / 2 + Mchi^2 * chi^2 / 2;
    \end{minted}
\end{example}

\subsection{Running the translator and producing output}
\label{sec:run-translator}
Once the model description is complete,
the translator is run to produce
{\CC} classes that implement the transport equations for it.
Provided your environment has been set up as described in
\S\ref{sec:environment} it should be possible to
invoke the translator simply by typing
\file{CppTransport} at the shell prompt.

The translator accepts a number of arguments.
Some of these perform simple housekeeping functions:
\begin{itemize}
    \item \option{{-}{-}help} \\
    Display brief usage information and a list of all
    available options

    \item \option{{-}{-}version} \\
    Show information about the version of {\CppTransport} being used.

    \item \option{{-}{-}license} \\
    Display licensing information.

    \item \option{{-}{-}no-colour} or \option{{-}{-}no-color} \\
    Do not produce colourized output. {\CppTransport} will normally detect
    the type of terminal in which it is running and adjust its
    output formatting appropriately. Where colour is available, it is used
    to add clarity.
    However, if you are redirecting its output to a file
    (or if this happens automatically as part of a batch
    environment), you may wish to suppress this behaviour.

    \item \option{{-}{-}verbose}, or abbreviate to \option{-v} \\
    Enable verbose output, giving more information about the different
    phases of translation and some statistics about
    the process.
\end{itemize}
Others affect the files read or written by the translator:
\begin{itemize}
    \item \option{{-}{-}include}, or abbreviate to \option{-I} \\
    Should be followed by a path to be added to the list
    of paths searched when looking for template files.
    For example, if extra templates have been written
    (or installed in a different location),
    this argument can be used to enable {\CppTransport} to find them.
    The templates should be stored in a directory named
    \file{templates} under this path.

    \item \option{{-}{-}no-search-env} \\
    Do not use the environment variable \envvar{CPPTRANSPORT_PATH}
    to determine a list of search paths;
    use only paths specified by \option{{-}{-}include} on the
    command line.

    \item \option{{-}{-}core-output} \\
    Followed by a path that specifies the file to which the
    customized \emph{core template} should be written.
    By default  the name
    \file{tag\_core.h} is used, where \file{tag} is the tag
    used to declare the \block{model} block.
    In most cases this default will be suitable, so it is not necessary
    to specify a filename explicitly.

    \item \option{{-}{-}implementation-output} \\
    Followed by a path that specifies the file to which the
    customized \emph{implementation template} should be written.
    By default  the name
    \file{tag\_implementation.h} is used, where \file{tag} is the tag
    used to declare the \block{model} block
    and \file{implementation}
    is the name of the integration implementation; in the
    current version this is always \file{mpi}.
    In most cases this default will be suitable, so it is not necessary
    to specify a filename explicitly.
\end{itemize}
A final set of options influence the {\CC} code generated by the translator:
\begin{itemize}
    \item \option{{-}{-}no-cse} \\
    Disable \emph{common sub-expression elimination}, described in more detail
    in~\S\ref{sec:codegen} below.

    \item \option{{-}{-}annotate} \\
    Annotate the generated code with comments,
    including comments to indicate which
    template line corresponds to each output line.
    This option can be useful for debugging, but often generates large
    files.

    \item \option{{-}{-}unroll-policy} \\
    Followed by an integer corresponding to the maximum allowed size of an
    \emph{unrolled index set}, described in more detail in~\S\ref{sec:codegen}.

    \item \option{{-}{-}fast} \\
    Unroll all index sets, regardless of size.
\end{itemize}

These options may also be specified in the
\file{\textasciitilde{}/.cpptransport} configuration
file discussed on p.\pageref{page:config-files}.
Each option should be placed on a new line,
without the leading \option{{-}{-}}.
Options such as \option{{-}{-}include} that accept an argument
should be written in the format
\mintinline{text}{option = argument},
such as
\mintinline{text}{include = /usr/local/share/cpptransport}
as described above.
If an option appears both in the configuration file and on the
command line, then values specified on the command line
are preferred.

\begin{example}
    For double-quadratic inflation no special options are required
    (although see the discussion of \option{{-}{-}fast}
    in~\S\ref{sec:codegen} below, which can be used to improve the
    execution time for this model).
    To print status messages during the different phases of translation
    we can run the translator with the \option{{-}{-}verbose} or
    \option{-v} switch to product verbose output.
    This gives:
    \begin{minted}[xleftmargin=0pt,bgcolor=blue!10,linenos=false]{text}
        $ CppTransport -v dquad.model
        CppTransport: translating '...templates/canonical_core.h' into 'dquad_core.h'
        CppTransport: translation finished with 1216 macro replacements
        CppTransport: macro replacement took 0.172s, of which time spent tokenizing 0.00881s (symbolic computation 0.0223s, common sub-expression elimination 0.087s)
        CppTransport: translating '...templates/canonical_mpi.h' into 'dquad_mpi.h'
        CppTransport: translation finished with 8102 macro replacements
        CppTransport: macro replacement took 0.169s, of which time spent tokenizing 0.00547s (symbolic computation 0.0179s, common sub-expression elimination 0.0919s)
        CppTransport: 153 expression cache hits, 415 misses (time spent performing queries 0.0116s)
        CppTransport: processed 1 model in time 0.391s
    \end{minted}

    {\CppTransport} gives information about each file it translates.
    Here, the files being translated are the core template
    \file{canonical\_core}
    which becomes \file{dquad\_core.h},
    and the implementation template
    \file{canonical\_mpi}
    which becomes \file{dquad\_mpi.h}.
    Recall that the stem \file{dquad} used to construct these filenames
    is taken from the tag provided to the model block
    in~\S\ref{sec:model-block}.

    In the subsequent messages, {\CppTransport} informs us of the number
    of tokens (`macros') replaced while customizing each file,
    and also the time spent performing each step.
    Sometimes common sub-expression elimination becomes very time-consuming;
    in this case, see the discussion in~\S\ref{sec:codegen}.
\end{example}

\subsection{Using the code generation options}
\label{sec:codegen}
As explained above,
the translator's task is to produce customized output.
It does this by rewriting the template files according
to well-defined rules.

In order to perform this rewriting
the translator recognizes a large number of
\emph{tokens},
of the form
\token{NAME},
\token{CITEGUIDE},
\token{DESCRIPTION}
(and so on),
which are replaced with the corresponding
data from the model description file.
There are also tokens such as
\token{HUBBLE\_SQ}
and \token{EPSILON}
which are replaced with symbolic expressions
computed from the Lagrangian of the model---here,
these would be
expressions to compute the square of the Hubble rate $H^2$
and the slow-roll parameter $\epsilon = - \dot{H}/H^2$,
respectively.

\para{Unrolling index sets}
In addition to these simple rewriting rules, the translator
must be able to generate code that implements the transport equations
for the two- and three-point functions.
Writing these correlation functions as
\begin{equation}
\begin{split}
    \langle
        \delta \phi^\alpha(\vect{k}_1)
        \delta\phi^\beta(\vect{k}_2)
    \rangle_t
    & = (2\pi)^3 \delta(\vect{k}_1 + \vect{k}_2)
    \Sigma^{\alpha\beta} \\
    \langle
        \delta \phi^\alpha(\vect{k}_1)
        \delta \phi^\beta(\vect{k}_2)
        \delta \phi^\gamma(\vect{k}_3)
    \rangle_t
    & = (2\pi)^3 \delta(\vect{k}_1 + \vect{k}_2 + \vect{k}_3)
    \alpha^{\alpha\beta\gamma} ,
\end{split}
\end{equation}
their evolution equations become
\begin{equation}
\begin{split}
    \frac{\d \Sigma^{\alpha\beta}}{\d N}
    & =
    {u^\alpha}_\gamma \Sigma^{\gamma\beta}
    + {u^\beta}_\gamma \Sigma^{\alpha\gamma} \\
    \frac{\d \alpha^{\alpha\beta\gamma}}{\d N}
    & =
    {u^\alpha}_\delta \alpha^{\delta\beta\gamma}
    +
    {u^\alpha}_{\delta\epsilon} \Sigma^{\delta \beta} \Sigma^{\epsilon \gamma}
    + \text{cyclic} ,
\end{split}
\label{eq:transport-equations}
\end{equation}
where $\d N = H \, \d t$ represents the number of e-folds which elapse in a
cosmic time interval $\d t$.
Here, ${u^\alpha}_\beta$ and ${u^\alpha}_{\beta\gamma}$ are coefficient matrices
calculated internally by the translator and
depending on the wavenumbers $\vect{k}_1$, $\vect{k}_2$, $\vect{k}_3$.
These matrices are represented using further tokens
such as \token{U2\_TENSOR[AB]} (corresponding to ${u^\alpha}_\beta$)
and \token{U3\_TENSOR[ABC]} (corresponding to ${u^\alpha}_{\beta\gamma}$).
The labels \indexset{AB} and \indexset{ABC} represent the associated indices.
The translator understands enough of the Einstein summation convention
to represent (for example) the transport equation for the two-point function
as
\begin{minted}{text}
    dSigma[$A][$B] $=  $U2_TENSOR[AC] * Sigma[$C][$B];
    dSigma[$A][$B] $+= $U2_TENSOR[BC] * Sigma[$A][$C];
\end{minted}
It has been assumed that the two-point function $\Sigma^{\alpha\beta}$ is
encoded in an array-like object \mintinline{text}{Sigma[][]}
and the derivative is to be written into a separate
array-like object \mintinline{text}{dSigma[][]}.
In particular, given these expressions the translator
understands that the free indices
\indexset{A} and \indexset{B}
label independent components of the
overall matrix equation,
and that the repeated index \indexset{C} is to be summed over.
The transport equation for the three-point function can
be represented similarly.

During translation these compact expressions must be unpacked into
valid {\CC} that performs the required calculations.
There are $N^2$ independent equations for the two-point function,
each of which entails a sum over one dummy index.
Therefore the overall size for this set of equations scales as $\Or(N^3)$.
For the three-point function there are $N^3$ independent equations,
but now each equation entails a sum over \emph{two} dummy indices.
Therefore the overall size scales as $\Or(N^5)$.
After unpacking we incur two types of cost.
One is \emph{execution time}:
no matter how they are expressed, the amount of work
involved in solving these evolution
equations will scale roughly like $\Or(N^3)$ or $\Or(N^5)$, respectively.
The other is \emph{space}:
if unpacked in the most literal fashion,
by simply writing out each component of Eq.~\eqref{eq:transport-equations}
sequentially,
the size of the generated {\CC} code will also scale roughly like
$\Or(N^3)$ or $\Or(N^5)$.

We cannot alter the power law in these scalings,
but it is possible to
make some limited tradeoffs between the space cost and execution time:
\begin{itemize}
    \item To obtain the fastest execution time, we can opt for the
    space-hungry strategy of writing out each equation explicitly.
    {\CppTransport} describes the indices being unpacked as an
    \emph{index set}, and refers to the process of writing them
    out sequentially as \emph{unrolling}.%
    	\footnote{The name is borrowed from a very similar
    	\href{https://en.wikipedia.org/wiki/Loop_unrolling}{loop optimization technique}.}

    Unrolling allows the {\CC} compiler to generate simple linear code
    that performs all the required computations without branches
    or jumps, which would be required by loops
    and may incur performance penalties.
    Also,
    because this strategy is completely explicit it maximizes
    the compiler's opportunities to optimize away redundant calculations.

    The downside is that generated {\CC} files become
    very large even for moderate $N$.
    If we require the three-point function
    then the dominant scaling comes from terms of the form
    ${u^\alpha}_{\delta\epsilon} \Sigma^{\delta\beta} \Sigma^{\epsilon^\gamma}$.
    We have been estimating the number of terms in each sum as $\sim N$,
    but it is usually $2N$ because we must account both for
    the fields and their canonical momenta.
    Therefore,
    assuming the dominant terms generate $\sim (2N)^5$ lines
    and supposing each of these lines to average $\sim 50$ characters,
    it follows that even $N = 20$ will generate an implementation file of
    size $\sim 10 \Gb$.
    Such large files require a prohibitively large amount of time and memory to compile.
    This places a practical upper limit on the maximum size of a {\CC} file.
    On typical hardware this limit is already much smaller than $1 \Gb$,
    making unrolling an unacceptable strategy except when $N$ is rather small.

    \item Alternatively, the indices can be unpacked into a
    {\CC} \mintinline{c++}{for}-loop.
    For example, the translator
    might unpack the line
    \mintinline{text}{dSigma[$A][$B] $= $U2_TENSOR[AC] * Sigma[$C][$B];}
    into
    \begin{minted}{c++}
        for(int A = 0; A < 2*N; ++A)
          {
            for(int B = 0; B < 2*N; ++B)
              {
                dSigma[A][B] = 0.0;
                for(int C = 0; C < 2*N; ++C)
                  {
                    dSigma[A][B] += U2_TENSOR[A][C] * Sigma[C][B];
                  }
              }
          }
    \end{minted}
    assuming that the array-like object
    \mintinline{c++}{U2_TENSOR[][]} has been initialized with the
    components of the tensor ${u^\alpha}_\beta$.

    The loop-based representation is considerably more economical with space,
    because its storage requirements do not grow with $N$.%
        \footnote{To be explicit, it is the storage requirements to
        express the \emph{algorithm itself} that are under discussion
        here. The storage requirements for the state variables
        \mintinline{c++}{Sigma} and \mintinline{c++}{dSigma}
        always scale with $N$.}
    For this reason
    it is the only viable approach for general $N$.
    The disadvantage is that we may forfeit opportunities
    for optimization.%
    	\footnote{Another disadvantage might come from
    	the extra overhead associated with a loop counter
    	and branch penalties. But the {\CC} compiler might
    	decide it it worthwhile to
    	optimize the loop by unrolling it anyway, in which
    	case these disadvantages disappear.
    	The disadvantage of masking zeros in
    	in ${u^\alpha}_\beta$ or ${u^\alpha}_{\beta\gamma}$
    	generally cannot be fixed, however, even by a good
    	optimizing compiler.}
    For example, it sometimes happens that
    certain components of ${u^\alpha}_\beta$
    or ${u^\alpha}_{\beta\gamma}$ are zero,
    and therefore the corresponding terms in the summation
    can be omitted.
    When all expressions are written explicitly
    it is easy for the compiler to make such
    optimizations.
    In the loop-based approach
    the compiler will normally be unable to
    skip particular iterations of the loop body,
    and therefore these terms will not be optimized away.
    The effect of these irrelevant operations
    can accumulate to a sizeable performance
    difference over many integrations;
    see Table~\ref{table:fast-v-unroll}.
\end{itemize}

{\CppTransport} attempts to find a compromise between these
strategies.
It will elect to unroll index sets where the result will not
be too large,
with
the distinction being set by the value assigned to
\option{{-}{-}unroll-policy}. The default is 1000.
Also,
the elements of a tensor such as ${u^\alpha}_\beta$
are stored in temporary arrays (such as
\mintinline{c++}{U2_TENSOR[][]} in the above example)
because this is required for
large index sets that `roll up' into
\mintinline{c++}{for}-loops.
Depending on the compiler settings, this use
of temporary arrays may inhibit some
optimization of redundant arithmetic.

The default unroll policy of 1000 is intended to give reasonable
performance for models with modest $N$, while simultaneously
allowing models with large $N$ to be handled.
However, we will see shortly
that if $N$ is not too large then global unrolling should be preferred.
In cases where it is known that the resulting {\CC} files will be
acceptable to the compiler it is possible to force global
unrolling using the command-line switch
\option{{-}{-}fast}. This instructs
{\CppTransport} to disregard the unroll policy limit and unroll
\emph{all} index sets.
In addition, the translator will no longer store the elements of tensors such as
${u^\alpha}_\beta$ in a temporary array, but instead cache them
in \mintinline{c++}{const} local variables.
This gives the compiler the best chance of removing unnecessary operations.

Table~\ref{table:fast-v-unroll}
shows
a comparison of execution times
for a particular $N=2$ model
with \option{{-}{-}fast}
and unrolling limits of $1000$ and $0$ (forcing roll-up of all
index sets).
It shows a significant advantage for \option{{-}{-}fast}.
This is a fairly general phenomenon;
testing has shown that models with $N=2$ or $N=3$ can often be unrolled effectively,
yielding a non-negligible performance improvement.

\begin{table}

    \begin{center}

        \small
    	\heavyrulewidth=.08em
    	\lightrulewidth=.05em
    	\cmidrulewidth=.03em
    	\belowrulesep=.65ex
    	\belowbottomsep=0pt
    	\aboverulesep=.4ex
    	\abovetopsep=0pt
    	\cmidrulesep=\doublerulesep
    	\cmidrulekern=.5em
    	\defaultaddspace=.5em
    	\renewcommand{\arraystretch}{1.5}

        \rowcolors{2}{gray!25}{white}

        \begin{tabular}{lrrrr}

            \toprule
            \semibold{setting}
                & \semibold{core}
                & \semibold{implementation}
                & \semibold{CPU/configuration}
                & \semibold{CPU total} \\
            \option{{-}{-}fast}
                & 135 kb & 649 kb & 0.993 s & 47 m 17 s \\
            \option{{-}{-}unroll-policy 1000}
                & 145 kb & 211 kb & 1.60 s & 1 h 16 m 20 s \\
            \option{{-}{-}unroll-policy 0}
                & 118 kb & 89 kb & 2.41 s & 1 h 54 m 16 s\\
            \bottomrule

        \end{tabular}

    \end{center}

    \caption{\label{table:fast-v-unroll}Comparison of
    generated code size and execution time
    for 2856 bispectrum configurations and
    an axion+quadratic model
    $V = m^2 \phi^2 / 2 + \Lambda^4 ( 1 - \cos 2\pi f^{-1} \chi )$;
    see Elliston et al. for a description of the parameters
    and initial conditions~\cite{Elliston:2011dr}.
    We use the first set of parameters described in {\S}5.1.2 of that
    reference.
    Timings are averages of 3 runs using OS X 10.11.4
    and the Apple Clang compiler 7.3.0
    on an Ivy Bridge i7-3770 machine.
    Each {\CppTransport} job used 7 worker processes.}

\end{table}

\para{Common sub-expression elimination}
To keep its generated files as small as possible,
{\CppTransport} uses a second strategy called
\emph{common sub-expression elimination}.
The automated symbolic calculations performed internally by the translator
are not automatically simplified,
and therefore the results resemble
those from {\Mathematica}
before application of
\texttt{Simplify[]}
or
\texttt{FullSimplify[]}.
These expressions often share common building blocks, such as the
Hubble rate $H$ or the slow-roll parameter $\epsilon$, which
{\CppTransport} tries to factor out intelligently.
Even when this has been done there may be further common pieces that
can be extracted.
For example, after common sub-expression elimination,
{\CppTransport} would translate the expression $(A+B+1)^2/(A+B)$ into
{\CC} of the form
\begin{minted}{c++}
    const auto temp_1 = A + B;
    const auto temp_2 = temp_1 + 1.0;
    const auto temp_3 = temp_2 * temp_2;
    const auto temp_4 = temp_3 / temp_1;
\end{minted}
The local variable \mintinline{c++}{temp_4} would be used to represent
the value of the expression.

This procedure is generally effective at minimizing the size of the generated
code, and therefore making the compiler's job as straightforward as possible.
However, the task of finding common sub-expressions is expensive
in the same way that
{\Mathematica}'s \texttt{Simplify[]} or
\texttt{FullSimplify[]} operations can be expensive.
For more complex models it is usually
the most time-consuming step in
the translation process, by a considerable margin.
If desired, {\CppTransport} provides the command-line switch
\option{{-}{-}no-cse}
to disable common sub-expression elimination.
This will dramatically speed up translation, but leaves the compiler
with a harder job because the same task has effectively been transferred
to it.

Normally it is advisable to leave common sub-expression elimination
enabled unless there is a particular difficulty with performing it
for a model.

\section{Building and running an integration task}
\label{sec:build-and-run}

\subsection{Coupling a model to the runtime system}
\label{sec:couple-to-runtime}
Once customized core and header files
have been produced, they can be used to perform
calculations. This involves
connecting the translated
files to other components of
{\CppTransport},
especially those that are needed to carry out integration tasks.
To do so we create a short {\CC} program;
for the double-quadratic example this could be called
\file{dquad.cpp}.
A simple implementation takes the form:
\begin{minted}{c++}
// include implementation header generated by translator
#include "dquad_mpi.h"

int main(int argc, char* argv[])
  {
    // set up a task_manager instance to control this process
    transport::task_manager<> mgr(argc, argv);

    // set up an instance of the double quadratic model
    std::shared_ptr< transport::dquad_mpi<> > model = mgr.create_model< transport::dquad_mpi<> >();

    // hand off control to the task manager
    mgr.process();

    return(EXIT_SUCCESS);
  }
\end{minted}
This code involves the following steps:
\begin{enumerate}
    \item First, the implementation header file produced in~\S\ref{sec:build-translator}
    is included using \mintinline{c++}{#include "dquad_mpi.h"}.
    Nothing else is needed to
    use the {\CppTransport} runtime system;
    any necessary library files are automatically included by the implementation header.

    \item The only function provided is \mintinline{c++}{main()}.
    It has three reponsibilities:
    \begin{enumerate}
        \item \semibold{Create a \emph{task manager} instance.}
        The task manager is a
        class provided by the runtime system. It is responsible for coordinating
        what happens during execution.
        For example: if we are running a parallel computation under {\MPI},
        each copy of the executable may be either the master process
        or a worker. It is the responsibility of the task manager
        to decide which is correct and behave appropriately.

        If it is the master process, the task
        manager builds a list of work using options specified
        in the configuration file or on the command line.
        It scatters these tasks to the workers and coordinates their activity.
        On the other hand, if it is a worker, the task
        manager waits for tasks to be issued by the master
        process and arranges for them to be carried out.

        The task manager class is called
        \mintinline{c++}{task_manager<>}.
        It shares a common feature with most other {\CppTransport}
        components:
        it lives inside the namespace
        \mintinline{c++}{transport}.
        This prevents any conflict between symbols
        defined in user code and those used internally by
        {\CppTransport}.
        As for objects defined in any namespace,
        each {\CppTransport} component should be prefixed
        by the namespace name and two colons, as in
        \mintinline{c++}{transport::task_manager}.%
            \footnote{It is possible pull all symbols defined
            within the \mintinline{c++}{transport}
            namespace
            into the global
            namespace with the \mintinline{c++}{using} directive,
            as in for example
            \mintinline{c++}{using namespace transport;}.
            However, this practice is not recommended because it risks
            conflicts between user-space symbols
            and those belonging to {\CppTransport} itself.}
        The meaning of the brackets
        \mintinline{c++}{<>}
        is explained in the
        \emph{Advanced usage}
        panel on p.\pageref{advanced:data-type}.

        The \mintinline{c++}{task_manager<>} constructor requires the
        arguments \mintinline{c++}{argc} and \mintinline{c++}{argv}
        provided to \mintinline{c++}{main()}.
        It will process these internally.
        The options understood by the task manager are described
        in XXX.

        \item \semibold{Create an instance of the implementation class.}
        Second, we need an instance of the implementation class
        generated by the translator.
        As explained in~\S\ref{sec:template-block},
        this class will be
        called \mintinline{c++}{tag_mpi}
        if we use the \file{canonical\_mpi} template.
        For us this will be \mintinline{c++}{dquad_mpi}.
        Like \mintinline{c++}{task_manager<>} its name should be followed
        by angle brackets \mintinline{c++}{<>}.

        To create the instance we use the \mintinline{c++}{task_manager<>} method
        \mintinline{c++}{create_model()}.
        This is a templated method that requires the name of the
        instance class to be provided between angle brackets;
        here, this is
        \mintinline{c++}{< transport::dquad_mpi<> >}.
        The method itself takes no arguments.
        It returns a
        \href{http://en.cppreference.com/w/cpp/memory/shared_ptr}{shared
        \emph{smart pointer}}
        to the implementation class instance.
        The use of
        \mintinline{c++}{create_model()} is necessary, rather than
        constructing an instance directly, in order that the {\CppTransport}
        runtime system is aware of the model and can find it when needed
        for computations.

        Notice that there is no need to explicitly deallocate the pointer
        \mintinline{c++}{model}.
        It is deallocated automatically when the
        smart pointer that manages it is destroyed.

        \item \semibold{Pass control to the task manager.}
        It is possible to create instances of as many implementation classes
        as are required.
        Each one should be constructed using
        \mintinline{c++}{create_model()}.
        When all implementation classes have been
        instantiated, control should be passed to the task
        manager via its
        \mintinline{c++}{process()} method.

        When running as the master,
        \mintinline{c++}{process()} will distribute tasks to the workers.
        When running as a worker it will await instructions from
        the master.
    \end{enumerate}

    \item Finally, the \mintinline{c++}{process()} method returns when all work
    has been exhausted. At this point the process should terminate,
    so we return \mintinline{c++}{EXIT_SUCCESS}.
\end{enumerate}
Most {\CppTransport} executables will contain a
\mintinline{c++}{main()} function of almost exactly this form.
The general case differs only by providing extra functions
to generate tasks and derived products,
which will be explained in \S\ref{sec:add-integration-task}
and \S\ref{sec:derived-products}
below.

\begin{advanced}{Custom integration data types}
    \label{advanced:data-type}
    Like most {\CppTransport} components,
    \mintinline{c++}{task_manager<>}
    is a \semibold{template class}. This is indicated by the
    angle brackets
    \mintinline{c++}{<>} following the object name.
    A \emph{template} is a class or other object that can be customized
    by providing a list of data types
    such as \mintinline{c++}{double} (or other parameters)
    between the brackets.
    For {\CppTransport}, the customization takes place in the integration engine,
    that is capable of integrating the transport equations using any
    suitable data type.
    If no type name is given, as in the example above,
    the integrator will default to
    \mintinline{c++}{double}.

    For almost all users the default choice is suitable,
    so there is no need to specify a type explicitly.
    As an alternative, however, is it possible use the single-precision type
    \mintinline{c++}{float}
    if the intention is to trade off some accuracy against speed.%
        \footnote{Although it is often true that
        \mintinline{c++}{float}s are half as long as
        \mintinline{c++}{double}s, this is a platform dependent statement.
        On some platforms there may be no difference between
        \mintinline{c++}{float} and \mintinline{c++}{double}.}
    For greater precision it is possible to use
    \mintinline{c++}{long double},
    or even a customized type
    from a library such as the GNU Multiple Precision Arithmetic Library
    \href{https://gmplib.org}{GMP}
    or the Class Library for Numbers \href{http://www.ginac.de/CLN}{CLN}.%
        \footnote{There are some caveats. Although {\CppTransport} can integrate
        using any required type, for storage it relies on the
        {\SQLite} engine which expects real numbers to be stored as
        64-bit IEEE floating-point numbers.
        On most platforms this corresponds to a \mintinline{c++}{double},
        and in fact the {\SQLite} API
        is written under this assumption.
        (For details, see
        \href{https://www.sqlite.org/c3ref/bind_blob.html}{the {\SQLite}
        documentation}.
        There is only a
        \mintinline{c++}{sqlite3_bind_double()}, but no
        comparable method for the other floating-point types.)
        This means that, no matter what precision was used during integration,
        {\CppTransport} always stores the results
        as \mintinline{c++}{double} precision.
        Using higher precision for intermediate values may still be useful,
        however, to reduce roundoff error in the integrator.}
    Using types with higher precision than \mintinline{c++}{double} will
    increase the computation time
    (eg. switching to \mintinline{c++}{long double}
    very roughly doubles the time required),
    whereas
    \mintinline{c++}{float} may require less stringent tolerances
    to prevent to integrator's stepsize becoming very small.
    Attempting to use types from GMP or CLN will likely require some
    template specializations to be provided. If so, this will manifest itself
    as missing symbols reported during the link step.

    It is possible to use the same core and implementation classes with
    different data types, just by changing the type name provided
    in the template specialization brackets
    \mintinline{c++}{<...>}.
    However, the current version of {\CppTransport} does not support mixing different
    types within the \emph{same} executable because the task manager
    needs to know which data type is in use, and therefore also requires
    a template specialization such as
    \mintinline{c++}{<double>}.
\end{advanced}

\subsection{Translate and build using a {\CMake} script}
It is possible to build {\CppTransport} executables by manually invoking the
compiler.
However, this
is not always convenient because it is necessary to locate the
{\Boost} and {\MPI}
libraries on which the runtime system depends.
The recommended way to build is using a {\CMake} build script.
When {\CppTransport} is installed, it provides some
{\CMake} tools that are intended to simplify this process.

The {\CMake} build script should be called \file{CMakeLists.txt} and
placed in the same directory as the
main {\CC} file---for the example of double-quadratic
inflation this is the file \file{dquad.mpi}
described above. A suitable script is:
\begin{minted}{cmake}
    CMAKE_MINIMUM_REQUIRED(VERSION 3.0)
    PROJECT(dquad)

    SET(CMAKE_MODULE_PATH ${CMAKE_MODULE_PATH} "~/.cpptransport-packages/share/cmake/")

    SET(CMAKE_CXX_FLAGS_RELEASE "-Ofast -DNDEBUG")
    SET(CMAKE_C_FLAGS_RELEASE "-Ofast -DNDEBUG")

    FIND_PACKAGE(CppTransport REQUIRED)

    INCLUDE_DIRECTORIES(${CPPTRANSPORT_INCLUDE_DIRS} ${CMAKE_CURRENT_BINARY_DIR})

    ADD_CUSTOM_COMMAND(
      OUTPUT ${CMAKE_CURRENT_BINARY_DIR}/dquad_core.h ${CMAKE_CURRENT_BINARY_DIR}/dquad_mpi.h
      COMMAND CppTransport --verbose --fast ${CMAKE_CURRENT_SOURCE_DIR}/dquad.model
      DEPENDS dquad.model
    )

    SET(HEADERS ${CMAKE_CURRENT_BINARY_DIR}/dquad_core.h ${CMAKE_CURRENT_BINARY_DIR}/dquad_mpi.h)
    ADD_CUSTOM_TARGET(Generator DEPENDS ${HEADERS})

    ADD_EXECUTABLE(dquad dquad.cpp)
    ADD_DEPENDENCIES(dquad Generator)
    TARGET_LINK_LIBRARIES(dquad ${CPPTRANSPORT_LIBRARIES})
    TARGET_COMPILE_OPTIONS(dquad PRIVATE -std=c++14 -mavx)
\end{minted}
The steps involved are:
\begin{enumerate}
    \item The lines
    \begin{minted}{cmake}
        CMAKE_MINIMUM_REQUIRED(VERSION 3.0)
        PROJECT(dquad)
    \end{minted}
    are required by {\CMake}.
    They specify the minimum version of the {\CMake} tool that is required
    (here version 3.0) and the name of the project being built.

    \item The line
    \begin{minted}{cmake}
        SET(CMAKE_MODULE_PATH ${CMAKE_MODULE_PATH} "~/.cpptransport-packages/share/cmake/")
    \end{minted}
    should be adjusted to point to the \file{share/cmake} directory
    installed under your installation prefix
    (see Fig.~\ref{fig:directory-structure}).
    This will allow {\CMake} to locate the build tools installed by
    {\CppTransport}.

    \item The lines
    \begin{minted}{cmake}
        SET(CMAKE_CXX_FLAGS_RELEASE "-Ofast -DNDEBUG")
        SET(CMAKE_C_FLAGS_RELEASE "-Ofast -DNDEBUG")
    \end{minted}
    set the compiler flags to be used when building in `Release' mode.
    Generally it is desirable to optimize {\CppTransport} to \emph{at least}
    \option{-O2} or similar, because the
    templated {\Boost}.{\odeint} steppers require optimization to produce
    acceptable results.
    Also, a high optimization setting will encourage the compiler
    to optimize the automatically-generated {\CC} produced by the translator.
    Clang and the Intel compiler produce good results using their
    \option{-Ofast} setting,
    and gcc produces good results using \option{-O3}.
    (For the Intel compiler, \option{-fast} is also a possibility.)
    The switch \texttt{-DNDEBUG}
    \href{http://en.cppreference.com/w/cpp/error/assert}{disables debugging code}
    associated with the
    \mintinline{c++}{assert()} macro.

    \item
    The next step is to detect the libraries and include files needed by
    {\CppTransport}. This is managed in a single line:
    \begin{minted}{cmake}
        FIND_PACKAGE(CppTransport REQUIRED)
    \end{minted}
    It was for this command to function correctly that we needed to
    adjust \cmakevar{CMAKE\_MODULE\_PATH} above.
    Specifically, this will detect the {\SQLite}, {\Boost} and {\MPI}
    libraries required by {\CppTransport}. It will also detect the
    libraries installed by {\CppTransport} itself.

    To make the header files required by these libraries available we use the
    line
    \begin{minted}{cmake}
        INCLUDE_DIRECTORIES(${CPPTRANSPORT_INCLUDE_DIRS} ${CMAKE_CURRENT_BINARY_DIR})
    \end{minted}
    The variable \cmakevar{CPPTRANSPORT\_INCLUDE\_DIRS}
    contains the \file{include} paths
    required by {\CppTransport} and its dependencies.
    The variable \cmakevar{CMAKE\_CURRENT\_BINARY\_DIR}
    adds the {\CMake} build directory to the include path,
    which is done to make the translated core and implementation header
    files available (see below).

    \item Next we must instruct {\CMake} to build the final executable.
    This is done in two stages:
    first, we arrange for the model description file
    \file{dquad.model} to be translated to the core and implementation headers
    \file{dquad\_core.h} and \file{dquad\_mpi.h};
    and second, we instruct the compiler to process the main
    file \file{dquad.cpp}
    with all the previously-determined include paths and library locations.

    \begin{enumerate}
        \item {\CMake} is instructed to invoke the {\CppTransport} translator
        using an \mintinline{cmake}{ADD_CUSTOM_COMMAND()} block,
        \begin{minted}{cmake}
            ADD_CUSTOM_COMMAND(
              OUTPUT
                ${CMAKE_CURRENT_BINARY_DIR}/dquad_core.h ${CMAKE_CURRENT_BINARY_DIR}/dquad_mpi.h
              COMMAND CppTransport --verbose --fast ${CMAKE_CURRENT_SOURCE_DIR}/dquad.model
              DEPENDS dquad.model
            )
        \end{minted}
        The
        \cmakevar{OUTPUT}
        line
        advises {\CMake} that this block gives a recipe for constructing
        the files \file{dquad\_core.h} and \file{dquad\_mpi.h}.
        The
        \cmakevar{DEPENDS}
        line advertises that this recipe
        depends on the file \file{dquad.model},
        and therefore should be re-run if it is changed.
        Finally, the
        \cmakevar{COMMAND} line gives the command
        to execute;
        it invokes
        the {\CppTransport} translator with the options
        \option{{-}{-}verbose} and \option{{-}{-}fast}.
        The {\CMake} variables
        \cmakevar{CMAKE\_CURRENT\_SOURCE\_DIR} and
        \cmakevar{CMAKE\_CURRENT\_BINARY\_DIR} refer
        to the source and build directories managed by {\CMake}.

        \item At this stage {\CMake}
        knows \emph{how} to generate the core and implementation
        header files, but it does not know that it \emph{should} do so.
        To instruct it that these files are required, we add a \emph{target}
        (a deliverable set of objects that {\CMake} can build):
        \begin{minted}{cmake}
            SET(HEADERS
              ${CMAKE_CURRENT_BINARY_DIR}/dquad_core.h ${CMAKE_CURRENT_BINARY_DIR}/dquad_mpi.h
            )
            ADD_CUSTOM_TARGET(Generator DEPENDS ${HEADERS})
        \end{minted}
        This tells {\CMake} that a target called
        \cmakevar{Generator} depends on the core and implementation header files.
        If we try to build this target, {\CMake} will invoke the recipe
        above in order to generate these files.

        \item Finally, we set up a second target
        \cmakevar{dquad} that consists of the
        finished executable and make this depend on the
        \cmakevar{Generator} target declared above.
        {\CMake} then knows that the files associated with \cmakevar{Generator}
        must be built before \cmakevar{dquad}.
        \begin{minted}{cmake}
            ADD_EXECUTABLE(dquad dquad.cpp)
            ADD_DEPENDENCIES(dquad Generator)
            TARGET_LINK_LIBRARIES(dquad ${CPPTRANSPORT_LIBRARIES})
            TARGET_COMPILE_OPTIONS(dquad PRIVATE -std=c++14 -mavx)
        \end{minted}
        The \mintinline{cmake}{TARGET_COMPILE_OPTIONS()} command adds
        extra compiler flags to the
        \cmakevar{dquad} target.
        The flag \cmakevar{-std=c++14} is required, because it
        enables certain {\CC}14 features that are used by the {\CppTransport}
        platform.
        Other code generation or optimization options can be specified here;
        an example is the switch \option{-mavx} which informs the compiler
        that it is allowed to generate code using the AVX instruction set
        extensions available on Intel since Sandy Bridge
        and on AMD since late 2011/early 2012.
        Where these instructions are available they
        can give a useful performance boost.

        Depending on your processor, even more recent instruction set extensions
        may be available such as AVX-2.
        These extensions have been available on Intel since Haswell,
        and
        on AMD they are currently implemented for the Carrizo platform.

        If using the Intel compiler to target Intel processors,
        the switch \option{-xHost} may be used
        to indicate that code generation should use all features of the
        machine being used to build.
        It is implied by the \option{-fast} optimization, which is more aggressive
        than \option{-Ofast}.
        Note, however, that \option{-xHost} should be used with caution if you
        plan to run executables in a heterogeneous cluster environment
        because different machines may support different instruction set enhancements.
        If the executable requires instructions that are not available on
        the host machine it will terminate with an error message.
    \end{enumerate}
\end{enumerate}
The {\CMake} script can be adapted for any {\CppTransport} executable.

\para{Build using {\CMake}}
The build process is the same as for {\CppTransport} itself.
First, starting from the directory containing the
\file{CMakeLists.txt} script,
create a build directory and move into it:
\begin{minted}{bash}
	mkdir build
	cd build
\end{minted}
Next, configure {\CMake} to build using the `Release' configuration.
\begin{minted}{bash}
	cmake .. -DCMAKE_BUILD_TYPE=Release
\end{minted}
Because it is typically unnecessary to install individual executables
the \cmakevar{CMAKE\_INSTALL\_PREFIX} option can be omitted.
However, if you wish to later install your executables to a standard location
such as \file{\textasciitilde{}/bin}
then you can specify a suitable prefix here.
Also,
if you wish to build with a compiler other than the default
then you should specify
\cmakevar{CMAKE\_C\_COMPILER}
and
\cmakevar{CMAKE\_CXX\_COMPILER}
as in~\S\ref{sec:build-translator}.

When configuration is complete, the build is initiated
by issuing the \mintinline{bash}{make} command.
If you then wish to install
to a different location, use \mintinline{bash}{make install}.
Once the executable has built you may wish to verify that it
function correctly by trying the following invocations:
\begin{minted}{bash}
	./dquad --version
	./dquad --models
	./dquad --help
\end{minted}

\subsection{Adding an integration task}
\label{sec:add-integration-task}
To make the executable \file{dquad} useful we must add tasks to generate
$n$-point functions, and also tasks to convert these raw $n$-point
functions into observables.
This is done by using the task manager's
\mintinline{c++}{add_generator()}
method to inform it that the executable includes specifications
for some number of
tasks.
The \mintinline{c++}{add_generator()}
method takes one argument, which should be a callable
object accepting a reference to
a \mintinline{c++}{transport::repository<>}
object as its single argument.
{\CppTransport} stores all information about
initial conditions, parameter choices, tasks,
derived products and any generated content
in disk-based databases called \emph{repositories}.
The \mintinline{c++}{repository<>} class manages
these databases and offers related services
to other
{\CppTransport} components.

The first step is always to build some number of integration tasks,
because all other tasks depend on the $n$-point functions that they compute.
In this section we illustrate the steps required to build, store and
execute a collection of integration tasks.

If you are not familiar with constructing callable objects
then a simple option is to use the {\CC}11 \emph{lambda} feature.
This is a shorthand way to notate functions.
First, declare a function
\mintinline{c++}{write_tasks()} that accepts
two arguments:
a \mintinline{c++}{repository<>} and a model pointer:
\begin{minted}{c++}
    void write_tasks(transport::repository<>& repo, transport::dquad_mpi<>* m);
\end{minted}
This function should be registered using the
\mintinline{c++}{add_generator()} method,
by inserting the lines
\begin{minted}{c++}
    // register task writer
    mgr.add_generator([=](transport::repository<>& repo) -> void { write_tasks(repo, model.get()); });
\end{minted}
immediately prior to the call to
\mintinline{c++}{mgr.process()}.
\begin{advanced}{Callable objects}
    The \mintinline{c++}{add_generator()} method accepts any callable,
    such as a \mintinline{c++}{std::function<>} object.
    It is not necessary to use lambdas
    if a different solution is preferable.
    For example, it is also possible supply an instance of any
    class that provides a call operator
    \mintinline{c++}{operator()}.
\end{advanced}

In this implementation, the argument of \mintinline{c++}{add_generator()}
is the function
\begin{center}
    \mintinline{c++}{[=](transport::repository<>& repo) -> void { write_tasks(repo, model.get()); }}
\end{center}
This is a lambda expression. It represents an object that behaves as a callable function,
taking a single \mintinline{c++}{repository<>} as an argument.
The function body is the code enclosed by braces
\mintinline{c++}{{ ... }}.
It calls the function \mintinline{c++}{write_tasks()},
passing on the \mintinline{c++}{repository<>} object
given as its own argument
and using the raw model pointer obtained from
\mintinline{c++}{model.get()}.
For the meaning of the prefix
\mintinline{c++}{[=]},
\href{http://en.cppreference.com/w/cpp/language/lambda}{see here}.

\para{Building a task}
The final step is to provide a definition
for \mintinline{c++}{write_tasks()}.
This should construct the integration tasks we want, and store them
in the \mintinline{c++}{repository<>} object it is passed.

Integration tasks package together all the information needed
to perform a computation of the 2- or 3-point functions.
This includes:
\begin{itemize}
	\item details of the model to be used, identified
	through the pointer to the model instance
	passed to \mintinline{c++}{write_tasks()}

	\item a choice for any parameters
	used in the Lagrangian,
	and a value for the Planck mass $\Mp$

	\item a choice for the initial values of the background
	fields (and optionally their derivatives)

	\item fixed start and end times for the integration, and a mesh
	of sample points between these times where samples will be recorded

	\item a mesh of wavenumber configurations
	(values of the wavenumber $k$ for the 2-point function,
	and configurations $\{ \vect{k}_1, \vect{k}_2, \vect{k}_3 \}$ for the
	3-point function)
	where the 2- and 3-point functions should be sampled
\end{itemize}
In addition, {\CppTransport} provides various options for customizing an
integration---for example, by changing the way initial conditions are handled.
These options will be described in~\S\ref{sec:int-options}
and are recorded as part of the integration task.

\para{Specifying parameters}
We work with the double-quadratic model as an example.
{\CppTransport} works in units where $c=\hbar=1$ but allows us
to measure the Planck scale $\Mp$ using whatever units we find
convenient.
Typically, however, `natural' units with $\Mp=1$ give good results
and in what follows we will make this choice.

The double-quadratic potential~\eqref{eq:double-quadratic-V}
requires us to specify the mass scales $M_\phi$ and $M_\chi$.
We will choose $M_\phi = 9 \times 10^{-5} \Mp$
and $M_\chi = 10^{-5} \Mp$.
A \emph{parameter package} consists of a model,
a choice for the Planck mass,
and choices for each of the parameters in the Lagrangian.
{\CppTransport} collects this information using a
\mintinline{c++}{transport::parameters<>} object.
Its constructor takes three arguments:
the value of the Planck mass;
a list of values for the model parameters
\semibold{in the same order they were declared in the model
description file};
and a pointer to the model instance.
With our choices we can construct a suitable parameter package using:
\begin{minted}{c++}
	void write_task::operator()(transport::repository<>& repo)
	  {
	    const double Mp   = 1.0;
	    const double Mphi = 9E-5 * Mp;
	    const double Mchi = 1E-5 * Mp;

	    transport::parameters<> params(Mp, {Mphi, Mchi}, model);
	  }
\end{minted}
To aid readability it can be helpful to use named temporary variables
that give meaning to numbers that are quoted directly.
We could have achieved the same effect by writing
the single-line construction
\begin{minted}{c++}
	transport::parameters<> params(1.0, {9E-5, 1E-5}, model);
\end{minted}
but it would then be more difficult to identify the meaning of the
numbers.

If it is more convenient,
the parameter list can be specified using any suitable iterable
container, such as
\mintinline{c++}{std::vector<double>}
or
\mintinline{c++}{std::list<double>}
rather than quoting it directly as the initialization list
\mintinline{c++}{{Mphi, Mchi}}.
If an incorrect number of parameters are passed
then {\CppTransport} will throw
a \mintinline{c++}{std::out_of_range} exception.

\para{Specifying initial conditions}
Next we combine the parameter package with a choice of initial conditions
to make an
\emph{initial conditions package}.
This information is stored in
a \mintinline{c++}{transport::initial_conditions<>}
object.
Its constructor accepts
three mandatory arguments:
a textual name, which will be used later to refer to this initial conditions package;
a parameter package, which specifies the model and parameters to be used;
and a list of initial values for the fields.
In an $N$-field model this list can contain either exactly $N$ or exactly $2N$ values:
\begin{itemize}
	\item if $N$ values are given, {\CppTransport} will interpret these as the initial
	conditions for the background fields
	\semibold{in the order they were declared in the model description}.
	It will infer initial conditions for the field derivatives using the slow-roll
	equation $3H \dot{\phi}^\alpha = \partial_\alpha V$,
	where $\partial_\alpha$ denotes the field derivative
	$\partial / \partial \phi^\alpha$.

	\item if $2N$ values are given, these are interpreted as $N$ initial
	conditions for the background fields $\phi^\alpha$
	(in the same order they were declared) followed by $N$ initial conditions
	for their derivatives $\d \phi^\alpha / \d N$ (in the same order as the fields).
	Here, $\d N = H \, \d t$ is a derivative with respect to e-folding number.
\end{itemize}
As for \mintinline{c++}{parameters},
the value list can be specified using any suitable container or by
quoting it directly as an initialization list.
If a number of values other than $N$ or $2N$ is given then
{\CppTransport} will raise
a \mintinline{c++}{std::out_of_range} exception.

In addition, the initial conditions package
should include
information about the time during inflation when these
initial field values are intended to apply.
This information can be specified in two ways:
\begin{itemize}
	\item as an initial time $\Ninit$ (specified in e-folds)
	together with the number of e-folds $\Npre$ from
	$\Ninit$ to the horizon-crossing time of a distinguished
	scale $\kstar$ (at time $\Nstar$) that is used as a reference.

	\item as an initial time $\Nzero$
	together with the horizon-crossing $\Nstar$
	associated with $\kstar$,
	and the desired number of e-folds $\Npre$
	from $\Ninit$ to $\Nstar$.
	This amounts to moving a set of initial conditions specified
	at $\Nzero$ to new initial conditions specified at
	$\Nstar - \Npre$.

	This version can be used to `settle' a set of field-only
	initial conditions onto the true dynamical attractor.
	If the slow-roll approximation holds to reasonable accuracy
	near the initial time then {\CppTransport}'s estimate
	of the field derivatives will normally be quite accurate.
	Nevertheless, there will be a period of adjustment while the
	numerical solution relaxes.
	This can lead to slight jitter if any $n$-point functions
	have initial conditions during this phase.

	If adaptive initial conditions are in use
	(this is normally the recommended configuration; see
	the discussion on p.\pageref{enum:adpative-ics}
	in~\S\ref{sec:int-options})
	then a customized initial condition will be computed for
	each $n$-point function.
	Provided $\Ninit$ is sufficiently early, this customization
	will automatically allow the initial conditions to relax onto
	the dynamical attractor.
	Manual settling is normally required only if
	$\Ninit$ if very close to the initial time for
	any $n$-point function, or if adaptive initial conditions
	are not being used.
\end{itemize}
For the purposes of illustration we will
set initial conditions for the double quadratic model
at $\phi = 10 \Mp$ and $\chi = 12.9 \Mp$
and allow {\CppTransport} to infer
values for the field derivatives.
We take the initial time to be $N=0$
(this is just a convention; any other value of $N$ could be used)
and set $\Nstar$ to occur at $N=12$.
To build an initial conditions package corresponding to these choices
we can use:
\begin{minted}{c++}
    const double phi_init = 10.0 * Mp;
    const double chi_init = 12.9 * Mp;

    const double N_init   = 0.0;
    const double N_pre    = 12.0;

    transport::initial_conditions<> ics("dquad", params, {phi_init, chi_init}, N_init, N_pre);
\end{minted}
Remember that when specified in this form, $\Nstar = \Ninit + \Npre$.
If we had used the second form, perhaps to arrange for some manual settling,
the last two parameters
\mintinline{c++}{N_init} and
\mintinline{c++}{N_pre}
would have been replaced by the \emph{three} parameters
\mintinline{c++}{N_0},
\mintinline{c++}{N_star}
and
\mintinline{c++}{N_pre},
corresponding to $\Nzero$, $\Nstar$ and $\Npre$.

\para{Selecting a mesh of time sample points}
The remaining task is to set up a series of sample points, both for
time and wavenumber configuration.
To assist in doing so, {\CppTransport} provides a mechanism
to construct arbitrary meshes that are unions
of ranges built using
linear or logarithmic spacing.
The building blocks of these meshes
are objects of type
\mintinline{c++}{transport::basic_range<>}. The constructor for
this object has the form
\begin{center}
    \mintinline{c++}{transport::basic_range<>(lo, hi, N, spacing);}
\end{center}
It constructs a range of \mintinline{c++}{N}+1 sample points between
\mintinline{c++}{lo}
and
\mintinline{c++}{hi}
(inclusive)
that divide the interval
[\mintinline{c++}{lo}, \mintinline{c++}{hi}] into $N$ parts.
The parameter \mintinline{c++}{spacing} should be one of:
\begin{itemize}
    \item \mintinline{c++}{transport::spacing::linear}: the sample points
    are spaced linearly

    \item \mintinline{c++}{transport::spacing::log_bottom}: the sample
    points are logarithmically spaced from the bottom of the interval

    \item \mintinline{c++}{transport::spacing::log_top}: the sample points are
    logarithmically spaced from the top of the interval
\end{itemize}
If $N=0$ the range consists of a single
value equal to \mintinline{c++}{lo}.

Any number of \mintinline{c++}{basic_range<>} ranges can be
composed to produce a composite range.
This produces an object of type
\mintinline{c++}{aggregate_range<>}.
If \mintinline{c++}{A}, \mintinline{c++}{B}, \mintinline{c++}{C}
are ranges (which may themselves be composite) then
the following are equivalent:
\begin{minted}{c++}
    transport::aggregate_range<> M = A + B + C;

    auto M = A + B + C;

    transport::aggregate_range<> M(A, B);
    M += C;

    transport::aggregate_range<> M(A);
    M.add_subrange(B);
    M.add_subrange(C);
\end{minted}
It is also possible to construct an empty
\mintinline{c++}{aggregate_range<>} by passing no arguments to its
constructor.
Often it assists readability to use the \mintinline{c++}{auto} type specifier,
which informs the compiler that it should deduce an appropriate type
for the given assignment.

The ability to construct arbitrary meshes
makes it possible to sample certain regions densely and others
sparsely. For example, it is possible to sample densely
in regions (either of time or wavenumber configuration)
that exhibit sharp features
while sampling sparsely elsewhere to keep the overall data volume
manageable.

For time sampling, a sensible starting point is to sample linearly
in $N$
to get a sense of how the correlation functions evolve.
Later, the sample mesh can be refined if required.
A reasonable starting point
might be 300 evenly spaced intervals between the minimum
and maximum values of $N$,
\begin{minted}{c++}
    const double N_end  = 60.0;

    transport::basic_range<> ts(N_init, N_end, 300, transport::spacing::linear);
\end{minted}

\begin{advanced}{Arbitrary value types}
    The \mintinline{c++}{basic_range<>} and
    \mintinline{c++}{aggregate_range<>} objects are templated
    and (if needed)
    can be used to construct a range of values for any numeric
    type. However, even if you are using a type other than
    \mintinline{c++}{double} in the integration engine,
    {\CppTransport} always measures times and wavenumbers
    using \mintinline{c++}{double}.
\end{advanced}

\para{Selecting a mesh of wavenumber samples}
Building a mesh of wavenumber samples is similar.
For the two-point function, a wavenumber configuration is fixed
by the magnitude $k$.
A set of samples can therefore be specified by a range
(possibly a composite, as above).
The wavenumber $k=1$ is \emph{defined} to exit the horizon at time
$N = \Nstar$,
as determined by the initial conditions package.
{\CppTransport} refers to wavenumbers normalized in this way as
\emph{conventionally normalized}. When producing derived products
it is possible to measure wavenumbers using a number of different
normalizations, as will be explained in~\S\ref{sec:examine-k-database}.

If $H$ is nearly constant then a general wavenumber $k$ will exit the
horizon roughly when $N = \Nstar + \ln k$.
This is a good rule-of-thumb when attempting to build a range
of $k$ that covers a given range of e-folds.
(When constructing a mesh of $k$s
it is often useful to make use of the logarithmic spacing option
in \mintinline{c++}{basic_range<>}.)
However, {\CppTransport} does \emph{not}
assume that $H$ is constant; it
calculates the horizon-exit time of each wavenumber exactly.

To begin, we will construct a range of wavenumbers that
sample horizon exit times between approximately
$\Nstar + 3.0$
and
$\Nstar + 8.0$:
\begin{minted}{c++}
    const double kt_lo = std::exp(3.0);
    const double kt_hi = std::exp(8.0);

    transport::basic_range<> ks(kt_lo, kt_hi, 50, transport::spacing::log_bottom);
\end{minted}

\para{Building 2- and 3-point function integration tasks}
With all of these elements in place, we can proceed to build
integration tasks.
Currently, {\CppTransport} offers two options.
While the integration
engine can compute the 3-point function for any model,
this calculation is expensive.
If the 3-point function is not required
(perhaps only a power-spectrum analysis is contemplated)
then it is much faster to omit it.
A task that computes \emph{only} the two-point function
is represented by an object of type
\mintinline{c++}{transport::twopf_task<>}:
\begin{minted}{c++}
    transport::twopf_task<> tk2("dquad.twopf", ics, ts, ks);
    tk2.set_adaptive_ics_efolds(5.0);
    tk2.set_description("Compute time history of the 2-point function from k ~ e^3 to k ~ e^9");
\end{minted}
Its constructor requires a name, an initial conditions package,
a range representing the time sample points,
and a range representing the wavenumber samples.
Setting a description is optional, but provides a convenient way to document
choices including the time- and wavenumber-sampling strategy.
The meaning of the
\mintinline{c++}{set_adaptive_ics_efolds()}
method will be explained in~\S\ref{sec:general-integration-options} below
(see p.\pageref{enum:adpative-ics}).

Alternatively, if we wish to compute the 3-point function,
a suitable task can be built using
\begin{minted}{c++}
    transport::threepf_cubic_task<> tk3("dquad.threepf", ics, ts, ks);
    tk3.set_adaptive_ics_efolds(5.0);
    tk3.set_description("Compute time history of the 3-point function on a cubic lattice from k ~ e^3 to k ~ e^9")
\end{minted}
This will sample the three-point function on a cubic lattice
$(k_1, k_2, k_3)$ built from the Cartesian product
\mintinline{c++}{ks} $\times$
\mintinline{c++}{ks} $\times$
\mintinline{c++}{ks},
after filtering out configurations that do not correspond to
a physical triangle.
Note that
this is only one way to construct a 3-point function task.
There are other ways to specify the wavenumber configurations
to be sampled, including use of
Fergusson--Shellard $(k_t, \alpha, \beta)$ parameters.
It is also possible to adjust the default
policies that determine which configurations
are regarded as physical triangles and whether all
configurations produced by the Cartesian product should
be retained for integration.
These features (and others) are described in~\S\ref{sec:threepf-options}.

To commit these tasks to the repository we use the
\mintinline{c++}{commit()} method:
\begin{minted}{c++}
    repo.commit(tk2);
    repo.commit(tk3);
\end{minted}

\subsection{Running tasks}
It is now possible to build the executable, enabling us to experiment
with creating repositories and running integration tasks.
The source code, as described above, is available from the
website \url{http://transportmethod.com} as
\file{dquad\_A.cpp}.
If the {\CMake} build directory was previously
configured correctly then there should be no need
to reconfigure.
To build, it is sufficient to execute
\mintinline{bash}{make}.

\subsubsection{Running executables under MPI and creating a repository}
The \file{dquad} executable can be invoked like any compiled object, by
passing its name to the shell.
Doing so without other arguments
will result in {\CppTransport} printing the error message
`Nothing to do: no repository specified'.
\begin{figure}
    \begin{center}
        \includegraphics[scale=0.7]{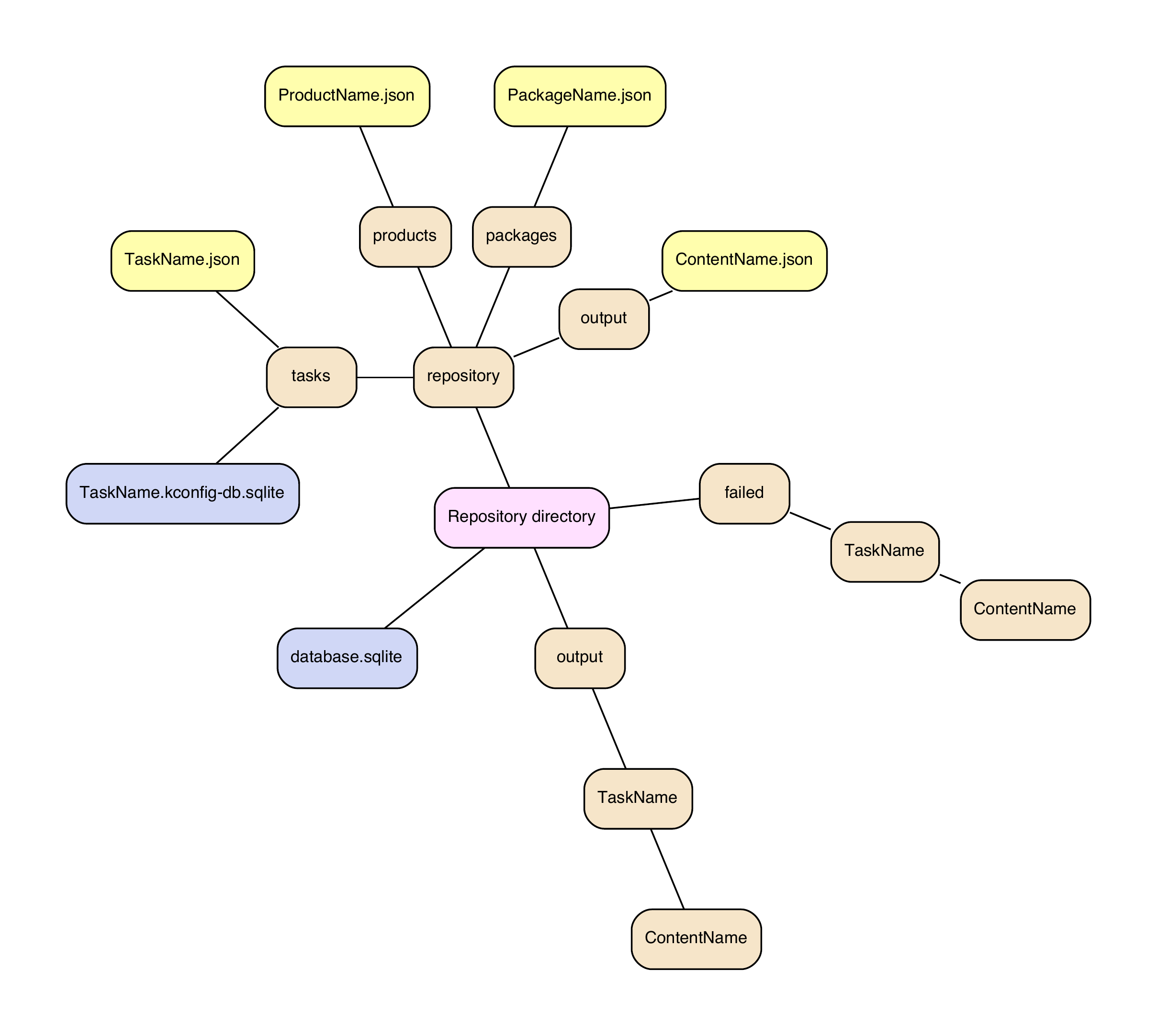}
    \end{center}
    \caption{\label{fig:repo-layout}Disk layout of a repository. The node
    labelled \emph{Repository directory} is the root directory whose name
    is passed to each {\CppTransport} executable.}
\end{figure}

\para{Repositories}
In order to carry out practical work it is necessary to specify a repository,
using the command line switch
\option{{-}{-}repo} or its abbreviation \option{-r}.
A repository
is a disk-based database managed by {\CppTransport},
distributed over a files in a predefined directory structure.
The specified repository
may already exist,
but if not a suitable directory layout will be created;
see Fig.~\ref{fig:repo-layout}.
Once created,
repositories are relocatable and can be moved to different filing-system
locations after creation, or even to a different machine.
The \emph{repository directory} is the name passed
as the argument of \option{{-}{-}repo}.
This directory contains up to four items:
\begin{itemize}
    \item \file{database.sqlite}. The is a SQLite database
    that contains summary information describing the relationship between
    all items in the database---%
    initial conditions packages, tasks, derived products and generated
    content.
    The database does not contain full information about these objects;
    this information is stored as
    \href{http://www.json.org}{JSON-format documents}
    elsewhere in the repository, in order that the information they contain
    is not hidden should it need to be recovered (or processed electronically)
    without {\CppTransport}.

    In addition, \file{database.sqlite} stores information about jobs that
    are currently running on the repository. This assists in automatically
    recovering data should there be a crash.

    \item \file{repository}. This is a directory containing the JSON documents
    that describe each repository object in detail.

    \item \file{output}. Output generated by tasks is placed in this folder.
    For a task named \emph{TaskName}, {\CppTransport} will generated a subdirectory
    also called \file{TaskName}.
    All content generated by \emph{TaskName}
    is placed in timestamped folders within this subdirectory.

    \item \file{failed}. If an error is encountered while generating output
    from a task, the log files and other content are placed within this folder
    for inspection.
    The organization is the same as for \file{output}.
\end{itemize}
The folders \file{output} and \file{failed}
are created only when needed. A freshly-created repository will contain only
\file{database.sqlite} and \file{repository}.

The \file{repository} folder contains further subfolders. As has been explained,
these house the JSON documents for each repository record.%
    \footnote{If necessary these can be edited by hand, although this practice
    is not recommended because it loses most of the advantages of a managed
    repository.}
\begin{itemize}
    \item \file{output}. Records describing each group of content generated
    by a task are stored in this folder
    in the format \file{ContentName.json}.

    \item \file{packages}. Contains records describing each initial conditions
    package. A package named \emph{PackageName} is stored as
    \file{PackageName.json}.

    \item \file{products}. Contains records describing each derived product.
    A product name \emph{ProductName} is stored as
    \file{ProductName.json}.

    \item \file{tasks}. Stores records describing each task
    \emph{TaskName} as \file{TaskName.json}.
    For integration tasks, this JSON document is accompanied by a
    SQLite database with filename
    \file{TaskName.kconfig-db.sqlite}
    storing the list of 2- and 3-point wavenumber configurations to be
    sampled, together with
    pre-computed information such as the corresponding time of horizon exit.
\end{itemize}
Repositories collect groups of related data products, ensuring that their
provenance is properly documented and that individual products do not become
orphaned. For example,
plots do not become separated from the datasets that were
used to produce them, and
computations of observables do not become separated from the raw $n$-point functions
and inflationary initial conditions on which they depend.

At the same time, repositories are intended to be a lightweight concept.
{\CppTransport} allows repositories to be created at will, and does not impose
limitations on their use.
At one extreme, it would be possible to write all integration tasks,
and all generated content, into the same repository.
This is probably not a good choice, partly because reporting on the
repository (see \S\ref{sec:HTML-analysis}) will take a long time as the repository
becomes large.
At the other extreme, every {\CppTransport} job could create a new
repository.
This strategy can work well in practice---%
especially if used in conjunction with the facility to attach notes
to repository records,
which
can be used to document
an evolving series of integrations.

\para{Launching {\CppTransport} using MPI}
If using a {\CppTransport} executable to create or
interrogate a repository, it can be launched as described above
in the same way as any other executable.
But to execute a task, {\CppTransport} expects to be run as a group
of related processes communicating within a managed {\MPI}
environment.
To carry out a task requires at least two processes,
but the task manager will make use of as many as are available.

To launch a {\CppTransport} executable under {\MPI},
use the following command:
\begin{minted}{bash}
    mpiexec -n 4 dquad --verbose --repo test-repo --create
\end{minted}
Replace the argument
\mintinline{bash}{-n 4} to \mintinline{bash}{mpiexec}
by the number of processes you wish to launch;
for running tasks it must be $\geq 2$.
It is seldom worth launching more processes than
there are physical cores to run them on,
except for some
products that support two
threads per core.
Intel calls this technology \emph{hyperthreading}
and it is available on certain i7 and Xeon
processors.
Such processors generally
identify themselves to the operating system with
two times their
physical core count.
Therefore, normally,
it is safe to use whatever number of
cores is reported by your machine.
In OS X, check Activity Monitor.
In Linux the Gnome System Monitor or equivalent performs
the same job.

In a cluster environment the argument
supplied to
\mintinline{bash}{-n} should match the number of
cores you request.
For example, an Open Grid Scheduler-like job submission
script requesting 36 cores
managed under {\OpenMPI}
might include the lines
\begin{minted}{bash}
#$ -pe openmpi 36
mpiexec -n 36 ...
\end{minted}
\begin{warning}
    If you are using {\CppTransport} on a network filing system
    such as NFS or Lustre (which is often the case when running on a cluster),
    you should add an extra command-line
    switch \option{{-}{-}network-mode}
    to the {\CppTransport} executable.

    In order to maximize performance, {\CppTransport}
    enables `write-ahead logging' mode in the underlying
    {\SQLite} database manager.
    This gives a significant performance improvement but is
    not compatible with network filing systems
    and must be disabled for reliable operation.
\end{warning}

\subsubsection{Examining the repository wavenumber configuration databases}
\label{sec:examine-k-database}
If the \mintinline{bash}{mpiexec} command given above succeeded,
{\CppTransport} will have created a repository called
\file{test-repo} in your current working directory.
The switch
\option{{-}{-}create} instructs it
to write any available tasks and derived products
into the repository by calling each object
registered with \mintinline{c++}{add_generator()}.
For the example of double-quadratic inflation this is the
function \mintinline{c++}{write_tasks()} described
in~\S\ref{sec:add-integration-task}.
The \option{{-}{-}create} option should be used only
once, the first time that task information needs to be
written to the repository.
If an attempt is made to commit a second task with the
same name as an existing task
then {\CppTransport} will report an error.

If verbose output is enabled using the switch
\option{{-}{-}verbose}
or
\option{-v}
then the constructors of
\mintinline{c++}{twopf_task<>}
and \mintinline{c++}{threepf_cubic_task<>}
will print brief summary information about the
tasks that have been constructed.
The constructor for \mintinline{c++}{tk2}
should print
\begin{minted}[bgcolor=blue!10]{text}
    dquad.twopf
    2pf configs: 51            Smallest k: 20.1
    Largest k: 2.98e+03        Earliest N_exit: N*+3.118
    Latest N_exit: N*+8.408    Inflation ends: N=67.78
\end{minted}
The information given is:
\begin{enumerate}
    \item the number of 2-point function configurations, here
    equal to 51 because the
    \mintinline{c++}{basic_range<>} object \mintinline{c++}{ts}
    was constructed with $N=50$ and therefore contains $N+1 = 51$
    points

    \item the smallest conventionally-normalized wavenumber sampled by
    the task, here
    $k = 20.1 \approx \e{3}$.

    \item the largest conventionally-normalized wavenumber sampled by
    the task, here
    $k = 2.9 \times 10^3 \approx \e{8}$.

    \item the earliest horizon exit time, relative to the distinguished
    time $\Nstar$. This will correspond to the smallest wavenumber,
    which is the largest physical scale.
    Here, that horizon exit time is $\approx 3.118$ e-folds after
    $\Nstar$.
    This is approximately what we expect from a mode with
    $k = \e{3}$, but shows already that
    the estimate $\Nexit \approx \Nstar + \ln k$
    is accurate only to a few percent.

    \item the latest horizon exit time, corresponding to the largest
    wavenumber or smallest physical scale.
    In this case, the error in the na\"{\i}ve estimate
    $\Nexit \approx \Nstar + \ln k$ has grown to $\sim 5\%$.
    Generally these estimates will become worse as the horizon
    exit time becomes farther from $\Nstar$.

    \item the time when inflation ends, if {\CppTransport}
    could detect it. By default, {\CppTransport} will search for
    1000 e-folds from the initial time
    and attempt to find the point where $\epsilon \equiv - \dot{H}/H^2 = 1$.
    If it does not find such a point within the 1000 e-fold search window
    then it will issue a warning, but this does not prevent
    successful calculation of the $n$-point functions.
\end{enumerate}
The constructor for \mintinline{c++}{tk3}
prints similar information, with the addition of the
number of bispectrum configurations that will be sampled:
\begin{minted}[bgcolor=blue!10]{text}
    dquad.threepf
    2pf configs: 51            3pf configs: 4017
    Smallest k: 20.1           Largest k: 2.98e+03
    Earliest N_exit: N*+3.118  Latest N_exit: N*+8.408
    Inflation ends: N=67.78
\end{minted}

For tasks that sample the 3-point function it is frequently
useful to inspect the list of wavenumber configurations that will
be computed.
There are two ways to do this.
The simplest, suitable for tasks that do not sample
too many configurations, is to run an HTML report
on the repository.
This writes details of the tasks into an easily-browsable
HTML document.
The second option is slightly less simple but
better suited to tasks that sample a large number of configurations.
The list of sample points is written into SQLite databases
held in the \file{repository/tasks} directory,
and
these can be inspected directly using a suitable tool.

\para{Option 1: examine configurations using an HTML report}
This option is generally preferred if the task samples
fewer than 5000 configurations.
The HTML report generator does not
include a wavenumber listing for
tasks with more than 5000 wavenumber configurations
because it makes the report too large: HTML documents
are not the best
way to examine such a large database.

\begin{figure}
    \begin{center}
        \includegraphics[scale=0.2]{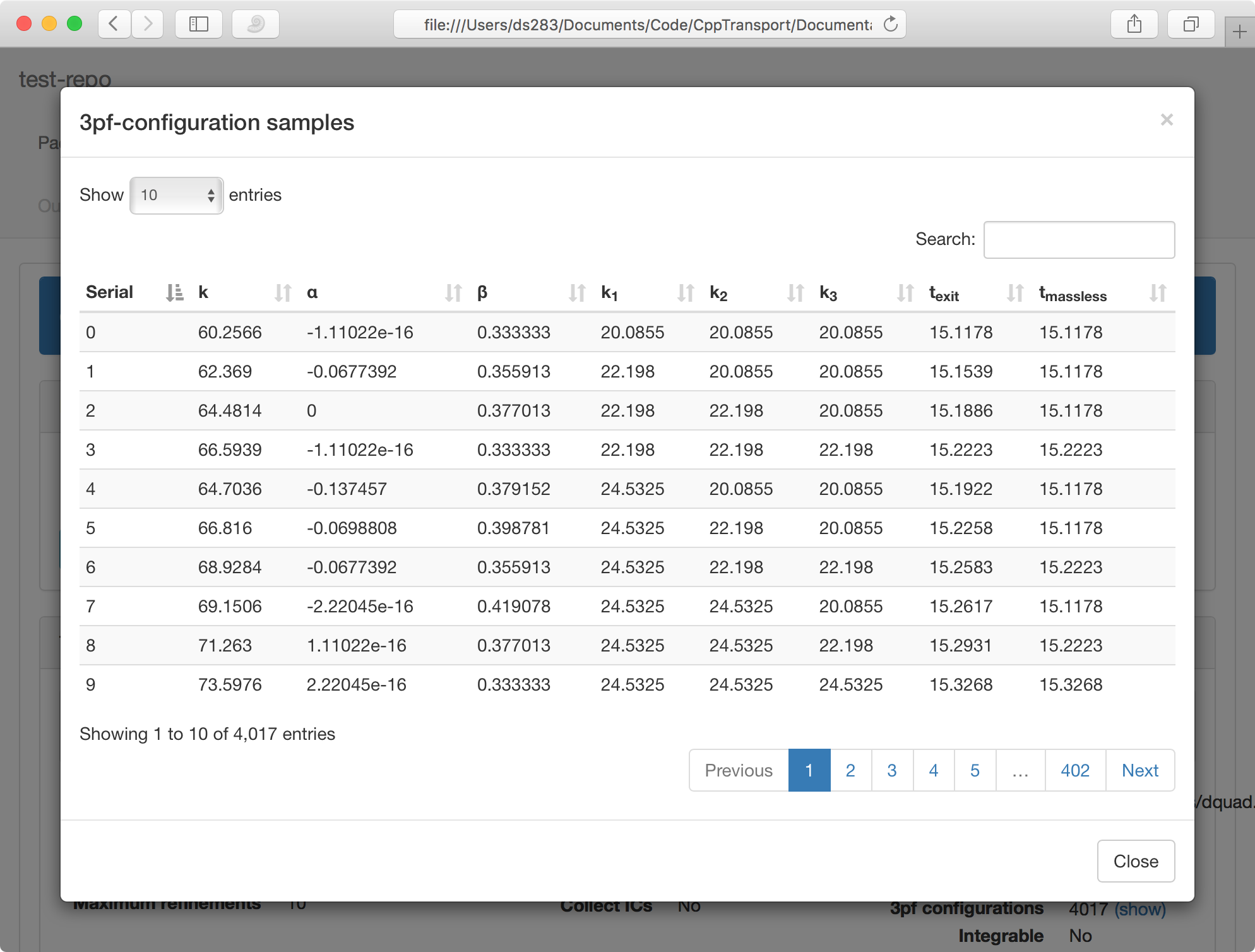}
    \end{center}
    \caption{\label{fig:threepf-config-report}Viewing the sampled 3-point function
    wavenumber configurations in an HTML report.}
\end{figure}

To produce a report for the \file{repo-test}
repository that has just been created,
execute
\begin{minted}{bash}
    ./dquad --repo test-repo --html test-report
\end{minted}
This will create a report in the directory
\file{test-report}.
In a desktop environment the report can usually
be viewed by opening the file
\file{test-report/index.html}, for example by a double-click.
The report is divided into a number of tabs, most of which will
be disabled at this stage because
the repository contains only integration tasks---there are no
tasks of other types, or any generated content.
However, it should be possible to see
details of the
\repoobject{dquad}
initial conditions package under the `Packages' tab,
and the
\repoobject{dquad.twopf} and
\repoobject{dquad.threepf}
tasks under the `Integration tasks' tab.
The report for each task will include
the number of 2- and 3-point wavenumber configurations
that are to be sampled.
Next to this number is a link labelled
`show'.
Clicking this link will display an overlay
listing the wavenumber configurations as a table;
see Fig.~\ref{fig:threepf-config-report}.

The drop-down menu in the top left can be used to adjust
the number of configurations displayed per page,
and the
position indicator in the bottom right can be used
to move through the available pages.
Clicking the arrows in the table header
will sort the table in ascending or descending order
on the corresponding column.
The information displayed for both
configurations of the 2- and 3-point functions
(`2pf configurations' and `3pf configurations') is:
\begin{itemize}
    \item \semibold{Serial.} This is a unique serial number identifying
    the configuration.

    \item \semibold{Wavenumber $k$.} For configurations of the 2-point function this is
    the conventionally-normalized magnitude $k$.
    For configurations of the 3-point function it is the
    conventionally-normalized
    value of $k_t = k_1 + k_2 + k_3$,
    which is the perimeter of the triangle formed by the momenta
    $\vect{k}_1$, $\vect{k}_2$, $\vect{k}_3$.

    \item \semibold{Horizon-exit time $\texit$.}
    For 2pf configurations this is the time
    (measured in e-folds)
    when $\kcom = aH$.
    Here, $\kcom$ is a \emph{comoving} wavenumber.
    Comoving wavenumbers are computed from conventionally-normalized
    wavenumbers by adjusting their normalization so that
    the conventional wavenumber $k=1$
    satisfies $\kcom = \astar \Hstar$
    at time $\Nstar$.
    For 3pf configurations
    there is no unique concept of horizon exit time
    because the wavenumber associated with each side
    of the momentum triangle can exit at different times.
    The time reported as $\texit$ is the average time in the
    sense $k_t = aH$, where here $k_t$ is comoving-normalized.

    If {\CppTransport} is unable to compute the horizon exit
    time for all configurations
    (presumably because for at least one configuration it occurs
    \emph{before} the initial time) it will issue an error:
    \begin{minted}[linenos=false,xleftmargin=0pt,bgcolor=blue!10]{text}
        dquad.twopf: extreme values of N did not bracket time of horizon exit; check whether range of N contains horizon exit times for all configurations
    \end{minted}
    To fix this it is usually necessary to move the initial time earlier
    while keeping the horizon exit time of the longest mode constant,
    or vice versa.

    \item \semibold{Massless time $\tmassless$.}
    \label{enum:massless-time}
    In order to apply suitable initial conditions,
    {\CppTransport} computes
    a \emph{massless time} $\tmassless$ for each 2pf configuration.
    This is defined to be the time when
    $(k/a)^2 = M^2$ where $M^2$ is the largest eigenvalue of the mass
    matrix $M_{\alpha\beta}$,
    or the time of horizon exit if it is earlier.

    For 2pf configurations the quoted time is the massless time
    computed according to this prescription.
    For 3pf configurations it is the earliest massless time
    associated with the individual wavenumbers
    $\vect{k}_1$, $\vect{k}_2$, $\vect{k}_3$.
    This is used when determining where to set initial conditions
    for a 3pf configuration.

    If {\CppTransport} is unable to compute the massless time
    for all configuration it will issue an error similar to that for
    the horizon-exit time:
    \begin{minted}[linenos=false,xleftmargin=0pt,bgcolor=blue!10]{text}
        dquad.twopf: extreme values of N did not bracket massless point; check whether range of N contains massless time for all configurations
    \end{minted}
\end{itemize}
In addition, the table of 3pf configurations includes some extra columns:
\begin{itemize}
    \item \semibold{Shape parameters $\alpha$, $\beta$.}
    Sometimes it is useful to measure the shape of the momentum triangle
    using the parameters $\alpha$, $\beta$ introduced by
    Fergusson \& Shellard~\cite{Fergusson:2006pr}.
    They are defined by
    \begin{equation}
    \begin{split}
        \alpha & = \frac{2(k_1 - k_2)}{k_t} \\
        \beta & = 1 - \frac{2k_3}{k_t} .
    \end{split}
    \label{eq:alpha-beta}
    \end{equation}
    These values are reported for each configuration.

    \item \semibold{Side lengths $k_1$, $k_2$, $k_3$.}
    The comoving side lengths are also reported.
\end{itemize}

\para{Option 2: inspect the SQLite databases directly}
If the table of sample configurations is large, or if
there is a requirement to inspect subsets of the list,
it is better to view the table in a dedicated SQL
database management tool.
Many such tools exist.
The simpler tools are intended to manage only {\SQLite}
databases.
\href{http://sqliteman.yarpen.cz}{\Sqliteman}
is an example of this type.
It is packaged with Ubuntu, and can be installed on OS X
using {\MacPorts} or {\Homebrew}.
More complex tools are capable of managing many different
types of database, and these are often more powerful
at the expensive of a more complex user interface.
The free tool
\href{http://dbeaver.jkiss.org}{\DBeaver} is an example
in this category.
Another is
\href{https://www.jetbrains.com/datagrip/}{\DataGrip};
this is a commercial product, but free licenses are available to
academic users.

The wavenumber configuration databases can be found in the
directory \file{repository/tasks}
within the repository.
The database for a task named \emph{TaskName} is
\file{TaskName.kconfig-db.sqlite}.
Opening this file in a database manager
will reveal
a table named
\mintinline{sql}{twopf_kconfig}
for a task sampling only the 2-point function,
and two tables
named
\mintinline{sql}{twopf_kconfig}
and
\mintinline{sql}{threepf_kconfig}
for a task that also samples the 3-point function.
These tables list the data described above, in addition
to some columns that are not displayed in the HTML table.
First, each wavenumber $k$ or $k_t$ is listed twice with
conventional and comoving normalizations.
Second, there are extra columns with names
beginning \mintinline{sql}{store_}.
These are used internally by {\CppTransport} and can be
ignored.

For 3pf configurations, the table is normalized in the sense
that the $k_1$, $k_2$, $k_3$ side lengths are not included
directly but refer to the serial number of the corresponding entry
in the 2pf configuration table.
This helps to ensure that the database remains internally
self-consistent.
To display a table that lists the side lengths explicitly
requires an SQL query:
\begin{minted}{sql}
    SELECT
      threepf_kconfig.serial          AS serial,
      threepf_kconfig.kt_conventional AS kt_conventional,
      threepf_kconfig.alpha           AS alpha,
      threepf_kconfig.beta            AS beta,
      threepf_kconfig.t_exit_kt       AS t_exit_kt,
      w1.conventional                 AS k1_conventional,
      w2.conventional                 AS k2_conventional,
      w3.conventional                 AS k3_conventional,
      threepf_kconfig.t_exit_kt       AS kt_exit_kt,
      threepf_kconfig.t_massless      AS t_massless
    FROM threepf_kconfig
      INNER JOIN twopf_kconfig AS w1 ON w1.serial = threepf_kconfig.wavenumber1
      INNER JOIN twopf_kconfig AS w2 ON w2.serial = threepf_kconfig.wavenumber2
      INNER JOIN twopf_kconfig AS w3 ON w3.serial = threepf_kconfig.wavenumber3
    ORDER BY serial;
\end{minted}

\subsubsection{Launch and track tasks from the command line}
If the set of 3pf sample configurations has been constructed correctly
the next step is to ask {\CppTransport} to carry out the tasks.
A {\CppTransport} executable can be instructed to
perform as many tasks as are desired,
in which case it will perform them sequentially.
To launch it with 4 processes, carrying out both the
\repoobject{dquad.twopf} and
\repoobject{dquad.threepf} tasks, we would use
\begin{minted}{bash}
    mpiexec -n 4 dquad -r test-repo --task dquad.twopf --task dquad.threepf
\end{minted}
While the job is in progress, executing the command
\begin{minted}{bash}
    ./dquad -r test-repo --inflight
\end{minted}
will show details of the tasks being processed:
\begin{minted}[linenos=false,xleftmargin=0pt,bgcolor=blue!10]{text}
    In-flight content:
    Name               Task           Type                      Initiated  Duration  Cores  Completion
    20160516T123513-1  dquad.threepf  integration    2016-May-16 12:35:13       33s      4          --
\end{minted}
For each `in flight' task,%
    \footnote{Only those tasks that
    are currently active are shown in this list. Although we specified
    two tasks on the command line, {\CppTransport} processes them sequentially and therefore
    only one at once will appear in the list.}
the information shown comprises:
\begin{itemize}
    \item The \semibold{group name}. {\CppTransport} refers to the output produced
    by each execution of a task as a \emph{content group}.
    For an integration task the content group will consist of a database
    containing
    various tables for the $n$-point functions and associated data products.
    For the post-processing tasks currently available---%
    to compute $n$-point functions of $\zeta$, and to take inner products
    with the bispectrum---%
    the content group will consist of
    further databases containing these quantities.
    For output tasks the content group will contain plots, tables
    or Python scripts.

    Each content group is given a unique name derived from its timestamp.
    The format is \repoobject{yyyymmddThhmmss}
    where \repoobject{yyyy} is replaced by the year,
    \repoobject{mm} by the month,
    and so on.
    The capital \repoobject{T} separates the date
    from the current time.

    If two tasks happen to be initiated close together, their time stamps
    may clash. In this case {\CppTransport} will append
    \repoobject{-N} to the group name, where $N$ is a unique number.
    The repository database is designed to be safe when used
    by multiple processes, so name collisions will not occur even if
    different {\CppTransport} jobs attempt simultaneously
    to generate a content group
    in the same repository.

    \item The \semibold{task name} and \semibold{task type}. This identifies the task to which the content group belongs.

    \item The \semibold{job start time} and \semibold{duration}.
    For long-running jobs, this enables you to keep track of the total elapsed time.

    \item The \semibold{number of cores} used by the job.

    \item An \semibold{estimated time of completion}.
    For long-running jobs, {\CppTransport}
    will attempt to estimate when the job will complete on the assumption that the
    items it has processed so far are typical.
    {\CppTransport} keeps track of the time required to process each item of work---for
    example, to integrate a single configuration of the 2- or 3-point function,
    or to generate a plot.
    The completion time is estimated assuming the time taken
    to process each remaining
    work item will be the current average time-per-item.
    This estimate is often accurate but can be misleading if the remaining
    work items are atypically expensive to compute.

    The time-to-completion estimate is first generated after 5 minutes.
    After this, it is updated at intervals of 10\% of the total number
    of work items.
    If verbose mode is enabled then {\CppTransport} will simultaneously
    print a brief advisory message summarizing progress so far.
    For the double quadratic example,
    on most modern hardware, both 2- and 3-pf integrations will complete
    before {\CppTransport} generates any progress update.
\end{itemize}
If verbose mode is enabled, {\CppTransport} will emit brief updates
as it works through the list of tasks.
For the
\repoobject{dquad.twopf}
and
\repoobject{dquad.threepf}
tasks of
double quadratic example
its output will be similar to:
\begin{minted}[linenos=false,xleftmargin=0pt,bgcolor=blue!10]{text}
    Task manager: processing task 'dquad.twopf' (1 of 2)
    Committed content group '20160516T123513' for task 'dquad.twopf' at 2016-May-16 13:35:13
    Task manager: processing task 'dquad.threepf' (2 of 2)
    Task manager: 2016-May-16 13:40:29
    Items processed: 3229  In flight: 496         Remaining: 292
    Complete: 80.4%        Mean CPU/item: 0.249s  Assignment: 60s
    Estimated completion: 2016-May-16 13:41:46 (1m 17.2s from now)
    Task manager: 2016-May-16 13:41:26
    Items processed: 3826  In flight: 181         Remaining: 10
    Complete: 95.2%        Mean CPU/item: 0.277s  Assignment: 60s
    Estimated completion: 2016-May-16 13:41:44 (18.6s from now)
    Task manager: 2016-May-16 13:41:36
    All work items processed: 2016-May-16 13:41:36
    Committed content group '20160516T123513-1' for task 'dquad.threepf' at 2016-May-16 13:42:06
    Task manager: processed 2 database tasks in wallclock time 6m 53s | time now 2016-May-16 13:42:06
\end{minted}
Notice that---in this case---the 2pf task \repoobject{dquad.twopf} completed so quickly that
the content group from the second task
\repoobject{dquad.threepf}
has the same timestamp;
the timestamp itself is measured using
\href{https://en.wikipedia.org/wiki/Coordinated_Universal_Time}{UTC}.
As explained above, to keep the names unique {\CppTransport}
has renamed the second group
\repoobject{20160504T202210-1}.

To confirm that output from these groups has been safely written to the repository,
use the \option{{-}{-}status} switch:
\begin{minted}{bash}
    ./dquad -r test-repo --status
\end{minted}
This causes {\CppTransport} to display a short summary of the tasks available
in the repository, and the number of content groups associated with each task.
In this case the output should be:
\begin{minted}[linenos=false,xleftmargin=0pt,bgcolor=blue!10]{text}
    Available tasks:
    Task           Type                  Last activity  Outputs
    dquad.threepf  integration    2016-May-16 12:42:06        1
    dquad.twopf    integration    2016-May-16 12:35:13        1
\end{minted}
If any tasks are still in flight then \option{{-}{-}status} will
additionally display the
same information shown by \option{{-}{-}inflight}.

\subsection{What happens while an integration task is in progress}
\label{sec:what-happens}
Now let us consider what happened while each integration task was in progress.
At the outset, when {\CppTransport} comes to process each new task,
it acquires a unique content group name derived from the current time.
A folder with this name is created in the
\file{output} directory of the repository, under a subdirectory
corresponding to the name of the owning task.

\begin{figure}
    \begin{center}
        \includegraphics[scale=0.65]{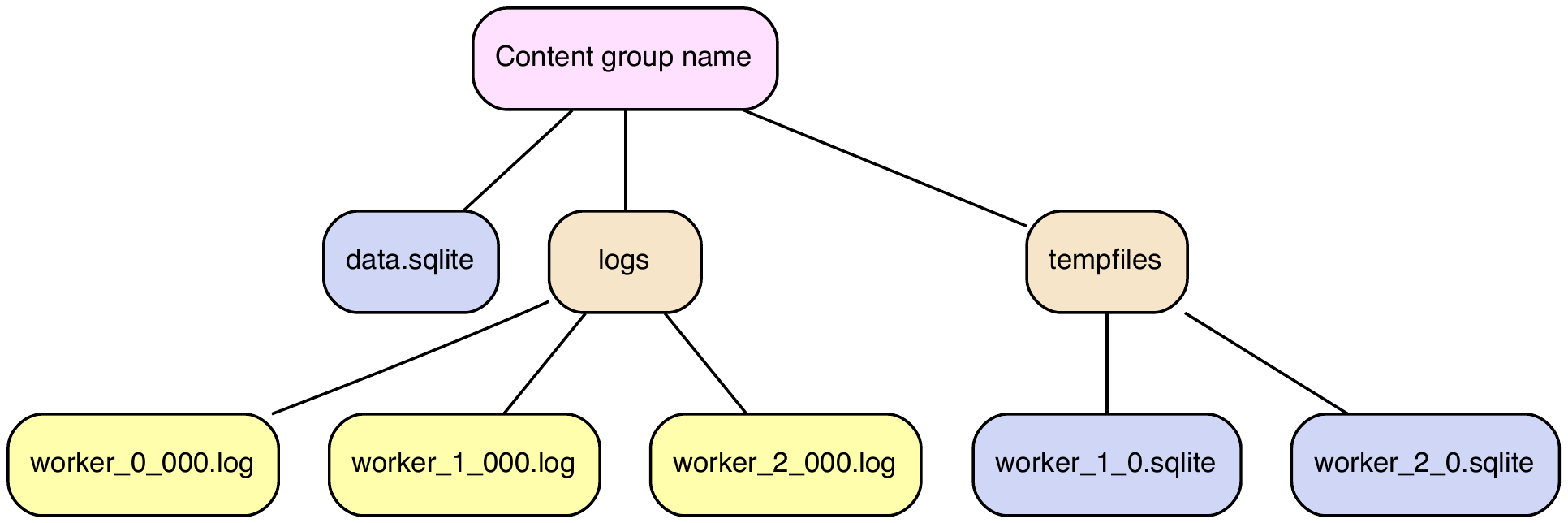}
    \end{center}
    \caption{\label{fig:integration-content}Directory structure for an integration
    content group.
    The \file{tempfiles} directory is present only while this content group
    is in flight; it is removed when the content group is committed to the
    repository.}
\end{figure}

Each content group directory has the structure shown in Fig.~\ref{fig:integration-content}.
There are three items at the top level:
\begin{itemize}
    \item A {\SQLite} database named \file{data.sqlite}.
    This is the main data container that stores the output produced by the job.
    It is just a normal {\SQLite} database, so it can be inspected
    using the database management tools described in~\S\ref{sec:examine-k-database}.

    \item A directory named \file{logs}.
    This contains fairly verbose logs describing the activity of each process
    while the job was active.
    Each process generates a logfile
    named \file{worker\_N\_yyy.log}
    where $N$ is a number identifying the process---the master has $N=0$ and
    the workers have $N > 0$---and \file{yyy} is a unique suffix
    that is usually \file{000}.

    Often it is unnecessary to inspect the logs, but they provide useful information
    if an integration task is not behaving as expected.

    \item A directory named \file{tempfiles}.
    This is used only while the task is in flight, and contains temporary {\SQLite}
    databases into which each worker writes its results.
    The master process aggregates data from these databases into the main container
    \file{data.sqlite}.
    This has the effect that \emph{only the master process} writes to the main container.

    The temporary databases are removed when their contents are merged with the main
    database.
    Once the task is complete and the content group has been committed to the repository,
    the \file{tempfiles} directory is removed.
\end{itemize}
A typical integration will involve the following sequence of steps.
\begin{enumerate}
    \item The task manager running on the master process begins work on a new
    integration task. It sends messages to the worker processes, instructing them
    to prepare for integration activity associated with this task.

    \item Once all workers have signalled that they are ready,
    the task manager issues a small number of work items to each worker.
    For integration tasks these work items are individual wavenumber
    configurations.
    The workers process these configurations and report the time taken.
    The results of each integration are held in memory on each worker.

    \item The task manager on the master process attempts to balance the workload
    of each worker by using the reported times to estimate capacity.
    It estimates the number of configurations each worker can process in 60 seconds
    and then issues a corresponding number of work items, modified if necessary to
    prevent any one worker consuming an unbalanced fraction of
    the queue.
    The effect is that slower cores will be issued with fewer work items,
    and faster cores will be issued with more.

    As the workers process each group of work items, they report the time taken
    back to the master. This information is used to update the estimate
    of each worker's capacity
    and to adjust future work allocations.

    \item The workers process integrations as they arrive, retaining the
    results in memory.
    To prevent memory requirements rising unboundedly, each worker
    is given a fixed capacity.
    When the accumulated integration products outgrow this capacity
    they are flushed to a temporary database in the
    \file{tempfiles} directory.
    (This strategy is adopted for performance reasons. It is faster
    to write the database to disk in one go, rather than
    writing the data piecemeal at the end of each integration.)

    The worker process then sends a message to the master,
    asking for this temporary database to be aggregated into the main
    container.

    \item The master process continues to issue new work and aggregate
    temporary databases.
    For large databases and some choices of filing system
    (especially slow network storage)
    the aggregation time may become lengthy, in which case
    the task manager will adjust the number of work items
    it allocates to each worker in an attempt to prevent
    workers waiting for new work allocations while an aggregation is
    in progress.

    \item Eventually all work items have been processed
    and their results aggregated into the main container.
    At this point the task manager performs an integrity check
    that attempts to detect any missing items.
    It then generates a repository record for the content group.
    If the integrity check was successful the group is marked
    as complete.
    Only complete groups can be used to generate subsequent
    derived products.
    However,
    incomplete groups are retained in the repository because their
    data can be re-used.
    Finally, the \file{tempfiles} folder is removed.
\end{enumerate}

\subsection{Using checkpoints and recovery to minimize data loss}
\label{sec:checkpointing}
The amount of memory available to each worker can be adjusted using
the command-line option
\option{{-}{-}caches}.
It should be followed by an argument representing the cache size
as an integer number of megabytes.
The default cache size is 500 Mb.

Generally, it is advantageous to keep the cache size fairly large.
A small cache size will encourage the workers to flush their
temporary databases frequently, which can generate a large amount
of disk activity.
If the master process is busy with aggregations then
some workers may stall while they wait for new work items to arrive.
A large cache mitigates this, at least if the main container
does not become too large, because a small number of
large aggregations tends to be faster than a large number of small
aggregations.

On the other hand, the total cache memory
allocated to cache should not become larger than the physical
memory available on a machine.
This will usually cause some physical memory to be swapped
out to disk, potentially slowing down the progress of the job.

\begin{advanced}{Estimating cache requirements with paired tasks}
    For paired post-integration tasks it should be remembered that
    each task uses an independent cache.
    While the data generated by post-integration tasks is usually
    smaller than an integration task, the cache size should be
    chosen to prevent out-of-memory problems.

    With the default cache size of 500 Mb, a 4-process job could
    consume up to 1.5 Gb for integration-only tasks
    or 3 Gb for a paired set.
    An 8-process job would consume up to 3.5 Gb for
    integration-only or 7 Gb when paired.
    If running on laptop- or desktop-class hardware these numbers
    should be adjusted to ensure that memory requirements stay within
    bounds.
    Similar calculations apply if running in a cluster environment.
\end{advanced}

If the cache is large then a significant amount of data can
accumulate in memory before being written to disk.
In ordinary execution this does not an issue.
However, if problems occur then this situation is
undesirable:
in the event of a crash, all data will be lost
and would need to be expensively regenerated.
The same applies to long-running tasks on a cluster,
where jobs may be terminated by the scheduler,
or the machine restarted,
in a way that is beyond the control of individual users.

To mitigate these risks {\CppTransport} offers the option
to set checkpoints for each task
at fixed time intervals.
At the first opportunity after a checkpoint,
a worker will flush its cache to a temporary database
and ask the master process to aggregate it.
If the job is subsequently terminated---by a crash,
machine restart, or other cause---then any information
stored in the main container can be recovered
and used to \emph{seed} another task.
In this way the CPU effort spent performing successful
integrations is not wasted.

To set checkpoints, the command-line option
\option{{-}{-}checkpoint} is used. It should be followed
by a time interval measured as an integer number of minutes.
Exactly what interval is appropriate will depend on the
typical time taken to integrate a configuration,
but could perhaps be 30 to 60 minutes.
Unless each integration takes substantially longer
than this, the result is that no more than $\sim$ 1 hour
of work would need to be regenerated following a crash.

\begin{advanced}{Setting default checkpoints for a task}
    In addition to the global checkpointing interval set by
    \option{{-}{-}checkpoint}, it is possible to set default
    checkpoint intervals for each task
    using the \mintinline{c++}{set_default_checkpoint()}
    method of an integration task; see~\S\ref{sec:general-integration-options}.
    Each task may set a different default checkpoint interval.

    If an explicit checkpoint interval is given on the
    command line, it overrides
    any task-specific defaults.
\end{advanced}

\para{Example}
We can simulate a crash by manually terminating
the \file{dquad} executable.
Start a new instance of the
\repoobject{dquad.threepf} task by typing
\begin{minted}{bash}
    mpiexec -n 2 dquad -v -r test-crash --create --checkpoint 1 --task dquad.threepf
\end{minted}
This starts a {\CppTransport} job using only two
processes---a master and a single worker. It will therefore proceed
more slowly, giving time to interrupt it.
The checkpoint interval has been set to 1 minute, and
the job output will be written into a new repository called
\file{test-crash}.
Monitor progress using a separate terminal and the
\option{{-}{-}inflight} option.
After the job has been in progress for a few minutes use
\texttt{Ctrl-C}
to terminate it.

Even after termination, using \option{{-}{-}inflight}
will appear to show that the task is still underway;
by itself the repository database has no way to know
that a crash has occurred:
\begin{minted}[bgcolor=blue!10]{text}
    $ ./dquad -r test-crash --inflight
    In-flight content:
    Name             Task           Type                      Initiated  Duration  Cores  Completion
    20160505T112925  dquad.threepf  integration    2016-May-05 11:29:25    3m 41s      2          --
\end{minted}
To deal with this situation we should use the command-line
switch \option{{-}{-}recover} to inform the repository that
the integration is no longer live, and recovery should be attempted.
{\CppTransport} informs us about each content group it is able to recover:
\begin{minted}[bgcolor=blue!10]{text}
    $ ./dquad -v -r test-crash --recover
    Committed content group '20160506T093816' for task 'dquad.threepf' at 2016-May-06 10:40:44
    Warning: Content group '20160506T093816' has missing content
\end{minted}
The recovery process may take some time.
{\CppTransport} must ensure that the database is left in
a consistent state and this can entail relatively costly comparison between
the various tables held in the database container.

Currently,
recovery is a global repository operation. When
\option{{-}{-}recovery} is specified, {\CppTransport}
will perform recovery for all tasks that are currently registered as in-flight.
If you have multiple tasks running against the same repository and need to
perform recovery, you may wish to to wait until no more active tasks remain
or manually terminate the remaining active processes.

\para{Seeding an integration from a recovered content group}
Normally, after successful recovery, the next step would be
to restart the task, recycling
any successful results from the recovered content group.
To do this we instruct {\CppTransport} to use the recovered group as a \emph{seed},
using the \option{{-}{-}seed} switch
followed by the name of the content group to use:
\begin{minted}[bgcolor=blue!10]{text}
    $ mpiexec -n 4 dquad -v -r test-crash --seed 20160506T093816 --task dquad.threepf
    Committed content group '20160506T094106' for task 'dquad.threepf' at 2016-May-06 10:42:38
    Task manager: processed 1 database task in wallclock time 1m 31.9s | time now 2016-May-06 10:42:38
\end{minted}
{\CppTransport} will copy any successful work items from the seed content group,
and then organize the worker processes to compute whatever items remain outstanding.
There is no limit to how many content groups can be chained together
by this process of recovery followed by seeding.

Once the seeded integration completes,
{\CppTransport}'s \option{{-}{-}status} report shows (as expected)
that two content groups are now attached to the
\repoobject{dquad.threepf} task:
\begin{minted}[bgcolor=blue!10]{text}
    $ ./dquad -r test-crash --status
    Available tasks:
    Task           Type                  Last activity  Outputs
    dquad.threepf  integration    2016-May-06 09:42:38        2
    dquad.twopf    integration    2016-May-06 09:38:16        0
\end{minted}
To obtain more information on these groups, use
the \option{{-}{-}info} switch followed by the task name:
\begin{minted}[bgcolor=blue!10]{text}
    $ ./dquad -r test-crash --info dquad.threepf
    dquad.threepf -- integration task
    Created: 2016-May-06 09:38:16      Last update: 2016-May-06 09:42:38
    Runtime version: 2016.1
    Task type: threepf
    Initial conditions package: dquad
    k-config database: repository/tasks/dquad.threepf.kconfig-db.sqlite

    Task description
    Compute time history of the 3-point function on a cubic lattice from k ~ e^3 to k ~ e^9

    Content group                   Created           Last update  Complete        Size
    20160506T093816    2016-May-06 09:40:42  2016-May-06 09:40:44        No    4e+02 Mb
    20160506T094106    2016-May-06 09:41:06  2016-May-06 09:42:38       Yes  9.1e+02 Mb
\end{minted}
The \option{{-}{-}info} switch can be used with the name of any repository object.
It prints similar information to
that shown by an HTML report, but can be more convenient for
quick inspection of individual records.

Here, the table at the bottom gives a short summary of the two
content groups. Notice that the interrupted group is marked as incomplete.
The reported size corresponds to the disk space occupied by the
data container \file{data.sqlite}.
For the final group (here \repoobject{20160506T094106}) this is just under
1 Gb, showing that the generated datasets can become large
even with a modest number of bispectrum configurations
and a few hundred time sample points.

{\CppTransport} tracks the relationship between different content groups.
Inspecting the repository record for \repoobject{20160506T094106}
shows that \repoobject{20160506T093816} was used as a seed:
\begin{minted}[bgcolor=blue!10]{text}
    $ ./dquad -r test-crash --info 20160506T094106
    20160506T094106 -- integration content
    Created: 2016-May-06 09:41:06      Last update: 2016-May-06 09:42:38
    Runtime version: 2016.1            Task: dquad.threepf
    Type: threepf                      Locked: No
    Tags: --

    Complete: Yes
    Workgroup: 1
    Seeded: Yes
    Data type: double
    Seed group: 20160506T093816
    Has statistics: Yes
    Initial conditions: No
    Size: 9.1e+02 Mb
    Container: output/dquad.threepf/20160506T094106/data.sqlite

    Metadata
    Wallclock time: 1m 20s
    Total time: 6m 22.3s
    Min integration: 0.0329s
    Max integration: 4.55s
    Configurations: 3188
    Failures: 0
\end{minted}

When complex relationships exist
between different content groups,
it becomes difficult to navigate the repository
using \option{{-}{-}info} from the command line.
In such cases it is often more convenient to generate an HTML report
that summarizes the repository contents.
We used these reports in~\S\ref{sec:examine-k-database}
to examine the wavenumber configurations
sampled as part of
each integration task, but they have many other uses.

\begin{figure}
    \begin{center}
        \includegraphics[scale=0.2]{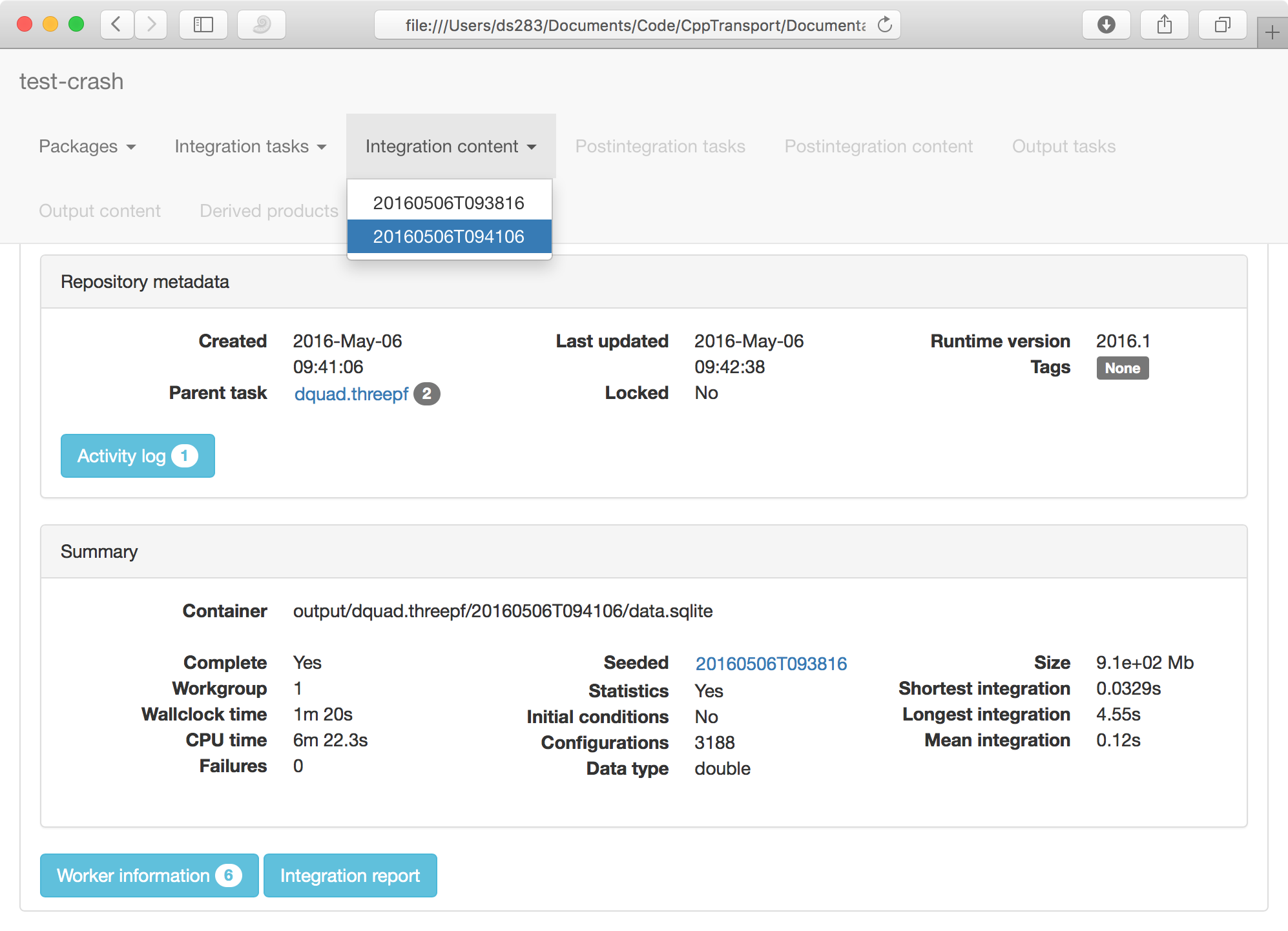}
    \end{center}
    \caption{\label{fig:seeded-group-report}HTML report for seeded
    content group.
    Notice that group relationships are hyperlinked, so it is possible
    to click on the name of the attached seed group to view its
    record.
    The same is true for the task names and any other repository objects.}
\end{figure}

Generating a report is done as in~\S\ref{sec:examine-k-database},
\begin{minted}{bash}
    ./dquad --repo test-repo --html test-report
\end{minted}
Now open \file{test-report/index.html} in a web browser
and
use the `Integration content' tab to view records for each
content group.
For the seeded group, the record includes the name of
seed but this is also hyperlinked---clicking on its name will take
you directly to its record.
Likewise, it is possible to click the task name to view
its details.
See Fig.~\ref{fig:seeded-group-report}.

\subsection{How is the integration time spent?}

\subsubsection{Using HTML reports to analyse integration performance}
\label{sec:HTML-analysis}
It is often useful to understand what determines the total execution time
for an integration task,
especially if performance is different from
what would be expected.
This can happen in two ways.
First, the combination of batch size, aggregation frequency and
number of processes can accidentally trigger an excessive number of stalls,
where the workers spend a long time idle because the master is too busy
to issue them with work.
To detect this kind of scenario the most helpful tool is the process
Gantt chart described in~\S\ref{sec:gantt} below.

Second, the integration time per configuration can vary significantly.
The total execution time therefore depends strongly
on the range of configurations included in each task.
A typical use of {\CppTransport} would be to sparsely sample a range
of configurations $(k_1, k_2, k_3)$
in an exploratory step, before increasing
the number of sample points to perform some specific science
analysis---for example, to predict the CMB temperature bispectrum
from a specific inflationary model.
Before stepping up to a more dense set of sample configurations
it is very helpful to be aware of the
way in which
integration time depends on configuration.
The information needed to understand this scaling
is included in an HTML report.

\begin{figure}
    \begin{center}
        \includegraphics[scale=0.18]{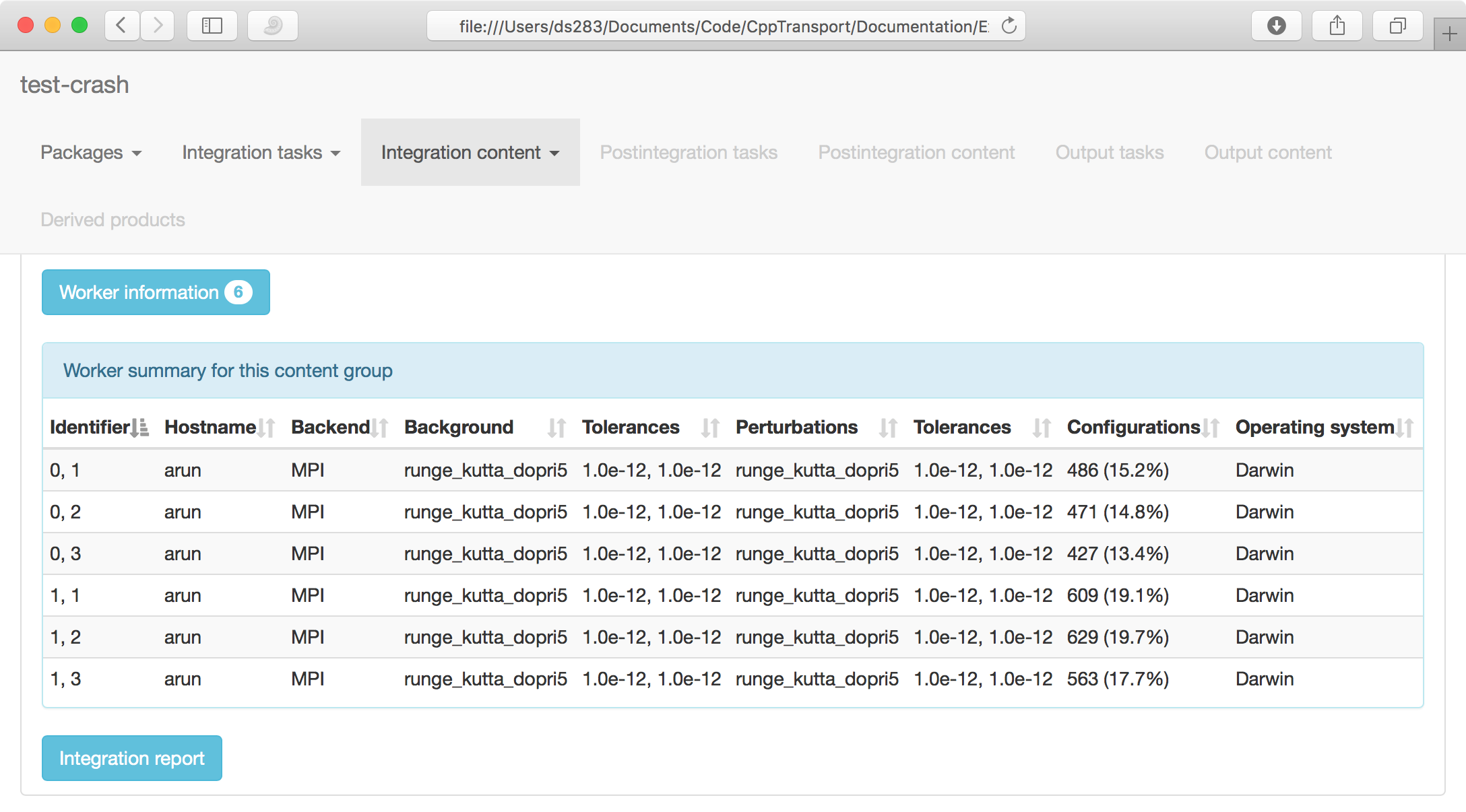}
    \end{center}
    \caption{\label{fig:worker-info}Worker information table displayed as
    part of a content group record.}
\end{figure}

\para{Worker information table}
Under the `Integration content' tab, the report for each content group
includes two blue buttons near the bottom.
One is labelled `Worker information', and clicking it will toggle
a table listing each worker process that has contributed to the
content group; see Fig.~\ref{fig:worker-info}.
The table identifies each worker by a pair of coordinates
$(w, n)$,
where $w$ is the \emph{workgroup number}
and $n$ is the \emph{worker number}.
When a new content group is created its workers are assigned
workgroup number 0.
If this group is subsequently used to seed another group then the workers
in the second group are assigned workgroup number 1, and so on.
The worker number identifies a unique worker within the workgroup.

For each worker, the table lists:
\begin{itemize}
    \item the hostname of the machine on which this worker was running.
    This can help identify problematic machines, especially in a cluster environment.

    \item the type of backend in use. In current versions of {\CppTransport},
    as explained in~\S\ref{sec:template-block}, this will be the
    {\MPI} implementation.

    \item the steppers used for integration of the background and perturabtions,
    and the tolerances applied in each case.
    The tolerances are given in the format
    (abserr, relerr).

    \item the number of configurations processed by the worker,
    both in total and as a percentage of the task.

    \item the operating system on which {\CppTransport} was running.
\end{itemize}

\begin{figure}
    \begin{center}
        \includegraphics[scale=0.65]{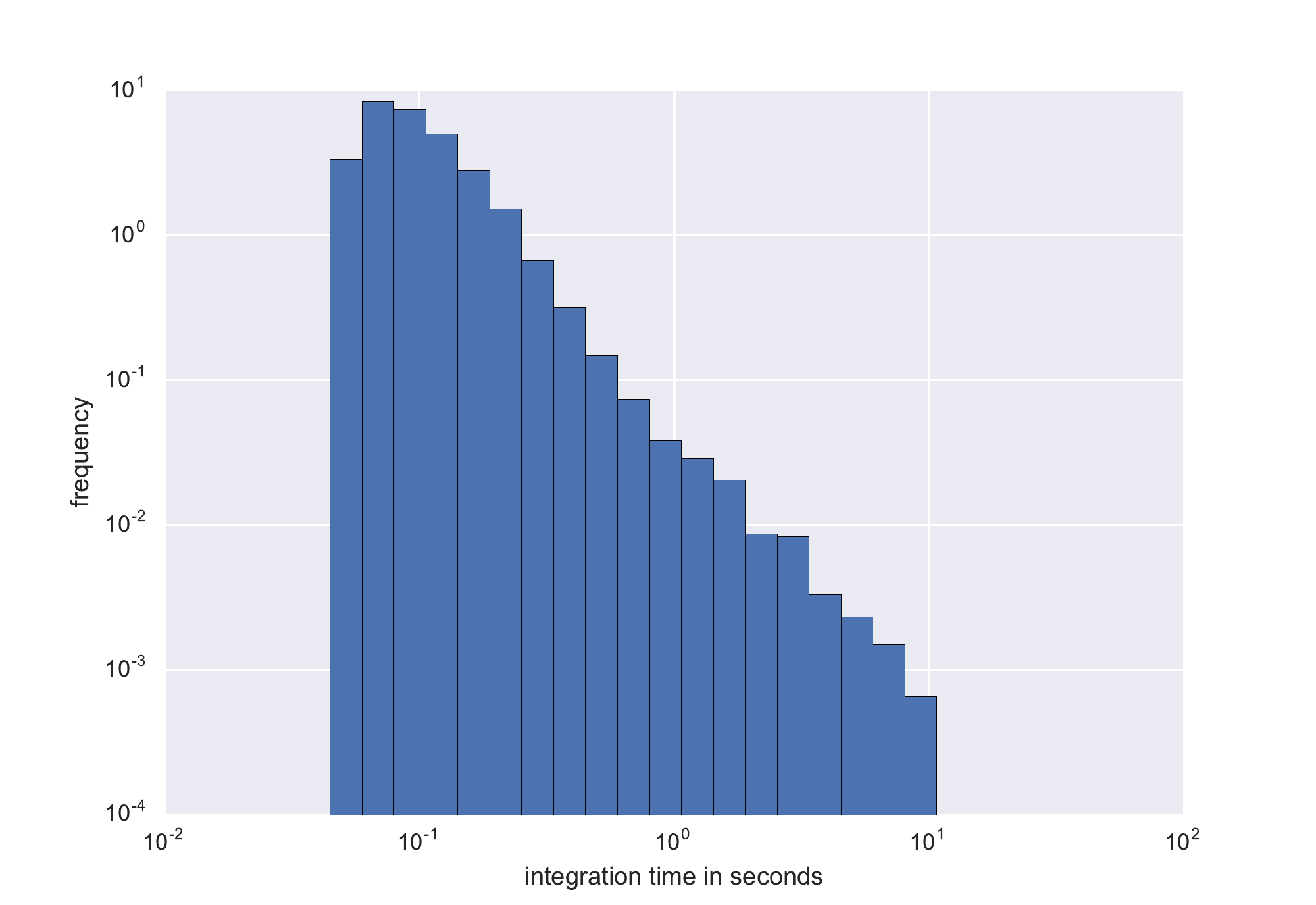}
    \end{center}
    \caption{\label{fig:time-distribution}Distribution of integration
    times for the three-point function in the double quadratic model
    constructed in~\S\ref{sec:add-integration-task}.}
\end{figure}

\begin{figure}
    \hspace{-8mm}
    \begin{tabular}{l@{\hspace{-10mm}}l}
        \includegraphics[scale=0.45]{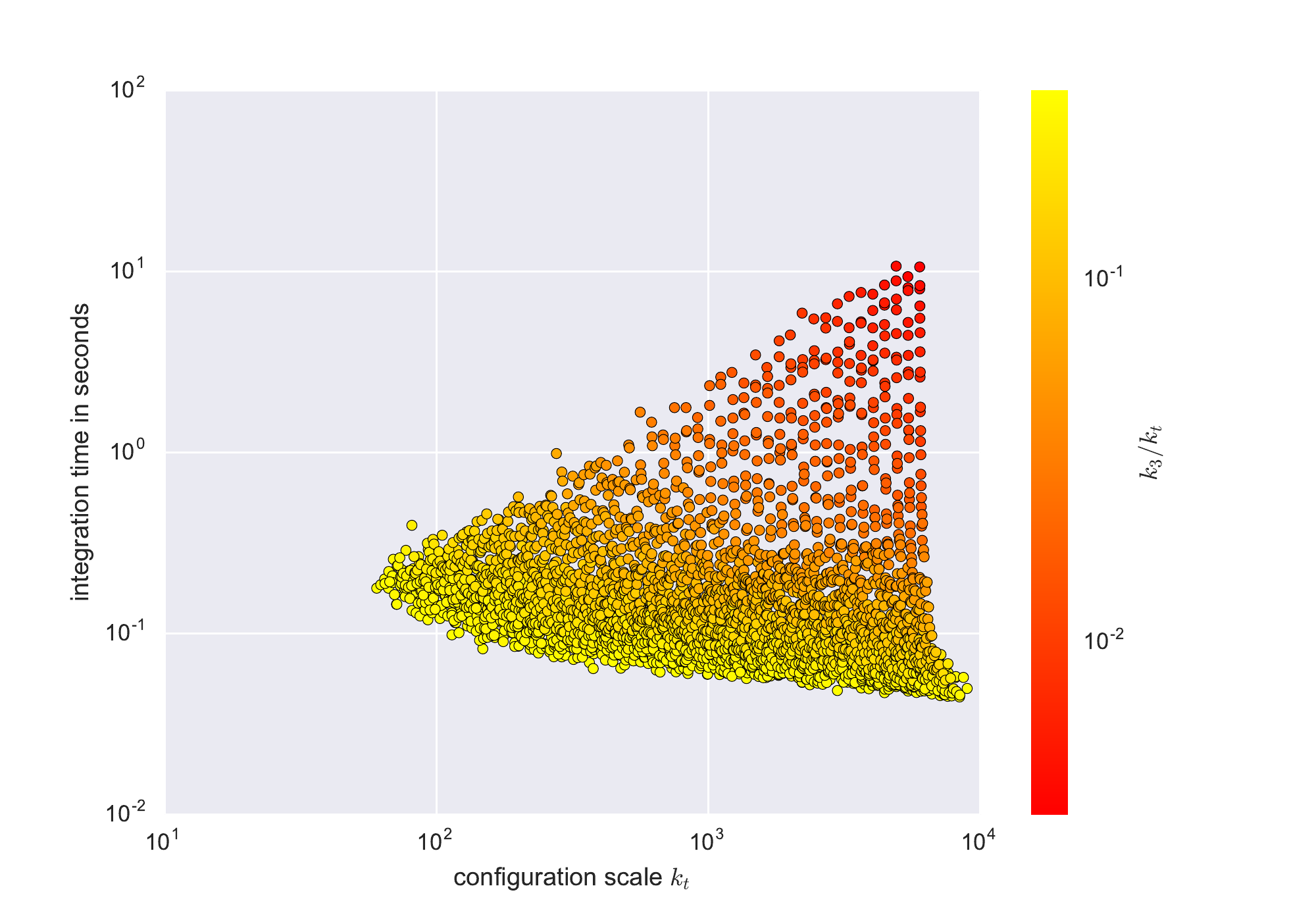} &
        \includegraphics[scale=0.45]{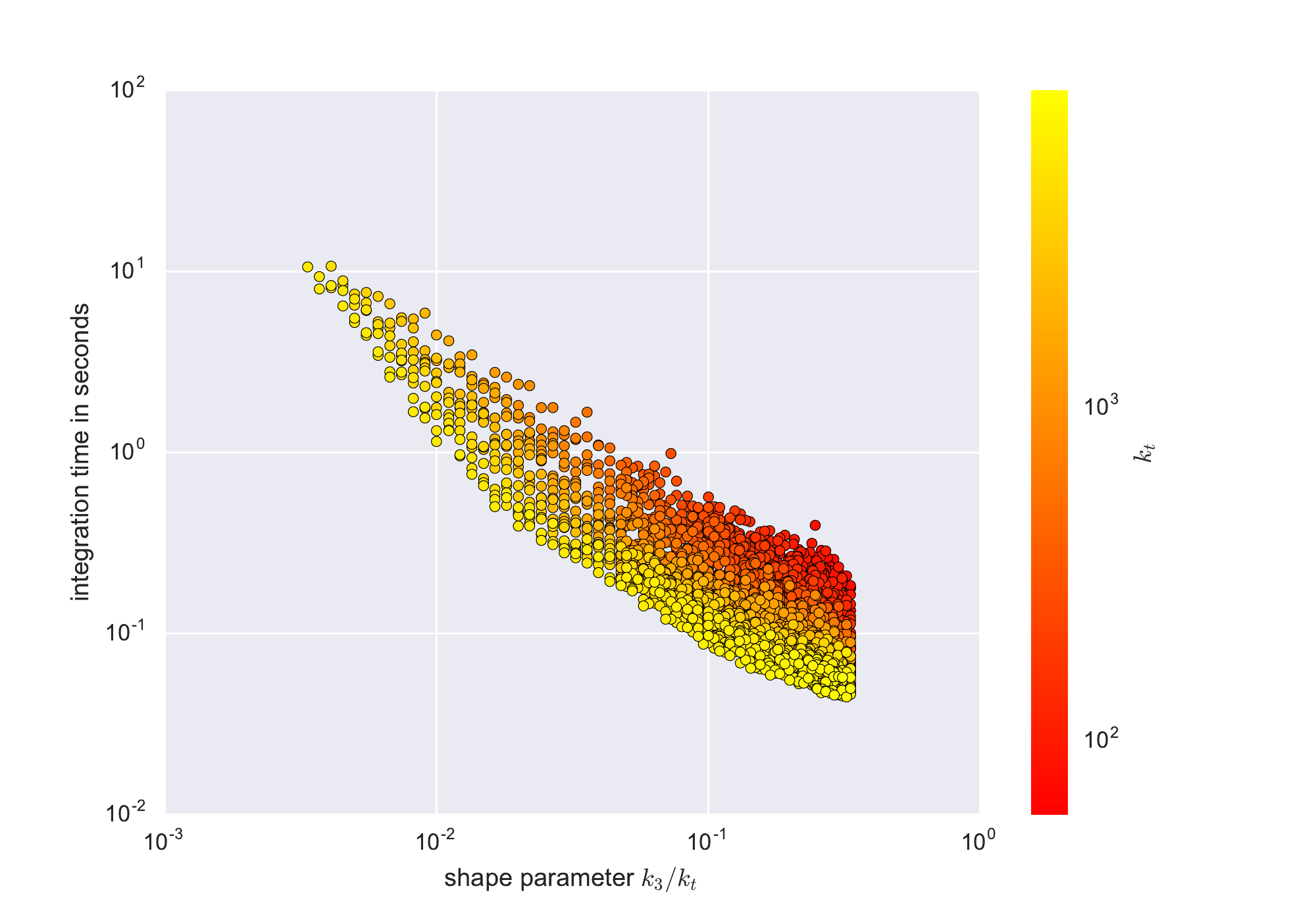} \\
        \includegraphics[scale=0.45]{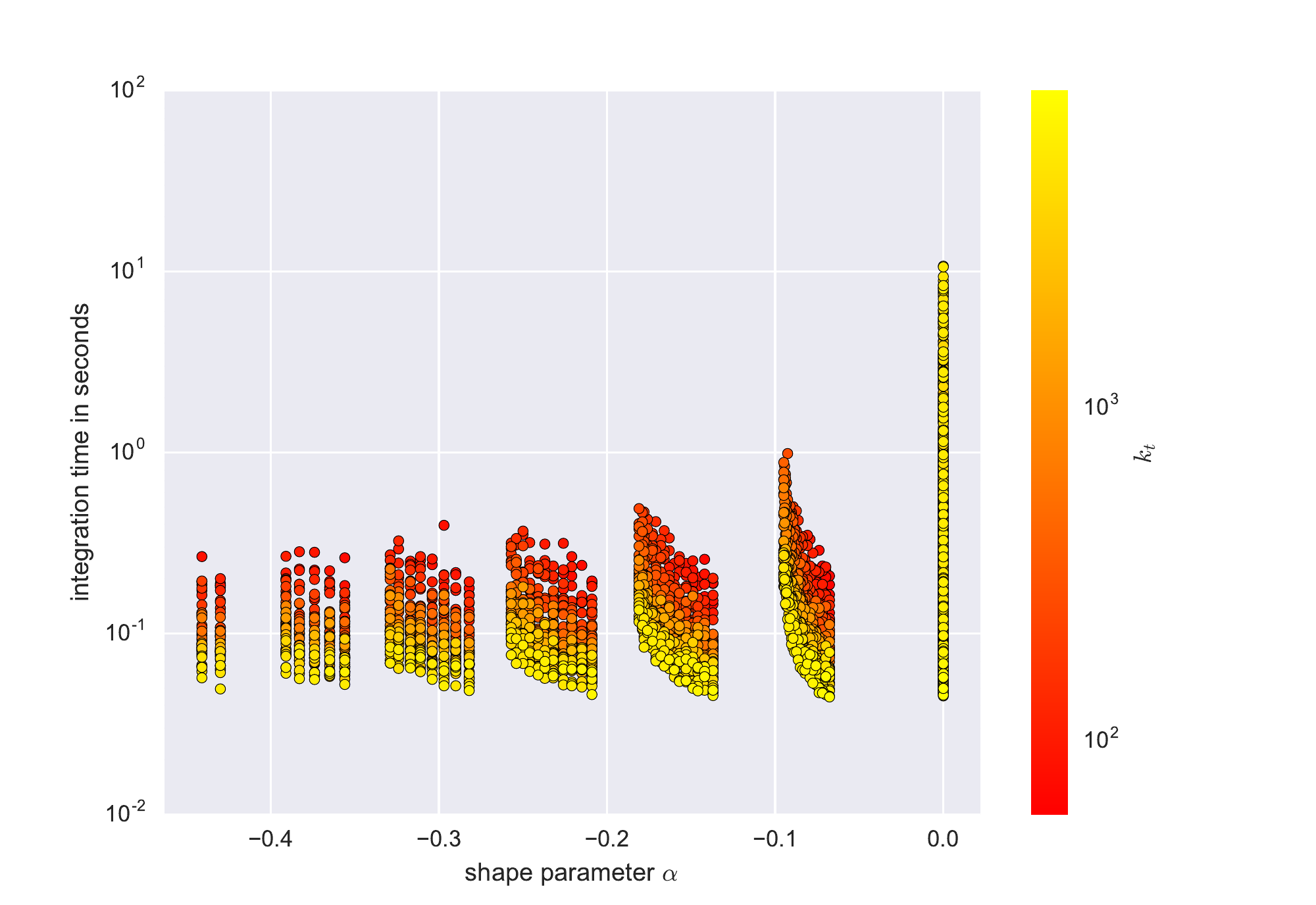}
    \end{tabular}
    \caption{\label{fig:time-scaling}Dependence of integration time on
    the parameters
    $k_t$ (top left), $k_3 / k_t = (1-\beta)/2$ (top right) and $\alpha$ (bottom left)
    in the double quadratic model of~\S\ref{sec:add-integration-task};
    for definitions, see Eq.~\eqref{eq:alpha-beta}.}
\end{figure}

\para{Integration report}
The second blue button is labelled `Integration report'.
Clicking it toggles a set of summary plots.
\begin{itemize}
    \item For any kind of integration task the report includes a
    bar chart showing the number of configurations processed by each
    worker, and a histogram showing the distribution of integration times
    within the task.
    For the 3-point function of the double quadratic model
    this distribution is shown in Fig.~\ref{fig:time-distribution};
    it is roughly a power law.
    This is fairly typical for
    {\CppTransport} tasks that sample a range of different configurations:
    most configurations integrate relatively quickly
    (here $\lesssim 0.1 \second$),
    but there is a heavy tail of
    rare configurations
    that take much longer
    (here, a few $\times$ $10^0 \second$).
    These expensive configurations are usually those that probe the
    squeezed limit of the 3-point function.

    \item For tasks that sample the 3-point function, {\CppTransport}
    includes extra plots showing configuration dependence of the
    integration time
    (Fig.~\ref{fig:time-scaling}).
    In smooth models where the bispectrum does not exhibit features,
    and assuming use of adaptive initial conditions (\S\ref{sec:general-integration-options}),
    the strongest dependence is usually on the squeezing parameter
    $k_3 / k_t$, where $k_3$
    is taken to be the shortest side of the momentum triangle.
    This is clearly visible in the top two plots of Fig.~\ref{fig:time-distribution}.
    In the $k_t$ plot (top-left panel)
    there are a range of integration times for each $k_t$, with the range
    widening as $k_t$ increases.
    However, the colour-bar shows that configurations generating the longest
    integration times have small $k_3/k_t$.
    The $k_3/k_t$ plot (top-right panel)
    gives the same conclusion in a different form.
    The dependence on $k_3/k_t$ is an approximate power law where the
    squeezing is significant.
\end{itemize}


\subsubsection{Generating a Gantt chart of worker activity}
\label{sec:gantt}
The integration report shows how the intrinsic computation cost scales
with configuration, and helps identify regions where it is cheap or
expensive to add
configurations.
These scalings are set partly by the nature of the problem,
partly by detailed microphysical interactions
determined by each specific model,
and partly by the integration scheme employed by
{\CppTransport} and its use of `transport' or evolution
equations.

These intrinsic scalings are one factor
contributing
to the overall numerical efficiency achieved by {\CppTransport}.
The remainder is determined by how effectively the master process is able to
feed tasks to its workers.
{\CppTransport} tries to adjust its scheduling strategy to respond to environmental
pressures. It is frequently successful but under some circumstances the master
process may become overwhelmed,
meaning that new batches of work are not allocated to workers
in a timely fashion.
This leads to workers idling while they wait for new work, described
as a `stall'.
For this reason it can be useful to check
what the workers are doing, and in particular whether progress is being
inhibited by stalls.

\begin{figure}
    \begin{center}
        \includegraphics[scale=0.5]{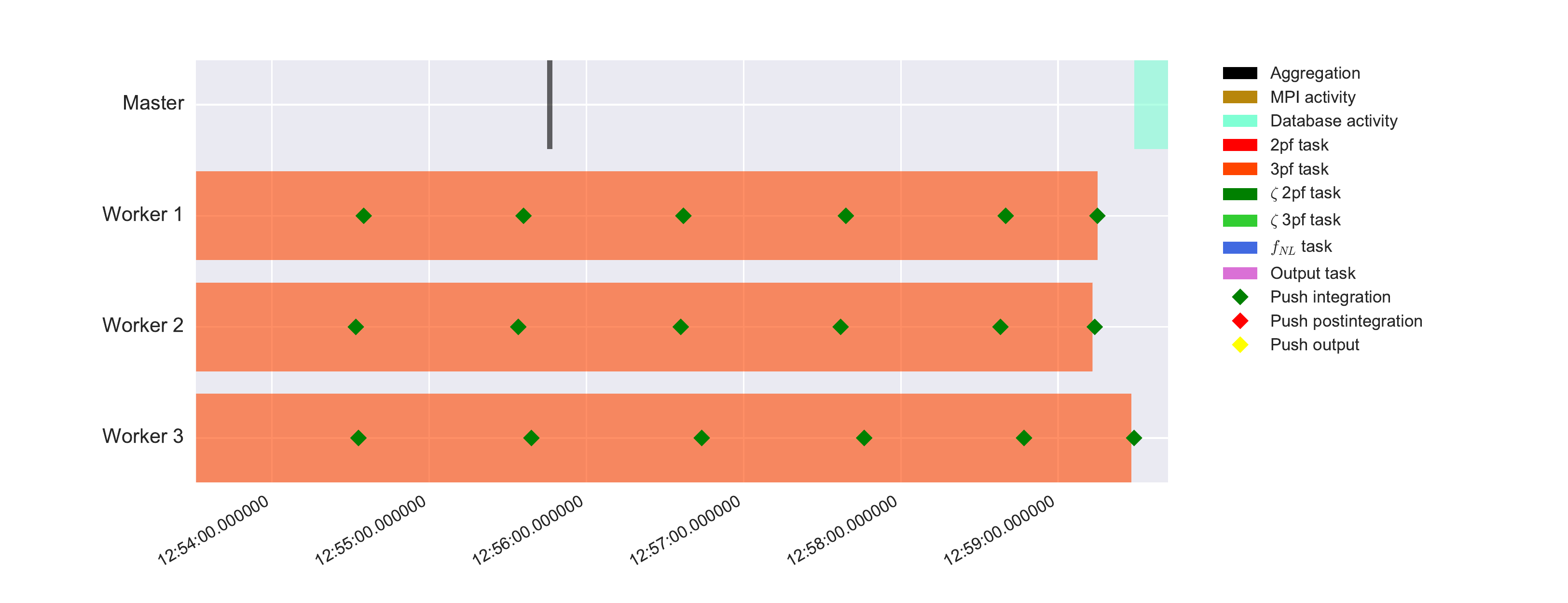}
    \end{center}
    \caption{\label{fig:gantt}Gantt chart
    generated by {\CppTransport}'s \option{{-}{-}gantt} option,
    showing worker activity
    for the task \repoobject{dquad.threepf}
    and \option{{-}{-}checkpoint 1}.
    The diamonds indicate where each worker flushed its cache
    to a temporary database.
    Dark grey bars on the master timeline show where the
    master process was aggregating temporary databases.
    (In many cases the time spent aggregating was too short to register
    on the plot. This test was performed on a machine with an SSD; on
    a machine with a hard disk, aggregation times are usually a few times longer.)
    The aquamarine bar on the far right-hand side
    shows database activity associated
    with checking integrity and performing routine maintenance.}
\end{figure}

To detect stalls, the most useful tool is the process Gantt chart.
Specifying the option \option{{-}{-}gantt} on the command line
will cause
{\CppTransport} to produce a Gantt-like timeline showing the activity of
each worker.
Bars are colour-coded according to the type of activity, and a legend is included
identifying each colour.
The \option{{-}{-}gantt} option
should be followed by a filename, and the output format will be determined by
its extension.
Any format supported by the {\Matplotlib}
backend can be used, of which the most useful
choices are usually SVG (\file{.svg} extension),
PDF (\file{.pdf}), or PNG (\file{.png}) if a bitmap format is desired.
Alternatively, specifying a \file{.py} extension will produce a Python script
to generate the Gantt chart, but without executing it. This is useful if further
customization is required.
For example, by producing a Python script it is possible to manually increase
the figure width if the timeline is too compressed for easy reading.
An example chart for the
\repoobject{dquad.threepf} task is shown in Fig.~\ref{fig:gantt}.

If more detail is required then {\CppTransport} can produce an activity
journal in JSON format using the
\option{{-}{-}journal} command-line argument.
It accepts a single parameter that is interpreted as the filename of the
output journal.
To use the journal for practical analysis it will probably be necessary
to write a custom code that parses each record and generates a suitable output.

\para{Avoiding stalls}
{\CppTransport} makes default choices that usually prevent significant stalls.
Stalls are more likely in the following circumstances:
\begin{itemize}
    \item if the repository is located on a slow filing system.
    Besides allocating work to each worker process, the master process is
    responsible for aggregating their results into the main data container
    \file{data.sqlite}.
    The time taken for each aggregation is dominated by the time required
    for a bulk \mintinline{sql}{INSERT} of rows from each temporary container
    into the main database.

    Depending how many configurations are being computed, and how many time
    sample points are being retained,
    the main database may become very large.
    {\CppTransport} has been optimized so that
    \mintinline{sql}{INSERT} performance is roughly independent of the database
    size,
    but its performance may begin to degrade if the database becomes extremely
    large (bigger than several tens of $\Gb$).
    On a fast internal SSD
    {\CppTransport} typically achieves $\sim 2\times 10^5$
    \mintinline{sql}{INSERT}s per second.
    On a conventional hard disc, either internal or attached via a fast interface
    such as USB or Thunderbolt,
    it achieves $\sim 5 \times 10^4$
    \mintinline{sql}{INSERT}s per second.
    Network filing systems are often somewhat slower.
    If you are forced into this situation then it is preferable
    to use the fastest available disk
    (eg. a Lustre-type system rather than NFS)
    in order to prevent aggregations from becoming too slow.

    With a slower filing system it can be a good strategy to
    set a checkpoint interval (or adjust the cache capacity)
    so that aggregations are distributed over the lifetime of the job.
    This means that master can be carrying out productive activity while the
    workers are busy with calculations.
    The downside of this approach is that a worker may finish its current
    allocation while the master is performing aggregation.
    If this happens the worker will stall until the master is free to
    allocate new work.

    To mitigate this effect, {\CppTransport} monitors how long is spent
    performing aggregations and adjusts the amount of work it offers
    to each worker.
    Initially it will attempt to allocate $60\second$ of work,
    but if aggregations are taking a long time
    it will try to allocate \emph{up to}
    ten times the current mean aggregation time.
    This reduces the probability that a worker will finish its current
    allocation while the master is performing aggregation.%
        \footnote{In practice it may allocate less work in order to balance the
        number of work items assigned to each worker process.}
    However, conversely, it increases the risk that the allocation becomes unbalanced
    in the sense that some workers run out of work items while other workers
    still have a long queue.

    Aggregation times are typically not long enough to be problematic.
    When writing to a hard disk {\CppTransport} can typically aggregate a $\sim 5 \Gb$
    main database in just a few minutes.
    Problems are likely to occur only in extreme scenarios, where a database grows to
    many tens of $\Gb$ while simultaneously being used over NFS.

    \item if a large number of workers are used with a slow filing system.
    Using many workers will generate a large number of aggregation requests,
    and if the filing system is too slow these can overwhelm the master process.
    There are possible mitigations:
    \begin{itemize}
        \item if you are not using a network filing system, ensure that
        you are not passing the command-line option
        \option{{-}{-}network-mode}.
        If used this disables {\SQLite}'s write-ahead log mode that confers
        a significant performance advantage.

        \item ensure that you are not defining the
        \mintinline{c++}{CPPTRANSPORT_STRICT_CONSISTENCY} macro during compilation.
        If used, this roughly doubles aggregation times.
        See~\S\ref{sec:strict-consistency}.

        \item experiment with different buffer sizes or checkpoint intervals,
        which will change how aggregation events
        are distributed over the lifetime of the job.
        In some scenarios it is faster to perform a small number of large aggregations
        near the end.
        In other scenarios it can be faster to perform frequent aggregations
        throughout the job, because this makes use of time when the master
        process would otherwise be idle.
    \end{itemize}
\end{itemize}

\subsection{Using the {\SQLite} data container}
\label{sec:using-integration-container}
In~\S\S\ref{sec:post-integration-tasks}--\ref{sec:derived-products}
we will consider post-integration tasks and show how they can be used
to generate observables from the raw $n$-point functions produced by
an integration task.
In many cases the facilities offered by these
tasks will be quite adequate to carry out a
standard analysis of an inflationary model.
However, in some circumstances, a nonstandard observable may be
needed.
In this case it is necessary to extract data for the raw $n$-point functions.
Another reason for needing this data is simply that you may
wish to produce your own customized plots.

\subsubsection{Table definitions}
This data can be retrieved from the main {\SQLite} container
\file{data.sqlite}
associated with each content group.
As explained above, this is done using SQL queries---yielding
a powerful method to easily extract appropriate subsets
of the data, formatted in any convenient way.

The {\SQLite} container for any integration task contains the following tables
and columns:
\begin{sqltablelist}
\begin{enumerate}
    \item \mintinline{sql}{time_samples} --- contains the list of time sample points. \\
    \label{sqltable:time-samples}
    \begin{tabular}{p{2.5cm}p{11.2cm}}
        \mintinline{sql}{serial} & unique serial number for this sample point \\
        \mintinline{sql}{time} & numerical value of time, measured in $N$, using
        the user-defined conventions established by the
        initial conditions package.
    \end{tabular}

    \item \mintinline{sql}{twopf_samples} --- contains the list of sample points for the 2-point function \\
    \label{sqltable:twopf-samples}
    \begin{tabular}{p{2.5cm}p{11.2cm}}
        \mintinline{sql}{serial} & unique serial number for this sample configuration \\
        \mintinline{sql}{conventional} & conventionally-normalized wavenumber $k$
        corresponding to this configuration \\
        \mintinline{sql}{comoving} & comoving-normalized wavenumber $k$
        corresponding to this configuration \\
        \mintinline{sql}{t_exit} & time of horizon exit for this wavenumber in the sense $k=aH$ \\
        \mintinline{sql}{t_massless} & mass-point for this wavenumber in the
        sense $(k/a)^2 = M^2$, where $M^2$ is the largest eigenvalue of the mass
        matrix
    \end{tabular}

    \item \mintinline{sql}{backg} --- contains the time evolution of the background fields \\
    \label{sqltable:backg}
    \begin{tabular}{p{2.5cm}p{11.2cm}}
        \mintinline{sql}{unique_id} & unique identifier labelling this row \\
        \mintinline{sql}{tserial} & reference to serial number
        \mintinline{sql}{time_samples.serial} identifying the time sample
        point for this row \\
        \mintinline{sql}{page} & the value of each background field is stored in the column
        \mintinline{sql}{coordN}, where \mintinline{sql}{N} is a 0-based integer
        labelling the fields in the same order that they were declared in
        the model description file. However, {\SQLite} allows only a limited number
        of columns per table. To accommodate this {\CppTransport} will store the roughly
        the first 2000 entries with \mintinline{sql}{page} equal to $0$ and then
        start a new row with \mintinline{sql}{page} equal to $1$, and so on.
        The same strategy is used for all data tables \\
        \mintinline{sql}{coordN} \ldots & values of the background fields in the order
        declared in the model description file.
        All fields come first as a block, followed by their $N$ derivatives as a second block.
    \end{tabular}

    \item \mintinline{sql}{twopf_re} --- contains real part of field-space 2-point function \\
    \begin{tabular}{p{2.5cm}p{11.2cm}}
        \mintinline{sql}{unique_id} & unique identifier labelling this row. Not used
        unless strict data consistency checks are enabled
        by defining the macro
        \mintinline{c++}{CPPTRANSPORT_STRICT_CONSISTENCY} during
        compilation. See~\S\ref{sec:strict-consistency}. \\
        \mintinline{sql}{tserial} & reference to serial number
        \mintinline{sql}{time_samples.serial} identifying the time sample point for this row \\
        \mintinline{sql}{kserial} & reference to serial number
        \mintinline{sql}{twopf_samples.serial} identifying the wavenumber sample point for this row \\
        \mintinline{sql}{page} & page number, defined as for \mintinline{sql}{backg.page}; see
        \mintinline{sql}{eleN} below \\
        \mintinline{sql}{eleN} \ldots & \semibold{dimensionless components} of the real part of the field-space 2-point
        function, packed in a `right-most' ordering defined as follows.
        Set $\langle \delta X^a(\vect{k}_1) \delta X^b(\vect{k}_2) \rangle = (2\pi)^3 \delta(\vect{k}_1 + \vect{k}_2)
        \Sigma^{ab}$, where $a$, $b$ range over the fields and their
        canonical momenta
        (which for the purposes of the 2-point function are the same as the
        derivatives with respect to $N$).
        Then $N$ = $D \times \#a + \#b$, if $D$ is the dimension of phase
        space (two times the number of fields)
        and $\#$ is the map that assigns a numerical
        value to a species in following order:
        fields first, in the order defined in the model description file,
        followed by momenta, in the same order as the fields.

        The quantity stored is $\Re(k^3 \Sigma^{ab})$,
        where $k = |\vect{k}_1| = |\vect{k}_2|$
        is the common magnitude of the momenta.
        This is independent of the comoving normalization, making the results
        easy to remap into whatever units are convenient.
    \end{tabular}

    \item \mintinline{sql}{tensor_twopf} --- contains the tensor 2-point function \\
    \label{sqltable:tensor-twopf}
    \begin{tabular}{p{2.5cm}p{11.2cm}}
        \mintinline{sql}{unique_id} & unique identifier labelling this row. Not used
        unless strict data consistency checks are enabled
        by defining the macro
        \mintinline{c++}{CPPTRANSPORT_STRICT_CONSISTENCY} during
        compilation. See~\S\ref{sec:strict-consistency}. \\
        \mintinline{sql}{tserial} & reference to serial number
        \mintinline{sql}{time_samples.serial} identifying the time sample point for this row \\
        \mintinline{sql}{kserial} & reference to serial number
        \mintinline{sql}{twopf_samples.serial} identifying the wavenumber sample point for this row \\
        \mintinline{sql}{page} & page number, defined as for \mintinline{sql}{backg.page}; see
        \mintinline{sql}{eleN} below \\
        \mintinline{sql}{eleN} \ldots & \semibold{dimensionless components}
        of the real part of the tensor two-point function
        for an individual helicity state $\gamma_s$ (where $s = +, \times$)
        and their derivatives $\pi_s = \d \gamma_s / \d N$,
        in the following order:
        $N = 0$ is $\gamma_s \gamma_s$,
        $N = 1$ is $\gamma_s \pi_s$,
        $N = 2$ is $\pi_s \gamma_s$
        and $N = 3$ is $\pi_s \pi_s$.

        If $\langle \gamma_s(\vect{k}_1 \gamma_{s'}(\vect{k}_2) \rangle =
        (2\pi)^3 \delta(\vect{k}_1 + \vect{k}_2) \Sigma_{ss'}$,
        where $s$ and $s'$ label helicities,
        then the quantity stored is
        $\Re(k^3 \Sigma_{ss'})$.
    \end{tabular}

    \item \mintinline{sql}{integration_statistics} --- contains integration metadata \\
    \label{sqltables:statistics}
    \begin{tabular}{p{2.5cm}p{11.2cm}}
        \mintinline{sql}{kserial} & reference to serial number
        \mintinline{sql}{twopf_samples.serial} or
        \mintinline{sql}{threepf_samples.serial} (depending whether the data is for
        a 2pf or 3pf task) to which this row corresponds \\
        \mintinline{sql}{workgroup} & workgroup number for the worker that processed this configuration \\
        \mintinline{sql}{worker} & worker number for the worker that processed this configuration \\
        \mintinline{sql}{integration_time} & time required to perform integration, in nanoseconds \\
        \mintinline{sql}{batch_time} & time required to transfer the integration data into storage, in nanoseconds \\
        \mintinline{sql}{steps} & number of steps taken by the integrator. Can be used to check whether
        the integrator is performing an excessive amount of work, which normally indicates that a different
        algorithm might give better performance \\
        \mintinline{sql}{refinements} & number of times the density of the time-sample mesh needed to be
        increased. Not used in the current version of {\CppTransport}.
    \end{tabular}
\end{enumerate}
\end{sqltablelist}
In addition, tasks sampling the 3-point function include the following extra tables:
\begin{sqltablelist}
\begin{enumerate}
    \setcounterref{enumi}{sqltables:statistics}
    \item \label{sqltable:threepf-samples}
    \mintinline{sql}{threepf_samples} --- contains the list of sample points for the 3-point function \\
    \begin{tabular}{p{2.5cm}p{11.2cm}}
        \mintinline{sql}{serial} & unique serial number for this sample configuration \\
        \mintinline{sql}{wavenumber1} & reference to \mintinline{sql}{twopf_samples.serial} identifying
        the $k_1$ mode for this configuration \\
        \mintinline{sql}{wavenumber2} & reference to \mintinline{sql}{twopf_samples.serial} identifying
        the $k_2$ mode for this configuration \\
        \mintinline{sql}{wavenumber3} & reference to \mintinline{sql}{twopf_samples.serial} identifying
        the $k_3$ mode for this configuration \\
        \mintinline{sql}{kt_conventional} & conventionally-normalized value of $k_t$ for this configuration \\
        \mintinline{sql}{kt_comoving} & comoving-normalized value of $k_t$ for this configuration \\
        \mintinline{sql}{alpha} & value of $\alpha$ for this configuration \\
        \mintinline{sql}{beta} & value of $\beta$ for this configuration \\
        \mintinline{sql}{t_exit_kt} & horizon-exit time for this configuration, in the sense
        $k_t / 3 = aH$ \\
        \mintinline{sql}{t_massless} & massless point for this configuration, given by the earliest
        of the massless points for $k_1$, $k_2$, $k_3$
    \end{tabular}

    \item \mintinline{sql}{twopf_im} --- contains the imaginary part of the field-space 2-point function \\
    The schema and storage conventions
    for this table are the same as for the real part of the 2-point function,
    \mintinline{sql}{twopf_re}.

    \item \mintinline{sql}{threepf_momentum} \\
    \mintinline{sql}{threepf_deriv} --- contains
    the real part of the field-space 3-point function \\
    These tables share the same schema. The difference is that
    \mintinline{sql}{threepf_momentum} stores correlation functions defined using
    the canonical momenta (rescaled by $a^3$), whereas
    \mintinline{sql}{threepf_deriv} stores correlation functions defined using
    derivatives with respect to $N$.
    These have a nontrivial relationship that is not easy to reconstruct without
    access to model-dependent information, and specifically the $A$, $B$ and
    $C$ tensors defined by Dias et al.~\cite{}.
    To mitigate difficulties with reconstructing one or other of these
    sets of correlation functions {\CppTransport} stores both, at the expense
    of roughly doubling the storage requirements for each 3-point function task.

    If the correlation functions produced by {\CppTransport} are to be used as
    input for a separate-universe type calculation,
    or are to be compared with the corresponding \emph{outputs},
    then it is the correlation functions involving derivatives that are required.
    \\
    \begin{tabular}{p{2.5cm}p{11.2cm}}
        \mintinline{sql}{unique_id} & unique identifier labelling this row. Not used
        unless strict data consistency checks are enabled
        by defining the macro
        \mintinline{c++}{CPPTRANSPORT_STRICT_CONSISTENCY} during
        compilation. See~\S\ref{sec:strict-consistency}. \\
        \mintinline{sql}{tserial} & reference to serial number
        \mintinline{sql}{time_samples.serial} identifying the time sample point for this row \\
        \mintinline{sql}{kserial} & reference to serial number
        \mintinline{sql}{twopf_samples.serial} identifying the wavenumber sample point for this row \\
        \mintinline{sql}{page} & page number, defined as for \mintinline{sql}{backg.page}; see
        \mintinline{sql}{eleN} below \\
        \mintinline{sql}{eleN} \ldots & \semibold{dimensionless components}
        of the real part of the field-space 3-point
        function, packed in a `right-most' ordering as for the 2-point function.
        That is, if
        $\langle \delta X^a(\vect{k}_1) \delta X^b(\vect{k}_2) X^c(\vect{k}_3) \rangle =
        (2\pi)^3 \delta(\sum_i \vect{k}_i) \alpha^{abc}$,
        then
        $N = D^2 \times \#a + D \times \#b + \#c$.

        The quantity stored is the \emph{shape function}
        $\Re(k_1^2 k_2^2 k_3^2 \alpha^{abc})$.
        As for the other data tables, this object is independent of the
        comoving normalization and can be easily remapped into
        any choice of wavenumber normalization.
    \end{tabular}
\end{enumerate}
\end{sqltablelist}

\subsubsection{Strict consistency checking}
\label{sec:strict-consistency}
In rare circumstances it may be useful to enforce strict consistency checks
within the database.
These checks ensure that there is only entry for each time- and wavenumber-configuration
combination,
and also ensure that the serial numbers
recorded in each data table
(such as \mintinline{sql}{backg}, \mintinline{sql}{twopf_re} and so on)
refer to valid entries in the sample tables
\mintinline{sql}{time_samples},
\mintinline{sql}{twopf_samples}
and
\mintinline{sql}{threepf_samples}.
The price paid for such checks is substantially reduced
\mintinline{sql}{INSERT} performance during aggregation, by at least a factor of
two but sometimes more.

To enable strict database checks, edit the main header file
\file{transport.h}
to define the macro
\mintinline{c++}{CPPTRANSPORT_STRICT_CONSISTENCY}, or define it by hand
before including any {\CppTransport} headers.
When enabled, this forces {\CppTransport} to use the
\mintinline{sql}{unique_id} fields in each data table.
These contain a unique integer encoding the
time-sample, wavenumber-sample and page number for the row.
The extra time cost arises because for each
\mintinline{sql}{INSERT} {\SQLite} must check
the existing database to ensure there are no conflicts.
In a worst-case scenario
this means that \mintinline{sql}{INSERT} performance could degrade
in proportion to the size of the main database, although typically
this is not realized because {\CppTransport} takes special measures to avoid it.
Nevertheless, if these checks are performed then
\emph{some} degradation must be expected as the size of the database
grows.

Normally it is best to leave strict checks disabled unless you need them for some special
purpose. For example, this situation might arise if
you want uniqueness to be enforced via
\mintinline{sql}{PRIMARY KEY} constraints after the database has been constructed;
{\SQLite} does not allow primary keys to be added later.
Alternatively you could consider post-processing the database to add your
own unique identifiers and creating a
\mintinline{sql}{UNIQUE} index.

\section{Options for integration tasks}
\label{sec:int-options}

\subsection{General options}
\label{sec:general-integration-options}
This section documents the options that can be used when customizing
integration tasks.
Typically options are set using member functions provided by
each task object, such as
the \mintinline{c++}{twopf_task} and \mintinline{c++}{threepf_cubic_task}
objects used in~\S\ref{sec:add-integration-task}.

\begin{itemize}
    \item \mintinline{c++}{set_name(std::string)} \\
    Sets the task name to the supplied string

    \item \mintinline{c++}{set_description(std::string)} \\
    Sets an (optional) description field. Reports
    for this task (either obtained
    at the command line by the \option{{-}{-}info} switch, or using the
    HTML report generator) include this description field where it is
    present.

    \item
    \label{enum:adpative-ics}
    \mintinline{c++}{bool get_adaptive_ics()} \\
    \mintinline{c++}{set_adaptive_ics(bool)} \\
    Get or set the current adaptive initial conditions state.

    {\CppTransport} sets initial conditions for each $n$-point function
    by working in the universal, massless approximation.
    This will be valid provided the initial time is set substantially earlier
    than the massless point for each wavenumber
    participating in the correlation function,
    as defined on p.\pageref{enum:massless-time}
    in~\S\ref{sec:examine-k-database}.
    The massless time is not later than the horizon exit time for the corresponding
    wavenumber, but can be earlier if there is a massive mode in the spectrum.

    Generally, increasing the number of e-folds of massless evolution
    will improve {\CppTransport}'s estimate of the correlation functions
    at horizon exit.
    The accuracy is these estimates is the principal factor determining accuracy
    of the evolution in the superhorizon regime.
    Unfortunately, subhorizon evolution is very expensive.
    In this phase the individual wavefunctions oscillate with exponentially
    increasing frequency
    as we move deeper into the subhorizon epoch.
    The correlation functions themselves do not oscillate,
    and therefore to ensure accuracy the stepper must adjust its
    step size so that counterbalancing oscillatory contributions cancel
    to an acceptable precision.
    The maximum acceptable step size decreases rapidly on subhorizon
    scales.

    In normal circumstances
    this means we wish to have just enough e-folds of massless evolution,
    but not more.
    Usually it is sufficient to have between $3$ and $5$ e-folds of massless evolution,
    with $4$ a good default choice.
    To make this possible
    {\CppTransport} supports two ways to set initial conditions for each $n$-point
    function---\emph{uniform} and \emph{adaptive}.
    With uniform initial conditions
    {\CppTransport} will compute an initial condition for each $n$-point function
    at the initial time specified in the initial conditions package
    (\S\ref{sec:add-integration-task})
    and evolve it from there.
    The advantage is that each correlation function gets the same initial time,
    so it is possible to compare their values at any subsequent point.
    The disadvantage is that this initial time may give some $n$-point functions
    a very large number of e-folds of massless evolution.
    Often this makes the corresponding integrations impracticably expensive.

    The alternative is adaptive initial conditions.
    Using adaptive initial conditions, {\CppTransport} will compute a customized
    initial condition a fixed number of e-folds before the massless point
    for the correlation function in question.
    (As explained on p.\pageref{sqltable:threepf-samples}, the massless point for
    a bispectrum configuration is defined to be the earliest massless point
    among all participating wavenumbers.)
    This choice is often dramatically faster than using uniform initial conditions.
    In many cases it is the only feasible way to perform the calculation.
    The disadvantage is that not all correlation functions will have
    sample data available at early times, because sampling
    begins at different times for different configurations.
    {\CppTransport} will only keep sample points for which all configurations have
    data available.

    \semibold{Adaptive initial conditions are the recommended setting
    for all {\CppTransport} tasks.}
    However, because they impose more stringent requirements on the initial
    conditions they are not enabled by default.
    This can be done by passing the value \mintinline{c++}{true} to
    \mintinline{c++}{set_adaptive_ics()},
    in which case the number of massless e-folds will default to 4,
    or by calling
    \mintinline{c++}{set_adaptive_ics_efolds()}
    with an alternative value.

    If the initial time specified as part of the initial conditions package
    does not allow this number of e-folds prior to the
    earliest massless point, then the task will throw
    a \mintinline{c++}{transport::runtime_exception} exception.

    \item \mintinline{c++}{double get_adaptive_ics_efolds()} \\
    \mintinline{c++}{set_adaptive_ics_efolds(double N)} \\
    Get or set the number of e-folds of massless evolution used by adapative
    initial conditions.

    \item \mintinline{c++}{double get_astar_normalization()} \\
    \mintinline{c++}{set_astar_normalization(double)} \\
    \label{method:set_astar_normalization}
    For concrete calculations {\CppTransport} must convert conventionally-normalized
    wavenumbers to comoving-normalized ones.
    This involves a choice of scale factor $\astar$ at the distinguished
    time $\Nstar$ when the conventionally-normalized scale $k=1$ leaves the
    horizon.

    Normally it is not necessary to be aware of what choice is made for
    $\astar$, because as explained in~\S\ref{sec:using-integration-container}
    the database stores \emph{dimensionless} correlation functions.
    The normalization of these correlation functions does not depend on
    $\astar$
    and therefore they are simple to remap into whatever wavenumber
    normalization is convenient for the problem at hand.

    By default {\CppTransport} sets
    $\astar = \exp(0) = 1$.
    The current implementation of the
    transport integrator uses dimensionless
    correlation functions, and therefore the numerical
    solutions are independent of its value.
    As a result this setting can mostly be ignored, unless
    you want a specific normalization for the purpose of producing
    plots or tables.

    \item \mintinline{c++}{bool get_collect_initial_conditions()} \\
    \mintinline{c++}{set_collect_initial_conditions(bool)} \\
    Get or set the current collection state for initial conditions.

    To assist comparison of the results produced by {\CppTransport} with
    other methods of computing $n$-point functions,
    it is possible to have {\CppTransport} write extra
    table sinto the {\SQLite} container that gives the field-values
    at horizon exit for each configuration.
    For tasks computing the 2-point function only,
    {\CppTransport} will write a table called
    \mintinline{sql}{horizon_exit_values}
    that records the
    field values at horizon exit for the corresponding wavenumber.
    For tasks computing the 3-point function {\CppTransport}
    writes two tables:
    for each configuration,
    \mintinline{sql}{horizon_exit_values}
    records field values at horizon-exit for the
    wavenumber which exits
    earliest.
    In addition \mintinline{sql}{kt_horizon_exit_values}
    contains the field values at horizon-exit for the
    average scale $k_t/3$.

    Note that these are \emph{not} the initial conditions used
    by {\CppTransport} internally.

    \item \mintinline{c++}{boost::optional<unsigned int> get_default_checkpoint()} \\
    \mintinline{c++}{set_default_checkpoint(unsigned int)} \\
    \mintinline{c++}{unset_default_checkpoint()} \\
    Get or set the current default checkpoint interval.

    A default checkpoint interval can be set on a per-task basis.
    If no \option{{-}{-}checkpoint} switch is given on the command line then
    the default checkpoint interval is applied, otherwise the value
    supplied to \option{{-}{-}checkpoint} is used.
    The current checkpoint interval can be cleared
    using
    \mintinline{c++}{unset_default_checkpoint()}.

    If there is no currently-set default interval then
    \mintinline{c++}{get_default_checkpoint()} returns an empty
    \mintinline{c++}{boost::optional}
    (see \href{http://www.boost.org/doc/libs/1_60_0/libs/optional/doc/html/index.html}{here}).
    Otherwise, it returns a
    \mintinline{c++}{boost::optional} containing the current
    checkpoint interval in minutes.

\end{itemize}

\subsection{Two-point function tasks}
\label{sec:twopf-options}
Two-point function tasks need few specific options, and those that exist are supplied
through the constructor. Its general form is
\begin{minted}[linenos=false,xleftmargin=0pt]{c++}
    twopf_task(
        std::string name,
        initial_conditions ics,
        range time_samples,
        range k_samples,
        bool adaptive_ics=false
    )
\end{minted}
This constructs a task with the given name and initial conditions,
and the specified time- and wavenumber sample points.
The \mintinline{c++}{range} objects can be instances of either
\mintinline{c++}{basic_range}
or
\mintinline{c++}{aggregate_range}
(\S\ref{sec:add-integration-task}).
If the boolean
\mintinline{c++}{adaptive_ics} is supplied then it determines whether adaptive initial conditions
are automatically enabled.
If \mintinline{c++}{true} the task will default to using 4 massless e-folds, but this can be
changed by calling \mintinline{c++}{set_adaptive_ics_efolds()}.

\subsection{Three-point function tasks}
\label{sec:threepf-options}
{\CppTransport} provides two different interfaces for
specifying three-point function tasks.
The differences relate to the way in which the sample configurations are specified.

\subsubsection{Cubic $(k_1, k_2, k_3)$ mesh}
One possibility is to build a set of bispectrum configurations from
the Cartesian product
\mintinline{c++}{ks} $\times$
\mintinline{c++}{ks} $\times$
\mintinline{c++}{ks}
built out of some range of wavenumbers
\mintinline{c++}{ks}.
This option was used to build the task
\repoobject{dquad.threepf} used as an example in~\S\ref{sec:build-and-run}.

To generate a task in this way one should use the
\mintinline{c++}{threepf_cubic_task} constructor. Its signature is
\begin{minted}[linenos=false,xleftmargin=0pt]{c++}
    threepf_cubic_task(
        std::string name,
        initial_conditions ics,
        range time_samples,
        range ks,
        bool adaptive_ics=false,
        StoragePolicy storage_policy = DefaultStoragePolicy,
        TrianglePolicy triangle_policy = DefaultTrianglePolicy
    )
\end{minted}
The last three parameters are optional, and if omitted their
default values will be used.
The range \mintinline{c++}{ks} is used to construct a Cartesian mesh
of wavenumbers as explained above, and the remaining parameters
share their meaning with the constructor for two-point function tasks
described in~\S\ref{sec:twopf-options}.
For details of the
\mintinline{c++}{StoragePolicy} and
\mintinline{c++}{TrianglePolicy} concepts
see~\S\ref{sec:storage-policy} and~\S\ref{sec:triangle-policy} respectively.

There is no need to specify a set of wavenumbers at which to sample the 2-point function.
Samples will automatically be recorded at points corresponding to each wavenumber appearing
on an external leg of the 3-point function.
In this case that means the 2-point function will be sampled at the wavenumbers
in \mintinline{c++}{ks}.

\subsubsection{Fergusson--Shellard $(k_t, \alpha, \beta)$ mesh}
Sampling from a wavenumber cube \mintinline{c++}{ks}$^3$
is convenient for some purposes but awkward for others.
In particular, if the intention is to sample a series of wavenumber configurations
probing the squeezed limit of the bispectrum then there may be more efficient
ways to proceed.
A good choice is to work with parameters that directly specify the configuration
shape.
For this purpose {\CppTransport} uses the $(\alpha, \beta)$ parameters
introduced by Fergusson \& Shellard~\cite{Fergusson:2006pr}; see Eq.~\eqref{eq:alpha-beta}.

To build a task from a grid of $k_t$, $\alpha$ and $\beta$ combinations one should use
the \mintinline{c++}{threepf_alphabeta_task} constructor:
\begin{minted}[linenos=false,xleftmargin=0pt]{c++}
    threepf_alphabeta_task(
        std::string name,
        initial_conditions ics,
        range time_samples,
        range kts,
        range alpha,
        range betas,
        bool adaptive_ics=false,
        StoragePolicy storage_policy = DefaultStoragePolicy,
        TrianglePolicy triangle_policy = DefaultTrianglePolicy
    )
\end{minted}
As for \mintinline{c++}{threepf_cubic_task} the last three parameters are optional.
The sample configurations are built from restricting the Cartesian product
of the $(k_t, \alpha, \beta)$ parameters
\mintinline{c++}{kts} $\times$ \mintinline{c++}{alphas} $\times$ \mintinline{c++}{betas}
to valid triangle configurations.

As for cubic tasks, the sample points for the 2-point function are not specified separately.
{\CppTransport} will compute the wavenumbers appearing on each external leg of the
3-point function and use these to build a list of sample points for the 2-point function.
For tasks specified using combinations of $(k_t, \alpha, \beta)$ this means that the sample
points can be irregularly spaced if the sampled set of bispectrum configurations
is sparse.

\subsubsection{Specifying a storage policy}
\label{sec:storage-policy}
Sometimes the grids
\mintinline{c++}{ks} $\times$ \mintinline{c++}{ks} $\times$ \mintinline{c++}{ks}
or
\mintinline{c++}{kts} $\times$ \mintinline{c++}{alphas} $\times$ \mintinline{c++}{betas}
will include combinations that are not needed for the task at hand.
Rather than wastefully compute these, {\CppTransport} allows them to be omitted by
specifying a \emph{storage policy}.

If supplied, a storage policy should be an instance of a callable object
that accepts a reference to a \mintinline{c++}{transport::threepf_kconfig}
object and returns a valid \mintinline{c++}{transport::storage_outcome}:
\begin{minted}[linenos=false,xleftmargin=0pt]{c++}
    transport::storage_outcome (StoragePolicy)(const transport::threepf_kconfig& config)
\end{minted}
The \mintinline{c++}{transport::threepf_kconfig} data structure specifies
the configuration that is to be inspected. Its fields are public data members:
\begin{itemize}
    \item \mintinline{c++}{serial} \\
    The proposed unique serial number for this configuration.

    \item \mintinline{c++}{k1_serial}, \mintinline{c++}{k2_serial}, \mintinline{c++}{k3_serial} \\
    The serial numbers for the wavenumbers $k_1$, $k_2$, $k_3$ associated with each external leg.

    \item \mintinline{c++}{k1_conventional}, \mintinline{c++}{k2_conventional}, \mintinline{c++}{k3_conventional} \\
    The conventionally-normalized wavenumbers associated with each external leg.

    \item \mintinline{c++}{k1_comoving}, \mintinline{c++}{k1_comoving}, \mintinline{c++}{k3_comoving} \\
    The comoving-normalized wavenumbers associated with each external leg.

    \item \mintinline{c++}{kt_conventional}, \mintinline{c++}{kt_comoving} \\
    The conventional- and comoving-normalized $k_t$ associated with this configuration.

    \item \mintinline{c++}{alpha}, \mintinline{c++}{beta} \\
    The $\alpha$ and $\beta$ values associated with this configuration.

\end{itemize}
The configuration presented for inspection is guaranteed to be a valid triangle as determined by the
active \mintinline{c++}{TrianglePolicy} (\S\ref{sec:triangle-policy}).
The callable should inspect the and determine whether it should be accepted or rejected.
The default storage policy will accept all configurations.

To accept this configuration the policy should return
\mintinline{c++}{transport::storage_outcome::accept}.
To reject, it should return either:
\begin{itemize}
    \item \mintinline{c++}{transport::storage_outcome::reject_remove} \\
    When constructing this configuration, it may have been necessary to assign
    unique serial numbers to new configurations for the two-point function.
    This will happen if at least one leg of the bispectrum corresponds to a wavenumber
    that has not yet been allocated a unique number.

    If \mintinline{c++}{reject_remove} is specified then {\CppTransport} will
    reject the configuration and deallocate any serial numbers that were speculatively
    assigned to `new' two-point function configurations.

    \item \mintinline{c++}{transport::storage_outcome::reject_retain} \\
    Alternatively, \mintinline{c++}{reject_retain} allows these serial numbers to be
    retained.

    A typical use case for this functionality occurs when you wish to
    subsample a cubic grid.
    Suppose the range \mintinline{c++}{ks} contains a set
    $\{ k_1, k_2, \ldots, k_m \}$ of wavenumbers, ordered so that
    $k_i < k_j$ if $i < j$.
    Therefore $k_m$ corresponds to the shortest physical scale that will be sampled.
    The set of isosceles triangles containing this mode
    is given by the configurations
    $(k_m, k_i, k_i)$ for all $i$, and a suitable storage policy picking out
    this subset might be
    \begin{minted}{c++}
        struct StoragePolicy
          {
            transport::storage_outcome operator()(const transport::threepf_kconfig& data)
              {
                if(data.k1_serial == m) return(transport::storage_outcome::accept);
                else                    return(transport::storage_outcome::reject_retain);
              }
          };
    \end{minted}
    Each configuration will be presented to
    \mintinline{c++}{StoragePolicy} for inspection in the order
    determined by the Cartesian product \mintinline{c++}{ks}$^3$,
    so by returning
    \mintinline{c++}{reject_retain} we can ensure that the
    serial numbers assigned to the 2-point function configurations
    will match the ordering in \mintinline{c++}{ks}.
\end{itemize}

\subsubsection{Specifying a triangle policy}
\label{sec:triangle-policy}
{\CppTransport} also allows specification of a \emph{triangle policy}.
This is used to determine which members of the grids
\mintinline{c++}{ks} $\times$ \mintinline{c++}{ks} $\times$ \mintinline{c++}{ks}
and
\mintinline{c++}{kts} $\times$ \mintinline{c++}{alphas} $\times$ \mintinline{c++}{betas}
are considered valid triangles.

{\CppTransport}'s default triangle policy imposes two conditions: first,
it requires $k_1 + k_2 + k_3 \geq 2 \max \{ k_1, k_2, k_3 \}$.
This is the geometrical condition that the sides
$\{ k_1, k_2, k_3 \}$ can form a triangle
and is equivalent to demanding that the sum of the two shortest sides is longer than the
longest side.
If this is not satisfied then the `triangle' cannot be closed.

Second, it imposes the condition
$k_1 \geq k_2 \geq k_3$.
This is intended to prevent unnecessary repeat calculations.
If we are interested only in connected correlation functions
then we are free to reorder the fields inside any
equal-time
correlation function
$\langle \delta X^a(\vect{k}_1) \delta X^b(\vect{k}_2) X^c(\vect{k}_3) \rangle$
because the failure of any pair of $X$s to commute will produce
contact terms.
Because {\CppTransport} solves the correlation functions for
all index combinations $(a,b,c)$, this reordering property
implies that it is only necessary to compute those configurations
satisfying $k_1 \geq k_2 \geq k_3$.
The remaining configurations can be determined by
permuting the index labels.

For some purposes, however, it is convenient to drop this restriction
or to adopt a different ordering. One use case is to produce
bispectrum shape plots in the $(\alpha, \beta)$ plane.
The restriction $k_1 \geq k_2 \geq k_3$ implies that
{\CppTransport} will produce correlation functions that cover only
$1/3! = 1/6$ of the allowed $(\alpha, \beta)$ space,
and in principle the remaining
regions can be determined by suitable transformations
of the region that is computed.
However, this transformation will map a rectangular mesh
in the computed region
to a non-rectangular mesh in the other regions.
Many plotting libraries require a uniform rectangular mesh
of points in order to produce surface or 3-dimensional plots,
making these non-rectangular transformed meshes unacceptable.
In this case, instead of splining the transformed regions and resampling
them on a rectangular mesh, it can be simpler
to absorb the factor of 6 computation cost and have
{\CppTransport} compute these points directly.

A triangle policy is an instance of a callable object that
determines whether a given configuration forms a triangle.
The `cubic' and `$\alpha\beta$' constructors require different types of policy:
\begin{itemize}
    \item for a `cubic' constructor, the triangle policy should accept 6 arguments
    and return a \mintinline{c++}{bool} indicating whether the configuration is valid:
    \begin{minted}[linenos=false,xleftmargin=0pt]{c++}
        bool (TrianglePolicy)(unsigned i, unsigned j, unsigned k, double k1, double k2, double k3)
    \end{minted}
    The arguments \mintinline{c++}{i}, \mintinline{c++}{j}, \mintinline{c++}{k}
    refer to the index of the $k_1$, $k_2$, $k_3$ values (respectively)
    within the original range \mintinline{c++}{ks}.
    The arguments \mintinline{c++}{k1}, \mintinline{c++}{k2}, \mintinline{c++}{k3}
    give these (conventionally-normalized) wavenumber values directly.

    The default {\CppTransport} policy rejects triangles according to the following
    criteria:%
        \footnote{In reality the policies do not perform literal \mintinline{c++}{<}
        or \mintinline{c++}{>} comparisons because of possible issues with
        floating point arithmetic. These subtleties are ignored here, and the
        policy is presented as if it were implemented purely using its
        mathematical definition. The same applies to the default
        policy for `$\alpha\beta$' tasks.}
    \begin{minted}{c++}
        // impose ordering k1 > k2 > k3
        if(i < j) return false;
        if(j < k) return false;

        // impose triangle
        double max = std::max(std::max(k1, k2), k3);
        return(k1 + k2 + k3 - 2.0*max >= 0.0);
    \end{minted}

    \item for an `$\alpha\beta$' task, the triangle policy should accept just two
    arguments corresponding to the values of $\alpha$ and $\beta$,
    \begin{minted}[linenos=false,xleftmargin=0pt]{c++}
        bool (TrianglePolicy)(double alpha, double beta)
    \end{minted}
    The default policy rejects triangles as follows:
    \begin{minted}{c++}
        // beta should lie between 0 and 1
        if(beta < 0.0) return false;
        if(beta > 1.0) return false;

        // alpha should lie between 1-beta and beta-1
        if(beta - 1.0 - alpha > 0.0) return false;
        if(alpha - (1.0 - beta) > 0.0) return false;

        // impose k1 > k2 > k3
        if(alpha < 0) return false;
        if(beta - (1.0 + alpha)/3.0 < 0.0) return false;
    \end{minted}
    In addition, it requires that none of the `squeezing' measures
    $1-\beta$,
    $|1+\alpha-\beta|$ and
    $|1-\alpha+\beta|$ becomes too small.
    By default the cutoff is taken to be $10^{-8}$.

\end{itemize}

\section{Adding postintegration tasks}
\label{sec:post-integration-tasks}
Although integration tasks are always the first step in using {\CppTransport},
for most purposes the $n$-point functions they produce need to be reprocessed
into observables.
At a minimum we normally require predictions for the correlation functions of
the primordial curvature perturbation
$\zeta$.
It is $\zeta$ that eventually determines the statistics of the
observed density perturbation---although it may happen
(where an adiabatic limit is not acheived by the end of inflation)
that
these observable statistics
also depend on post-inflationary
physics that is not handled by {\CppTransport}.
In such cases the $\zeta$ correlation functions produced by
the {\CppTransport} platform should be used as initial
conditions for a Boltzmann solver (or equivalent)
that is capable of following their evolution through the
subsequent radiation- and matter-dominated phases.

\subsection{$\zeta$ tasks for the two- and three-point functions}
Post-processing or `postintegration' tasks accept output
from integration tasks (or other postintegration tasks)
and convert it into some other form.
To produce $\zeta$ correlation functions we need to specify
post-processing tasks that can perform the gauge transformation
from field space into $\zeta$.

Returning to the example of double-quadratic inflation,
we can `connect' suitable two- and three-point function
tasks for $\zeta$ to the existing integration tasks
\mintinline{c++}{tk2} and
\mintinline{c++}{tk3}.
It is only necessary to add the following lines at the end of
the
\mintinline{c++}{write_tasks()} function:
\begin{minted}{c++}
    transport::zeta_twopf_task<> ztk2("dquad.twopf-zeta", tk2);
    ztk2.set_description("Convert the output from dquad.twopf into a zeta 2-point function");

    transport::zeta_threepf_task<> ztk3("dquad.threepf-zeta", tk3);
    ztk3.set_description("Convert the output from dquad.threepf into zeta 2- and 3-point functions");

    repo.commit(ztk2);
    repo.commit(ztk3);
\end{minted}
The constructor for
\mintinline{c++}{zeta_twopf_task} accepts a name and an instance of a
\mintinline{c++}{twopf_task}, whereas
the constructor for
\mintinline{c++}{zeta_threepf_task} accepts a name and an instance of
a generic
\mintinline{c++}{threepf_task}.
The 3-point function task may
have been constructed using either
\mintinline{c++}{threepf_cubic_task} or
\mintinline{c++}{threepf_alphabeta_task}.
It is not possible to use
a \mintinline{c++}{twopf_task} with
\mintinline{c++}{zeta_threepf_task}
or a \mintinline{c++}{threepf_task} with
\mintinline{c++}{zeta_twopf_task}.

After constructing
\mintinline{c++}{ztk2} and
\mintinline{c++}{ztk3}
they are committed to the
repository
database using \mintinline{c++}{repo.commit()},
as for any repository object.
The original
\mintinline{c++}{commit()}
calls for
\mintinline{c++}{tk2} and
\mintinline{c++}{tk3} can be removed because
{\CppTransport} will realize that
these integration tasks must be stored in the
repository database in order for
\mintinline{c++}{ztk2} and
\mintinline{c++}{ztk3} to make sense.

Build the \file{dquad} executable as before and invoke it
with an instruction to execute the
new \repoobject{dquad.threepf-zeta} task:
\begin{minted}{bash}
    mpiexec -n 4 dquad -v -r test-zeta --create --task dquad.threepf-zeta
\end{minted}
As usual, the \mintinline{bash}{-n} argument of
\mintinline{bash}{mpiexec} should be adjusted to use a suitable
number of processes on your machine.
{\CppTransport} will print short updates as it carries out each
task:
\begin{minted}[linenos=false,xleftmargin=0pt,bgcolor=blue!10]{text}
    Task manager: processing task 'dquad.threepf' (1 of 2)
    Committed content group '20160516T130625' for task 'dquad.threepf' at 2016-May-16 14:12:44
    Task manager: processing task 'dquad.threepf-zeta' (2 of 2)
    Committed content group '20160516T131244' for task 'dquad.threepf-zeta' at 2016-May-16 14:13:50
    Task manager: processed 2 database tasks in wallclock time 7m 24.5s | time now 2016-May-16 14:13:50
\end{minted}
Although we only asked for the
\repoobject{dquad.threepf-zeta} task,
{\CppTransport} has scheduled execution of both
\repoobject{dquad.threepf} and \repoobject{dquad.threepf-zeta}.
This happens because it is aware that
\repoobject{dquad.threepf-zeta} needs content
from \repoobject{dquad.threepf} before it can be processed.
If no content groups are available for
\repoobject{dquad.threepf} then it is added to the list of tasks
to be processed.
The same applies for any chain of dependent tasks;
if a task specified on the command line depends on output from
some other task for which no content groups are available,
{\CppTransport} will automatically schedule execution of these tasks.

If we now re-run the same command
(removing the unnecessary option \option{{-}{-}create}),
{\CppTransport} will notice that
content for \repoobject{dquad.threepf} is already available
and assume that it should be used to feed
the new execution of
\repoobject{dquad.threepf-zeta}:
\begin{minted}[bgcolor=blue!10]{text}
    $ mpiexec -n 4 dquad -v -r test-zeta --create --task dquad.threepf-zeta
    Committed content group '20160516T131956' for task 'dquad.threepf-zeta' at 2016-May-16 14:21:02
    Task manager: processed 1 database task in wallclock time 1m 6.97s | time now 2016-May-16 14:21:02
\end{minted}
Checking the number of content groups attached to each task using
\option{{-}{-}status} shows that
\repoobject{dquad.threepf-zeta} now has 2 groups, while
\repoobject{dquad.threepf} has only one:
\begin{minted}[bgcolor=blue!10]{text}
    $ ./dquad -r test-zeta/ --status
    Available tasks:
    Task                Type                      Last activity  Outputs
    dquad.threepf       integration        2016-May-16 13:12:44        1
    dquad.threepf-zeta  postintegration    2016-May-16 13:21:02        2
    dquad.twopf         integration        2016-May-16 13:06:25        0
    dquad.twopf-zeta    postintegration    2016-May-16 13:06:25        0
\end{minted}

\subsection{Applying tags to control which content groups are used}
\label{sec:apply-tags}
When several suitable content groups are available
to feed into a postintegration task,
{\CppTransport} must decide which one should be used.
By default it will use the most recently-generated group, but this
not always the appropriate choice.
For example, we might have several content groups attached to the
task \repoobject{dquad.threepf}, perhaps generated using
different steppers or different numerical tolerances.
To control which content groups are used by postintegration tasks
{\CppTransport} allows \emph{tags} to be attached to each group.
These are short, descriptive strings that label some property
of the group.
(Of course, these labels may be useful in their own right in addition
to their ability to control selection of content groups.)

Using the \option{{-}{-}tag} switch,
tags can be attached to content groups at the time of generation.
Alternatively
they may be applied later
using \option{{-}{-}add-tag}
(see~\S\ref{sec:add-remove-tags}).
For example, suppose we generate two content groups
for the same integration task,
one
using the
Dormand--Prince $4^{\mathrm{th}}$/$5^{\mathrm{th}}$-order
stepper and another using the
Fehlberg $7^{\mathrm{th}}$/$8^{\mathrm{th}}$-order stepper.
This could be done
by compiling two different executables using different
model description files, or by editing the model description file
and rebuilding.
Using \option{{-}{-}tag},
we can tag each of these content groups appropriately:
\begin{minted}{bash}
    $ mpiexec -n 4 dquad -v -r test-tags --create --task dquad.threepf --tag stepper-dopri5
    $ mpiexec -n 4 dquad -v -r test-tags --task dquad.threepf --tag stepper-fehlberg78
\end{minted}
It can be checked that the tags have been correctly applied
by using \option{{-}{-}info} with the content group
name, or by generating an HTML report
and viewing the group records.
We could have specified multiple tags, if necessary, by repeating
the \option{{-}{-}tag} argument for each one.
There is no limit to the number of tags that can be applied.

To compare the results from each stepper we need to obtain
$\zeta$ predictions from each content group
by feeding them to
\repoobject{dquad.threepf-zeta}.
This can be done by specifying the appropriate tag (or group of tags)
when we launch the \repoobject{dquad.threepf-zeta} job:
\begin{minted}{bash}
    $ mpiexec -n 4 dquad -v -r test-tags --tag stepper-dopri5 --task dquad.threepf-zeta
\end{minted}
This time, the option
\option{{-}{-}tag} has two meanings:
first, as for an integration task,
the resulting content group will be
tagged with whatever labels we specify, here
`\grouptag{stepper-dopri5}'.
Second, when searching for content groups to use as a source,
{\CppTransport} will restrict attention to those with matching tags.
If multiple tags are specified then {\CppTransport} will attempt to match
them all,
and if no suitable content groups are found
then the task will fail with an error:%
    \footnote{If a task has no content groups, but tags are specified on the
    command line, then {\CppTransport} will schedule execution of the corresponding
    task. The generated content group will be labelled with the given tags.
    But if content groups \emph{are} present, simply without matching tags, then
    {\CppTransport} will issue an error.
    The presumption is that tags
    have been specified with an intention to pick out
    a content group with specific properties.}
\begin{minted}[bgcolor=blue!10]{text}
    $ mpiexec -n 4 dquad -v -r test-tags --tag no-matching-tag --task dquad.threepf-zeta
    Repository error: no matching content groups for task 'dquad.threepf'
\end{minted}

This process can be repeated.
For example, in~\S\ref{sec:derived-products}
we will see how to build derived products and generate them using output tasks.
Each output task draws its data from a collection of integration and
postintegration tasks such as \repoobject{dquad.threepf}
and \repoobject{dquad.threepf-zeta}.
By specifying a collection of tags while executing the output task we can cause it
to select input data that satisfy specific labels
such as
\grouptag{stepper-dopri5}
or
\grouptag{stepper-fehlberg78}.

\para{Selecting a content group by name}
Alternatively, it is possible to select a content group by name.
To do this,
specify the name as a tag.
The resulting content groups will
themselves be tagged with the original group name,
meaning that in principle it is possible to use this feature
to follow chains
of groups back to an original integration content group.
In practice, however, this kind of dependency tracking is best performed
using HTML reports. These can generate dependency diagrams that summarize the
interrelation between content groups more efficiently than a large number of tags.

The option to specify content groups by name remains useful
when it is necessary to force a particular group to be used.

\subsection{Paired $\zeta$ tasks}
It sometimes happens that the output from an integration task is of limited
interest by itself.
For example, if our interest is to determine the observational viability
of a particular
inflationary model then we only require predictions for $\zeta$,
at least if a suitable adiabatic limit is reached during the inflationary
phase.
In these circumstances
the content group generated by an integration task
will be used only for post-processing by other postintegration tasks.
This is such a common occurrence that {\CppTransport} provides a short-cut
enabling a pair of coupled integration/postintegration tasks
to be processed simultaneously.
Tasks coupled in this way are said to be \emph{paired}.

To pair
the postintegration tasks
\mintinline{c++}{ztk2} and
\mintinline{c++}{ztk3} with their parent integration tasks, use
the
\mintinline{c++}{set_paired()} method.
The {\CC} code used to generate these tasks could be replaced by
\begin{minted}{c++}
    transport::zeta_twopf_task<> ztk2("dquad.twopf-zeta", tk2);
    ztk2.set_description("Convert the output from dquad.twof into a zeta 2-point function");
    ztk2.set_paired(true);

    transport::zeta_threepf_task<> ztk3("dquad.threepf-zeta", tk3);
    ztk3.set_description("Convert the output from dquad.threepf into zeta 2- and 3-point functions");
    ztk3.set_paired(true);

    repo.commit(ztk2);
    repo.commit(ztk3);
\end{minted}
Once tasks have been paired in this way:
\begin{itemize}
    \item asking {\CppTransport} to execute the \emph{integration} task will generate
    a content group for it, as usual. This content group can be used
    in the same way as any other.

    \item asking {\CppTransport} to execute the \emph{postintegration} task
    will commence simultaneous processing of the integration/postintegration pair.
    The postintegration task will not look for an existing content group
    attached to the integration task, as described above;
    a new integration content group is generated every time.
    To mark that this content group is paired with a postintegration group,
    {\CppTransport} will add the suffix `-paired' to its name.

    Notice that this means it isn't possible to select a pre-existing
    content group to use with a paired postintegration task.
\end{itemize}
During simultaneous processing, the gauge transformation from field-space
to $\zeta$ will be calculated on-the-fly, without writing all the integration data
out to a database and then reading it back again as with unpaired
post-processing. This means that paired tasks normally execute
more quickly than unpaired tasks.

\para{Task options}
It is possible to specify independent checkpoint intervals for the tasks
in an integration/postintegration pair, but in the event of failure
the data stored in each container will be synchronized.
In practice this means that data in one container that has no counterpart
in the other will be discarded,
and therefore there is no real use case for unequal checkpoint intervals.

\subsection{Using $\zeta$ {\SQLite} data containers}
\S\ref{sec:using-integration-container}
described the database containers used to hold data products from
an integration task.
The data products from postintegration tasks are handled very similarly.
In particular, each content group will have the same physical layout
shown in Fig.~\ref{fig:integration-content}.
All $n$-point functions and supporting metadata
are aggregated into a main container
\file{data.sqlite},
and logs are left in the directory \file{logs}.

\subsubsection{$\zeta$ two-point function tasks}
The {\SQLite} container for a task generating the $\zeta$ two-point function will container
the following tables:
\begin{sqltablelist}
\begin{enumerate}
	\item \mintinline{sql}{time_samples} --- contains the list of time sample points. \\
	This table duplicates, and is inherited from, the one described in \S\ref{sec:using-integration-container}
	(see p.\pageref{sqltable:time-samples}).

	\item \mintinline{sql}{twopf_samples} --- contains the list of sample points for the 2-point function \\
	This table duplicates, and is inherited from, the one described in \S\ref{sec:using-integration-container}
	(see p.\pageref{sqltable:twopf-samples}).

	\item \mintinline{sql}{zeta_twopf} --- contains the real part of the $\zeta$ 2-point function \\
    \begin{tabular}{p{2.5cm}p{11.2cm}}
        \mintinline{sql}{unique_id} & unique identifier labelling this row. Not used
        unless strict data consistency checks are enabled
        by defining the macro
        \mintinline{c++}{CPPTRANSPORT_STRICT_CONSISTENCY} during
        compilation. See~\S\ref{sec:strict-consistency}. \\
        \mintinline{sql}{tserial} & reference to serial number \mintinline{sql}{times_samples.serial}
        identifying the time sample point for this row \\
        \mintinline{sql}{kserial} & reference to serial number \mintinline{sql}{twopf_samples.serial}
        identifying the wavenumber sample point for this row \\
        \mintinline{sql}{twopf} & \semibold{dimensionless value} of
        the real part of the $\zeta$ 2-point function,
        $\langle \zeta(\vect{k}_1) \zeta(\vect{k}_3) \rangle = (2\pi)^3 \delta(\vect{k}_1 + \vect{k}_2)
        \Pzeta$.

        The value stored is $\Re(k^3 \Pzeta)$.
    \end{tabular}

    \item \mintinline{sql}{gauge_xfm1} --- contains the linear gauge transformation $N_a$ \\
    The required Fourier-space gauge transformation has the form
    \begin{equation}
    	\zeta(\vect{k}) =
    	N_a(\vect{k})
    	\delta X^a(\vect{k})
    	+
    	\frac{1}{2}
    	\int \frac{\d^3 q}{(2\pi)^3} \frac{\d^3 r}{(2\pi)^3}
    	\delta(\vect{k} - \vect{q} - \vect{r})
    	N_{ab}(\vect{q},\vect{r})
    	\delta X^a(\vect{q})
    	\delta X^b(\vect{r})
    	+ \cdots ,
    	\label{eq:gauge-xfm}
    \end{equation}
	where $\delta X^a$, $\delta X^b$ (and so on) label fluctuations over the full phase space
	of fields and their derivatives with respect to $N$,
	and the omitted terms represented by `$\cdots$' are higher order
	$X$.
	Expressions for the transformation coefficients
	$N_a$ and $N_{ab}$ may be extracted from Dias et al.~\cite{Dias:2014msa}.
	In the notation of that reference,
	the expressions used by {\CppTransport} correspond to
	$\zeta_1^{\text{simple}}$
	and
	$\zeta_2^{\text{simple}}$,
	making $N_a$ independent of $\vect{k}$ in practice.

	The $\zeta$ post-processing tasks write these transformation coefficients into the
	data containers.
	Specifically, the table
	\mintinline{sql}{gauge_xfm1} is associated with production of the
	$\zeta$ power spectrum,
	for which only $N_a$ is required.
    Although the expression used for $N_a$ in current versions is
    momentum independent, {\CppTransport} does not assume this will always be the
    case and records the gauge transformation for each momentum configuration. \\
    \begin{tabular}{p{2.5cm}p{11.5cm}}
        \mintinline{sql}{unique_id} & unique identifier labelling this row. Not used
        unless strict data consistency checks are enabled
        by defining the macro
        \mintinline{c++}{CPPTRANSPORT_STRICT_CONSISTENCY} during
        compilation. See~\S\ref{sec:strict-consistency}. \\
    	\mintinline{sql}{tserial} & 	reference to serial number \mintinline{sql}{times_samples.serial}
        identifying the time sample point for this row \\
        \mintinline{sql}{kserial} & reference to serial number \mintinline{sql}{twopf_samples.serial}
        identifying the wavenumber sample point for this row \\
        \mintinline{sql}{page} & page number, defined as for \mintinline{sql}{backg.page}
        in~\S\ref{sec:using-integration-container}; see p.\pageref{sqltable:backg} \\
        \mintinline{sql}{eleM} \ldots & components of $N_a$ for this
        momentum configuration, packed with $M= \#a$.
        Here, as in~\S\ref{sec:using-integration-container}, `$\#$'
        is the map that takes a species label to its numeric equivalent
        defined by the order of declaration in the model description file.
        All fields come first, followed by all field derivatives in the same order.
        Notice that
		the gauge transformation~\eqref{eq:gauge-xfm} is taken to be written in
		terms of \emph{derivatives}, as in Ref.~\cite{Dias:2014msa}, and not canonical momenta.

		Notice that because $\zeta$ is dimensionless but the $\delta X^a$ have dimensions of
		mass, $N_a$ has dimensions of inverse mass. It therefore scales with the
		value assigned to the Planck scale $\Mp$.
    \end{tabular}
\end{enumerate}
\end{sqltablelist}

\subsubsection{$\zeta$ three-point function tasks}
In addition to the generic tables described above,
postintegration tasks that generate the $\zeta$ three-point function also
write the following tables to the data container:
\begin{sqltablelist}
\begin{enumerate}
	\item \mintinline{sql}{threepf_samples} --- contains the list of sample points for the 3-point function \\
	This table duplicates, and is inherited from, the one decribed in \S\ref{sec:using-integration-container}
	(see p.\pageref{sqltable:threepf-samples}).

	\item \mintinline{sql}{threepf} --- contains the real part of the $\zeta$ 3-point function \\
	\begin{tabular}{p{2.5cm}p{11.5cm}}
        \mintinline{sql}{unique_id} & unique identifier labelling this row. Not used
        unless strict data consistency checks are enabled
        by defining the macro
        \mintinline{c++}{CPPTRANSPORT_STRICT_CONSISTENCY} during
        compilation. See~\S\ref{sec:strict-consistency}. \\
        \mintinline{sql}{tserial} & reference to serial number \mintinline{sql}{times_samples.serial}
        identifying the time sample point for this row \\
        \mintinline{sql}{kserial} & reference to serial number \mintinline{sql}{twopf_samples.serial}
        identifying the wavenumber sample point for this row \\
        \mintinline{sql}{threepf} & \semibold{dimensionless value} of
        the real part of the $\zeta$ 3-point function.
        If this is taken to satisfy
        $\langle \zeta(\vect{k}_1) \zeta(\vect{k}_2) \zeta(\vect{k}_3) \rangle
        = (2\pi)^3 \delta(\vect{k}_1 + \vect{k}_2 + \vect{k}_3) \Bzeta$
        then the stored quantity is
        the dimensionless shape function
        $k_1^2 k_2^2 k_3^2 \Bzeta$. \\
        \mintinline{sql}{redbsp} & \semibold{dimensionless reduced bispectrum},
        defined by
        \begin{equation}
        	\frac{6}{5} \fNL(k_1, k_2, k_3) = \frac{\Bzeta(k_1, k_2, k_3)}
        		{\Pzeta(k_1) \Pzeta(k_2) +
        		 \Pzeta(k_1) \Pzeta(k_3) +
        		 \Pzeta(k_2) \Pzeta(k_3)} .
            \label{eq:reduced-bispectrum}
        \end{equation}
	\end{tabular}

	\item \mintinline{sql}{gauge_xfm2_123} \\
    \mintinline{sql}{gauge_xfm2_213} \\
    \mintinline{sql}{gauge_xfm2_312} --- contains the quadratic gauge transformation $N_{ab}$ \\
    For each bispectrum configuration there are three possible assignments
    for $N_{ab}$.
    {\CppTransport} writes them all to the data container, in three separate tables
    with the mapping: \\
    \begin{tabular}{p{2.5cm}p{11.5cm}}
    	\mintinline{sql}{gauge_xfm_123} & $N_{ab}(\vect{k}_2, \vect{k}_3)$ \\
    	\mintinline{sql}{gauge_xfm_213} & $N_{ab}(\vect{k}_1, \vect{k}_3)$ \\
    	\mintinline{sql}{gauge_xfm_312} & $N_{ab}(\vect{k}_1, \vect{k}_2)$
    \end{tabular}
	The columns in each table are: \\
	\begin{tabular}{p{2.5cm}p{11.5cm}}
        \mintinline{sql}{unique_id} & unique identifier labelling this row. Not used
        unless strict data consistency checks are enabled
        by defining the macro
        \mintinline{c++}{CPPTRANSPORT_STRICT_CONSISTENCY} during
        compilation. See~\S\ref{sec:strict-consistency}. \\
		\mintinline{sql}{tserial} & reference to serial number \mintinline{sql}{times_samples.serial}
        identifying the time sample point for this row \\
        \mintinline{sql}{kserial} & reference to serial number \mintinline{sql}{twopf_samples.serial}
        identifying the wavenumber sample point for this row \\
        \mintinline{sql}{page} & page number, defined as for \mintinline{sql}{backg.page}
        in~\S\ref{sec:using-integration-container}; see p.\pageref{sqltable:backg} \\
        \mintinline{sql}{eleM} \ldots & components of $N_{ab}$ for this
        momentum configuration and momentum assignment, packed with the right-most
        ordering $M = D \times \#a + \#b$.
        As for the linear gauge transformation, the expressions assume~\eqref{eq:gauge-xfm}
        is written in terms of derivatives with respect to e-folding $N$
        rather than canonical momenta.
	\end{tabular}
\end{enumerate}
\end{sqltablelist}
On superhorizon scales all the gauge transformation coefficients become independent of
momenta.

\subsection{Inner-product tasks to compute $\fNL$-like amplitudes}
\label{sec:inner-products}
Production of $\zeta$ correlation functions
is not the only kind of post-processing that could be
considered.
In principle it is possible to implement {\CppTransport} postintegration
tasks to compute almost any observable of interest,
but at present only one option is offered---an inner product between the
numerically-computed bispectrum and one of the standard bispectrum shapes.

To generate such inner products we use an
\mintinline{c++}{fNL_task}.
{\CppTransport} uses this name because, \emph{under certain circumstances},
the amplitudes they produce may be related to the $\fNL$ amplitudes
reported by surveys of the cosmic microwave background anisotropies
or the large-scale galaxy distribution.
However, this interpretation should be treated with caution.
As will be explained below,
there is no necessary relation between these inner products and
the quantities observed by some particular instrument,
and to obtain truly accurate results requires a more careful calculation.
For an example, see Ref.~\cite{Byrnes:2015dub}.

To compute an inner product
we must provide a source of 3-point function
data by attaching a \mintinline{c++}{zeta_threepf_task}.
However, {\CppTransport} places restrictions on the
source 3-point functions for which it can compute inner products.
In particular:
\begin{itemize}
	\item currently, inner products are supported only for
	3-point function tasks built from a linear grid
	(whether of cubic or $\alpha\beta$ type).
	\item the grid should be complete---no configurations should
	be dropped by the storage policy.
\end{itemize}

The inner product between two bispectra is defined to be
a sum over the corresponding dimensionless shape functions.
Given a bispectrum $B(k_1, k_2, k_3)$, its shape function is
$S \equiv (k_1 k_2 k_3)^2 B(k_1, k_2, k_3)$
and the inner product between two bispectra
$B_1$, $B_2$ satisfies~\cite{Babich:2004gb,Fergusson:2009nv}
\begin{equation}
	\iprod{B_1}{B_2} \equiv
	\sum_{\text{triangles}}
	S_1(k_1, k_2, k_3) S_2(k_1, k_2, k_3) .
	\label{eq:inner-product}
\end{equation}
Sometimes the configurations in this sum are weighted
to approximate the sensitivity of a particular instrument,
but {\CppTransport} chooses to weight all configurations equally.
With this choice
the formal sum over triangles can be understood as an integral
\begin{equation}
	\iprod{B_1}{B_2} \equiv
	\int \d k_1 \, \d k_2 \, \d k_3 \;
	S_1(k_1, k_2, k_3) S_2(k_1, k_2, k_3) .
\end{equation}

For the reasons explained above
this choice is not unique,
and even when we have settled on a weighting for the
integrand
the definition must be completed by
specifying UV and IR limits.
The numerical value of a typical inner product
$\iprod{B_1}{B_2}$
(and any conclusions that are drawn from it)
will depend on the choices that are made.
However, this arbitrariness is not really a cause for concern, because
the purpose of the inner product is not mathematical but physical.
Present-day experiments achieve only low signal-to-noise
in measurements of any individual bispectrum configuration.
Results with high signal-to-noise can be obtained only by
supposing a relationship between the amplitude of different
groups of configurations
and averaging
to obtain to cumulative contribution of the entire group.
Our choice of inner product
is useful if it happens that the averaging
involved in Eq.~\eqref{eq:inner-product}
can be regarded as a proxy for the averaging over
configurations performed by a realistic experiment.

\subsubsection{The standard templates}
{\CppTransport} knows about a number of standard `templates',
or suppositions for the relationship between the amplitude of
nearby bispectrum configurations.
These are:
\begin{itemize}
    \item \semibold{The local template.} This is defined by
    \begin{equation}
        \Bzeta^\text{local} =
            2
            \Big(
                \Pzeta(\vect{k}_1) \Pzeta(\vect{k}_2)
                + \Pzeta(\vect{k}_1) \Pzeta(\vect{k}_3)
                + \Pzeta(\vect{K}_2) \Pzeta(\vect{k}_3)
            \Big) .
    \end{equation}

    \item \semibold{The equilateral template.} This satisfies~\cite{Creminelli:2005hu}
    \begin{equation}
    \begin{split}
        \Bzeta^\text{equi} =
        6
        \Big(
        &
            {- \Pzeta(\vect{k}_1)} \Pzeta(\vect{k}_2)
            - \Pzeta(\vect{k}_1) \Pzeta(\vect{k}_3)
            - \Pzeta(\vect{k}_2) \Pzeta(\vect{k}_3)
        \\
        & -
            2 \big[ \Pzeta(\vect{k}_1) \Pzeta(\vect{k}_2
                \Pzeta(\vect{k}_3) \big]^{2/3}
        \\
        & +
            \big[
                \Pzeta(\vect{k}_1)
                \Pzeta^2(\vect{k}_2)
            \big]^{1/3}
            \Pzeta(\vect{k}_3)
            + \text{5 cyclic permutations}
        \Big)
    \end{split}
    \end{equation}

    \item \semibold{The orthogonal template.} This satisfies~\cite{Senatore:2009gt}
    \begin{equation}
    \begin{split}
        \Bzeta^\text{ortho} =
        6
        \Big(
        &
            {- 3\Pzeta(\vect{k}_1)} \Pzeta(\vect{k}_2)
            - 3\Pzeta(\vect{k}_1) \Pzeta(\vect{k}_3)
            - 3\Pzeta(\vect{k}_2) \Pzeta(\vect{k}_3)
        \\
        & -
            8 \big[ \Pzeta(\vect{k}_1) \Pzeta(\vect{k}_2
                \Pzeta(\vect{k}_3) \big]^{2/3}
        \\
        & +
            3 \big[
                \Pzeta(\vect{k}_1)
                \Pzeta^2(\vect{k}_2)
            \big]^{1/3}
            \Pzeta(\vect{k}_3)
            + \text{5 cyclic permutations}
        \Big)
    \end{split}
    \end{equation}
\end{itemize}
For any template $\Bzeta^i$
{\CppTransport} defines an associated amplitude parameter
$\fNL^i$,
\begin{equation}
    \fNL^i = \frac{5}{3} \frac{\iprod{B}{\Bzeta^i}}{\iprod{\Bzeta^i}{\Bzeta^i}}
    \qquad
    \text{(no sum on $i$)}
    \label{eq:fNL-amplitudes}
\end{equation}
When the inner products in~\eqref{eq:fNL-amplitudes}
are a good estimate for the response of
the quadratic estimator used for practical
parameter estimation in a given experiment, the
corresponding $\fNL^i$ should be a good estimate
for the amplitudes
$\fNLlocal$, $\fNLequi$, $\fNLortho$
(and so on)
measured by that experiment, assuming the
true primordial bispectrum is $B$.
On the other hand, if the inner product
$\iprod{\cdot}{\cdot}$ is not a good estimate
for the response of a practical estimator
then $\fNL^i$ computed in this way may have little or
nothing to do with the response of a real experiment.
In particular, this can happen if
$B$ is too far from scale-invariant.
The 3-dimensional inner product~\eqref{eq:inner-product}
will weight configurations near the upper limit
$\kmax$ like $\kmax^3$,
whereas a realistic CMB estimator working in $\ell$ space
would
weight such configurations like $\ellmax^2$.
This difference in dimensionality
can mean that
$\iprod{\cdot}{\cdot}$ overestimates the signal-to-noise
contributed by the high-$k$ configurations.

\subsubsection{Building an inner-product task}
To compute an inner product and associated $\fNL$-parameter
for the double quadratic model we will need a new
3-point function task build from a linear grid.
The output from this task should be attached to a
$\zeta$ task, and the output from \emph{that}
should be attached to the inner-product task.

To specify the template the constructor for
\mintinline{c++}{fNL_task<>} requires a third
argument.
The possible values technically belong to
{\CppTransport}'s visualization toolkit rather
than the underlying base platform,
and therefore
live in the
\mintinline{c++}{vis_toolkit} namespace rather than
the
\mintinline{c++}{transport} namespace.
\begin{itemize}
    \item \mintinline{c++}{vis_toolkit::bispectrum_template::local} -- local template
    \item \mintinline{c++}{vis_toolkit::bispectrum_template::equilateral} -- equilateral template
    \item \mintinline{c++}{vis_toolkit::bispectrum_template::orthogonal} -- orthogonal template
\end{itemize}

\begin{minted}{c++}
    transport::basic_range<> ks_linearspaced(kt_lo, kt_hi, 50, transport::spacing::linear);

    transport::threepf_cubic_task<> tk3_linear("dquad.threepf-linear", ics, ts, ks_linearspaced);
    tk3_linear.set_adaptive_ics_efolds(4.0);
    tk3_linear.set_description("Compute time history of the 3-point function on a linear grid");

    transport::zeta_threepf_task<> ztk3_linear("dquad.threepf-linear-zeta", tk3_linear);
    ztk3_linear.set_description("Convert the output from dquad.threepf-linear into zeta" " 2 and 3-point functions");

    transport::fNL_task<> fNL_local("dquad.fNL-local", ztk3_linear, transport::bispectrum_template::local);
    fNL_local.set_description("Compute inner product of double-quadratic bispectrum" " with local template");

    repo.commit(fNL_local);
\end{minted}
The resulting executable can be built in the usual way,
and launched using \mintinline{bash}{mpiexec}.
We need only specify the final
task \repoobject{dquad.fNL-local},
because {\CppTransport} will realise that content groups
for \repoobject{dquad.threepf-linear}
and \repoobject{dquad.threepf-linear-zeta}
are required in order to generate suitable input.

\begin{figure}
    \begin{center}
        \includegraphics[scale=0.5]{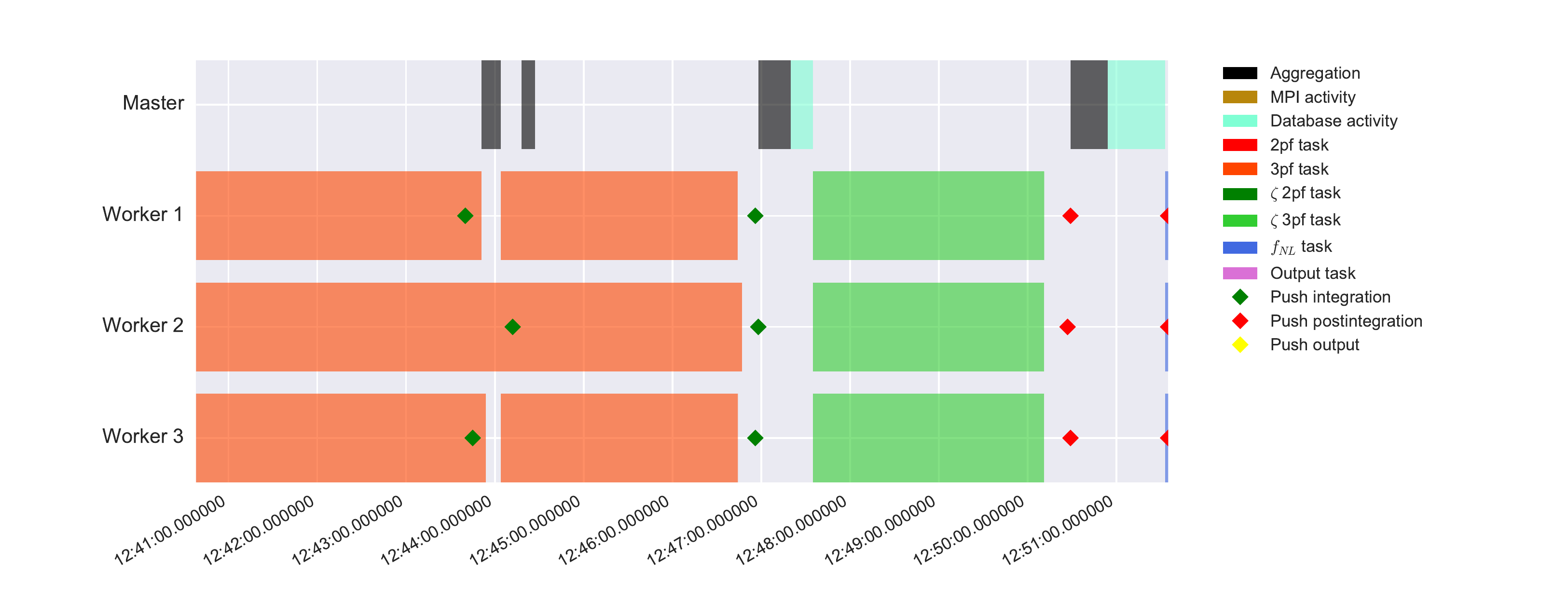}
    \end{center}
    \begin{center}
    	\includegraphics[scale=0.5]{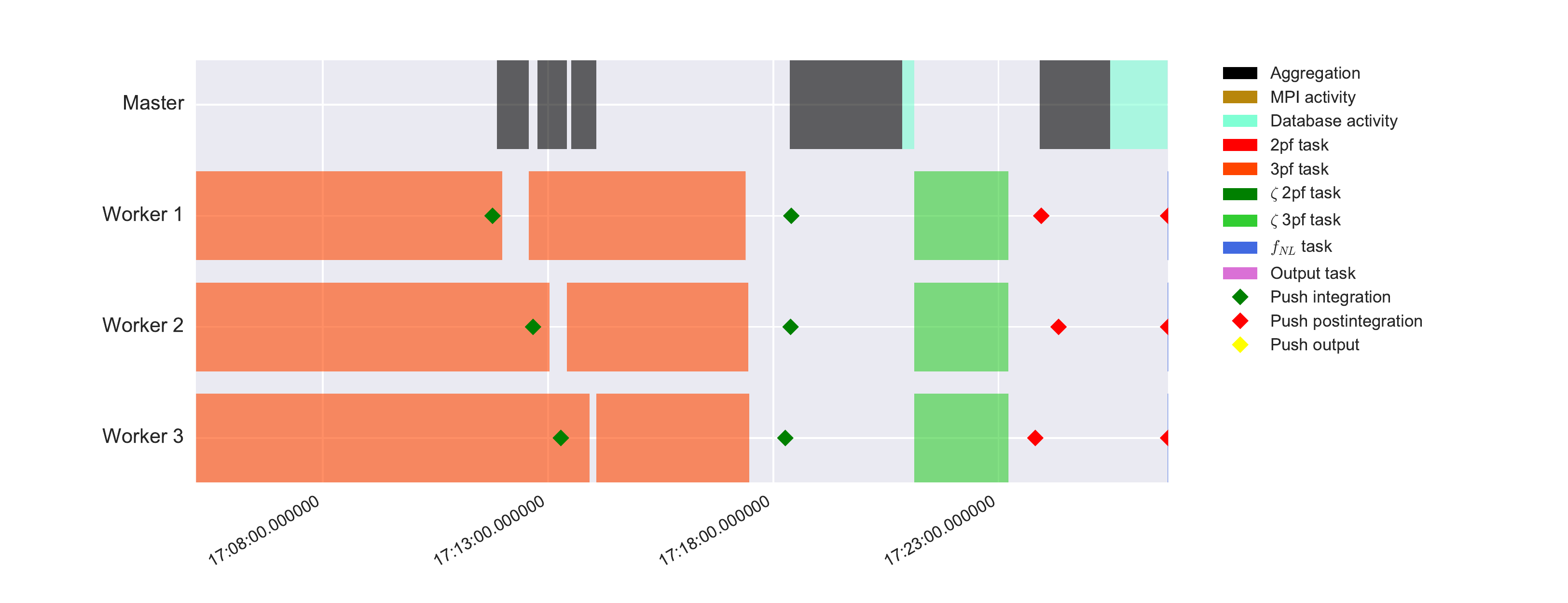}
    \end{center}

    \caption{\label{fig:gantt-fNL}Process Gantt charts for the inner-product task \repoobject{dquad.fNL-local}.
    Top chart: using a machine with solid-state storage.
    Bottom chart: using a machine with an external USB3 hard disk.
    Notice that aggregation is more time consuming using disks.
    Aggregation was performed using {\SQLite}'s write-ahead logging mode.}
\end{figure}

With linear spacing in $k$ we generate many more configurations,
roughly 12,000.
Although each configuration integrates fairly quickly, the
total number means these tasks will take a few minutes to process:
\begin{minted}[bgcolor=blue!10]{text}
    $ mpiexec -n 4 dquad -v -r test-fNL --create --task dquad.fNL-local --gantt gantt-fNL.pdf
    Task manager: processing task 'dquad.threepf-linear' (1 of 3)
    Task manager: 2016-May-16 14:32:36
    Items processed: 9073   In flight: 634          Remaining: 2344
    Complete: 75.3%         Mean CPU/item: 0.0985s  Assignment: 60s
    Estimated completion: 2016-May-16 14:34:26 (1m 50.7s from now)
    Task manager: 2016-May-16 14:32:47
    Items processed: 9707   In flight: 1196         Remaining: 1148
    Complete: 80.5%         Mean CPU/item: 0.0975s  Assignment: 60s
    Estimated completion: 2016-May-16 14:34:11 (1m 24.2s from now)
    Task manager: 2016-May-16 14:33:50
    Items processed: 10971  In flight: 868          Remaining: 212
    Complete: 91%           Mean CPU/item: 0.0969s  Assignment: 60s
    Estimated completion: 2016-May-16 14:34:30 (40.4s from now)
    Task manager: 2016-May-16 14:34:08
    All work items processed: 2016-May-16 14:34:08
    Committed content group '20160516T132659' for task 'dquad.threepf-linear' at 2016-May-16 14:35:01
    Task manager: processing task 'dquad.threepf-linear-zeta' (2 of 3)
    Committed content group '20160516T133501' for task 'dquad.threepf-linear-zeta' at 2016-May-16 14:39:12
    Task manager: processing task 'dquad.fNL-local' (3 of 3)
    Committed content group '20160516T133912' for task 'dquad.fNL-local' at 2016-May-16 14:39:13
    Task manager: processed 3 database tasks in wallclock time 12m 14.7s | time now 2016-May-16 14:39:13
\end{minted}
On a slow filing system you may find that
{\CppTransport}'s
periodic progress updates
show an adjustment of the target work assignment.
For this task the final data container reaches $\sim 3.5 \Gb$, and aggregation
may take a few seconds per temporary database on an SSD,
or more if writing to a conventional disk.
As explained above,
when aggregation times are non-negligible
{\CppTransport} schedules larger batches of work in an attempt
to avoid wasteful stalls.
That process Gantt chart for this job is shown
in Fig.~\ref{fig:gantt-fNL}.

\subsubsection{Example: Using the {\SQLite} data container to produce a plot}
In~\S\ref{sec:derived-products}
we will see how to generate a derived product
that plots the time history of
$\fNLlocal$ in this model.
As for other types of data-generating task, however, it is also possible to extract
raw data from the {\SQLite} database
and use that directly.

\begin{figure}
    \begin{center}
        \includegraphics[scale=0.65]{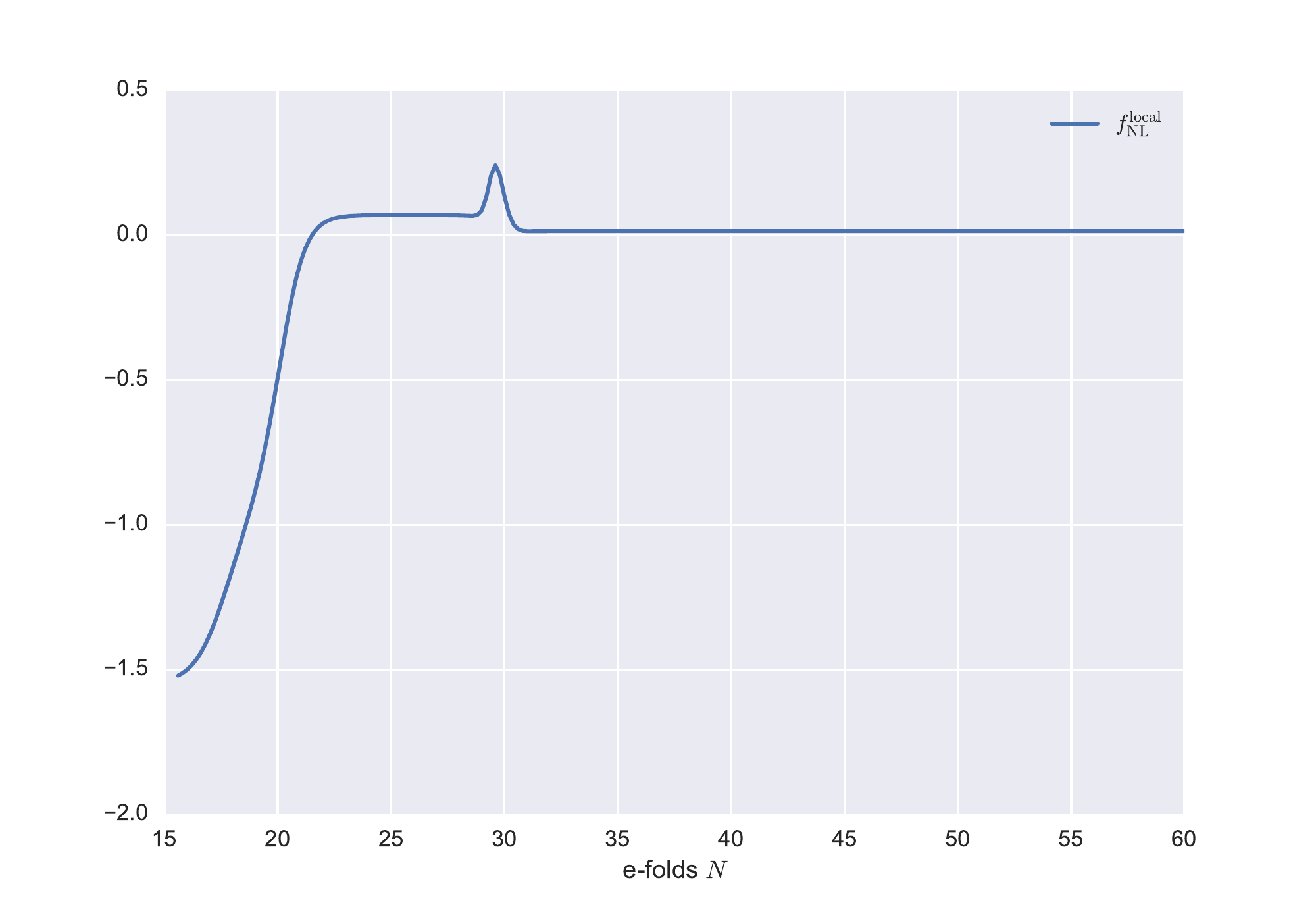}
    \end{center}
    \caption{\label{fig:dquad-fNL-plot}Evolution of the inner-product
    amplitude $\fNLlocal$ in the double-quadratic inflation model.}
\end{figure}

Each inner-product task will write the following tables
into its container:
\begin{sqltablelist}
\begin{enumerate}
    \item \mintinline{sql}{time_samples} -- contains the list of time sample points.
	This table duplicates, and is inherited from, the one described in \S\ref{sec:using-integration-container}
	(see p.\pageref{sqltable:time-samples}).

	\item \mintinline{sql}{fNL_local} \\
	\mintinline{sql}{fNL_equi} \\
	\mintinline{sql}{fNL_ortho} --- contain the inner products needed to compute $\fNL^i$ \\
	Typically only one of these tables will be present, as appropriate. Each table contains the columns: \\
	\begin{tabular}{p{2.5cm}p{11.5cm}}
	   \mintinline{sql}{tserial} & reference to serial number \mintinline{sql}{time_samples.serial}
	   identifying the time sample point for this row \\
	   \mintinline{sql}{BB} & contains the inner product $\iprod{B}{B}$, where $B$ is the numerical
	   bispectrum produced by the $\zeta$ task used as a source \\
	   \mintinline{sql}{BT} & contains the inner product $\iprod{B}{T}$, where $B$ is as above and
	   $T$ is the appropriate template \\
	   \mintinline{sql}{TT} & contains the inner product $\iprod{T}{T}$
	\end{tabular}
\end{enumerate}
\end{sqltablelist}
As an example, consider what would be required to compute
$\fNL^i$ from this database and plot it.
The first step is to construct an SQL query that computes the
required combination
of the columns \mintinline{sql}{BT}
and \mintinline{sql}{TT}.
We also need a column representing the time sample point.
A simple way to achieve this is
\begin{minted}{sql}
    SELECT
      tserial           AS tserial,
      (5 / 3) * BT / TT AS fNL
    FROM fNL_local;
\end{minted}
This is an example of the way in which short SQL queries can be used
to perform simple data analysis without
the need to write programs.
For example, if we wished it would be possible to add
a new column representing the (square of the) correlation `cosine'
between $B$ and $T$,
\begin{equation}
    \cos^2 (B,T) = \frac{\iprod{B}{T}^2}{\iprod{B}{B} \iprod{T}{T}} .
\end{equation}
To do this we would change the query:
\begin{minted}{sql}
    SELECT
      tserial             AS tserial,
      (5 / 3) * BT / TT   AS fNL,
      BT * BT / (BB * TT) AS cos_square
    FROM fNL_local;
\end{minted}
We would usually prefer to plot e-foldings $N$ on the $x$-axis rather than the
time serial number.
This requires us to join the
\mintinline{sql}{fNL_local} and
\mintinline{sql}{time_samples} tables,
by splicing together rows from each table
to form a bigger table with more columns.
This is accomplished by the SQL
\mintinline{sql}{INNER JOIN} clause:
\begin{minted}{sql}
    SELECT
      time_samples.time                                           AS time,
      (5 / 3) * fNL_local.BT / fNL_local.TT                       AS fNL,
      fNL_local.BT * fNL_local.BT / (fNL_local.BB * fNL_local.TT) AS cos_square
    FROM fNL_local
      INNER JOIN time_samples ON fNL_local.tserial = time_samples.serial
    ORDER BY time;
\end{minted}
Most database managers will allow the results of this query to be saved
in any suitable format.
`Comma Separated Value' or CSV is usually a good choice that has
widespread support among third party tools.
Once the data has been exported we can plot it as we wish,
by using a suitable tool such as
\href{http://www.gnuplot.info}{Gnuplot}.
Alternatively, the following Python script will produce a simple plot
from a CSV file.
\begin{minted}{python}
    import numpy as np
    import matplotlib.pyplot as plt
    import seaborn as sns
    import csv

    with open('fNL_local.csv') as f:
        time = []
        fNL_local = []

        reader = csv.DictReader(f)
        for row in reader:
            time.append(float(row['time']))
            fNL_local.append(float(row['fNL']))

        time_data = np.array(time)
        fNL_local_data = np.array(fNL_local)

    plt.figure()
    plt.plot(time_data, fNL_local_data, label='$f_{\mathrm{NL}}^{\mathrm{local}}$')

    plt.xlabel(r'e-folds $N$')
    plt.legend(frameon=False)
    plt.savefig('fNL_local.pdf')
\end{minted}
The resulting plot is shown in Fig.~\ref{fig:dquad-fNL-plot}.

\section{Generating derived products using output tasks}
\label{sec:derived-products}

We are now in a position to consider the final step in
performing an analysis with {\CppTransport}---using its visualization
toolkit to produce derived products.
{\CppTransport} can convert its raw $n$-point functions into a large
number of derived quantities, some of which are observables.
In many cases these quantities will be sufficient to undertake
a simple model analysis, or comparison with summary observables
such as the power spectrum amplitude or spectral index.
For more advanced uses it may be
preferable or necessary to export raw data from the database
and process it using a dedicated pipeline.

The current implementation of the visualization toolkit can
produce
good quality two-dimensional line and scatter plots
with minimal manual intervention.
Outputs are built from a series of building blocks:
\begin{itemize}
    \item \semibold{derived \emph{lines}} are simple functions of time
    or configuration that can be extracted from the output of one or more tasks.
    Examples might be
    the values of raw correlation functions,
    the spectral index of the power spectrum,
    or the tensor-to-scalar ratio---all as functions of time or configuration.
    Many such quantities can be constructed using just one task but others require
    two, such as the tensor-to-scalar ratio.
    (The tensor amplitude is computed by
    a \mintinline{c++}{twopf_task} or
    \mintinline{c++}{threepf_task}, while the $\zeta$ amplitude
    is computed by the
    \mintinline{c++}{zeta_*} versions of these.)

    \item \semibold{derived \emph{products}} are groups of derived lines,
    presented either as a table or as a plot.
    A single derived product can include lines that draw data from many different
    integration or postintegration tasks.
    However, only groups of lines that share a common type of $x$-axis can be
    aggregated into a single derived product.

    \item \semibold{output tasks} are collections of derived products.
    When an output task is executed, {\CppTransport}
    produces fresh copies of each derived product it contains.
    The standard rules described in~\S\ref{sec:apply-tags}
    are used to select which content groups are used as data sources.
\end{itemize}
{\CppTransport} supports two types of derived line:
\emph{time series}, which capture the evolution of some quantity as a function
of e-foldings $N$, and
\emph{wavenumber series},
which capture the dependence on some aspect of the wavenumber configuration.
This could be the scale $k$ or $k_t$, but also shape parameters such as
$\alpha$ or $\beta$, or the squeezing ratios $k_i/k_t$.
It can automatically take derivatives of wavenumber series in order to
measure spectral indices.

\subsection{Selecting which data to plot using SQL query objects}
To build a derived line requires selecting among the time and wavenumber samples
stored by {\CppTransport} during integration,
or subsequently generated by a postintegration task.
Because {\CppTransport}'s internal data storage services are built on top of
SQL databases, a natural and efficient
means to perform this selection is to use
SQL expressions as building blocks.

{\CppTransport} provides three classes that encapsulate SQL expressions suitable
for selecting a group of time of configuration sample points.
\mintinline{c++}{SQL_time_query} is used to select time samples,
and
\mintinline{c++}{SQL_twopf_query}
or \mintinline{c++}{SQL_threepf_query}
are used to select suitable configurations from the sample points
for 2- or 3-point function wavenumber configurations.
To use them, supply a suitable SQL expression based on the columns
of the tables
\mintinline{sql}{time_samples},
\mintinline{sql}{twopf_samples}
and
\mintinline{sql}{threepf_samples},
as appropriate.
Each of these accepts a single argument, which is the SQL
expression to use.

\para{Selecting times}
For example, to build a trivial query that selects all configurations
we require an expression that is always true, such as
\mintinline{sql}{1=1}.
For example:
\begin{minted}{c++}
	SQL_time_query all_times("1=1");
	SQL_twopf_query all_twopfs("1=1");
	SQL_threepf_query all_threepfs("1=1");
\end{minted}
Alternatively, to select the latest time recorded in the database
we could use
\begin{minted}{sql}
	tserial IN (SELECT MAX(tserial) FROM time_samples)
\end{minted}
The function \mintinline{sql}{MAX} is provided by SQL.
There is also a \mintinline{sql}{MIN} function that can be used to select
the earliest time available in the database,
or we can combine these to get two sample points---both the earliest \emph{and}
latest times,
\begin{minted}{sql}
	tserial IN (SELECT MAX(tserial) FROM time_samples UNION SELECT MIN(tserial) FROM time_samples)
\end{minted}
To pick a specific time, it is possible to look up the corresponding serial
number (using an HTML report as described in~\S\ref{sec:examine-k-database}
for wavenumber configurations, but checking the time sample points instead)
and specify it directly.
Alternatively you can specify a range of times using a query such as
\mintinline{sql}{time > 10 AND time < 30}.

\para{Selecting wavenumber configurations}
To select wavenumber configurations is equally easy.
For 2-point function configurations we could select a range of wavenumbers
using an expression such as
\begin{minted}{sql}
	conventional > 10 AND conventional < 50
\end{minted}
The available column names are those
that apply to the corresponding SQL table, as described in~\S\ref{sec:using-integration-container}.
This means that we could also select configurations based on their horizon-exit
time via the column
\mintinline{sql}{t_exit}
or their massless point
\mintinline{sql}{t_massless}.
So, for example, we could select all wavenumbers $k$ that leave the horizon in
some specified e-folding interval using
\begin{minted}{sql}
	t_exit > 0.0 AND t_exit < 10.0
\end{minted}
The same facilities are available
for 3-point function configurations,
for which it is possible to select on
the columns
\mintinline{sql}{kt_conventional},
\mintinline{sql}{kt_comoving},
\mintinline{sql}{alpha},
\mintinline{sql}{beta},
\mintinline{sql}{t_exit_kt}
or
\mintinline{sql}{t_massless}.
If you wish to select on a squeezing ratio such as $k_3/k_t$ then
this has to be done by rewriting the expression in terms of
$\alpha$ and $\beta$, eg. the SQL expression
\begin{minted}{sql}
	(1-beta)/2 < 1E-3
\end{minted}
will select all configurations with squeezing ratio
$k_3/k_t < 10^{-3}$.

\subsection{Example: plotting the evolution of the background fields}
\label{sec:example-backg-evolve}
\begin{figure}
	\begin{center}
		\includegraphics[scale=0.75]{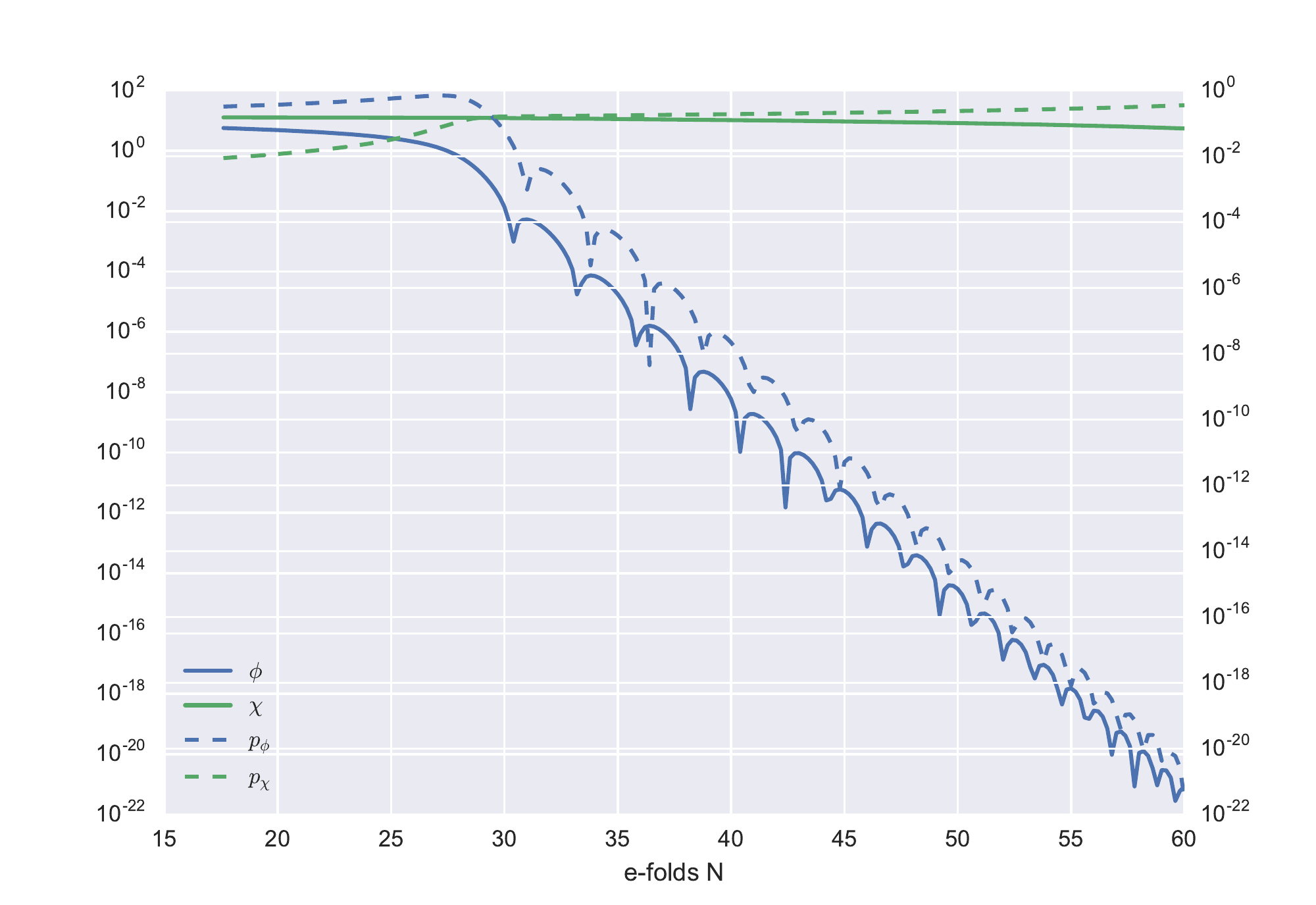}
	\end{center}
	\caption{\label{fig:background-plot}Time evolution of background fields $\phi$, $\chi$ and their
	derivatives $p_\phi$, $p_\chi$ generated by the produce
	\repoobject{dquad.product.bg\_plot}.}
\end{figure}

To see how these queries are used in practice,
consider building a plot showing the time evolution of the background
fields and their derivatives.
To keep our function
\mintinline{c++}{write_tasks()} from becoming excessively long, we break it
into two parts: one that builds the sample points and initial conditions,
and another that builds tasks and derived products.
Later, we will add further tasks and products; these can go in separate functions.
Our prototypes are now
\begin{minted}{c++}
	void write_tasks(transport::repository<>& repo, transport::dquad_mpi<>* model);

	void write_zeta_products(transport::repository<>& repo, transport::initial_conditions<>& ics, transport::range<>& ts, transport::range<>& ks);
\end{minted}
The function \mintinline{c++}{write_tasks()} becomes
\begin{minted}{c++}
	void write_tasks(transport::repository<>& repo, transport::dquad_mpi<>* model)
	  {
	    const double Mp = 1.0;
	    const double Mphi = 9E-5 * Mp;
	    const double Mchi = 1E-5 * Mp;

	    transport::parameters<> params(Mp, { Mphi, Mchi }, model);

	    const double phi_init = 10.0 * Mp;
	    const double chi_init = 12.9 * Mp;

	    const double N_init = 0.0;
	    const double N_pre = 12.0;
	    const double N_end = 60.0;

	    transport::initial_conditions<> ics("dquad", params, { phi_init, chi_init }, N_init, N_pre);

	    transport::basic_range<> ts(N_init, N_end, 300, transport::spacing::linear);

	    const double kt_lo = std::exp(3.0);
	    const double kt_hi = std::exp(8.0);

	    transport::basic_range<> ks_logspaced(kt_lo, kt_hi, 50, transport::spacing::log_bottom);
	    transport::basic_range<> ks_linearspaced(kt_lo, kt_hi, 50, transport::spacing::linear);

	    write_zeta_products(repo, ics, ts, ks_logspaced);
	  }
\end{minted}
and the new function
\mintinline{c++}{write_zeta_products()} is
\begin{minted}{c++}
void write_zeta_products(transport::repository<>& repo, transport::initial_conditions<>& ics, transport::range<>& ts, transport::range<>& ks)
  {
	constexpr unsigned int num_fields = 2;

    transport::threepf_cubic_task<> tk3("dquad.threepf", ics, ts, ks);
    tk3.set_adaptive_ics_efolds(54.0);
    tk3.set_description("Compute time history of the 3-point function on a cubic lattice" " from k ~ e^3 to k ~ e^9");

    transport::zeta_threepf_task<> ztk3("dquad.threepf-zeta", tk3);
    ztk3.set_description("Convert the output from dquad.threepf into zeta 2-" " and 3-point functions");

    vis_toolkit::SQL_time_query all_times("1=1");

    vis_toolkit::background_time_series<> bg_fields(tk3, vis_toolkit::index_selector<1>(num_fields).all(), all_times);

    vis_toolkit::time_series_plot<> bg_plot("dquad.product.bg_plot", "background.pdf");
    bg_plot.set_legend_position(vis_toolkit::legend_pos::bottom_left);
    bg_plot += bg_fields;

    transport::output_task<> out_tk("dquad.output.zeta");
    out_tk += bg_plot;

    repo.commit(out_tk);
  }
\end{minted}
The steps involved here are:
\begin{enumerate}
	\item We set up a \mintinline{c++}{threepf_cubic_task} and couple it to a
	\mintinline{c++}{zeta_threepf_task} as described above.
	These tasks will collaborate to produce the raw $n$-point functions in the model.

	\item We set up an \mintinline{c++}{SQL_time_query} containing the trivial
	SQL query \mintinline{sql}{1=1}. This will select all sample times stored in the database,
	so we will see the evolution of the background fields over the entire integration.

	However, remember that with adaptive initial conditions, {\CppTransport} will not begin
	storing samples until all $n$-point functions are available. This means that the
	first time stored in the database will be some time later than the initial time
	$\Ninit = 0$ specified in the initial conditions package
	\mintinline{c++}{ics}.

	\item We set up a time-series line for the background fields using the
	\mintinline{c++}{vis_toolkit::background_time_series<>} class.
	Its constructor has signature
	\begin{minted}{c++}
		background_time_series(integration_task tk, index_selector<1> selector, SQL_time_query query)
	\end{minted}
	The first argument identifies the integration task that will supply the data.
	The second argument is a new type of object,
	an \mintinline{c++}{index_selector<>}.
	This determines which fields to plot.
	The final argument is an \mintinline{c++}{SQL_time_query}.
	As explained above, this will determine which time sample points to include.
	Our choice \mintinline{c++}{all_times} will use them all.

	The constructor for the \mintinline{c++}{index_selector}
	accepts a single argument, corresponding to the number of fields in the model.
	It takes a template argument between angle brackets,
	here \mintinline{c++}{<1>},
	to indicate the number of indices being selected from.
	For example: the background fields have one index,
	the two-point function $\Sigma^{ab}$ has two indices, and
	the three-point function $\alpha^{abc}$ has three indices.

	By default all components are selected, but to be explicit we have used the
	\mintinline{c++}{all()} method to indicate that they should all be included.
	This will give us a plot showing the field expectation values
	$\phi(N)$, $\chi(N)$
	and also their derivatives $p_\phi \equiv \d \phi / \d N$
	and $p_\chi = \d \chi / \d N$.

	\item Once the derived line has been created we build a plot
	to contain it. This is the object
	\mintinline{c++}{time_series_plot<>}.
	Its constructor takes two arguments:
	a \emph{derived product name} (which should be unique)
	and a filename.
	The format of the output file is inferred by
	{\Matplotlib} from the extension of the filename.
	Any format that can be written by {\Matplotlib} may be used,
	and typically PDF (\file{.pdf}) or SVG (\file{.svg})
	are good for plots.

	As an alternative it is possible to specify a filename with extension
	\file{.py}.
	This causes {\CppTransport} to write a Python script that will produce
	the plot, but it does not execute it---the script \emph{is} the derived
	product.
	This gives an opportunity to apply custom formatting or to apply
	adjustments that {\CppTransport} cannot make automatically.

	\item By default, the plot includes a legend
	that is placed in the upper right-hand corner.
	In this case it will overlay the lines, so we elect to move it to the bottom
	left-hand corner using the
	\mintinline{c++}{set_legend_position()} method.

	\item The data line \mintinline{c++}{bg_fields} is added to the plot
	by writing
	\mintinline{c++}{bg_plot += bg_fields}.
	If we had more than one line to add we could include them all in a summation
	on the right-hand side.

	\item Finally, we build an output task called
	\repoobject{dquad.output.zeta}. We add the plot to the task
	by writing
	\mintinline{c++}{out_tk += bg_plot}.
	The output task is committed to the database.
	Notice that there is no need to commit the derived products on which the
	task depends,
	or the tasks on which those derived products depend,
	because {\CppTransport} will automatically commit these dependencies if they are
	required.
\end{enumerate}
To test, build the executable and launch the task
\repoobject{dquad.output.zeta} using
\begin{minted}{bash}
	$ mpiexec -n 4 dquad -v -r test-output --create --task dquad.output.zeta
\end{minted}
As usual, {\CppTransport} will schedule execution of any tasks needed to produce
content groups that feed later tasks in the chain:
\begin{minted}[linenos=false,xleftmargin=0pt,bgcolor=blue!10]{text}
    Task manager: processing task 'dquad.threepf' (1 of 2)
    Committed content group '20160516T135141' for task 'dquad.threepf' at 2016-May-16 15:00:18
    Task manager: processing task 'dquad.output.zeta' (2 of 2)
    Committed content group '20160516T140133' for task 'dquad.output.zeta' at 2016-May-16 15:01:43
    Task manager: processed 2 database tasks in wallclock time 10m 1.8s | time now 2016-May-16 15:01:43
\end{minted}
Fig.~\ref{fig:background-plot} shows the generated plot, using the
\option{seaborn} plot style.
We will discuss its features below and explain how it can be customized.

{\CppTransport} employs a similar directory structure to that
used for integration and postintegration tasks.
Generated content is placed in the \file{output} folder of the repository,
under a subfolder with name corresponding to the output task.
Inside this folder, each content group
is placed in a directory with name given by the content group name.
Each derived product is generated inside this directory, with the filename
specified when the derived product was set up.
There is also the usual \file{logs} folder that contains detailed logging
from the master and worker processes while the task was in progress.

\para{Reporting on output content groups}
Asking {\CppTransport} to report on \repoobject{dquad.output.zeta} shows the derived products
that are generated as part of the task, and lists each associated content group:
\begin{minted}[bgcolor=blue!10]{text}
    $ ./dquad -r test-output/ --info dquad.output.zeta
dquad.output.zeta -- output task
    Created: 2016-May-16 13:51:40      Last update: 2016-May-16 14:01:43
    Runtime version: 2016.1

    Derived product                            Type     Matches tags  Filename
    dquad.product.bg-plot                      2d plot  --            background.pdf

    Content group                   Created           Last update
    20160516T140133    2016-May-16 14:01:33  2016-May-16 14:01:43
\end{minted}
If we instead report on the content group generated by this task, {\CppTransport} will summarize
the content groups that were used to feed data into its derived products:
\begin{minted}[bgcolor=blue!10]{text}
    $ ./dquad -r test-output/ --info 20160516T140133
    20160516T140133 -- derived content
    Created: 2016-May-16 14:01:33      Last update: 2016-May-16 14:01:43
    Runtime version: 2016.1            Task: dquad.output.zeta
    Locked: No                         Tags: --

    Summary of content groups used
    Content group    Task                Type                        Last update
    20160516T135141  dquad.threepf       integration        2016-May-16 14:00:18
\end{minted}
The table aggregates all dependencies, but sometimes it is more useful
to see indivudally how each derived product depends on other content groups.
{\CppTransport} can provide this information as a
\emph{provenance report},
obtained by using
the command-line option \option{{-}{-}provenance} rather than
\option{{-}{-}info}.

\begin{figure}
	\begin{center}
		\includegraphics[scale=0.4]{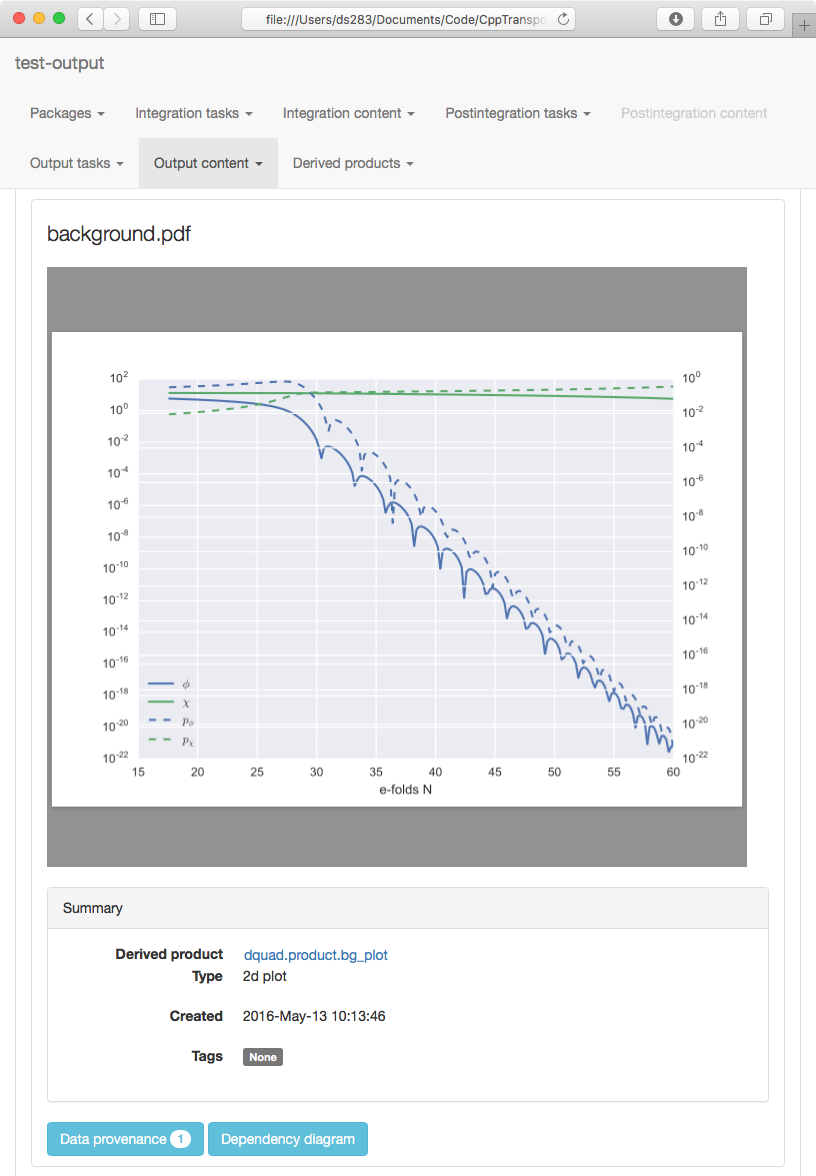}
	\end{center}
	\caption{\label{fig:bg-plot-screenshot}The HTML report for content groups
	generated by output tasks will include the output where possible.
	Click the `Data provenance' button to display a hyperlinked list of content
	groups showing the data sources for each product.
	The `Dependency diagram' shows how the data sources depend on each other,
	on particular integration and postintegration tasks, and ultimately
	on a set of initial conditions packages.}
\end{figure}

\begin{figure}
	\begin{center}
		\includegraphics[scale=0.5]{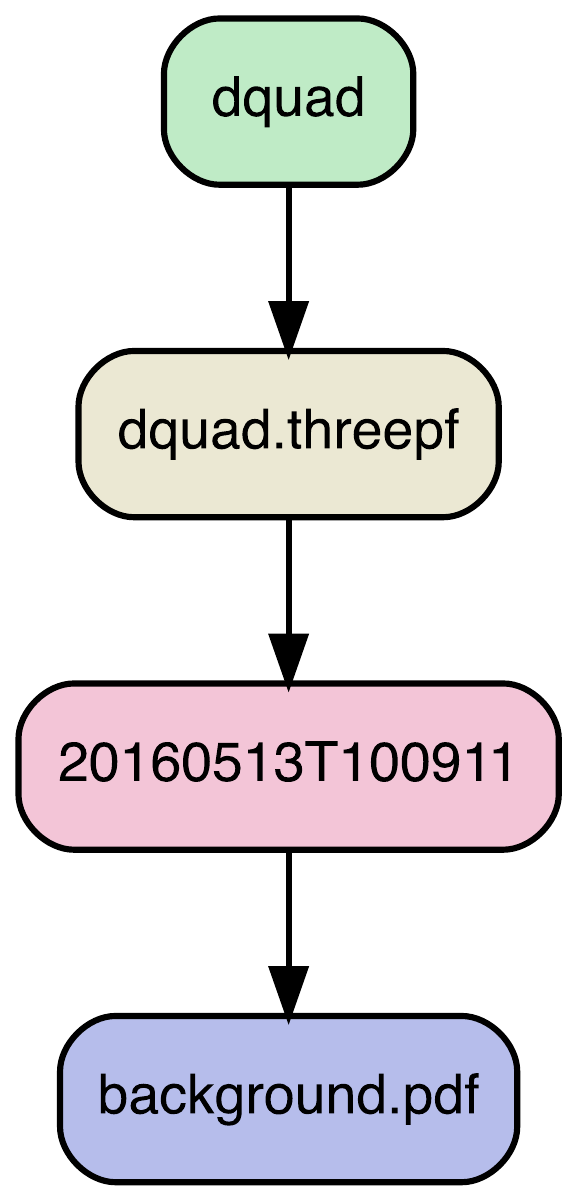}
	\end{center}
	\caption{\label{fig:bg-dependencies}Dependency diagram for
	the background field plot \repoobject{dquad.products.bg\_plot}.
	In this case the chain of dependencies is very straightforward, but
	more complex cases are possible. Initial conditions packages are coloured
	green, tasks are coloured yellow, content groups are red and
	derived products are blue.}
\end{figure}

The same information is embedded in an HTML report,
and often these are the easiest way to navigate the network of
dependencies.
Content groups generated by output tasks are listed
under the `Output content' tab, and where possible the report
includes any generated plots.
This is often a good way to view the products generated by an output
task, especially if there are more than a few.
See Fig.~\ref{fig:bg-plot-screenshot}.
The report for each derived product includes a table of
dependencies similar to that provided by the
\option{{-}{-}provenance} option.
Also, if {\Graphviz} is installed, it will produce a dependency diagram
showing the interrelation of
initial conditions packages,
tasks and content groups that
yielded the data for this product.
An example for the background fields plot is shown in Fig.~\ref{fig:bg-dependencies}.

\subsection{Derived products: plots and tables}
The current implementation of the visualization toolkit supports four principal
types of derived product:
\begin{itemize}
	\item The products \mintinline{c++}{time_series_plot<>}
	and \mintinline{c++}{wavenumber_series_plot<>}
	are two-dimensional graphs
	suitable for plotting time-series and wavenumber-series
	data, respectively.

	\item The products \mintinline{c++}{time_series_table<>}
	and \mintinline{c++}{wavenumber_series_table<>}
	are suitable for tabulating time- and wavenumber-series data.
\end{itemize}
The constructors for these derived products have a standard signature
consisting of two arguments:
a unique name, and an output filename.
We have already seen this in action for the
\mintinline{c++}{time_series_plot<>} used to plot background field data
above.

\subsubsection{Standard options}
\label{sec:product-standard-options}
Each derived product can be customized by applying options.
The \mintinline{c++}{set_*} methods can be chained together for more
economical expression.
\begin{itemize}
	\item \mintinline{c++}{bool get_log_x()} \\
	\mintinline{c++}{set_log_y(bool f)} \\
	Determine whether logarithmic $x$-axis values are used, if this makes sense
	for the product. Off by default.

	\item \mintinline{c++}{bool get_log_y()} \\
	\mintinline{c++}{set_log_y(bool f)} \\
	Determine whether logarithmic $y$-axis values are used, if this makes sense
	for the product. Enabled by default for plots.

	\item \mintinline{c++}{bool get_abs_y()} \\
	\mintinline{c++}{set_abs_y(bool f)} \\
	Determine whether absolute $y$-values are used. Enabled by default for plots.

	If a logarithmic $y$-axis is in use, {\CppTransport} will detect whether
	absolute values are needed for lines that have negative values.

	\item \mintinline{c++}{bool get_use_LaTeX()} \\
	\mintinline{c++}{set_use_LaTeX()} \\
	\label{method:use_LaTeX}
	Determine whether {\LaTeX}-format labels are used
	in preference to plain text.
	Notice that this has nothing to do with whether {\LaTeX}
	is involved in typesetting. Enabled by default for plots.
\end{itemize}

\subsubsection{Plot-specific options}
\label{sec:plot-options}
In addition, there are a number of options that are specific to plots,
ie. the objects
\mintinline{c++}{time_series_plot<>}
and
\mintinline{c++}{wavenumber_series_plot<>}.
As usual, the \mintinline{c++}{set_*} methods can be chained.

\para{Axis handling}
The following options influence axes:
\begin{itemize}
    \item \mintinline{c++}{bool get_reverse_x()} \\
    \mintinline{c++}{set_reverse_x(bool)} \\
    Determine whether $x$-axis is reversed. Off by default.

    \item \mintinline{c++}{bool get_reverse_y()} \\
    \mintinline{c++}{set_reverse_y(bool)} \\
    Determine whether $y$-axis is reversed. Off by default.
\end{itemize}

\para{Labelling}
The following options determine labelling:
\begin{itemize}
    \item \mintinline{c++}{bool get_x_label()} \\
    \mintinline{c++}{set_x_label(bool)} \\
    Determine whether a label is added to the $x$-axis. Enabled by default.

    \item \mintinline{c++}{std::string get_x_label_text()} \\
    \mintinline{c++}{set_x_label_text(std::string)} \\
    \mintinline{c++}{clear_x_label_text()} \\
    Set or clear
    text for $x$-axis label, if it is in use. Does not imply labelling is enabled.

    If labelling is enabled but no text is explicitly supplied, {\CppTransport}
    will use a default label.
    In this case, \mintinline{c++}{get_x_label_text()} will return an empty string.

    \item \mintinline{c++}{bool get_y_label()} \\
    \mintinline{c++}{set_y_label(bool)} \\
    Determine whether a label is added to the $y$-axis. Off by default.

    On the example of Fig.~\ref{fig:background-plot}
    we could switch on labelling on the $y$-axis by adding
    \begin{minted}{c++}
        bg_fields.set_y_label(true);
    \end{minted}
    before adding the \mintinline{c++}{bg_fields} lines to
    \mintinline{c++}{bg_plot}.
    It is not necessary to specify a label; {\CppTransport} will
    apply default labels based on quantities represented by each axis.
    If we wish, however, it is possible to specify a label
    using the \mintinline{c++}{set_y_label_text} method described below.
    This sets the label for the first axis only.

    \item \mintinline{c++}{std::string get_y_label_text()} \\
    \mintinline{c++}{set_y_label_text(std::string)} \\
    \mintinline{c++}{clear_y_label_text()}
    Set or clear
    text for $y$-axis label, if it is in use. Does not imply labelling is enabled.

    If labelling is enabled but no text is explicitly supplied, {\CppTransport}
    will use a default label.
    In this case, \mintinline{c++}{get_y_label_text()} will return an empty string.

    \item \mintinline{c++}{bool get_title()} \\
    \mintinline{c++}{set_title(bool)} \\
    Determine whether a title is added to the plot. Off by default.

    \item \mintinline{c++}{std::string get_title_text()} \\
    \mintinline{c++}{set_title_text(std::string)} \\
    \mintinline{c++}{clear_title_text()} \\
    Set or clear
    text for plot title, if it is in use. Does not imply titling is enabled.

    If no title text has been supplied,
    \mintinline{c++}{get_title_text()} will return an empty string.

    \item \mintinline{c++}{bool get_legend()} \\
    \mintinline{c++}{set_legend(bool)} \\
    Determine whether a legend is included on the plot. Enabled by default.

    \item \mintinline{c++}{vis_toolkit::legend_pos get_legend_position()} \\
    \mintinline{c++}{set_legend_position(vis_toolkit::legend_pos)} \\
    Determine position of legend, if one is enabled. By default the legend is
    position in the top right corder of the plot

    The allowed values are: \\
    \begin{tabular}{p{6.5cm}p{7cm}}
        \mintinline{c++}{vis_toolkit::legend_pos::top_left} & legend in top left corner \\
        \mintinline{c++}{vis_toolkit::legend_pos::top_right} & legend in top right corner \\
        \mintinline{c++}{vis_toolkit::legend_pos::bottom_left} & legend in bottom left corner \\
        \mintinline{c++}{vis_toolkit::legend_pos::bottom_right} & legend in bottom right corner \\
        \mintinline{c++}{vis_toolkit::legend_pos::right} & legend position to right of plot \\
        \mintinline{c++}{vis_toolkit::legend_pos::centre_right} & legend in right side of plot, vertically centred  \\
        \mintinline{c++}{vis_toolkit::legend_pos::centre_left} & legend in left side of plot, vertically centred \\
        \mintinline{c++}{vis_toolkit::legend_pos::upper_centre} & legend in top of plot, horizontally centred \\
        \mintinline{c++}{vis_toolkit::legend_pos::lower_centre} & legend in bottom of plot, horizontally centred \\
        \mintinline{c++}{vis_toolkit::legend_pos::centre} & legend in centre of plot \\
    \end{tabular}

    \item \mintinline{c++}{bool get_typeset_with_LaTeX()} \\
    \mintinline{c++}{set_typeset_with_LaTeX(bool)} \\
    Determine whether {\Matplotlib} is
	asked to offload typesetting responsibilities
	to {\LaTeX}. This produces higher quality results but is slower, and requires
	a {\LaTeX} toolchain to be installed on
	the current user's
	\mintinline{bash}{PATH}.
	{\LaTeX} is needed for some typesetting choices because {\Matplotlib}'s
	built-in typesetting engine does not support the full range of
	{\LaTeX} commands.

	Disabled by default. Notice that this setting is independent of
	whether {\LaTeX}-format labels have been enabled using
	\mintinline{c++}{set_use_LaTeX()}.

	\item \mintinline{c++}{bool get_dash_second_axis()} \\
	\mintinline{c++}{set_dash_second_axis(bool)} \\
	Determine whether lines associated with a second vertical axis
	are distinguished by being dashed.
	{\CppTransport} understands that different derived products have different
    units and scales,
    and where necessary it will add a second axis to handle lines
    with two different intrinsic scales.
    An example is the plot of background fields already produced
    in Fig.~\ref{fig:background-plot}.
    Here, although the axes are not labelled in the default configuration,
    the left-hand vertical axis refers to the value of the fields
    $\phi$, $\chi$
    and the right-hand axis refers to the value of their
    derivatives.

    To disable dashing on our example plot Fig.~\ref{fig:background-plot}
    we should add
    \begin{minted}{c++}
        bg_fields.set_dash_second_axis(false);
    \end{minted}
    before adding \mintinline{c++}{bg_fields} to the plot
    \mintinline{c++}{bg_plot}.

    Enabled by default.
\end{itemize}

\subsubsection{Table-specific options}
\label{sec:table-options}
For the table objects
\mintinline{c++}{time_series_table<>}
and
\mintinline{c++}{wavenumber_series_table<>}
there is an option to select the table format:
\begin{itemize}
    \item \mintinline{c++}{vis_toolkit::table_format get_format()} \\
    \mintinline{c++}{set_format(vis_toolkit::table_format f)} \\
    Determine the format of the table. The allowed options are: \\
    \begin{tabular}{p{6.5cm}p{7cm}}
        \mintinline{c++}{vis_toolkit::table_format::justified} & aligned in columns, padded with spaces \\
        \mintinline{c++}{vis_toolkit::table_format::csv} & comma-separated value \\
        \mintinline{c++}{vis_toolkit::table_format::tsv} & tab-separated value
    \end{tabular} \\
    The default format is determined from the filename
    extension. If this is
    \file{.csv} or \file{.tsv}
    then the format is set to CSV or TSV respectively.
    The other allowed formats are
    \file{.txt}, \file{.data}
    and \file{.dat}. These use justified output.
\end{itemize}

\subsection{Available derived lines}
\label{sec:derived-lines}

In this section we document
the different derived lines provided by the visualization toolkit.
Each line is represented by a suitable object whose constructor
typically requires:
\begin{itemize}
    \item one or more \semibold{task objects} that specify the data sources
    for the line
    \item for lines representing tensor objects such as the two-point
    function $\Sigma^{ab}$ or the three-point function $\alpha^{abc}$,
    an \semibold{index selector} that specifies which components should be
    plotted
    \item for both time-series and wavenumber-series objects,
    two \semibold{SQL query expressions}---one for time configurations, and
    another for wavenumber configurations.
    \begin{itemize}
        \item for \emph{time series}, the time query determines
        the range of points sampled on the $x$-axis. If the wavenumber-configuration
        query generates multiple points then these are used to produce different
        lines on the plot.

        \item for \emph{wavenumber series} the situation is reversed. The
        wavenumber configuration query determines the points sampled on the
        $x$-axis, and if the time query generates multiple points then these
        are plotted as multiple lines.
    \end{itemize}
\end{itemize}

\subsubsection{Standard options}
\label{sec:derived-standard-options}

In addition, each line has a number of standard options. These settings are available
for all derived lines, although not every line will make use of each option.
\begin{itemize}
    \item \mintinline{c++}{vis_toolkit::dot_type get_dot_meaning()} \\
    \mintinline{c++}{set_dot_meaning(vis_toolkit::dot_type)} \\
    Determines whether the phase space is taken to be built from the
    fields and their derivatives with respect to $N$, or the fields
    and their canonical momenta.

    The allowed values for \mintinline{c++}{vis_toolkit::dot_type} are: \\
    \begin{tabular}{p{6.5cm}p{7cm}}
        \mintinline{c++}{vis_toolkit::dot_type::derivatives} &
            use derivatives \\
        \mintinline{c++}{vis_toolkit::dot_type::momenta} &
            use momenta
    \end{tabular}

    This setting affects only derived products containing
    field space correlation functions.

    \item \mintinline{c++}{vis_toolkit::klabel_type get_klabel_meaning()} \\
    \mintinline{c++}{set_klabel_meaning(vis_toolkit::klabel_type)} \\
    Determines whether conventional or comoving momenta are used.

    The allowed values for \mintinline{c++}{vis_toolkit::klabel_type} are: \\
    \begin{tabular}{p{6.5cm}p{7.5cm}}
        \mintinline{c++}{vis_toolkit::klabel_type::conventional} &
            use conventional normalization \\
        \mintinline{c++}{vis_toolkit::klabel_type::comoving} &
            use comoving normalization
    \end{tabular}

    The comoving normalization is the one set by the value of
    $\astar$; see p.\pageref{method:set_astar_normalization}.

    \item \mintinline{c++}{vis_toolkit::axis_value get_current_x_axis_value()} \\
    \mintinline{c++}{set_current_x_axis_value(vis_toolkit::axis_value)} \\
    Determines how the $x$-axis is constructed.

    {\CppTransport} can use several different quantities
    for the $x$-axis scale: \\
    \begin{tabular}{p{6.5cm}p{7cm}}
        \mintinline{c++}{vis_toolkit::axis_value::efolds} &
            use e-folds $N$ \\
        \mintinline{c++}{vis_toolkit::axis_value::k} &
            use wavenumber $k$ (or $k_t$ for 3-point function data) \\
        \mintinline{c++}{vis_toolkit::axis_value::efolds_exit} &
            use e-folds between $\Nstar$ and horizon exit of
            wavenumber $k$ (or $k_t/3$ for 3-point function data) \\
        \mintinline{c++}{vis_toolkit::axis_value::alpha} &
            use the shape parameter $\alpha$ \\
        \mintinline{c++}{vis_toolkit::axis_value::beta} &
            use the shape parameter $\beta$ \\
        \mintinline{c++}{vis_toolkit::axis_value::squeeze_k1} \\
        \mintinline{c++}{vis_toolkit::axis_value::squeeze_k2} \\
        \mintinline{c++}{vis_toolkit::axis_value::squeeze_k3} &
            use the squeezing ratio $k_i/k_t$
    \end{tabular}

    Not all derived lines support all $x$-axis types.
    Time series lines support only
    \mintinline{c++}{efolds},
    and wavenumber series associated with configurations of the 2-point
    function support only
    \mintinline{c++}{k} and
    \mintinline{c++}{efolds_exit}.
    Wavenumber series associated with configurations of the 3-point function
    support all types except
    \mintinline{c++}{efolds}.
    {\CppTransport} will raise an exception if you attempt to
    select an unsupported $x$-axis value.

    Derived products can contain only lines with the same $x$-axis type.
    If you attempt to add lines with different types to the same
    product then
    {\CppTransport} will raise an exception.

    \item \mintinline{c++}{clear_label_text()} \\
    \mintinline{c++}{set_label_text(std::string latex, std::string non_latex)} \\
    Set or clear a customized label for this line.

    Two labels are required: one in {\LaTeX} format
    and one in plain text format.
    The {\LaTeX}-format label is used
    if {\LaTeX} labels have been enabled by
    \mintinline{c++}{set_use_LaTeX()}
    (see~\S\ref{sec:product-standard-options}).
    Otherwise, the plain text label is used.

    Labels are used as column headings (if the derived product is a table)
    or displayed on the legend (if the derived product is a plot).
    If no label is set then {\CppTransport} will use a suitable default.

    \item \mintinline{c++}{bool get_label_tags()} \\
    \mintinline{c++}{set_label_tags(bool)} \\
    \label{method:get_label_tags}
    Determine whether identifying `tags' are added to each label.
    For time series, the tag identifies the wavenumber configuration
    corresponding to each line (if relevant).
    For wavenumber series it identifies the time sample point.

    Tags are useful where a single derived line generates multiple physical
    lines and you wish to distinguish them.
\end{itemize}

\subsubsection{Options for 2-point correlation functions}
The lines representing 2-point correlation functions are:
\begin{itemize}
    \item \mintinline{c++}{twopf_time_series}
    \item \mintinline{c++}{twopf_wavenumber_series}
    \item \mintinline{c++}{tensor_time_series}
    \item \mintinline{c++}{tensor_wavenmber_series}
    \item \mintinline{c++}{zeta_twopf_time_series}
    \item \mintinline{c++}{zeta_twopf_wavenumber_series}
\end{itemize}
These lines
have additional options:
\begin{itemize}
    \item \mintinline{c++}{set_dimensionless(bool)} \\
    Determine whether the dimensionful or dimensionless correlation
    function is plotted.
    The dimensionless correlation function $\DimlessSigma$ is defined
    in terms of the dimensionful correlation function $\Sigma$
    via
    \begin{equation}
        \DimlessSigma = \frac{k^3}{2\pi^2} \Sigma .
    \end{equation}
    Enabled by default.
\end{itemize}
The field-space two-point function lines
\mintinline{c++}{twopf_time_series}
and \mintinline{c++}{twopf_wavenumber_series} have an additional option that
allows them to be switched between
real and imaginary values, where available.
\begin{itemize}
    \item \mintinline{c++}{vis_toolkit::twopf_type get_type()} \\
    \mintinline{c++}{set_type(vis_toolkit::twopf_type)} \\
    Determine whether real or imaginary values are used.

    The allowed values
    of \mintinline{c++}{vis_toolkit::twopf_type} are: \\
    \begin{tabular}{p{6.5cm}p{7.5cm}}
        \mintinline{c++}{vis_toolkit::twopf_type::real} & use real values \\
        \mintinline{c++}{vis_toolkit::twopf_type::imaginary} & use imaginary values
    \end{tabular}
    Real values are used by default.
    Notice that only datasets generated by three-point function tasks
    include imaginary values for the 2-point function.
\end{itemize}

\subsubsection{Options for 3-point correlation functions}
The lines representing 3-point correlation functions are:
\begin{itemize}
    \item \mintinline{c++}{threepf_time_series}
    \item \mintinline{c++}{threepf_wavenumber_series}
    \item \mintinline{c++}{zeta_threepf_time_series}
    \item \mintinline{c++}{zeta_threepf_wavenumber_series}
\end{itemize}
These lines have additional options:
\begin{itemize}
    \item \mintinline{c++}{set_dimensionless(bool)} \\
    Determine whether the dimensionful or dimensionless correlation
    function is plotted.
    The dimensionless correlation function $\DimlessB^{abc}$
    is sometimes called the shape function.
    It is defined
    in terms of the dimensionful 3-point function $\alpha$
    by
    \begin{equation}
        \DimlessB = (k_1 k_2 k_3)^2 \alpha .
    \end{equation}
    Enabled by default.

    \item \mintinline{c++}{bool get_use_kt_label()} \\
    \mintinline{c++}{set_use_kt_label(bool)} \\
    If tags are
    being attached to labels
    (see p.\pageref{method:get_label_tags}),
    determine whether the $k_t$ value is used as part of the label.
    Enabled by default.

    \item \mintinline{c++}{bool get_use_alpha_label()} \\
    \mintinline{c++}{set_use_alpha_label(bool)} \\
    If tags are
    being attached to labels
    (see p.\pageref{method:get_label_tags}),
    determine whether the $\alpha$ shape parameter is used as part of the label.
    Disabled by default.

    \item \mintinline{c++}{bool get_use_beta_label()} \\
    \mintinline{c++}{set_use_beta_label(bool)} \\
    If tags are
    being attached to labels
    (see p.\pageref{method:get_label_tags}),
    determine whether the $\beta$ shape parameter is used as part of the label.
    Disabled by default.
\end{itemize}
If your configurations vary only with scale $k_t$ then the default tag
will be correct.
However, if you are varying more than one parameter, or keeping $k_t$
constant while varying $\alpha$ or $\beta$,
you should considering changing which quantities are
used to generate the tag.

In addition, the `reduced bispectrum'~\eqref{eq:reduced-bispectrum}
is defined by a ratio of 2- and 3-point correlation functions.
It is represented by the lines
\begin{itemize}
    \item \mintinline{c++}{zeta_reduced_bispectrum_time_series}
    \item \mintinline{c++}{zeta_reduced_bispectrum_wavenumber_series}
\end{itemize}
They have the same options given above except for
\mintinline{c++}{set_dimensionless()}.

\subsubsection{Time series}
\label{sec:time-series}
This section lists the available time series lines and the
signature of their constructors.
As a shorthand, the namespace
\mintinline{c++}{vis_toolkit}
is omitted for the names of the lines themselves,
and for types appearing in their constructors
such as \mintinline{c++}{vis_toolkit::SQL_time_query}
and
\mintinline{c++}{vis_toolkit::index_selector<>}.
We also use the following abbreviations for task types: \\
\begin{tabular}{p{3.5cm}p{10.5cm}}
    \mintinline{c++}{integration_task} & any integration task \\
    \mintinline{c++}{threepf_task} & any 3-point function integration task \\
    \mintinline{c++}{zeta_task} & any $\zeta$ postintegration task \\
    \mintinline{c++}{zeta_threepf_task} & a $\zeta$ 3-point function task \\
    \mintinline{c++}{fNL_task} & any $\fNL$ postintegration task
\end{tabular} \\
Each time series requires an
\mintinline{c++}{vis_toolkit::SQL_time_query} to select the
time sample points.
Each sample is taken at fixed wavenumber configuration,
selected by a
\mintinline{c++}{vis_toolkit::SQL_twopf_query} or
\mintinline{c++}{vis_toolkit::SQL_threepf_query}
that must also be supplied.
Where this query picks out multiple configurations, they are included
as separate columns in a table or lines in a plot.

\begin{itemize}
    \item \mintinline{c++}{background_time_series<>} \\
    Time evolution of the background fields and their derivatives.
    \begin{minted}{c++}
        background_time_series(
            integration_task,
            index_selector<1>,
            SQL_time_query
        )
    \end{minted}

    \item \mintinline{c++}{twopf_time_series<>} \\
    Time evolution of the field-space two-point correlation function $\Sigma^{ab}$
    (or its dimensionless counterpart).
    \begin{minted}{c++}
        twopf_time_series(
            integration_task,
            index_selector<2>,
            SQL_time_query,
            SQL_twopf_query
        )
    \end{minted}

    \item \mintinline{c++}{threepf_time_series<>} \\
    Time evolution of the field-space three-point correlation function $\alpha^{abc}$
    (or its dimensionless counterpart).
    \begin{minted}{c++}
        threepf_time_series(
            threepf_task,
            index_selector<3>,
            SQL_time_query,
            SQL_twopf_query
        )
    \end{minted}

    \item \mintinline{c++}{tensor_time_series<>} \\
    Time evolution of the tensor
    two-point correlation function $\Sigma_{ss'}$
    (or its dimensionless counterpart).
    \begin{minted}{c++}
        tensor_time_series(
            integration_task,
            index_selector<2>,
            SQL_time_query,
            SQL_twopf_query
        )
    \end{minted}

    \item \mintinline{c++}{r_time_series<>} \\
    Time evolution of the tensor-to-scalar ratio $r$.
    \begin{minted}{c++}
        r_time_series(
            zeta_task,
            SQL_time_query,
            SQL_twopf_query
        )
    \end{minted}

    \item \mintinline{c++}{zeta_twopf_time_series<>} \\
    Time evolution of the $\zeta$ two-point correlation function $\Sigma_\zeta$
    (or its dimensionless counterpart).
    \begin{minted}{c++}
        zeta_twopf_time_series(
            zeta_task,
            SQL_time_query,
            SQL_twopf_query
        )
    \end{minted}

    \item \mintinline{c++}{zeta_threepf_time_series<>} \\
    Time evolution of the $\zeta$ three-point correlation function $\alpha_\zeta$
    (or its dimensionless counterpart).
    \begin{minted}{c++}
        zeta_threepf_time_series(
            zeta_threepf_task,
            SQL_time_query,
            SQL_twopf_query
        )
    \end{minted}

    \item \mintinline{c++}{zeta_reduced_bispectrum_time_series<>} \\
    Time evolution of the $\zeta$ reduced bispectrum, defined by
    Eq.~\eqref{eq:reduced-bispectrum}.
    \begin{minted}{c++}
        zeta_reduced_bispectrum_time_series(
            zeta_threepf_task,
            SQL_time_query,
            SQL_twopf_query
        )
    \end{minted}

    \item \mintinline{c++}{fNL_time_series<>} \\
    Time evolution of an inner product amplitude $\fNL^i$;
    see Eq.~\eqref{eq:fNL-amplitudes}.
    \begin{minted}{c++}
        fNL_time_series(
            fNL_task,
            SQL_time_query
        )
    \end{minted}

    \item \mintinline{c++}{background_line<>} \\
    Time evolution of a background quantity. Currently-implemented options are
    $H$, $\epsilon$ and $aH$.
    \begin{minted}{c++}
        background_line(
            integration_task,
            SQL_time_query,
            background_quantity
        )
    \end{minted}
    The allowed values of \mintinline{c++}{vis_toolkit::background_quantity}
    are: \\
    \begin{tabular}{p{7cm}p{7.5cm}}
        \mintinline{c++}{vis_toolkit::background_quantity::epsilon} &
            compute $\epsilon = - \dot{H}/H^2$ \\
        \mintinline{c++}{vis_toolkit::background_quantity::Hubble} &
            compute Hubble parameter $H$ \\
        \mintinline{c++}{vis_toolkit::background_quantity::aH} &
            compute comoving Hubble scale $aH$
    \end{tabular}

    \item \mintinline{c++}{u2_line<>} \\
    Time evolution of components of the tensor $u_{ab}$.
    \begin{minted}{c++}
        u2_line(
            integration_task,
            index_selector<2>,
            SQL_time_query,
            SQL_twopf_query
        )
    \end{minted}

    \item \mintinline{c++}{u3_line<>} \\
    Time evolution of components of the tensor $u_{abc}$.
    \begin{minted}{c++}
        u3_line(
            threepf_task,
            index_selector<2>,
            SQL_time_query,
            SQL_threepf_query
        )
    \end{minted}
    Notice that because $u_{abc}$ is a function of a bispectrum configuration
    it can be computed only for 3-point function integrations.

    \item \mintinline{c++}{largest_u2_line<>} \\
    Time evolution of the largest component of the momentum--field block
    of $u_{ab}$, which is numerically equivalent to the field-space
    mass matrix in Hubble units $M_{\alpha\beta}/H^2$.
    The other components of $u_{ab}$ have fixed magnitudes or depend only
    on $\epsilon$, and so are excluded.
    \begin{minted}{c++}
        largest_u2_line(
            integration_task,
            index_selector<2>,
            SQL_time_query,
            SQL_twopf_query
        )
    \end{minted}

    \item \mintinline{c++}{largest_u3_line<>} \\
    Time evolution of the largest component of $u_{abc}$.
    Unlike $u_{ab}$, \emph{all} components are included.
    \begin{minted}{c++}
        largest_u3_line(
            threepf_task,
            index_selector<3>,
            SQL_time_query,
            SQL_threepf_query
        )
    \end{minted}

\end{itemize}

\subsubsection{Wavenumber series}
\label{sec:wavenumber-series}
For each quantity that is wavenumber-configuration dependent there is a corresponding
wavenumber series.
We describe these using the same conventions used for time series
in~\S\ref{sec:time-series}.

Like the corresponding time series, each wavenumber series
requires
a \mintinline{c++}{vis_toolkit::SQL_time_query}
and one
or other of
\mintinline{c++}{vis_toolkit::SQL_twopf_query} or
\mintinline{c++}{vis_toolkit::SQL_threepf_query}.
For wavenumber series, the twopf- or threepf-query is used to select the
sample points that will appear on the $x$-axis
and the time query selects the particular time at which we wish to sample.
If multiple times are included, they generate separate columns in a table
or lines on a plot.

\para{Spectral indices}
All wavenumber series have an extra option to compute the spectral index,
using the current $x$-axis type;
the definition is
\begin{equation}
    n = \frac{\d \ln y}{\d \ln x} .
\end{equation}
If $y$ can be approximated as a slowly varying power law, then $n$
can be regarded as an estimate for the power-law index
$y \sim x^n$.
For example, if $x$ is a wavenumber and $y$ is a power spectrum, then
$n$ computed in this way is the usual spectral index.
Alternatively, if $y$ is a reduced bispectrum and
$x$ is a squeezing ratio $k_i/k_t$ then
$n$ gives the local approximate power law
$\fNL(k_1, k_2, k_3) \sim (k_i/k_t)^n$.

The calculation of spectral indices is enabled using
\mintinline{c++}{set_spectral_index(bool)}.

\begin{warning}
    When computing spectral indices associated with variation of the
    bispectrum configuration, as above, it is important to be
    careful about which quantities vary and which are held fixed.

    For example, to measure the spectral index associated with
    variations of the squeezing $k_i/k_t$ at fixed scale, we should keep
    $k_t$ fixed. The SQL query expression that selects those configurations
    to include should enforce this constraint.

    Likewise, to measure the spectral index associated with variations of
    scale $k_t$ we should usually keep the squeezing parameters $k_i/k_t$ fixed.
\end{warning}

\begin{warning}
    In the current version of {\CppTransport},
    spectral indices generated using this option should be treated
    with caution.
    The calculation is performed by fitting a spline to $y(x)$
    and differentiating this spline.

    In general this gives acceptable results, but there are some pitfalls.
    First, the spline and its derivative can lose accuracy near the edges of the
    fitted region.
    This tends to introduce some spurious jitter into the spectral index.
    Second, with a large number of sample points the spline tends to
    be overfit. {\CppTransport} tries to compensate for this by using a
    $p$-spline (a `penalized' spline that tries to smoothly interpolate
    between sample points, rather than strictly passing through each point),
    but it is not always successful.

    For these reasons it is wise to check the spectral index calculation using
    other methods before relying on the result.
\end{warning}

\begin{itemize}
    \item \mintinline{c++}{twopf_wavenumber_series<>} \\
    $k$-dependence of the field-space two-point correlation function $\Sigma^{ab}$
    (or its dimensionless counterpart).
    \begin{minted}{c++}
        twopf_wavenumber_series(
            integration_task,
            index_selector<2>,
            SQL_time_query,
            SQL_twopf_query
        )
    \end{minted}

    \item \mintinline{c++}{threepf_wavenumber_series<>} \\
    Configuration-dependence
    of the field-space three-point correlation function $\alpha^{abc}$
    (or its dimensionless counterpart).
    \begin{minted}{c++}
        threepf_wavenumber_series(
            threepf_task,
            index_selector<3>,
            SQL_time_query,
            SQL_twopf_query
        )
    \end{minted}

    \item \mintinline{c++}{tensor_wavenumber_series<>} \\
    $k$-dependence of the tensor
    two-point correlation function $\Sigma_{ss'}$
    (or its dimensionless counterpart).
    \begin{minted}{c++}
        tensor_wavenumber_series(
            integration_task,
            index_selector<2>,
            SQL_time_query,
            SQL_twopf_query
        )
    \end{minted}

    \item \mintinline{c++}{r_wavenumber_series<>} \\
    $k$-dependence of the tensor-to-scalar ratio $r$.
    \begin{minted}{c++}
        r_time_series(
            zeta_task,
            SQL_time_query,
            SQL_twopf_query
        )
    \end{minted}

    \item \mintinline{c++}{zeta_twopf_wavenumber_series<>} \\
    $k$-dependence of the $\zeta$ two-point correlation function $\Sigma_\zeta$
    (or its dimensionless counterpart).
    \begin{minted}{c++}
        zeta_twopf_wavenumber_series(
            zeta_task,
            SQL_time_query,
            SQL_twopf_query
        )
    \end{minted}

    \item \mintinline{c++}{zeta_threepf_wavenumber_series<>} \\
    Configuration-dependence
    of the $\zeta$ three-point correlation function $\alpha_\zeta$
    (or its dimensionless counterpart).
    \begin{minted}{c++}
        zeta_threepf_wavenumber_series(
            zeta_threepf_task,
            SQL_time_query,
            SQL_twopf_query
        )
    \end{minted}

    \item \mintinline{c++}{zeta_reduced_bispectrum_wavenumber_series<>} \\
    Configuration-dependence of the $\zeta$ reduced bispectrum, defined by
    Eq.~\eqref{eq:reduced-bispectrum}.
    \begin{minted}{c++}
        zeta_reduced_bispectrum_time_series(
            zeta_threepf_task,
            SQL_time_query,
            SQL_twopf_query
        )
    \end{minted}

\end{itemize}

\subsubsection{Integration cost analysis}

Finally, a special set of derived lines exist that do not represent
data or observables, but rather provide metadata about the
performance of the integration.
These can be used to provide more targeted versions of the
scatter plots generated by HTML reports
and discussed in~\S\ref{sec:HTML-analysis}.
(See Fig.~\ref{fig:time-scaling}.)
These automatically-generated plots include all configurations,
whereas the derived lines offer more control through the use of SQL
query expressions to restrict the configuration that are included.

\begin{itemize}
    \item \mintinline{c++}{cost_wavenumber<>} \\
    Supply the per-configuration integration cost, where this information is
    available.
    \begin{minted}{c++}
        cost_wavenumber(
            twopf_task,
            SQL_twopf_query,
            cost_metric
        )

        cost_wavenumber(
            threepf_task,
            SQL_threepf_task,
            cost_metric
        )
    \end{minted}
    The allowed values of \mintinline{c++}{vis_toolkit::cost_metric}
    are \\
    \begin{tabular}{p{6cm}p{8.5cm}}
        \mintinline{c++}{vis_toolkit::cost_metric::time} &
            measure integration time \\
        \mintinline{c++}{vis_toolkit::cost_metric::steps} &
            measure number of steps taken by stepper
    \end{tabular}

\end{itemize}

\subsection{Enabling or disabling indices using an \mintinline{c++}{index_selector<>}}
\label{sec:index-selector}

In~\S\S\ref{sec:example-backg-evolve}--\ref{sec:derived-lines}
it was explained that
for derived lines generated by objects with multiple indices, a
\mintinline{c++}{index_selector<>} object must be supplied
to select which indices should be included.

Constructing a selector requires specifying a template argument that fixes the
number of indices.
The constructor accepts a single argument giving the number
of fields.
For example, to build a selector for a 2-index object such as the field-space
correlation function $\Sigma^{ab}$, we could use
\begin{minted}{c++}
	vis_toolkit::index_selector<2> sel(num_fields);
\end{minted}
where \mintinline{c++}{num_fields} should be suitably defined.
(It could be extracted from the \mintinline{c++}{get_N_fields()} method of a
\mintinline{c++}{transport::model} object, which returns the number of
fields used in the model.)

Index selectors enable all components by default.
Once constructed, a selector provides methods
\mintinline{c++}{all()}
and
\mintinline{c++}{none()}
that explicitly enable or disable all components.
It is also possible to enable or disable individual
components by supplying a tuple of numbers representing
the component in question.
As in the mapping used for index assignment in {\SQLite} containers,
fields are assigned increasing integers beginning at 0
and with ordering inherited from the model description file.
The field derivatives or momenta
follow the fields, in the same order.

For example, in the double quadratic model the fields are
$\phi$ and $\chi$, with $\phi$ being declared before $\chi$
in the description file (see~\S\ref{sec:field-param-block}).
To plot only the components $\Sigma_{\phi\phi}$ and
$\Sigma_{\chi\chi}$ we could use the selector
\begin{minted}{c++}
    vis_toolkit::index_selector<2>().none().set_on( {0,0} ).set_on( {1,1} )
\end{minted}
On the other hand, to plot components \emph{except} the
momentum cross-terms
$\Sigma_{p_\phi p_\chi}$
and
$\Sigma_{p_\chi p_\phi}$
we could use
\begin{minted}{c++}
    vis_toolkit::index_selector<2>().all().set_off( {2,3} ).set_off( {3,2} )
\end{minted}

\subsection{Examples: double quadratic inflation}
To illustrate the use of these derived quantites, consider generating
plots for a set of standard observables in the double-quadratic model.
We will use the following standard queries to select time- and wavenumber-configuration
samples:
\begin{minted}[breakbytoken=false]{c++}
    // time query -- all sample points
    vis_toolkit::SQL_time_query all_times("1=1");

    // time query -- last time
    vis_toolkit::SQL_time_query last_time("serial IN (SELECT MAX(serial) FROM time_samples)");

    // twopf query -- all configuraitons
    vis_toolkit::SQL_twopf_query all_twopfs("1=1");

    // threepf query -- all equilateral configurations
    vis_toolkit::SQL_threepf_query all_equilateral("ABS(alpha) < 1E-5 AND ABS(beta-1.0/3.0) < 1E-5");

    // threepf query -- isosceles triangles
    vis_toolkit::SQL_threepf_query all_isosceles("ABS(alpha) < 1E-5");

    // threepf query -- largest and smallest equilateral triangles
    vis_toolkit::SQL_threepf_query large_small_equilateral("ABS(alpha) < 1E-5 AND ABS(beta-1.0/3.0) < 1E-5 AND wavenumber1 IN (SELECT MAX(serial) FROM twopf_samples UNION SELECT MIN(serial) FROM twopf_samples)");
\end{minted}

\subsubsection{$\zeta$ power spectrum}
\begin{figure}
    \begin{center}
        \includegraphics[scale=0.75]{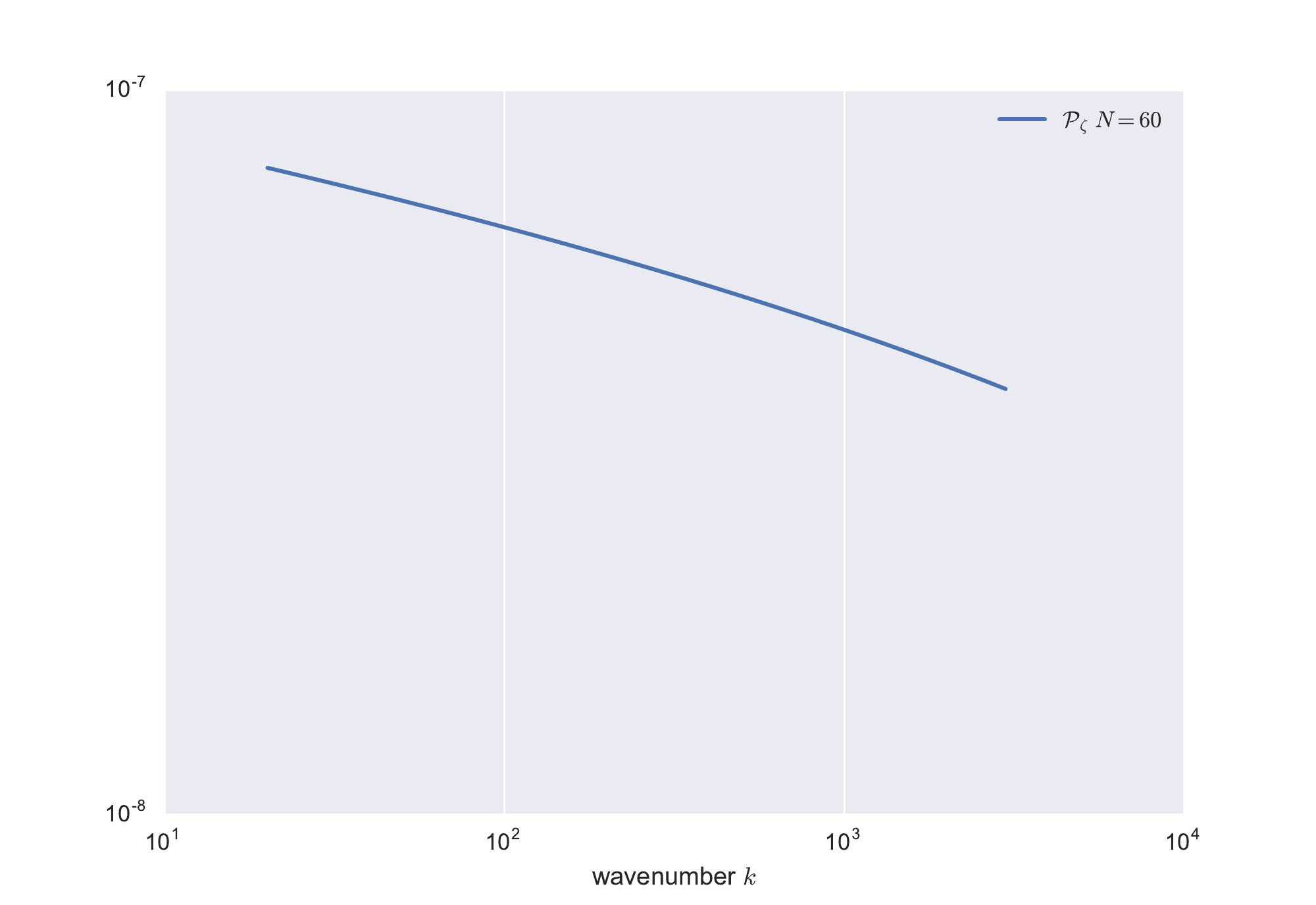}
    \end{center}
    \caption{\label{fig:twopf-plot}$\zeta$ 2-point function}
\end{figure}

\begin{minted}{c++}
    // 2. Zeta power spectrum

    vis_toolkit::zeta_twopf_wavenumber_series<> zeta_twopf(ztk3, last_time, all_twopfs);
    zeta_twopf.set_dimensionless(true);

    vis_toolkit::wavenumber_series_plot<> zeta_twopf_plot("dquad.product.zeta-twopf.plot", "twopf-plot.pdf");
    zeta_twopf_plot.set_log_x(true);
    zeta_twopf_plot += zeta_twopf;
\end{minted}
This produces the plot of Fig.~\ref{fig:twopf-plot}.

\subsubsection{Spectral index for $\zeta$ power spectrum}
\begin{figure}
    \begin{center}
        \includegraphics[scale=0.75]{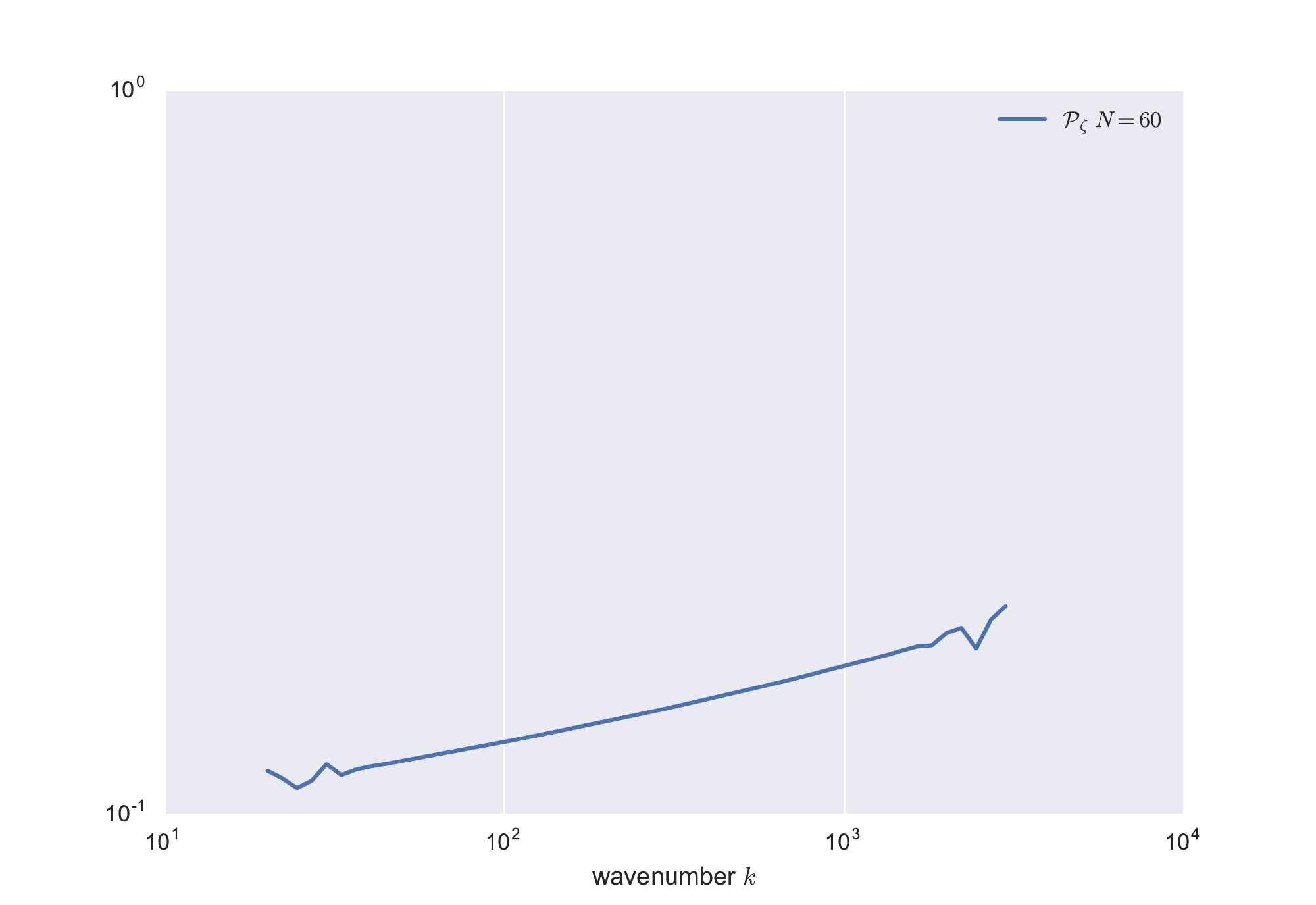}
    \end{center}
    \caption{\label{fig:twopf-index-plot}Spectral index of $\zeta$ two-point function}
\end{figure}

\begin{minted}{c++}
    // 3. Zeta power spectrum spectral index

    vis_toolkit::zeta_twopf_wavenumber_series<> zeta_twopf_index(ztk3, last_time, all_twopfs);
    zeta_twopf_index.set_dimensionless(true);
    zeta_twopf_index.set_spectral_index(true);

    vis_toolkit::wavenumber_series_plot<> zeta_twopf_index_plot("dquad.product.zeta-twopf.index-plot", "twopf-index-plot.pdf");
    zeta_twopf_index_plot.set_log_x(true);
    zeta_twopf_index_plot += zeta_twopf_index;
\end{minted}
This produces the plot of Fig.~\ref{fig:twopf-index-plot}.
Notice the jitter near the ends of the line;
see the discussion of spectral index calculations in~\S\ref{sec:wavenumber-series}.
The results could be improved by extending the range of wavenumbers being sampled,
to move these edge effects away from the region of interest.

\subsubsection{Reduced bispectrum on equilateral configurations}
\begin{figure}
    \begin{center}
        \includegraphics[scale=0.75]{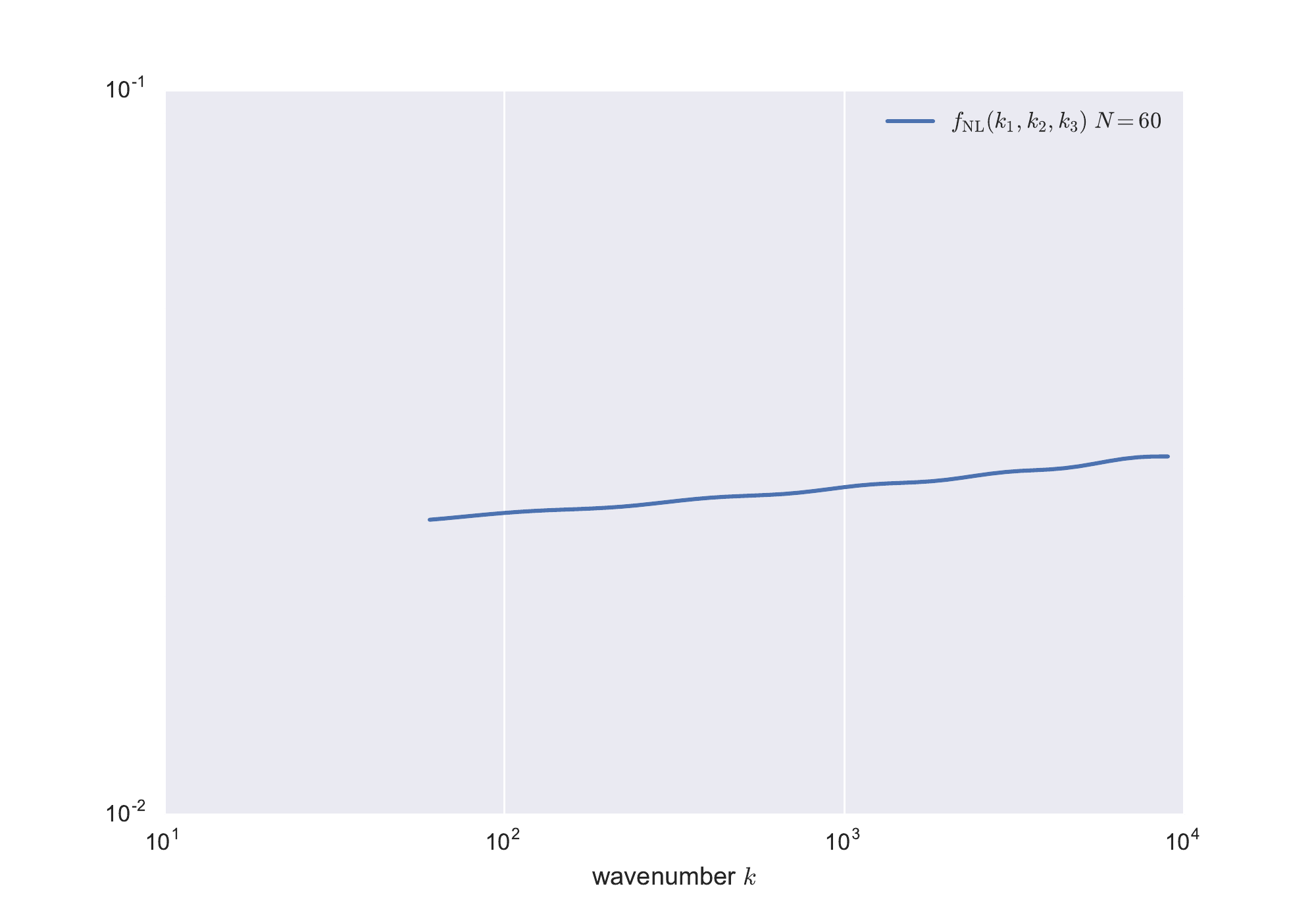}
    \end{center}
    \caption{\label{fig:equi-plot}Reduced bispectrum on equilateral configurations of
    varying scale $k_t$}
\end{figure}

\begin{minted}{c++}
    // 4. Reduced bispectrum on equilateral configurations

    vis_toolkit::zeta_reduced_bispectrum_wavenumber_series<> zeta_redbsp_equi(ztk3, last_time, all_equilateral);
    zeta_redbsp_equi.set_current_x_axis_value(vis_toolkit::axis_value::k);

    vis_toolkit::wavenumber_series_plot<> zeta_redbsp_equi_plot("dquad.product.zeta-redbsp.equi-plot", "equi-plot.pdf");
    zeta_redbsp_equi_plot.set_log_x(true);
    zeta_redbsp_equi_plot += zeta_redbsp_equi;
\end{minted}
This produces the plot of Fig.~\ref{fig:equi-plot}.
Notice the slight oscillations caused by settling of the heavy field
into its minimum.

\subsubsection{Spectral index of reduced bispectrum on equilateral configurations}
\begin{figure}
    \begin{center}
        \includegraphics[scale=0.75]{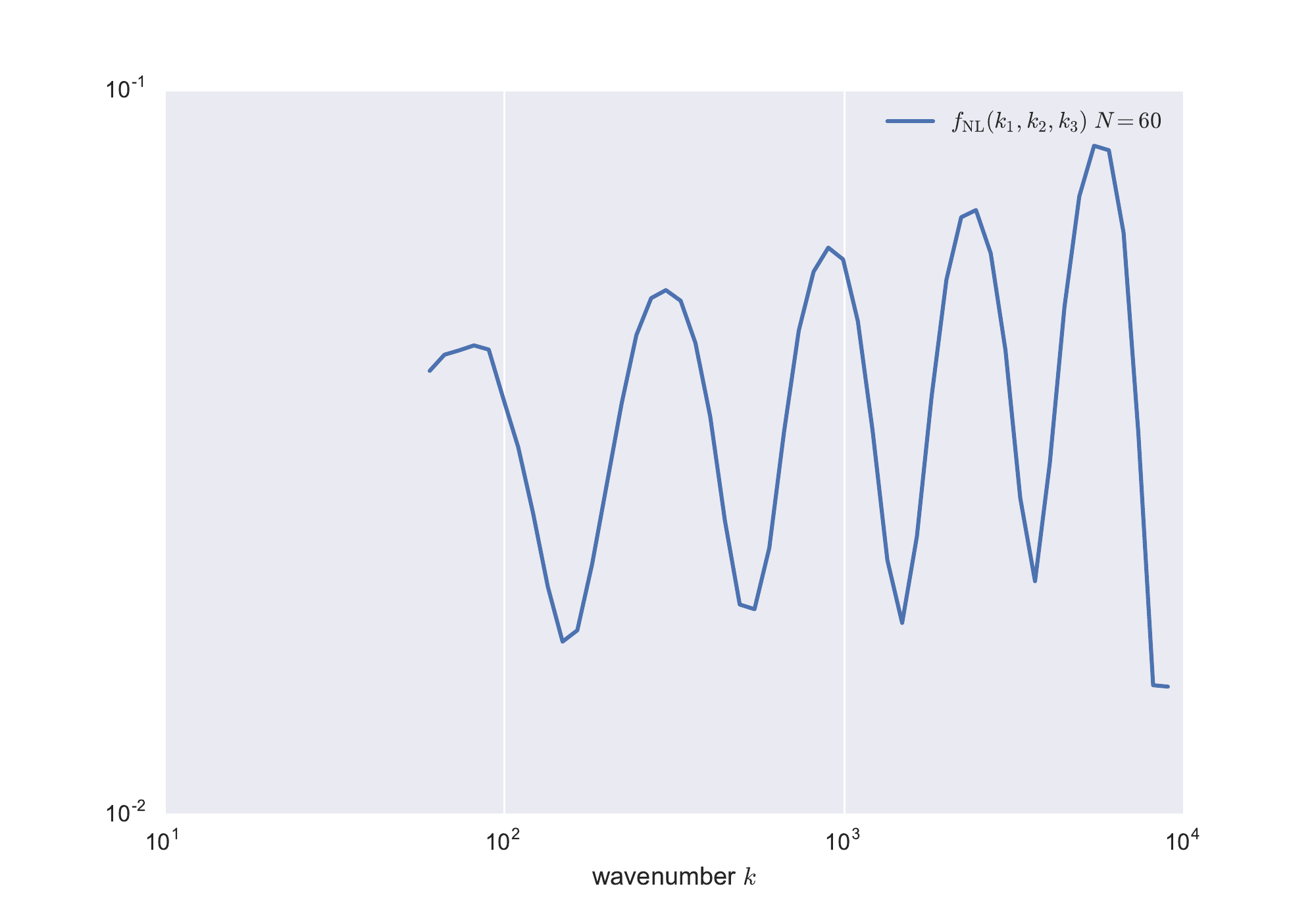}
    \end{center}
    \caption{\label{fig:equi-index-plot}Spectral index (with $k_t$) of reduced bispectrum
    on equilateral configurations}
\end{figure}

\begin{minted}{c++}
    // 5. Spectral index of reduced bispectrum on equilateral configurations

    vis_toolkit::zeta_reduced_bispectrum_wavenumber_series<> zeta_redbsp_equi_index(ztk3, last_time, all_equilateral);
    zeta_redbsp_equi_index.set_spectral_index(true).set_current_x_axis_value(vis_toolkit::axis_value::k);

    vis_toolkit::wavenumber_series_plot<> zeta_redbsp_equi_index_plot("dquad.product.zeta-redbsp.equi-index-plot", "equi-index.pdf");
    zeta_redbsp_equi_index_plot.set_log_x(true);
    zeta_redbsp_equi_index_plot += zeta_redbsp_equi_index;
\end{minted}
This produces the plot of Fig.~\ref{fig:equi-index-plot}.
The oscillatory structure is much more visible than it is in the amplitude.
In general, obtaining accurate spectral indices requires
higher accuracy in the integration.
This can be
achieved by increasing the number of e-folds of massless evolution,
and by decreasing the absolute and relative tolerances if needed.

\subsubsection{Squeezing dependence of reduced bispectrum: isosceles triangles}
\begin{figure}
    \begin{center}
        \includegraphics[scale=0.75]{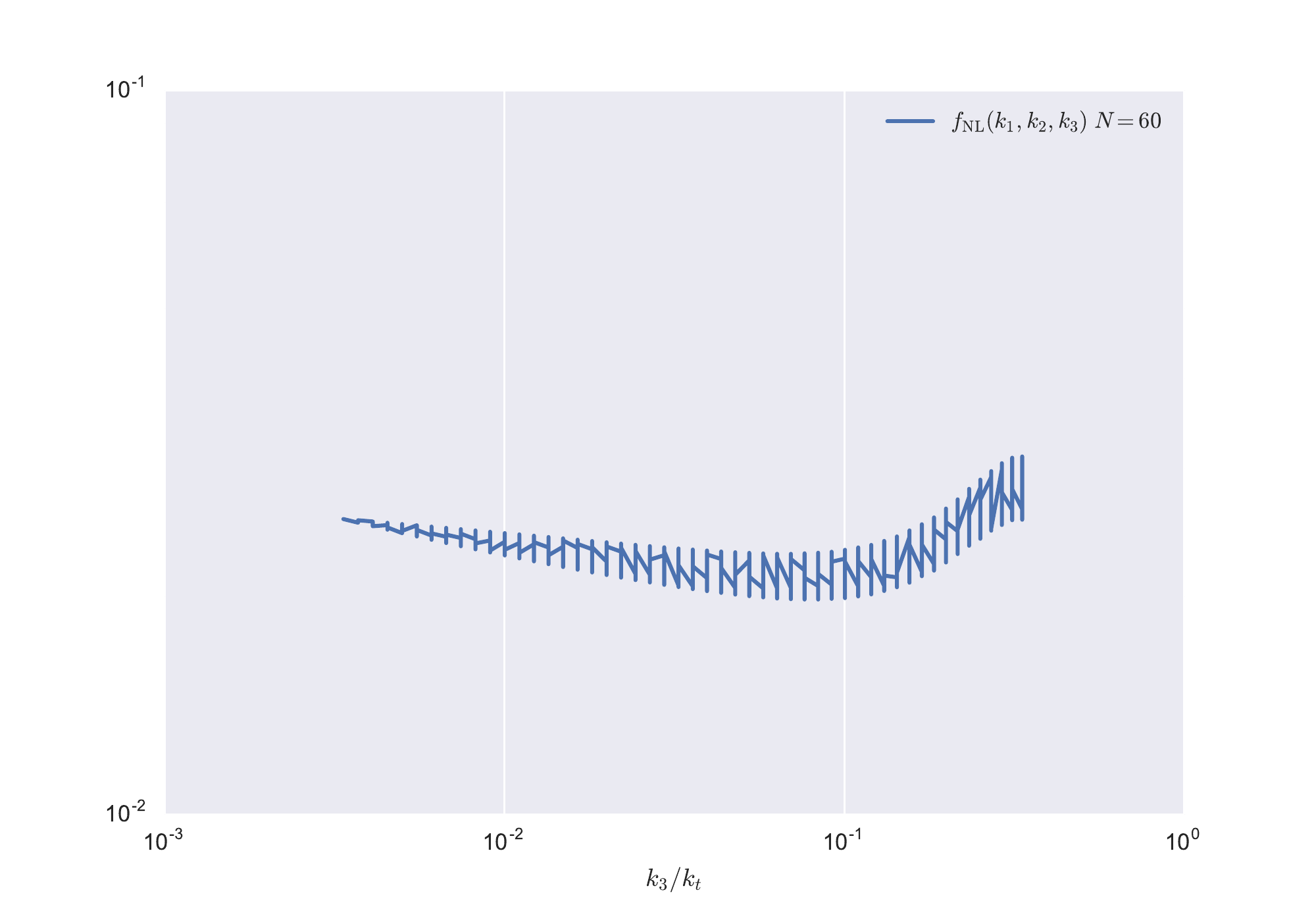}
    \end{center}
    \caption{\label{fig:squeeze-plot}Variation of reduced bispectrum with squeezing
    $k_3/k_t$}
\end{figure}

\begin{minted}{c++}
    // 6. Reduced bispectrum as a function of squeezing on isosceles triangles

    vis_toolkit::zeta_reduced_bispectrum_wavenumber_series<> zeta_redbsp_squeeze(ztk3, last_time, all_isosceles);
    zeta_redbsp_squeeze.set_current_x_axis_value(vis_toolkit::axis_value::squeeze_k3);

    vis_toolkit::wavenumber_series_plot<> zeta_redbsp_squeeze_plot("dquad.product.zeta_redbsp.squeeze-plot", "squeeze-plot.pdf");
    zeta_redbsp_squeeze_plot.set_log_x(true);
    zeta_redbsp_squeeze_plot += zeta_redbsp_squeeze;
\end{minted}
This produces the plot of Fig.~\ref{fig:squeeze-plot}.
The general trend is clear, but there are obvious jumps in amplitude
from configuration to configuration.
This happens because the plot is constructed from a cubic mesh.
Although only the squeezing ratio $k_3/k_t$ is plotted on the $x$-axis,
the configurations are also varying in $k_t$.
To get a smooth line---for example, suitable for computing a spectral index---it
would be necessary to switch to an $\alpha\beta$-type mesh
that would allow sampling from different values of $k_3/k_t$ at
\emph{fixed} $k_t$.

\subsubsection{Time evolution of 3pf correlation functions}
The most commonly used data products
are functions of wavenumber (or other configuration variables)
at fixed time. Often this fixed time will be the end of inflation,
although it may be earlier if the system converges to an adiabatic
limit characterized by conservation of $\zeta$ (as it does for the double quadratic model).

\begin{figure}
    \begin{center}
        \includegraphics[scale=0.75]{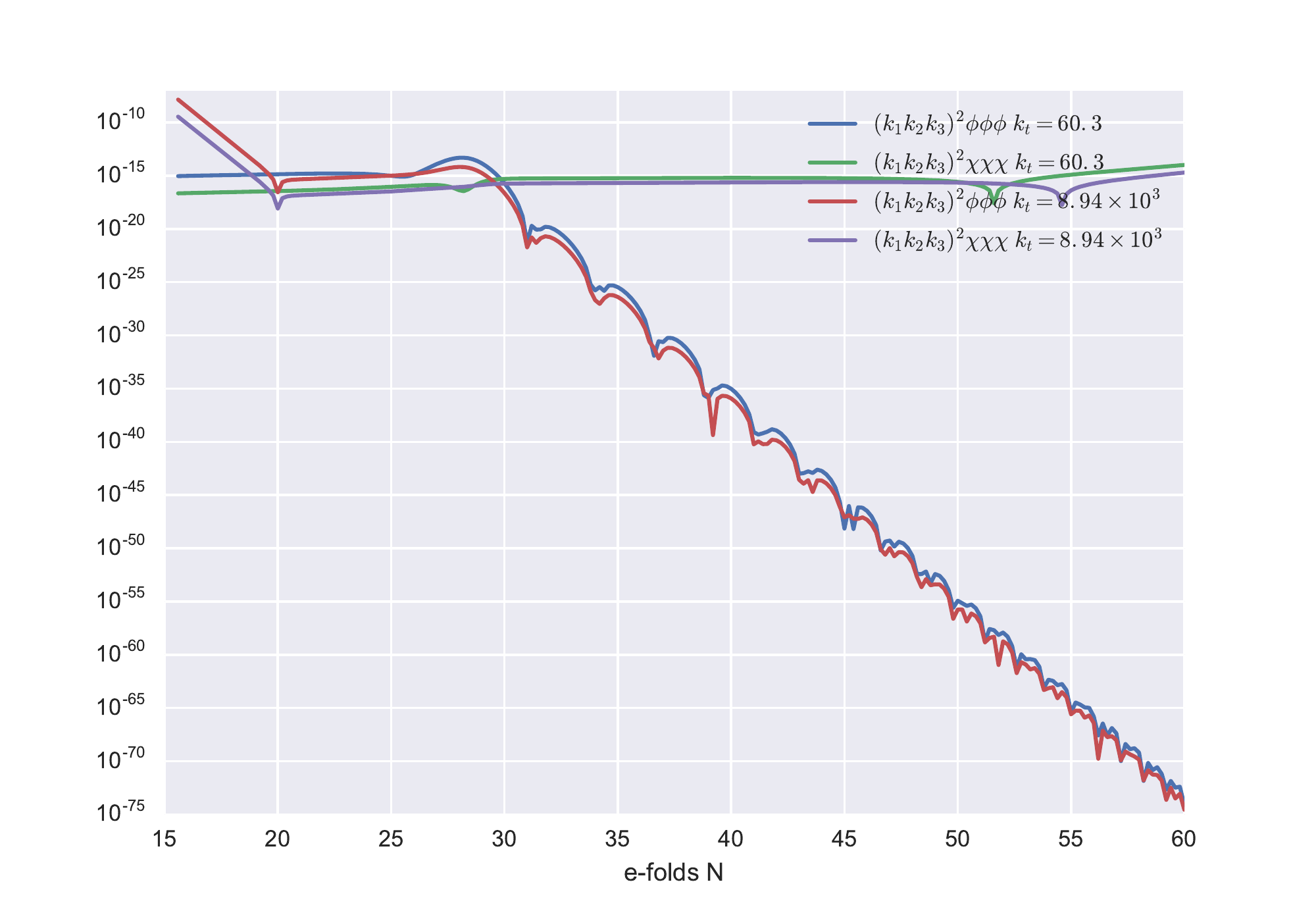}
    \end{center}
    \caption{\label{fig:threepf-time}Time evolution of the
    3-point functions $\langle \phi \phi \phi \rangle$
    and $\langle \chi \chi \chi \rangle$ for the largest and smallest values of $k_t$}
\end{figure}

However, it is also useful to plot the time \emph{evolution} of individual
quantities. This is especially useful to check for any anomalies in the integration
that might make the final data products inaccurate.
As a sanity check, it is useful to plot the evolution of the raw 3-point
functions for at least a few scales.
\begin{minted}{c++}
    // 7. Time evolution of some sample 3-point correlation functions

    vis_toolkit::threepf_time_series<> threepf_time(tk3, vis_toolkit::index_selector<3>(num_fields).none().set_on({ 0, 0, 0 }).set_on({ 1, 1, 1 }),
                                                    all_times, large_small_equilateral);

    vis_toolkit::time_series_plot<> threepf_time_plot("dquad.product.threepf-time", "threepf-time.pdf");
    threepf_time_plot += threepf_time;
\end{minted}
The resulting plot is shown in Fig.~\ref{fig:threepf-time}.
Notice the use of `tags' giving the $k_t$ value, which distinguish between the
different lines generated by the wavenumber configuration query;
see the discussion of label tagging in~\S\ref{sec:derived-standard-options}.

Such plots are an especially useful
diagnostic tool where they include the subhorizon
evolution, here visible as a steeply falling straight line at the left-hand
edge of the plot for the lines with $k_t = 8.94 \times 10^3$.
In this region the individual wavefunctions oscillate rapidly but the
\emph{correlation function} is smooth, as was explained in the discussion of
adaptive initial conditions in~\S\ref{sec:general-integration-options}.
In the absence of special features, the subhorizon evolution should be a smooth
decaying power law. Any noise in these lines, or the appearance of oscillations,
tends to indicate loss of accuracy during the integration.
The normal response should be to tighten the numerical tolerances
(\S\ref{sec:stepper-block}),
increase the number of e-folds of massless evolution
(\S\ref{sec:add-integration-task} and p.\pageref{enum:adpative-ics}), or both.

\subsubsection{$\fNL$ amplitude}
\begin{figure}
    \begin{center}
        \includegraphics[scale=0.75]{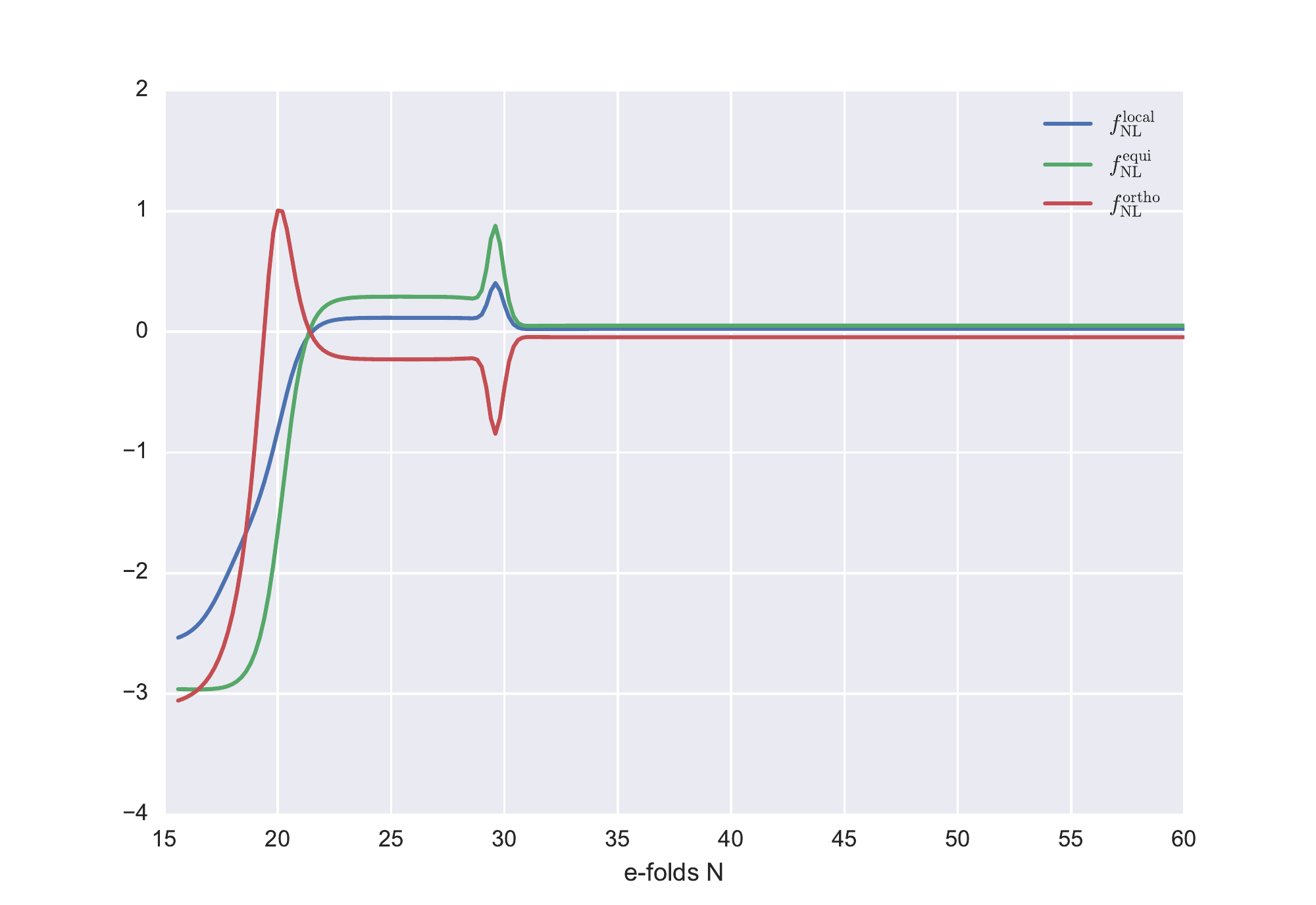}
    \end{center}
    \caption{\label{fig:fNL-auto-plot}Time evolution of
    inner-product amplitudes
    $\fNLlocal$, $\fNLequi$ and $\fNLortho$.
    As above, we caution that these quantities must be interpreted with
    care and are not necessarily related to the constraints reported
    from experiment.}
\end{figure}

Finally, consider setting up a plot of the amplitudes generated by
inner-products with the equilateral, orthogonal and local
templates.
We can collect these in a separate function; see Fig.~\ref{code:fNL-code}.
\begin{figure}
\begin{minipage}[t][15cm]{\textwidth}
\begin{minted}{c++}
void write_fNL_products(transport::repository<>& repo, transport::initial_conditions<>& ics,
                        transport::range<>& ts, transport::range<>& ks)
  {
    transport::threepf_cubic_task<> tk("dquad.threepf-linear", ics, ts, ks);
    tk.set_adaptive_ics_efolds(5.0);
    tk.set_description("Compute time history of the 3-point function on a linear grid");

    transport::zeta_threepf_task<> ztk("dquad.threepf-linear-zeta", tk);
    ztk.set_description( "Convert the output from dquad.threepf-linear into zeta 2 and 3-point functions");

    transport::fNL_task<> fNL_local("dquad.fNL-local", ztk, vis_toolkit::bispectrum_template::local);
    fNL_local.set_description( "Compute inner product of double-quadratic bispectrum with local template");

    transport::fNL_task<> fNL_equi("dquad.fNL-equi", ztk, vis_toolkit::bispectrum_template::equilateral);
    fNL_equi.set_description( "Compute inner product of double-quadratic bispectrum with equilateral template");

    transport::fNL_task<> fNL_ortho("dquad.fNL-ortho", ztk, vis_toolkit::bispectrum_template::orthogonal);
    fNL_ortho.set_description( "Compute inner product of double-quadratic bispectrum with orthogonal template");

    vis_toolkit::SQL_time_query all_times("1=1");

    vis_toolkit::fNL_time_series<> local(fNL_local, all_times);
    vis_toolkit::fNL_time_series<> equi(fNL_equi, all_times);
    vis_toolkit::fNL_time_series<> ortho(fNL_ortho, all_times);

    vis_toolkit::time_series_plot<> fNL_plot("dquad.product.fNL_plot", "fNL_plot.pdf");
    fNL_plot.set_log_y(false).set_abs_y(false);
    fNL_plot += local + equi + ortho;

    transport::output_task<> out_tk("dquad.output.fNL");
    out_tk += fNL_plot;

    repo.commit(out_tk);
  }
\end{minted}
\end{minipage}
\vspace{1cm}
\caption{\label{code:fNL-code}Function to generate $\fNL$ tasks and derived products}
\end{figure}
The resulting plot is shown in Fig.~\ref{fig:fNL-auto-plot} and can be
compared with the hand-generated version Fig.~\ref{fig:dquad-fNL-plot}.

\section{Managing repositories}
\label{sec:repo-management}

\subsection{Managing records}
\S\ref{sec:apply-tags} explained the system of `tags' used to control which
content groups are selected as data sources
for postintegration or output tags.
Tags can be set immediately, when each content group is generated,
but it is also possible to adjust them after
the group has been entered in the repository.

In addition to tags, {\CppTransport} provides a facility to attach
longer `notes' to each group.
These consist of free-format text and can be used for any purpose.
For example, they could be used to add persistent working notes
to an ongoing project,
to highlight features when circulating data to collaborators,
or to provide enriched documentation for archival purposes.

Finally, it is sometimes desirable to remove unneeded content groups from a repository.
Notice, however, that {\CppTransport} does \emph{not} offer an option to remove
or edit
the details of initial conditions packages, tasks, or derived products.
This is an intentional design choice.
These details form part of the documentation associated with each content group,
and if they were to be changed then some of this documentation would be lost
or become ambiguous.
If it is necessary to make adjustments to any of these definitions, it is
preferable to add a new definition with a different name or to write into a new
repository.

\subsubsection{Specifying which objects to modify}
\label{sec:specify-objects}
{\CppTransport} allows modifications to be applied to many different
repository records simultaneously.
To specify records use the \option{{-}{-}object} command-line argument,
followed by the name of a content group.
It is possible to use multiple \option{{-}{-}object} arguments to specify
multiple records.

Alternatively, to name provided to \option{{-}{-}object} can be
enclosed in braces
\mintinline{bash}+{+
$\cdots$
\mintinline{bash}+}+.
{\CppTransport} will interpret the name between the braces
as a regular expression and attempt to match it to any available content groups.
For example, to match any content group produced in 2016 we could write
\option{{-}{-}object \{2016.*\}}.

\subsubsection{Adding and removing tags}
\label{sec:add-remove-tags}

Adding tags is accomplished using the
\option{{-}{-}add-tag} command-line switch,
followed by the name of a tag. If that tag contains spaces then it should be
wrapped in quotation marks.
The given tag is added to all content groups that match
an argument provided to \option{{-}{-}object}.
To remove a tag from all such groups use
\option{{-}{-}delete-tag}.

\subsubsection{Adding and removing notes}
\label{sec:add-remove-notes}

To add notes, use
\option{{-}{-}add-note}
followed by the text to be added.
If it contains spaces, it should be wrapped in quotation marks.
The note is added to all content groups matching an argument provided
to \option{{-}{-}object}.

To remove a note, obtain a list of notes attached to the content group of your
choice using \option{{-}{-}info}.
Then use \option{{-}{-}delete-note} followed by the number of the note to be
removed.

\begin{warning}
    Notice that the given note is removed from every content group that matches
    an argument provided to \option{{-}{-}option}. If the note you intend to delete
    is not in the same position in every content group then you will need to carry out
    the removal in batches.
\end{warning}

\subsubsection{Deleting content groups}
\label{sec:delete-content}

To delete an unwanted content group use \option{{-}{-}delete}.
{\CppTransport} will not allow you to delete content groups that were used
as data sources for other groups that remain in the repository,
because this would disrupt its ability to provide a provenance for those groups.

\subsubsection{Lock and unlock groups}
\label{sec:lock-unlock}

Finally, groups can be \emph{locked} to prevent modifications. To do this use the
\option{{-}{-}lock} switch.
Locked groups cannot be altered or deleted until they are unlocked using
\option{{-}{-}unlock}.

\subsection{Summary of command-line options}
\label{sec:option-summary}
This section summarizes the command-line
options recognized by {\CppTransport} executables.
\vspace{3mm}

\noindent Housekeeping functions:
\begin{itemize}
	\item \option{{-}{-}help} \\
	Display brief usage information and a list of all available options.

	\item \option{{-}{-}version} \\
	Show version of {\CppTransport} used to build the model headers,
	and the version of the runtime system. These need not be the same,
	although {\CppTransport} requires the runtime system to be at least as recent
	as the version used to build the headers.

	\item \option{{-}{-}license} \\
	Display licensing information.

	\item \option{{-}{-}models} \\
	Show list of models understood by this executable.

	\item \option{{-}{-}no-colour} or \option{{-}{-}no-color} \\
	Do not produce colourized output.
	Normally {\CppTransport} will detect whether the terminal
	in which it is running can support colour.
	However, if you are redirecting {\CppTransport}'s output to a file
	then you may wish to manually suppress the use of colour.

	\item \option{{-}{-}include}, or abbreviate to \option{-I} \\
	Adds the following path to the list of paths searched for resources.
	Currently, the only resources needed by {\CppTransport}
	are those used by the HTML report generator.
\end{itemize}

\noindent Configuration options:
\begin{itemize}
	\item \option{{-}{-}verbose}, or abbreviate to \option{-v} \\
	Display extra status and update messages.

	\item \option{{-}{-}repo}, or abbreviate to \option{-r} \\
	Should be followed by a path identifying the repository to be used.
	If the repository does not exist then new, blank repository is created.

	\item \option{{-}{-}caches} \\
	\option{{-}{-}batch-cache} \\
	\option{{-}{-}datapipe-cache} \\
	Followed by a cache size in Mb. Sets the corresponding cache size
	(or both caches, if the option \option{{-}{-}caches} is used).
	The \emph{batching cache} is used to temporarily
	hold the data products from integration in memory before
	flushing them to disk; see~\S\ref{sec:what-happens}.
	The \emph{datapipe cache} stores data used to generate derived products.
	This normally requires database access, which can be time consuming on a slow
	filing system. Storing data in memory can give a significant performance
	boost if the same data is re-used.

	\item \option{{-}{-}network-mode} \\
	Disable use of the {\SQLite} write-ahead log. Must be used if the repository
	is stored on a network filing system such as NFS or Lustre, but should
	otherwise be omitted.
\end{itemize}

\noindent Job specification:
\begin{itemize}
	\item \option{{-}{-}create} \\
	Write records held by this executable into the repository (\S\ref{sec:examine-k-database}).

	\item \option{{-}{-}task} \\
	Followed by the name of task. Adds the named task to the list of work.

	\item \option{{-}{-}tag} \\
	Specify a tag to be attached to any content groups generated by this
	{\CppTransport} job.
	For postintegration or output tasks,
	filters the available content groups to those that share the specified tag.
	Can be repeated multiple times to specify more than one tag.

	\item \option{{-}{-}checkpoint} \\
	Set the checkpoint interval, measured in minutes.
	Overrides any default checkpoints set by individual tasks.

	\item \option{{-}{-}seed} \\
	Seed jobs using the specified content group.
\end{itemize}

\noindent Repository actions:
\begin{itemize}
	\item \option{{-}{-}object} \\
	Select objects to be modified.
	Regular expressions can be used between curly braces
	\mintinline{bash}+{+
	$\cdots$
	\mintinline{bash}+}+.

	\item \option{{-}{-}lock} \\
	Lock repository records (preventing modification or deletion)
	for content groups matching the object specification list.

	\item \option{{-}{-}unlock} \\
	Unlock repository records matching the object specification list.

	\item \option{{-}{-}add-tag} \\
	Add the specified tag to content groups matching the object specification
	list. Can be repeated multiple times to add more than one tag.

	\item \option{{-}{-}delete-tag} \\
	Remove the specified tag from content groups matching the object specification
	list. Can be repeated to delete multiple tags.

	\item \option{{-}{-}add-note} \\
	Add the specified note (which should be quoted if it contains spaces)
	to any content groups
	matching the object specification list. Can be repeated to add multiple notes.

	\item \option{{-}{-}delete-note} \\
	Specifies a note to remove by number (check the repository record using
	\option{{-}{-}info} to obtain a list of notes).
	Can be applied to multiple content groups, but will remove the same numbered
	note from each list.

	\item \option{{-}{-}delete} \\
	Remove content groups matching the object specification list, provided no
	other content groups depend on them.

	Notice that operations can be chained. For example,
	\option{{-}{-}unlock} and
	\option{{-}{-}delete} can be specified
	at the same time, in which case unlocking is performed
	\emph{before} deletion.
	The same applies to other operations such as
	adding or removing tags and notes.
	If \option{{-}{-}lock} is specified then the record is locked
	only after all other operations have been processed.
\end{itemize}

\noindent Repository reporting and status:
\begin{itemize}
	\item \option{{-}{-}record} \\
	Perform recovery on the repository; see~\S\ref{sec:checkpointing}.

	\item \option{{-}{-}status} \\
	Print brief report showing repository status.
	Includes available tasks and the number of content groups
	attached to each task, in addition to the details of any
	in-flight jobs.

	\item \option{{-}{-}inflight} \\
	Similar to \option{{-}{-}status}, but shows details of in-flight jobs only.

	\item \option{{-}{-}info} \\
	Report on a specified repository record.
	Matches any objects whose names begin with the specified string, so it is
	not necessary to write the name out exactly.
	Alternatively, a regular expression can be provided by wrapping
	it in curly braces
	\mintinline{bash}+{+ $\cdots$ \mintinline{bash}+}+.

	\item \option{{-}{-}provenance} \\
	Report on the provenance of a specified output content group.
	The provenance report shows all content groups that contributed
	to each derived product generated as part of the group.
	Name matching is as for \option{{-}{-}info}.

	\item \option{{-}{-}html} \\
	Write a HTML-format report on the contents of the repository to
	the specified folder.
\end{itemize}

\noindent Plotting options:
\begin{itemize}
	\item \option{{-}{-}plot-style}, or abbreviate to \option{-p} \\
	Select a plotting style. See the discussion in~\S\ref{sec:environment}.

	\item \option{{-}{-}mpl-backend} \\
	Force {\CppTransport} to use a specified {\Matplotlib} backend.
	See the discussion in~\S\ref{sec:environment}.
\end{itemize}

\noindent Journaling options:
\begin{itemize}
	\item \option{{-}{-}gantt} \\
	Write a process Gantt chart, showing the activities of each process
	in a multiprocess MPI job, to the specified file.
	Any output format supported by {\Matplotlib} may be used,
	selected by its extension.
	Alternatively the extension \file{.py} may be specified to
	obtain the Python script suitable for generating the plot.

	\item \option{{-}{-}journal} \\
	Write a (very detailed) JSON-format journal showing the MPI
	communication between workers. Mostly of value when debugging.
\end{itemize}

\section{Acknowledgments}
It is a pleasure to acknowledge a longstanding collaboration
with
Mafalda Dias,
Jonathan Frazer
and David Mulryne.

Development of {\CppTransport} has been supported
by an ERC grant:
\begin{itemize}
    \item \emph{Precision tests of the inflationary
    scenario},
    funded by
    the European Research Council under the European Union's
    Seventh Framework Programme (FP/2007--2013) and ERC Grant Agreement No. 308082.
\end{itemize}
In addition,
some development of {\CppTransport} has been supported by
other funding sources.
Portions of
the work described in this document have been supported by:
\begin{itemize}
    \item The UK
    Science and Technology Facilities Council via grants
    ST/I000976/1 and ST/L000652/1,
    which funded the science programme at the University of Sussex Astronomy
    Centre from April 2011--March 2014
    and April 2014--March 2017, respectively.
    \item The Leverhulme Trust via a Philip Leverhulme Prize.
    \item The National Science Foundation Grant No. PHYS-1066293
    and the hospitality of the Aspen Center for Physics.
    \item The hospitality of the Higgs Centre for Theoretical
    Physics at the University of Edinburgh,
    and the Centre for Astronomy \& Particle Physics
    at the University of Nottingham.
\end{itemize}

\appendix

\section{Third-party software used by {\CppTransport}}

The {\CppTransport} sources incorporate portions of the following open source projects:

\begin{itemize}
    \item The GinacPrint common subexpression elimination algorithm made available by Doug Baker.
    \begin{center}
        \url{http://www.ginac.de/pipermail/ginac-list/2010-May/001631.html}
    \end{center}
    License: GPL-2 \\
    This is incorporated in the source files
    \begin{center}
    \file{translator/backends/infrastructure/language\_concepts/cse.cpp} \\
    \file{translator/backends/infrastructure/language\_concepts/cse.h}
    \end{center}
\end{itemize}
The {\CppTransport} translator and runtime system are linked to the following
libraries. The build process assumes they are available on the system,
but does not automatically install them.

The {\CppTransport} sources do not include source code (or derivatives of the source code)
from these libraries; they only link to them as external resources.
\begin{itemize}
    \item {\GiNaC} (used by translator) \\
    \url{http://www.ginac.de} \\
    License: GPL-2

    \item The {\Boost} libraries (used by translator and runtime system) \\
    \url{http://www.boost.org} \\
    License: Boost Software License

    \item {\SQLite} (used by runtime system) \\
    \url{https://www.sqlite.org} \\
    License: Public Domain (\url{https://www.sqlite.org/copyright.html})

    \item {\OpenSSL} (used by runtime system) \\
    \url{https://www.openssl.org} \\
    License: OpenSSL License (\url{https://www.openssl.org/source/license.html})
\end{itemize}
In addition, the {\CppTransport} build process automatically downloads
and installs the following libraries. They are statically linked to executables
constructed by (1) running the translator and (2) building the resulting code
using the provided runtime system.

The {\CppTransport} sources do not include source code (or derivatives of the source code)
from these libraries. Compiled executables using the provided runtime system
link to them only as external resources.
\begin{itemize}
    \item {\SPLINTER} \\
    \url{https://github.com/bgrimstad/splinter} \\
    License: Mozilla public license (see \file{thirdparty/License/SPLINTER.txt})

    \item {\JsonCpp} \\
    \url{https://github.com/open-source-parsers/jsoncpp} \\
    License: MIT License (see \file{thirdparty/License/JsonCpp.txt})
\end{itemize}
{\CppTransport} also depends on the {\Eigen} library using the version
bundled as part of {\SPLINTER}.
\begin{itemize}
    \item {\Eigen} \\
    \url{http://eigen.tuxfamily.org/index.php?title=Main\_Page} \\
    License: Mozilla public license
    (for details, see files installed by the {\SPLINTER} build process)
\end{itemize}
Also, the {\CppTransport} platform bundles parts of the following open source
projects. These parts are included in the Git repository for {\CppTransport}
or the installation tarball in the \file{thirdparty/} directory.

The {\CppTransport}
sources do not include source code (or derivatives of the source code)
from this projects. It depends on them only as external resources that are
used by HTML reports.
\begin{itemize}
    \item {\jQuery} \\
    \url{https://jquery.com/download/} \\
    License: MIT License
    (see \file{thirdparty/License/jQuery.txt})

    \item Twitter Bootstrap \\
    \url{http://getbootstrap.com} \\
    License: MIT License
    (see \file{thirdparty/License/Bootstrap.txt})

    \item {\packagefont bootstrap-tab-history} \\
    \url{http://mnarayan01.github.io/bootstrap-tab-history/} \\
    License: Apache License
    (see \file{thirdparty/License/bootstrap-tab-history.txt})

    \item {\DataTables} \\
    \url{https://datatables.net} \\
    License: MIT License
    (see \file{thirdparty/License/DataTables.txt})

    \item {\packagefont Prism.js} \\
    \url{http://prismjs.com} \\
    License: MIT license
    (see \file{thirdparty/License/prism.txt})
\end{itemize}

\bibliographystyle{JHEP}
\bibliography{paper}

\end{document}